
\input epsf
\magnification = 1200
\def\lapp{\hbox{$ {     \lower.40ex\hbox{$<$}
                   \atop \raise.20ex\hbox{$\sim$}
                   }     $}  }
\def\rapp{\hbox{$ {     \lower.40ex\hbox{$>$}
                   \atop \raise.20ex\hbox{$\sim$}
                   }     $}  }
\def\barre#1{{\not\mathrel #1}}
\def\krig#1{\vbox{\ialign{\hfil##\hfil\crcr
           $\raise0.3pt\hbox{$\scriptstyle \circ$}$\crcr\noalign
           {\kern-0.02pt\nointerlineskip}
$\displaystyle{#1}$\crcr}}}
\def\upar#1{\vbox{\ialign{\hfil##\hfil\crcr
           $\raise0.3pt\hbox{$\scriptstyle \leftrightarrow$}$\crcr\noalign
           {\kern-0.02pt\nointerlineskip}
$\displaystyle{#1}$\crcr}}}
\def\ular#1{\vbox{\ialign{\hfil##\hfil\crcr
           $\raise0.3pt\hbox{$\scriptstyle \leftarrow$}$\crcr\noalign
           {\kern-0.02pt\nointerlineskip}
$\displaystyle{#1}$\crcr}}}
\def\larr#1{\vbox{\ialign{\hfil##\hfil\crcr
$\raise0.3pt\hbox{$\scriptstyle \leftarrow$}$\crcr\noalign
{\kern-0.02pt\nointerlineskip}
$\displaystyle{#1}$\crcr}}}

\def\svec#1{\skew{-2}\vec#1}
\def\tr{\,{\rm tr}\,}
\def\Tr{\,{\rm Tr}\,}

\def\g5{\gamma_5}
\def\ks{\barre k}

\def\leff{{\cal L}_{{\rm eff}}}
\def\lqcd{{\cal L}_{{\rm QCD}}}
\def\m{{\cal M}}
\def\su3{SU(3)}
\def\l{\Lambda}
\def\hen{{\rm high \, \, energy}}
\def\len{{\rm low \, \, energy}}
\def\zva{{\cal Z}[v,a,s,p]}
\topskip=0.60truein
\leftskip=0.18truein
\vsize=8.8truein
\hsize=6.5truein
\tolerance 10000
\hfuzz=20pt

\baselineskip 12pt plus 1pt minus 1pt
\pageno=0
\font\bigtenrm=cmr10 scaled\magstep2
\font\bigeigrm=cmr10 scaled\magstep1
\centerline{\bf {\bigtenrm CHIRAL DYNAMICS IN NUCLEONS AND NUCLEI}}
\vskip 40pt
\centerline{\bigeigrm{V. Bernard}$^{1)}$}
\vskip  4pt
\centerline{\it Centre de Recherche Nucleaire et Universit\'e Louis
  Pasteur de Strasbourg}
\centerline{{\it Physique Th{\'e}orique, BP 28,
F--67037 Strasbourg Cedex 2, France}}
\vskip 10pt
\centerline{\bigeigrm{N. Kaiser}$^{2)}$}
\vskip  4pt
\centerline{\it Technische Universit\"at M{\"u}nchen, Physik Department T30 }
\centerline{\it James-Franck-Stra{\ss}e, D--85747 Garching, Germany}
\vskip 10pt
\centerline{\bigeigrm{Ulf-G. Mei{\ss}ner}$^{3)}$}
\vskip  4pt
\centerline{\it   Universit\"at Bonn, Institut f{\"u}r
 Theoretische Kernphysik,}
\centerline{\it Nussallee 14--16, D--53115 Bonn, Germany}
\vskip 1truecm
\centerline{ email: $^{1)}$bernard@crnhp4.in2p3.fr,
$^{2)}$nkaiser@physik.tu-muenchen.de}
\centerline{ $^{3)}$meissner@pythia.itkp.uni-bonn.de}
\vskip 3truecm
\baselineskip 12pt plus 1pt minus 1pt
\centerline{ ABSTRACT:}
\smallskip
\centerline{
We review the implications of the spontaneous chiral
symmetry breaking in QCD}
\centerline{for processes involving one, two or more nucleons.}
\vfill
\centerline{Commissioned article for Int. J. Mod. Phys. E}
\vskip 20pt
\noindent CRN 95/3

\noindent TK 95 1 \hfill February 1995
\eject
\baselineskip 12pt plus 1pt minus 1pt
\noindent {\bf CONTENTS}
\bigskip
\item{1.} INTRODUCTION \hfill 2
\medskip
\item{2.} CHIRAL SYMMETRY IN QCD \hfill 8
\smallskip
2.1. \quad Elementary introduction to chiral symmetry \hfill 8

2.2. \quad Three--flavor QCD \hfill 9

2.3. \quad Chiral perturbation theory \hfill 11

2.4. \quad Modelling the pion \hfill 21
\medskip
\item{3.} THE PION--NUCLEON SYSTEM \hfill 26
\smallskip
3.1. \quad Effective Lagrangian \hfill 26

3.2. \quad Extreme non--relativistic limit \hfill 33

3.3. \quad Renormalization \hfill 38

3.4. \quad Low--energy constants and the role of the $\Delta(1232)$
\hfill 42

3.5. \quad Aspects of pion--nucleon scattering \hfill 46

3.6. \quad The reaction $\pi N \to \pi \pi N$ \hfill 50

3.7. \quad The pion--nucleon vertex \hfill 54
\medskip
\item{4.} NUCLEON STRUCTURE FROM ELECTROWEAK PROBES \hfill 58
\smallskip
4.1. \quad Electromagnetic form factors of the nucleon  \hfill 58

4.2. \quad Nucleon Compton scattering \hfill 63

4.3. \quad Axial properties of the nucleon  \hfill 76

4.4. \quad Threshold pion photo-- and electroproduction \hfill 79

4.5. \quad Two--pion production \hfill 88

4.6. \quad Weak pion production \hfill 94
\medskip
\item{5.} THE NUCLEON--NUCLEON INTERACTION \hfill 102
\smallskip
5.1. \quad General considerations  \hfill 102

5.2. \quad Nucleon--nucleon potential    \hfill 106

5.3. \quad More than two nucleons \hfill  112

5.4. \quad Three--body interactions between nucleons, pions and photons
\hfill 114

5.5. \quad Exchange currents  \hfill 118
\medskip
\item{6.} THREE FLAVORS, DENSE MATTER AND ALL THAT \hfill 122
\smallskip
6.1. \quad Flavor SU(3), baryon masses and $\sigma$--terms  \hfill 122

6.2. \quad Kaon---nucleon scattering  \hfill 130

6.3. \quad The pion in matter  \hfill 134

6.4. \quad Miscellaneous omissions \hfill 138
\medskip
\noindent \quad APPENDICES \hfill 142
\vfill
\eject
\noindent{\bf I. \quad INTRODUCTION}
\medskip
Effective field theories (EFTs) have become a popular tool in particle
and nuclear physics. An effective field theory differs from a conventional
renormalizable ("fundamental") quantum field theory in the following
respect. In EFT, one only works at low energies (where "low" is defined
with respect to some scale specified later) and expands the theory in
powers of the energy/characteristic scale. In that case,
renormalizability at all scales
is not an issue and one has to handle strings of
non--renormalizable interactions. Therefore, at a given order
in the energy expansion, the theory is specified by a finite
number of coupling (low--energy) constants (this allows  e.g.
for an order--by--order renormalization). All observables are
parametrized in terms of these few constants and thus there is a host
of predictions for many different processes. Obviously, at some high
energy this effective theory fails and one has to go over to a better
$\hen$ theory (which again might be an EFT of some fundamental
theory). The trace of this underlying $\hen$ theory are the particular
values of the $\len$ constants. The EFT presumably studied in most detail
is chiral perturbation theory (CHPT). The central topic
of this review will be the application of this framework when nucleons
(baryons) are present, with particular emphasis on processes with
exactly one nucleon in the initial and one nucleon in the final
state. Before elaborating on these particular aspects of CHPT,
it is useful to make some general comments concerning the applications of
EFTs.
\medskip
EFTs come into play when the underlying fundamental theory contains
massless (or very light) particles. These induce poles and cuts
and conventional Taylor expansions in powers of momenta fail. A
typical example is QED where gauge invariance protects the photon
from acquiring a mass. One photon exchange involves a propagator
$\sim 1/t$, with $t$ the invariant four--momentum transfer squared.
Such a potential can not be Taylor expanded. A classic example to deal
with such effects is the work of
Euler and Heisenberg [1.1] who considered the scattering of light
by light at very low energies, $\omega \ll m_e$, with $\omega$
the photon energy and $m_e$ the electron mass. To
calculate the scattering amplitude, one does not need full QED but
rather integrates out the electron from the theory. This leads to an
effective Lagrangian of the form
$$\leff = {1 \over 2} ({\vec E}^2 - {\vec B}^2)
 +{e^4 \over 360 \pi^2 m_e^4} \biggl[({\vec E}^2 - {\vec B}^2)^2
 + 7 ({\vec E} \cdot {\vec B})^2 \biggr] + \ldots
 \eqno(1.1) $$
 which is nothing but a derivative expansion since ${\vec E}$ and
 ${\vec B}$ contain derivatives of the gauge potential. Stated
 differently, since the photon energy is small, the electromagnetic
 fields are slowly varying.
 From eq.(1.1)
 one reads off that corrections to the leading term are suppressed
by powers of $(\omega/m_e)^4$. Straightforward calculation leads
to the cross section $\sigma  (\omega ) \sim \omega^6 / m_e^8$. This
can, of course, also be done using full QED [1.2], but the EFT caculation
is much simpler. The results of [1.2] nicely agree with the ones
making use of the EFT for $\omega \ll m_e$.
\medskip
A similar situation arises in QCD which is a non--abelian gauge
theory of colored quarks and gluons,
$$ \eqalign{
\lqcd &= -{1 \over 4 g^2} G_{\mu \nu}^a G^{\mu \nu,a}
  + \bar q i \gamma^\mu D_\mu q  - \bar q \m q \cr
&=
\,\,\,\,\,\,\,\,\,\,\,\,\,\,\,\,\,\,\,\,\,\,\,\,\,\,
\lqcd^0
\,\,\,\,\,\,\,\,\,\,\,\,\,\,\,\,\,\,\,\,\,\,\,\,\,\,\,\,\,
+  \lqcd^{\rm I} \cr}
\eqno(1.2)$$
with $\m = {\rm diag} (m_u,m_d,m_s, \ldots)$ the quark mass matrix.
For the full theory, there is a conserved charge for every quark
flavor separately since the quark masses are all different. However,
for the first three flavors ($u,d,s)$ it is legitimate to set the
quark masses to zero since they are small on a typical hadronic scale
like {\it e.g.} the $\rho$--meson mass. The absolute values of the
running quark masses at 1 GeV are $m_u \simeq 5$ MeV,
$m_d \simeq 9$ MeV,
$m_s \simeq 175$ MeV,
{\it i.e.} $m_u/M_\rho \simeq 0.006$, $m_d/M_\rho \simeq 0.012$ and
$m_s/M_\rho \simeq 0.23$ [1.3]. If one sets the quark masses to zero, the
left-- and right--handed quarks defined by
$$
q_L = {1 \over 2} ( 1 - \gamma_5) \,q, \,\,\,\,\,
q_R = {1 \over 2} ( 1 + \gamma_5)\, q
\eqno(1.3) $$
do not interact with each other and the whole theory admits an $U(3)
\times U(3)$ symmetry. This is further reduced by the axial anomaly,
so that the actual symmetry group of three flavor massless QCD is
$$ G= \su3_L \times \su3_R \times U(1)_{L+R} \eqno(1.4) $$
The U(1) symmetry related to baryon number conservation will not be
discussed in any further detail. The conserved charges which come along
with the chiral $\su3 \times \su3$ symmetry generate the corresponding
Lie algebra. In the sixties and seventies, manipulations of the
commutation relations between the conserved vector (L+R) and
axial--vector (L-R) charges were called "PCAC relations" or
"current algebra calculations" and lead to a host of $\len$ theorems
and predictions [1.4]. These rather tedious manipulations have
nowadays been replaced by EFT methods, in particular by CHPT (as will
be discussed later on). Let us come back to QCD. One quickly realizes
that the ground state does not have the full symmetry $G$, eq.(1.4). If
that were the case, every known hadron would have a partner of the
same mass but with opposite parity. Clearly, this is in contradiction
with the observed particle spectrum. Further arguments that the chiral
symmetry is not realized in the Wigner--Weyl mode are given in section 2.
The physical ground state must therefore
be asymmetric under the chiral $\su3_L \times \su3_R$ [1.5].
In fact, the chiral symmetry is spontaneously broken down (hidden) to
the vectorial subgroup of isospin and hypercharge, generated by the
vector currents,
$$ H = \su3_{L+R} \times U(1)_{L+R} \eqno(1.5) $$
As mandated by Goldstone's theorem [1.6], the spectrum of massless QCD
must therefore contain $N_f^2 - 1 = 9-1 = 8$ massless bosons
with quantum numbers $J^P = 0^-$ (pseudoscalars) since the axial
charges
do not annihilate the vacuum. Reality is a bit more complex. The
quark masses are not exactly zero which gives rise to an explicit
chiral symmetry breaking (as indicated by the term $\lqcd^I$ in
eq.(1.2)). This is in agreement with the observed particle spectrum --
there are no massless strongly interacting particles. However, the
eight lightest hadrons are indeed pseudoscalar mesons. These are the
pions ($\pi^\pm\, , \,\, \pi^0$), the kaons ($K^\pm\,,\,\, {\bar K}^0\, ,
\,\, K^0)$ and the eta ($\eta$). One observes that $M_\pi \ll M_K
\approx M_\eta$ which indicates that the masses of the quarks in  the
$SU(2)$ subgroup (of isospin) should be considerably smaller than the
strange quark mass. This expectation is borne out by actual calculation
of quark mass ratios. Also, from the relative
size of the quark masses $m_{u,d} \ll m_s$ one expects the chiral
expansion to converge much more rapidly in the two--flavor case than
for $\su3_f$. These basic features of QCD can now be explored
in a similar fashion as outlined before for the case of QED.
\medskip
As already noted, the use of EFTs in the context of strong interactions
preceeds QCD. The Ward identities related to the spontaneously broken
chiral symmetry were explored in great detail in the sixties in the
context of current algebra and pion pole dominance [1.4,1.7].  The work
of Dashen and Weinstein [1.8], Weinberg [1.9] and Callan, Coleman, Wess
and Zumino [1.10] clarified the relation between current algebra
calculations and the use of effective Lagrangians (at tree level).
However, only with Weinberg's [1.11] seminal paper in 1979 it became
clear how one could systematically generate loop corrections to the
tree level (current algebra) results. In fact, he showed that these
loop corrections are suppressed by powers of ($E/\l)^2$, with $E$
a typical energy (four--momentum) and $\l$ the scale below which
the EFT can be applied (typically the mass of the first non--Goldstone
resonance, in QCD $\l \simeq M_\rho$). The method was systematized
by Gasser and Leutwyler for $SU(2)_f$ in Ref.[1.12] and for $\su3_f$
in Ref.[1.13] and has become increasingly popular ever since. The basic
idea of using an effective Lagrangian instead of the full theory is
based on a universality theorem for $\len$ properties of field
theories containing massless (or very light) particles. Consider
a theory (like QCD) at low energies. It exhibits the following properties:
\medskip
\item{$\bullet$}${\cal L}$ is symmetric under some Lie group $G$ (in QCD:
$G = \su3_L \times \su3_R$).
\medskip
\item{$\bullet$}The ground state $|0>$ is symmetric under
$ H \subset G$
(in QCD: $H = SU(3)_V$). To any broken generator of $G$ there appears
a massless Goldstone boson (called "pion") with the corresponding
quantum numbers ($J^P = 0^-$ in QCD).
\medskip
\item{$\bullet$}The Goldstone bosons have a finite transition amplitude
to decay into the vacuum (via the current associated with the broken
generators). This matrix element carries a scale $F$, which is of
fundamental importance for the $\len$ sector of the theory (in QCD:
$<0| A_\mu^a | \pi^b> = i p_\mu \delta^{ab} F$, with $F$ the pion decay
constant in the chiral limit).
\medskip
\item{$\bullet$}There exists no other massless (strongly interacting)
particles.
\medskip
\item{$\bullet$}Explicit symmetry breaking (like the quark mass term
in QCD) can be treated in a perturbative fashion.
\medskip
\item{$\bullet$}Matter fields (such as the spin--1/2 baryons) can
be incorporated in the EFT according to the strictures of
non--linearly realized chiral symmetry. However, special care has to
be taken about their mass terms (see below).
\medskip
\noindent Now any
theory with these properties looks the same (in more than
two space-time dimensions). This means that to leading order the solution
to the Ward identities connected to the broken symmetry is unique
and only contains the scale $F$. Thus, the EFT to lowest order is
uniquely fixed and it is most economical to formulate it in terms
of the Goldstone fields [1.14].
In fact, one collects the pions in a matrix--valued function (generally
denoted '$U$') which transforms linearly under the full action of $G$.
In QCD, a popular choice is $U(x)
= \exp [i  \lambda^a  \pi^a (x)
/F]$  with $\lambda^a\, ( \,\, a = 1, \ldots, 8)$ the Gell--Mann
matrices and $U'(x) = R U(x) L^\dagger$
under chiral $\su3_L \times \su3_R$
(with $L,R$ an element of $\su3_{L,R}$). Accordingly, the pion fields
transform in a highly non--linear fashion. This is a
characteristic feature of EFTs.
\medskip
The inclusion of the lowest--lying baryon octet in the EFT of the
strong interactions again preceeds QCD, see e.g.[1.4,1.7--1.11]. However, the
first systematic analysis of QCD Green functions and current
matrix--elements due to Gasser, Sainio and ${\rm {\check S}}$varc is
much more recent [1.15]. They showed that the fully relativistic
treatment of the spin--1/2 matter fields (the nucleons) spoils the
exact one--to--one correspondence between the loop expansion and the
expansion in small momenta and quark masses. This can simply be
understood from the fact that the nucleon mass $m$ does not vanish in the
chiral limit and thus an extra scale is introduced into the
problem. Stated differently, nucleon four--momenta can
 never be small.\footnote{$^*$}{In ref.[1.15], the two-flavor case was
considered. However, the  problems related to the non--vanishing
mass in the chiral limit generalize straightforwardly to flavor
SU(3).  In this introduction,
we therefore casually switch between the terms 'nucleon' and 'baryon'.}
This problem was overcome by Jenkins and Manohar [1.16] who used
methods borrowed from heavy quark EFT to eliminate the troublesome
baryon mass term. This amounts to considering the baryons as very heavy,
static sources. Consequently, all the mass dependence is shuffled into
a string of interaction vertices with increasing powers of $1/m$ and a
consistent power counting scheme emerges. In this review, we wish to
summarize the developments which have taken place over the last few
years, with particular emphasis on the two--flavor sector and
processes with one nucleon line running through the pertinent Feynman
diagrams. To our opinion, these are the best studied processes from
the theoretical as well as from the experimental side. However, there
has also been considerable activity concerning processes involving two
(or more) nucleons starting from the work of Weinberg [1.17] plus
extensions to the three--flavor case, dense matter and much more.
To summarize the present state of the art, we believe that to
rigorously test the consequences of the spontaneous chiral symmetry
breaking of QCD in nucleon and nuclear studies, calculations to order
${\cal O}(E^4 / \Lambda^4)$ are mandatory in many cases. On the
experimental side, the advances in machine and detector technology
have lead, are leading and will lead to many more data of
unprecedented accuracy. These will serve as a good testing ground of the
chiral structure of QCD.
\medskip
Another non--perturbative method which is much used in studying baryon
properties at low energies is lattice gauge theory (LGT). To our opinion,
LGT has not yet
reached a sufficient accuracy to describe dynamical processes such as pion
production or Compton scattering in the non--perturbative regime. However,
we would like to stress that one should consider these methods as
complementary. For example, one hopes that in
the not too distant future LGT will significantly contribute
by supplying e.g. numerical values for the pertinent low--energy constants.
\medskip
The material is organized as follows. In section 2, we give an
elementary introduction to chiral symmetry, discuss three--flavor QCD
and give a brief account of CHPT for the meson sector. We also show
how one can model the Goldstone pion in a quark model
language. Section 3 contains the basic discussion of the pion--nucleon
Lagrangian, its construction, the extreme non--relativistic limit and
the renormalization procedure to order $E^3$. We give a complete list
of the numerical values of the low--energy constants for the
next--to--leading order effective Lagrangian ${\cal L}_{\pi N}^{(2)}$
and summarize to what extent these values can be understood from a
resonance exchange picture. The inclusion of the spin--3/2 decuplet,
i.e. the $\Delta(1232)$, as an active degree of freedom
 in the EFT is critically examined.  Applications to
pion--nucleon scattering and the reaction $\pi N \to \pi \pi N$ are
also discussed. Section 4 is devoted to the nucleon as probed by
electroweak currents. We discuss in detail such topics
as the electromagnetic form factors, Compton scattering, axial
properties and, furthermore, single and double pion production with real and
virtual photons as well as W--bosons. Together with section 3,
this is the main body of the
work presented in this review. Section 5 contains the extensions to
systems with two and more nucleons. Here, a complication arises due to
the appearance of IR divergences in reducible diagrams which leads to
a modification of the power counting scheme. This is discussed in some
detail and the pertinent method  of applying the chiral power counting
only to the irreducible diagrams together with the solution of a
Schr\"odinger or Lippmann--Schwinger equation to generate the S--matrix
 is then
applied to the potential between two, three and four nucleons. Since
the construction of the NN--potential from the chiral Lagrangian
involves a large number of low--energy constants,
it appears to be favorable for certain applications
to supply as much phenomenological input as possible, i.e. by
taking the two--body  pion--nucleon and the nucleon--nucleon
interaction suitably parametrized from phenomenology.
 The chiral machinery is then used to provide the
remaining three--body forces. As an example  pion--deuteron
scattering is discussed. Similarly, in the description of the meson exchange
currents it is argued that the nuclear short--range correlations
indeed suppress the badly known contact terms thus leading to a more
predictive scheme than for the NN--potential.
Section 6 contains extension to the
three--flavor sector, kaon--nucleon scattering, the
density--dependence of pion properties in matter and gives a summary
of topics not treated in detail. Many of these developments are only
in their infancy and we therefore have decided more to highlight the
weak points than to give any details. However, the reader is supplied
with sufficiently many references on these topics to get a more
detailed (and eventually less biased) picture. The appendices contain
various technicalities such as a summary of the pertinent Feynman
rules or the definition of loop functions which are needed for actual
calculations. To keep the sections self--contained, the relevant
references are given at the end of each section.
\bigskip
\bigskip
\noindent{\bf REFERENCES}
\medskip
\item{1.1}H. Euler, {\it Ann. Phys.\/} (Leipzig) {\bf 26} (1936) 398;

H. Euler and  W. Heisenberg, {\it Z. Phys.\/} {\bf 98} (1936) 714.
\smallskip
\item{1.2}R. Karplus and M. Neuman, {\it Phys. Rev.\/} {\bf 83} (1951) 776.
\smallskip
\item{1.3}J. Gasser and H. Leutwyler, {\it Phys. Reports\/} {\bf C87}
 (1982) 77.  \smallskip
\item{1.4}S. L. Adler and R. F. Dashen, "Current Algebras and applications to
particle physics", Benjamin, New York, 1968.
\smallskip
\item{1.5}Y. Nambu and G. Jona--Lasinio, {\it Phys. Rev.\/} {\bf 122}
(1961) 345; {\bf 124} (1961) 246.
\smallskip
\item{1.6}J. Goldstone, {\it Nuovo Cim.\/} {\bf 19} (1961) 154;

J. Goldstone, A. Salam and S. Weinberg, {\it Phys. Rev.} {\bf 127}
(1962) 965.
\smallskip
\item{1.7}H. Pagels, {\it Phys. Reports\/} {\bf 16} (1975) 219.
\smallskip
\item{1.8}R. Dashen and M. Weinstein, {\it Phys. Rev.\/} {\bf 183}
(1969) 1261.
\smallskip
\item{1.9}S. Weinberg, {\it Phys. Rev.\/} {\bf
166} (1968) 1568.
\smallskip
\item{1.10}S. Coleman, J. Wess and B. Zumino,
{\it Phys. Rev.\/} {\bf 177} (1969) 2239;

C. G. Callan, S. Coleman, J. Wess and B. Zumino,
{\it Phys. Rev.\/} {\bf 177} (1969) 2247.
\smallskip
\item{1.11}S. Weinberg, {\it Physica} {\bf 96A} (1979) 327.
\smallskip
\item{1.12}J. Gasser and H. Leutwyler, {\it Ann. Phys. (N.Y.)\/}
 {\bf 158} (1984) 142.
\smallskip
\item{1.13}J. Gasser and H. Leutwyler, {\it Nucl. Phys.\/}
 {\bf B250} (1985) 465.
\smallskip
\item{1.14}H. Leutwyler, {\it Ann. Phys. (N.Y.)\/}
 {\bf 235} (1994) 165. \smallskip
\item{1.15}J. Gasser, M.E. Sainio and A. ${\rm {\check S}}$varc,
{\it Nucl. Phys.\/}  {\bf B307} (1988) 779.
\smallskip
\item{1.16}E. Jenkins and A.V. Manohar, {\it Phys. Lett.} {\bf 255}
(1991) 558. \smallskip
\item{1.17}S. Weinberg, {\it Phys. Lett.} {\bf 251}
(1990) 288. \smallskip
\vfill
\eject
\noindent{\bf II. CHIRAL SYMMETRY OF THE STRONG INTERACTIONS}
\bigskip
In this section, we first discuss chiral symmetry on an elementary level.
We extend these considerations to three--flavor QCD and the formulation of
its effective low--energy field theory  in terms of the Goldstone
bosons related to the spontaneous chiral symmetry breaking. We also outline
briefly how the structure of the pion can be modeled in a
four--quark interaction cut--off theory of the Nambu--Jona-Lasinio type.
\medskip
\noindent{\bf II.1. ELEMENTARY INTRODUCTION TO CHIRAL SYMMETRY}
\medskip
Before discussing full QCD, let us give a few very introductory remarks about
chiral symmetry. The reader familiar with this concept is invited to skip this
section. To be specific, consider a free and massless spin--1/2 (Dirac) field,
$$ {\cal L}[\Psi ] = i {\bar \Psi} \gamma_\mu \partial^\mu \Psi \eqno(2.1)$$
The state of a free relativistic fermion (of arbitrary mass) is completely
characterized by its energy $E$, its momentum $\svec p$ and its helicity
$\hat{h} ={\svec \sigma} \cdot {\svec p} / |p|$.
For massless fermions helicity is
identical to chirality with $\gamma_5 $ the chirality operator
(one speaks of chirality related to the greek word for "hand").
Let us decompose the spinor into a right-- and a left--handed component,
$$\eqalign{ \Psi & = {1 \over 2} (1-\gamma_5) \Psi +
{1 \over 2} (1+\gamma_5) \Psi   \cr
& = P_L \Psi + P_R \Psi \cr & = \Psi_L + \Psi_R \cr} \eqno(2.2)$$
Obviously, the operators $P_{L,R}$ are projectors,
$$P_L^2 = P_L,P_R^2 = P_R,P_L \cdot P_R = 0 , P_L + P_R = 1  \eqno(2.3)$$
with the property\footnote{*}{For a massive fermion, the $P_{L,R}$ are still
projectors but do not yield exactly the helicity.}
$$ {1 \over 2}{\hat h} \Psi_{L,R} = \pm {1 \over 2} \Psi_{L,R}    \eqno(2.3)$$
This shows that the states $\Psi_{L,R}$ are helicity eigenstates.
 In terms of these fields, the Lagrangian (2.1) takes the form
$$ {\cal L}[\Psi_L, \Psi_R ] =
 i {\bar \Psi}_L \gamma_\mu \partial^\mu \Psi_L +
 i {\bar \Psi}_R \gamma_\mu \partial^\mu \Psi_R
 \eqno(2.4)$$
One notices that the  left-- and right--handed fermion modes do not
communicate. Stated differently, one can apply separate $U(1)_{L,R}$
transformations which leave the Lagrangian invariant,
$$ \Psi_L \to {\rm e}^{i \epsilon_L} \Psi_L , \quad
\Psi_R \to {\rm e}^{i \epsilon_R} \Psi_R
\eqno(2.5)$$
leading to conserved left-- and right--handed currents,
$$J_\mu^I = {\bar \Psi}_I \gamma_\mu \Psi_I , \quad I = L,R \eqno(2.6)$$
with
$$\partial_\mu J^{\mu,I} =  0 , \quad I = L,R    \eqno(2.7)$$
Equivalently, one can construct conserved vector and axial--vector currents,
$$\eqalign{ V_\mu & = {\bar \Psi} \gamma_\mu \Psi , \quad \partial_\mu V^\mu =
0
\cr A_\mu & = {\bar \Psi} \gamma_\mu \gamma_5 \Psi ,
\quad \partial_\mu A^\mu = 0 \cr} \eqno(2.8)$$
since $J_{L,R} = (V \pm A) / 2$. To reiterate, chiral symmetry means that
for massless fermions chirality is a constant of motion.
A fermion mass term explicitely breaks
this symmetry since it mixes the left-- and right--handed components,
$$ {\bar \Psi} M \Psi = {\bar \Psi}_L M \Psi_R +
{\bar \Psi}_R M \Psi_L    \eqno(2.9)$$
To make chiral symmetry a viable concept for massive fermions, the
corresponding eigenvalues of the mass matrix have to be small compared to a
typical energy  scale of the system under consideration. As an example, we will
now consider the case of three--flavor Quantumchromodynamics (QCD).
\goodbreak \bigskip
\noindent{\bf II.2. THREE--FLAVOR QCD}
\medskip
The standard model of the strong, electromagnetic and weak interactions
involves three generations of fermion doublets,
alas six different quark flavors.
{}From these six quark types, three are labelled 'light' ($u,d,s$)
and the other three 'heavy' ($c,b,t$). Here light and heavy refers to a typical
hadronic scale $M_H \sim 1$ GeV. In fact, $m_c > M_H$ and $m_{b,t} >> M_H$
whereas typical values of the light quark masses at a renormalization point of
1 GeV are [2.1]
$$ m_d = 5 \pm 2 \, {\rm MeV}, \quad  m_d = 9 \pm 3 \, {\rm MeV}, \quad
 m_s = 175 \pm 55 \, {\rm MeV}   \eqno(2.10)$$
Note that there exist some controversy about these values, for reviews with
detailed references see e.g. [2.2,2.3,2.4].
 In the three--flavor sector, the QCD
Lagrangian takes the form
$$ {\cal L}_{\rm QCD} = -{1 \over 2 g^2} G_{\mu \nu}^a G^{\mu \nu ,a} +
{\bar q} i \gamma^\mu (\partial_\mu - i G_\mu ) q - {\bar q} {\cal M} q
- {\Theta \over 16 \pi^2} G_{\mu \nu}^a  {\tilde G}^{\mu \nu ,a}  \eqno(2.11)$$
with $q^T (x) = ( u(x) , d(x) , s(x) )$, $G_\mu$ the gluon field, $G_{\mu \nu}$
the corresponding field strength tensor and ${\tilde G}_{\mu \nu ,a} = {1 \over
2} \epsilon_{\mu \nu \alpha \beta} G^{\alpha \beta}_a$ its dual. The last term
in (2.11) is related to the strong CP--problem. In what follows, we will set
$\Theta = 0$. The quark mass matrix can be chosen to be diagonal,
$$ {\cal M} = {\rm diag}( m_u , m_d , m_s )    \eqno(2.12)$$
In (2.11), we have not made explicit the generators related to the local
$SU(3)_{colour}$  transformations. From the chiral symmetry point of view we
rewrite (2.11) as
$$ {\cal L}_{\rm QCD} = {\cal L}_{\rm QCD}^0 - {\bar q} {\cal M} q
\eqno(2.13)$$
and $ {\cal L}_{\rm QCD}^0$ is invariant under the global transformations of
the group
$$ {\cal G} = SU(3)_L \times SU(3)_R \times U(1)_V \times U(1)_A \eqno(2.14)$$
Projecting onto left-- and right--handed quark fields, $q_{L,R} = (1/2)(1 \mp
\gamma_5 ) q$, these transform under the chiral group $SU(3)_L \times SU(3)_R$
as
$$q_L \to {\rm e}^{i T^a \alpha^a_L} \, q_L , \quad
q_R \to {\rm e}^{i T^a \alpha^a_R} \, q_R , \quad  a= 1, \ldots ,8
\eqno(2.15)$$
with the generators $T^a$ (a=1,$\ldots$,8) given in terms of the Gell-Mann
SU(3) matrices via $T^a = \lambda^a / 2$ with $\Tr (T^a T^b) = \delta^{ab} /
2$. In what follows, we will not be concerned with the vectorial $U(1)$
symmetry related to the baryon current ${\bar q} \gamma_\mu q$ and the
anomalous $U(1)_A$ current. It is believed that the axial $U(1)$ is broken by
instanton effects [2.5]. To the global $SU(3)_L \times SU(3)_R$ symmetry of
${\cal L}_{\rm QCD}^0$ one associates $16 = 2(N_f^2 -1 )$ conserved currents,
$$\eqalign{ V_\mu^a & = {\bar q} \gamma_\mu T^a q , \quad \partial_\mu
V^{\mu ,a} =0 \cr
 A_\mu^a & = {\bar q} \gamma_\mu \gamma_5 T^a q , \quad \partial_\mu
A^{\mu ,a} =0 \cr} \eqno(2.16)$$
with the corresponding conserved charges
$$\eqalign{ Q_V^a & = \int d^3 x V_0^i (x) , \quad {dQ^a_V \over dt} = 0 \cr
 Q_A^a & = \int d^3 x A_0^i (x) , \quad {dQ^a_A \over dt} = 0 \cr}
\eqno(2.17)$$
Of course, in the presence of quark mass terms, this symmetry is explicitely
broken.

One might now ask the question whether this chiral symmetry is also manifest in
the ground state or the particle spectrum of QCD? In fact, there are numerous
indications that this is not the case. The realization of the chiral symmetry
in the Wigner mode (i.e. all generators defined in (2.17)  annihilate the
vacuum) would lead to degenerate hadron doublets of opposite parity in plain
contradiction to the observed spectrum. Furthermore, in the Wigner phase the
vector--vector and axial-vector--axial-vector correlators  in the ground state
 would be equal,
$$<0| V_\mu^a(x) V_\nu^b(y) |0>
\, = \, <0| A_\mu^a(x) A_\nu^b(y) |0> \quad . \eqno(2.18)$$
These correlators can be extracted from $\tau$ decay data, $\tau \to \nu_\tau +
n \pi$ $(n=1,2, \ldots)$ with $n$ even (odd) containing the information about
the VV (AA) correlation function.  As shown in refs.[2.6], the VV correlator
strongly peaks around $s \simeq 0.5$ GeV$^2 \simeq M_\rho^2$ whereas the AA
correlator has a broad maximum around $s \simeq 1.5$ GeV$^2 \simeq M_{A_1}^2$.
{}From that and the approximate flavour SU(3) symmetry of the hadron spectrum
we
conclude that the chiral symmetry is spontaneously broken down to its vectorial
subgroup,
$$SU(3)_L \times SU(3)_R \to SU(3)_V              \eqno(2.19)$$
with the appearance of $N_f^2 -1 = 8$ massless pseudoscalar mesons,
the Goldstone bosons [2.7]. These are the analog to the spin waves in a
ferromagnet which underwent spontaneous magnetization (thus breaking the
rotational symmetry of the magnet Hamiltonian). In nature, however, these
Goldstone bosons are not exactly massless but acquire a small mass due to
the explicit symmetry breaking from the quark masses, $M_P^2 \sim {\cal M}$,
where $P$ is a generic symbol for the three pions, the four kaons
 and the eta. From
the sytematics of the hadron spectrum, $M_\eta \simeq M_K \gg M_\pi$ we can
immediately conclude that $m_s \gg m_d \simeq m_u$ since the pions do not
contain any strange quarks. These Goldstone bosons are in fact the lightest
observed hadrons and they saturate the pertinent Ward identities of the strong
interactions at low energies. To calculate QCD Green functions in the
non--perturbative regime, one therefore makes use of an effective field theory
(EFT) with the pseudoscalar mesons as the relevant degrees of freedom. The
essential feature which makes this EFT amenable to a systematic perturbative
expansion is the fact that the interaction between the Goldstone bosons at low
energies is weak. To be more precise, consider the elastic scattering process
$\pi^+ \pi^0 \to \pi^+ \pi^0 $ (for massless pions) [2.8]
$$T( \pi^+ \pi^0 \to \pi^+ \pi^0 ) = {t \over F_\pi^2}     \eqno(2.20)$$
with $t$ the invariant four--momentum transfer squared.
Indeed, as $t$ approaches zero,
the Goldstone boson interaction vanishes. This fact is at the heart of the
systematic low energy expansion in terms of small momenta and quark masses -
chiral perturbation theory (CHPT) - as discussed in some detail in the next
section. For a more detailed account see e.g. the monograph [2.9], the original
papers by Gasser and Leutwyler [2.10] or the review [2.3].
\medskip
\noindent{\bf II.3. \quad CHIRAL PERTURBATION THEORY}
\medskip
In this section, we briefly review how to construct the effective
chiral Lagrangian of the strong interactions at next--to--leading order,
following closely the work of Gasser and Leutwyler [2.10]. It is most
economical to use the external field technique since it avoids
any complication related to the non--linear transformation properties of
the pions. The basic objects to consider are currents and densities
with external fields coupled to them [2.11] in accordance with the
symmetry requirements. The associated Green functions automatically obey
the pertinent Ward identities and higher derivative terms can be
constructed systematically. The S--matrix elements for processes
involving physical mesons follow then via standard LSZ reduction.
To be specific, consider the vacuum--to--vacuum transition
amplitude in the presence of external fields
$$
{\rm e}^{i\zva} = <0 \, \, {\rm out} | 0 \, \, {\rm in}>_{v,a,s,p}
\eqno(2.21)$$
based on the QCD Lagrangian
$${\cal L} = \lqcd^0 + \bar q (\gamma^\mu v_\mu (x) + \gamma^5 \gamma^\mu
a_\mu (x))q - \bar q (s(x) - ip(x)) q
\eqno(2.22) $$
The external vector
$(v_\mu)$, axial--vector ($a_\mu)$, pseudoscalar $(p)$ and scalar
$(s)$ fields are hermitean $3 \times 3$ matrices in flavor space.
The quark mass matrix $\m$ (2.12)
is contained in the scalar field $s(x)$. The Green functions of
massless QCD are obtained by expanding the generating functional
around $v_\mu = a_\mu = s= p =0$.
For the real world, one has to expand around
$v_\mu = a_\mu = p =0\, , \, \, s(x) = \m$.
The Lagrangian ${\cal L}$ is invariant even under {\it local} $\su3 \times
\su3$
chiral transformations if the quark and external fields transform as
follows:
$$\eqalign{
q'_R &= Rq \, \, \, \, ; \, \, \, \, q'_L = Lq \cr
v'_\mu + a'_\mu & = R(v_\mu + a_\mu) R^\dagger
+ i R \partial_\mu R^\dagger \cr
v'_\mu - a'_\mu &= L(v_\mu - a_\mu) L^\dagger
+ i L \partial_\mu L^\dagger \cr
s' + i  p' &= R(s + i p) L^\dagger \cr}
\eqno(2.23) $$
with $L,R$ elements of $\su3_{L,R}$ (in general, these are elements
of $U(3)_{L,R}$, but we already account for the axial anomaly to be
discussed later). The path integral representation of ${\cal Z}$
reads:
$$ {\rm e}^{i \zva} = \int [D G_\mu] [Dq] [D \bar q] {\rm e}^{
\int i d^4x {\cal L}(q,\bar q, G_{\mu \nu}\, ; \, v,a,s,p)}
\eqno(2.24) $$
It allows one to make contact to the effective meson theory. Since
we are interested in processes were the momenta are small (the low
energy sector of the theory), we can expand the Green functions
in powers of the external momenta.
This amounts to an expansion in derivatives of the external fields.
This low energy expansion is not a simple
Taylor expansion since the Goldstone bosons generate poles at $q^2 =0$
(in the chiral limit) or $q^2 = M_\pi^2$ (for finite quark masses).
The $\len$ expansion involves two small parameters, the external
momenta $q$ and the quark masses $\m$ (or the Goldstone masses $M_\pi, M_K,
M_\eta$).
 One expands in powers of these
with the ratio $\m /q^2$ fixed. The effective meson Lagrangian  to carry
out this procedure follows from the $\len$ representation of the
generating functional
$$ {\rm e}^{i \zva} = \int [DU]
{\rm e}^{
\int i d^4x {\cal L}_{{\rm eff}}(U \, ; \, v,a,s,p)}
\eqno(2.25) $$
where the matrix $U$ collects the pseudoscalar Goldstone fields. The low energy
expansion is now obtained from a perturbative expansion of the meson
EFT,
$$ \leff = {\cal L}_2 + {\cal L}_4 +  \ldots
\eqno(2.26)$$
where the subscript ($n = 2$, 4, $\ldots$) denotes the low energy dimension
(number of derivatives and/or quark mass terms). Let us now discuss the
 various terms in this expansion.
The leading term (called ${\cal L}_2$) in the
$\len$ expansion (2.26) can easily be written down in terms of the mesons
which  are described by a unitary $3 \times 3$ matrix in flavor space,
$$ U^\dagger U = 1 \, \,  , \, \, \, \, {\rm det} \, U = 1 \eqno(2.27) $$
The matrix $U$ transforms linearly under chiral symmetry, $U' =
R U L^\dagger$. The lowest order Lagrangian consistent with Lorentz
invariance, chiral symmetry, parity, G--parity and charge conjugation
reads [2.10]
$$
{\cal L}_2 = {1 \over 4} F^2 \biggl\lbrace {\rm Tr} [ \nabla_\mu U^\dagger
\nabla^\mu U + \chi^\dagger U + \chi U^\dagger] \biggr\rbrace
\eqno(2.28) $$
The covariant derivative $\nabla_\mu U$ transforms linearly  under
chiral $\su3 \times \su3$ and contains the couplings to the external
vector and axial fields,
$$ \nabla_\mu U = \partial_\mu U - i (v_\mu + a_\mu) U + iU(v_\mu - a_\mu)
\eqno(2.29) $$
The field $\chi$ embodies the scalar and pseudoscalar externals,
$$\chi = 2 B (s + ip) \eqno(2.30) $$
There are two constants appearing in
eqs.(2.28,2.30). The scale $F$ is related to the axial vector currents,
$A_\mu^a = -F \partial_\mu \pi^a + \ldots$
and thus can be identified with the pion decay constant in the chiral
limit, $ F= F_\pi \lbrace 1 + {\cal O}(\m) \rbrace$, by direct
comparison with the matrix--element $<0|A_\mu^a|\pi^b> = i p_\mu
\delta^{ab} F$.\footnote{*}{Strictly speaking the axial-axial correlator
in the vacuum has a pion pole term with its residuum given by $F_\pi$.}
 The constant $B$, which appears in the field $\chi$,
is related to the explicit chiral symmetry breaking. Consider the
symmetry breaking part of the Lagrangian and expand it in powers of
the pion fields (with $p=0$, $s = \m$ so that
$\chi = 2 B \m$)
$${\cal L}_2^{SB} = {1 \over 2} F^2 B {\rm Tr}[\m(U + U^\dagger)]=
(m_u + m_d) B [F^2 - {\pi^2 \over 2} + {\pi^4 \over 24 F^2}
+ {\cal O}(\pi^6) ] \, + \,      \ldots
\eqno(2.31) $$
where the ellipsis denotes the contributions for the kaons and the eta.
The first term on the right hand side
of eq.(2.31) is obviously related to the vacuum energy, while the
second and third are meson mass and interaction terms, respectively.
Since $\partial H_{{\rm QCD}} / \partial m_q = \bar q q$ it follows
from (2.31) that
$$
<0| \bar u u|0> =
<0| \bar d d|0> =
<0| \bar s s|0> = - F^2 B \lbrace 1 + {\cal O}(\m) \rbrace
\eqno(2.32) $$
This shows that the constant $B$ is related to the vev's of the scalar
quark densities $<0| \bar q q| 0>$, the order parameter of the
spontaneous chiral symmetry breaking. The relation (2.32) is only
correct modulo higher order corrections in the quark masses as
indicated by the term ${\cal O}(\m)$. One can furthermore read off the
pseudoscalar mass terms from (2.31). In the case of isospin symmetry
($m_u = m_d = \hat m)$, one finds
$$\eqalign{
M_\pi^2&= 2 \hat m B \lbrace 1 + {\cal O}(\m)\rbrace
= \krig M_\pi^2  \lbrace 1 + {\cal O}(\m)\rbrace \cr
M_K^2&= ( \hat m + m_s) B \lbrace 1 + {\cal O}(\m)\rbrace
= \krig M_K^2    \lbrace 1 + {\cal O}(\m)\rbrace \cr
M_\eta^2&= {2 \over 3}( \hat m +2 m_s) B \lbrace 1 + {\cal O}(\m)\rbrace
= \krig M_\eta^2    \lbrace 1 + {\cal O}(\m)\rbrace \cr}
\eqno(2.33) $$
with $\krig M_P$ denoting the leading term in the quark mass expansion
of the pseudoscalar meson masses. For the $\krig M_P$, the Gell--Mann--Okubo
relation is exact, $4 \krig M_K^2 = \krig M_\pi^2 + 3 \krig M_\eta^2$.
 In the case of isospin breaking, which leads to $\pi^0 - \eta$
mixing, these mass formulae are somewhat more complicated (see e.g.
ref.[2.10]). Eq.(2.33) exhibits nicely the Goldstone character of the
pions -- when the quark masses are set to zero, the pseudoscalars
are massless and $\su3 \times \su3$ is an exact symmetry. For small
symmetry breaking, the mass of the pions is proportional to the square
root of the symmetry breaking parameter, {\it i.e.} the quark masses.
 From eqs.(2.31)
and (2.33) one can eliminate the constant $B$ and gets the celebrated
Gell--Mann--Oakes--Renner [2.12] relations
$$\eqalign{
F_\pi^2 M_\pi^2&= - 2 \hat m <0| \bar u u |0> + {\cal O}(\m^2) \cr
F_K^2 M_K^2&= -(\hat m + m_s) <0| \bar u u |0> + {\cal O}(\m^2) \cr
F_\eta^2 M_\eta^2&=
-{2 \over 3} (\hat m + m_s) <0| \bar u u |0> + {\cal O}(\m^2) \cr}
\eqno(2.34) $$
where we have used $F_P = F \lbrace 1 + {\cal O}(\m) \rbrace$ ($P=
\pi$, $K$, $\eta$), {\it i.e.} the differences in the physical
decay constants $F_\pi \ne F_K \ne F_\eta$ appear in the terms
of order ${\cal M}^2$ in eq.(2.34).
{}From this discussion we realize that to leading order the strong
interactions are characterized by two scales, namely $F$ and $B$.
Numerically, using the QCD sum rule value $<0| \bar u u |0> = (-225$ MeV)$^3$
 one has $F \simeq F_\pi \simeq 93 \, \,  {\rm MeV}$ and
$ B \simeq 1300 \, \, {\rm MeV}$.
The large value of the ratio $B/F \simeq 14$ has triggered some
investigations of alternative scenarios concerning the mode of
quark condensation [2.13].

One can now calculate tree diagrams using the effective Lagrangian
${\cal L}_2$ and derive with ease all so--called current algebra
predictions ($\len$ theorems). Current algebra is, as should have
become evident by now, only the first term in a systematic $\len$
expansion. Working out tree graphs using ${\cal L}_2$ can not be
the whole story -- tree diagrams are always real and thus unitarity
is violated. One has to include higher order corrections to cure this.
To do this in a consistent fashion, one needs a counting scheme.
The leading term in the $\len$ expansion of $\leff$ (2.26) was denoted
${\cal L}_2$ because it has dimension (chiral power) two. It contains
two derivatives or one power of the quark mass matrix. If one assumes
the matrix $U$ to be order one, $U = {\cal O}(1)$, a consistent
power counting scheme for local terms containing $U$, $\partial_\mu  U$,
$v_\mu$, $a_\mu$, $s$, $p$, $\ldots$ goes as follows. Denote by $q$
a generic small momentum (for an exact definition of 'small', see
below). Derivatives count as order $q$ and so do the external fields
which occur linearly in the covariant derivative $\nabla_\mu U$. For
the scalar and pseudoscalar fields, it is most convenient to book them
as order $q^2$. This can be traced back to the fact
that the scalar field $s(x)$ contains the quark mass matrix, thus
$s(x) \sim \m \sim M_\pi^2 \sim q^2$. With these rules, all terms
appearing in (2.28) are of order $q^2$, thus the notation ${\cal L}_2$
(notice that a term of order one is a constant since $U^\dagger U
= 1$ and can therefore be disregarded. Odd powers of $q$ clash with
parity requirements). To summarize, the building blocks of all terms
containing derivatives and/or quark masses have the following
dimension:
$$\eqalign{
& \partial_\mu U(x) \, , \,\, v_\mu (x) \, , \, \, a_\mu(x) =
{\cal O}(q) \cr
& s(x) \, , \, \, p(x) \, , \, \, F_{\mu \nu}^{L,R} (x) =
{\cal O}(q^2) \cr} \eqno(2.35) $$
where we have introduced the field strengths $F_{\mu \nu}^{L,R}$ for
later use. They are defined via
$$\eqalign{
F_{\mu \nu}^I&= \partial_\mu F_\nu^I - \partial_\nu F_\mu^I
- i [F_\mu^I, F_\nu^I] \, , \, \, \, I= L,R \cr
F_{\mu \nu}^R&= v_\mu + a_\mu \, \, ; \, \, \, \,
F_{\mu \nu}^L= v_\mu - a_\mu \cr}
\eqno(2.36) $$
As already mentioned, unitarity calls for pion loop graphs. Weinberg
[2.14] made the important observation that diagrams with $n$ ($n = 1$,
$2$, $\ldots$) meson loops are suppressed by powers of ${(q^2)}^n$
with respect to the leading term. His rather elegant argument
goes as follows. Consider the S--matrix for a reaction involving
$N_e$ external pions
$$ S = \delta(p_1 + p_2 + \ldots + p_{N_e}) M \eqno(2.37)$$
with $M$ the transition amplitude.  Now $M$ depends on the total
momentum flowing through the amplitude, on the pertinent coupling
constants $g$ and the renormalization  scale $\mu$ (the loop
diagrams are in general divergent and need to be regularized\footnote{*}{It is
advantegoeus to use dimensional regularization since it that case one avoids
the appearance of power--law divergences.}),
$$M = M (q \, , \, \, g \, , \, \, \mu) = q^D f(q/\mu \, , \, \, g)
\eqno(2.38) $$
with the total scaling dimension $D$ of $M$ given by [2.14]
$$ D = 2 + \sum_d N_d (d-2) + 2 N_L \quad . \eqno(2.39) $$
Here, $N_L$ is the number of pion loops and $N_d$ the number of vertices
with $d$ derivatives (or quark mass insertions).
The dominant graphs at low energy carry the smallest value of $D$. The
leading terms with $d=2$ scale like $q^2$ at tree level $(N_L =0)$,
like $q^4$ at one loop level ($N_L = 1$) and so on. Higher derivative
terms with $d=4$ scale as $q^4$ at tree level, as $q^6$ at one--loop
order {\it etc}. This power suppression of loop diagrams is at
the heart of the $\len$ expansion in EFTs like {\it e.g.} chiral
perturbation theory (CHPT).

Up to now, we have been rather casual with the meaning of the word
"small". By small momentum or small quark mass we mean this with
respect to some typical hadronic scale, also called the scale of
chiral symmetry breaking (denoted by $\l_\chi$). Georgi and Manohar
[2.15] have argued that a consistent chiral expansion is possible if
$ \l_\chi \le 4 \pi F_\pi \simeq 1 \, \, \,  {\rm GeV}$.
Their argument is based on the observation that under a change of
the renormalization scale of order one typical loop contributions
(say to the $\pi \pi$ scattering amplitude) will correspond to
changes in the effective couplings of the order $F_\pi^2 / \l_\chi^2
\simeq 1/(4 \pi)^2$. Setting $\l_\chi = 4 \pi F_\pi$ and cutting
the logarithmically divergent loop integrals at this scale, quantum
corrections are of the same order of magnitude as changes in the
renormalized interaction terms. The factor $(4 \pi)^2$ is generic for
one--loop integrals (in 3+1 dimensions).
Another type of argument is related to the non--Goldstone spectrum. Consider
$\pi \pi$ scattering in the $I=J=1$ channel. There, at $\sqrt s =
770$ MeV, one hits the $\rho$--resonance. This is a natural barrier
to the derivative expansion of the Goldstone mesons and therefore
serves as a cut off. The appearance of the $\rho$ signals the
regime of the non--Goldstone particles and describes new physics.
It is therefore appropriate to choose
$\l_\chi \simeq M_\rho \simeq 770 \, \, \, {\rm MeV}$,
which is not terribly different from the previous estimate. In summary,
small external momenta $q$ and small quark masses $\m$ means
$q/M_\rho \ll 1 \, \, {\rm and} \, \, \, \, \m/M_\rho \ll 1 \, \,.$

We have now assembeld all tools to discuss the generating functional
${\cal Z}$ at next--to--leading order, {\it i.e.} at ${\cal O}(q^4)$.
It consists of three different contributions:
(1) The anomaly functional is of order $q^4$ (it contains four
derivatives). We denote the corresponding functional by ${\cal Z}_A$.
The explicit construction was given by Wess and Zumino [2.16] and can
also be found in ref.[2.10]. A geometric interpretation is provided by
Witten [2.17].
(2) The most general effective Lagrangian of order $q^4$ which
is gauge invariant. It leads to the action ${\cal Z}_2 + {\cal Z}_4
= \int d^4x {\cal L}_2 + \int d^4x {\cal L}_4$.
(3) One loop graphs associated with the lowest order term,
${\cal L}_2$. These also scale as terms of order $q^4$.

Let us first discuss the anomaly functional ${\cal Z}_A$. It subsumes
all interactions which break the intrinsic parity and is responsible
{\it e.g.} for the decay $\pi^0 \to 2 \gamma$. It also generates
interactions between five or more Goldstone bosons [2.17]. In what
follows, we will not consider this sector in great detail (for a review,
see ref.[2.18]).

What is now the most general Lagrangian at order $q^4$? The building
blocks and their $\len$ dimensions were already discussed -- we can
have terms with four derivatives or with two derivatives and one quark mass
or with two quark masses (and, correspondingly, the other external fields).
In $\su3$, the only invariant tensors are $g_{\mu \nu}$ and
$\epsilon_{\mu \nu \alpha \beta}$, so one is left with (imposing
also P, G and gauge invariance) [2.10]
$$ {\cal L}_4 = \sum_{i=1}^{10} L_i P_i + \sum_{j=1}^2 H_j \tilde P_j
\eqno(2.40) $$
with
$$\eqalign{
P_1 &= \Tr (\nabla^\mu U^\dagger \nabla_\mu U)^2 , \quad
P_2  = \Tr (\nabla_\mu U^\dagger \nabla_\nu U)
       \Tr (\nabla^\mu U^\dagger \nabla^\nu U) \cr
P_3 &= \Tr (\nabla^\mu U^\dagger \nabla_\mu U
            \nabla^\nu U^\dagger \nabla_\nu U) , \quad
P_4  = \Tr (\nabla^\mu U^\dagger \nabla_\mu U)
       \Tr (\chi^\dagger U + \chi U^\dagger) \cr
P_5 &= \Tr (\nabla^\mu U^\dagger \nabla_\mu U)
           (\chi^\dagger U + \chi U^\dagger) , \quad
P_6  = \bigl[\Tr (\chi^\dagger U + \chi U^\dagger)\bigr]^2 \cr
P_7 &= \bigl[\Tr (\chi^\dagger U - \chi U^\dagger)\bigr]^2 , \quad
P_8  = \Tr (\chi^\dagger U
            \chi^\dagger U +
            \chi U^\dagger
            \chi U^\dagger) \cr
P_9 &=
    - i \Tr (F_{\mu \nu}^R \nabla^\mu U \nabla^\nu U^\dagger )
    - i \Tr (F_{\mu \nu}^L \nabla^\mu U^\dagger \nabla^\nu U) \cr
P_{10} &= \Tr(U^\dagger F^R_{\mu \nu} U F^{L, \mu \nu}) \cr
\tilde P_1 &=
        \Tr(F^R_{\mu \nu}  F^{R, \mu \nu}
         +  F^L_{\mu \nu}  F^{L, \mu \nu} ) , \quad
\tilde P_2  = \Tr(\chi^\dagger \chi) \cr}
\eqno(2.41) $$
For the two flavor case, not all of these terms are independent. The
pertinent $q^4$ effective Lagrangian is discussed in ref.[2.10]. The
first ten terms of (2.40) are of physical relevance for the $\len$
sector, the last two are only necessary for the consistent renormalization
procedure discussed below. These terms proportional to $\tilde P_j
(j = 1$, $2)$ do not contain the Goldstone fields and are therefore
not directly measurable at low energies.
The constants $L_i \, \, (i=1 , \ldots , 10)$ appearing in (2.40) are
the so--called low--energy constants. They are not fixed by the
symmetry and have the generic structure
$$ L_i = L_i^r \, + \, L_i^{\rm inf}   \eqno(2.42)$$
These constants serve to renormalize the infinities of the pion loops
$(L_i^{{\rm inf}})$
and the remaining finite pieces $(L_i^r)$ have to be
fixed phenomenologically or to be estimated by some model (see below). It
should be noted that a few of the low--energy constants are in fact finite.
At next--to--leading order, the strong interactions dynamics is
therefore determined in terms of twelve parameters -- $B$, $F$, $L_1,$
$\ldots$, $L_{10}$ (remember that we have disregarded the singlet
vector and axial currents). In the absence of external fields, only
the first three terms in (2.40) have to be retained.

Finally, we have to consider the loops generated by the lowest order
effective Lagrangian. These are of dimension $q^4$ (one loop
approximation) as mandated by Weinberg's scaling rule. To evaluate
these loop graphs one considers the neighbourhood of the solution $\bar
U (x)$ to the classical equations of motion. In terms of the
generating functional, this reads
$$ {\rm e}^{i {\cal Z}} =
{\rm e}^{i \int d^4 x [{\cal L}_2(\bar U)
+ {\cal L}_4 (\bar U)]} \int [DU]
{\rm e}^{i \int d^4 x [{\cal L}_2( U)
- {\cal L}_2 (\bar U)]}
\eqno(2.43) $$
The bar indicates that the Lagrangian is evaluated at the classical
solution. According to the chiral counting, in the second factor of
(2.43) only the term ${\cal L}_2$ is kept. This leads to
$${\cal Z} = \int d^4x (
\bar {\cal L}_2 +
\bar {\cal L}_4) + {i \over 2} {\rm ln} \, {\rm det} D
\eqno(2.44) $$
The operator $D$ is singular at short distances. The ultraviolet
divergences contained in ${\rm ln} \, {\rm det} D$
can be determined via the
heat kernel expansion. Using dimensional regularization, the
UV divergences in four dimensions take the form
$$ -{1 \over (4 \pi)^2} {1 \over d-4} {\rm Sp}({1 \over 2} \hat \sigma^2
+{1 \over 12} \hat \Gamma_{\mu \nu} \hat \Gamma^{\mu \nu})
\eqno(2.45) $$
The explicit form of the operators $\hat \sigma$ and $\hat \Gamma_{\mu
\nu}$ can be found in ref.[2.10]. Using their explicit expressions, the
poles in ${\rm ln} \, {\rm det} D$ can be absorbed by the following
renormalization of the $\len$ constants:
$$ \eqalign{
L_i&= L_i^r + \Gamma_i L , \, \, \, \, i = 1, \ldots, 10 \cr
H_j&= H_j^r + \tilde
\Gamma_j L , \, \, \, \, j = 1, 2        \cr}
\eqno(2.46)$$
with
$$\eqalign{
L & = {1 \over 16 \pi^2} \lambda^{d-4} \biggl\lbrace {1 \over d-4}
- {1 \over 2} [ {\rm ln}(4 \pi) + \Gamma'(1) + 1 ] \biggr\rbrace \cr
\Gamma_1 & = {3 \over 32} \, , \, \, \, \,
\Gamma_2   = {3 \over 16} \, , \, \, \, \,
\Gamma_3   = 0            \, , \, \, \, \,
\Gamma_4   = {1 \over 8} \, , \, \, \, \, \cr
\Gamma_5 & = {3 \over 8} \, , \, \, \, \,
\Gamma_6   = {11 \over 144} \, , \, \, \, \,
\Gamma_7   = 0            \, , \, \, \, \,
\Gamma_8   = {5 \over 48} \, , \, \, \, \, \cr
\Gamma_9 & = {1 \over 4} \, , \, \, \, \,
\Gamma_{10}   = -{1 \over 4} \, , \, \, \, \,
\tilde \Gamma_{1}   = -{1 \over 8} \, , \, \, \, \,
\tilde \Gamma_{2}  = {5 \over 24} \, , \, \, \, \, \cr}
\eqno(2.47) $$
and $\lambda$ is the scale of dimensional regularization. The $q^4$
contribution ${\cal Z}_4 + {\cal Z}_{{\rm 1-loop}}$ is finite at
$d = 4$ when expressed in terms of the renormalized coupling constants
$L_i^r$ and $H_i^r$. The next step consists in the expansion
of the differential operator $D$ in powers of the external fields. This
gives the explicit contributions of the one--loop graphs to a given
Green function. The full machinery is spelled out in Gasser and
Leutwyler [2.10]. In general, one groups the loop contributions into
tadpole and unitarity corrections. While the tadpoles contain one
vertex and one loop, the unitarity corrections contain one loop
and two vertices. The tadpole contributions renormalize the
couplings of the effective Lagrangian. Both of these loop contributions
also depend on the scale of dimensional regularization. In contrast,
physical observables are $\lambda$--independent. For actual calculations,
however, it is sometimes convenient to choose a particular value
of $\lambda$, say, $\lambda = M_\eta$ or $\lambda = M_\rho$.
To one--loop order, the generating functional therefore takes the form
$${\cal Z} =
{\cal Z}_2 +
{\cal Z}_4 +
{\cal Z}_{{\rm one-loop}} +
{\cal Z}_{{\rm anom}}
\eqno(2.48) $$
and what remains to be done is to determine the values of the
renormalized $\len$ constants, $L_i^r (i=1$, $\ldots$, $10)$.
These  are in principle calculable from QCD, they depend
on $\Lambda_{{\rm QCD}}$ and the heavy quark masses
$$ L_i^r = L_i^r( \Lambda_{{\rm QCD}} \, ; \, \, m_c \, , \,  m_b\, ,
\, m_t)    \eqno(2.49)$$
In practice, such a calculation is not feasible. One therefore resorts
to phenomenology and determines the $L_i^r$ from data. However, some
of these constants are not easily extracted from empirical information.
Therefore, one uses constraints from the large $N_c$ world.
Using this and experimental information from $\pi \pi$ scattering,
$F_K / F_\pi$, the electromagnetic radius of the pion and so on, one
ends up with the values for the $L_i^r (\lambda = M_\eta )$ given in table 1
(large $N_c$ arguments are used to estimate $L_1 \, , \, L_4$ and $L_6$
). For comparison, we also give the values at $\lambda = M_\rho$.
 More
accurate data will allow to further pin down this quantity. In the case
of $SU(2)$, one can define scale--independent couplings $\bar{\ell}_i$
 ($i = 1, \ldots , 7$). These are discussed in ref.[2.10].
Can one now understand the values of the $L_i^r$ from some underlying
principles? Already in their 1984 paper, Gasser and Leutwyler [2.10]
made the following observation. They considered an effective theory
of $\rho$ mesons coupled to the pseudoscalars. Eliminating the heavy
field by use of the equations of motion in the region of momenta
much smaller than the $\rho$ mass, one ends up with terms of order
$q^4$. The values of the corresponding $\len$ constants are given
in terms of $M_\rho$ and the $\rho$--meson coupling strengths to
photons and pions. This leads to a fair description of the SU(2) $\len$
constants.
This method has been generalized by Ecker {\it et al}. [2.19] and by
Donoghue {\it et al}. [2.20]. They consider the lowest order effective
theory of Goldstone bosons coupled to resonance fields (R). These
resonances are of vector (V), axial--vector (A), scalar (S) and
non--Goldstone pseudoscalar (P) type. For the latter category, only the
$\eta '$ is of practical importance. The form of the
pertinent couplings is dictated by chiral symmetry\footnote{*}{For the vectors
and axials, this naturally leads to the tensor--field formulation.}
 in terms of a
few coupling constants which can be determined from data (from
meson--meson and meson--photon decays). At low momenta, one
integrates out the resonance fields. Since their couplings to the
Goldstone bosons are of order $q^2$, resonance exchange produces terms
of order $q^4$ and higher. Symbolically, this reads
$$ \int [dR] {\rm exp}\biggl( i \int d^4 x \tilde \leff [U,R]\biggr) =
             {\rm exp}\biggl( i \int d^4 x \leff [U]\biggr)
\eqno(2.50) $$
So to leading order $(q^4)$, one only sees the momentum--independent
part of the resonance propagators,
$$ {1 \over M_R^2 - t} = {1 \over M_R^2} \biggl(1 + {t \over M_R^2 - t}
\biggr)      \eqno(2.51)    $$
and thus the $L_i^r (\lambda \simeq M_R)$  can be expanded in terms of the
resonance coupling constants and their masses. This leads to
$$L_i^r (\lambda) = \sum_{R=V,A,S,P} L_i^{\rm Res} + \hat L_i (\lambda)
\eqno(2.52)   $$
with $\hat L_i (\lambda)$ a remainder. For this scenario to make sense,
one has to choose $\lambda$ somewhere in the resonance region so that one
can neglect the remainder. A preferred
choice is $\lambda = M_\rho$ (as shown in Ref.[2.19], any value of $\lambda$
between 500 MeV and 1 GeV does the job).
In table 1, we show the corresponding values for all $\len$ constants estimated
from resonance exchange.
It is apparent that the resonances almost completely saturate
the $L_i^r$, with no need for additional contributions. This method
of estimating $L_i^r$ is sometimes called QCD duality or the QCD version
of VMD. In fact, it is rather natural that the higher lying hadronic
states leave their imprints in the sector of the light pseudoscalars
-- as already stated, the typical resonance mass is the scale of new
physics not described by the Goldstone bosons.
\bigskip
$$\hbox{\vbox{\offinterlineskip
\def\strut{\hbox{\vrule height  8pt depth  8pt width 0pt}}
\hrule
\halign{
\strut\vrule# \tabskip 0.1in &
\hfil#\hfil  &
\vrule# &
\hfil#\hfil &
\hfil#\hfil &
\hfil#\hfil &
\vrule# \tabskip 0.0in
\cr
\noalign{\hrule}
&  $i$
&& $L_i^r (M_\eta)$  & $L_i^r (M_\rho)$   & $L_i^{\rm RES}$
& \cr
\noalign{\hrule}
& 1   && $0.9 \pm 0.5$  & $0.7 \pm 0.5$ & 0.6 & \cr
& 2   && $1.6 \pm 0.4$  & $1.2 \pm 0.4$ & 1.2 & \cr
& $3^*$   && $-3.6 \pm 1.3$  & $-3.6 \pm 1.3$ & $-3.0$ & \cr
& 4   && $0.0 \pm 0.5$  & $-0.3 \pm 0.5$ & 0.0 & \cr
& 5   && $2.2 \pm 0.5$  & $ 1.4 \pm 0.5$ & 1.4 & \cr
& 6   && $0.0 \pm 0.3$  & $ -0.2 \pm 0.3$ & 0.0 & \cr
& $7^*$   && $-0.4 \pm 0.15$  & $-0.4 \pm 0.15$ & $-0.3$ & \cr
& 8   && $1.1 \pm 0.3$  & $  0.9 \pm 0.3$ & 0.9 & \cr
& 9   && $7.4 \pm 0.2$  & $  6.9 \pm 0.2$ & 6.9 & \cr
& 10  && $-5.7 \pm 0.3$  & $ -5.2 \pm 0.3$ & $-6.0$ & \cr
\noalign{\hrule}}}}$$
\smallskip
{\noindent\narrower Table 1:\quad Low--energy constants for $\su3_L \times
\su3_R$. The first two columns give the phenomenologically determined
values at $\lambda = M_\eta$ and $\lambda = M_\rho$. The $L_i^r (i=4$,
$\ldots$,
8) are from ref.[2.10], the $L^r_{9,10}$ from ref.[2.21] and the $L^r_{1,2,3}$
from the recent determination in ref.[2.22]. The '$*$' denotes
the constants which are not renormalized. The third column shows the
estimate based on resonance exchange [2.19].
\smallskip}
\goodbreak
\medskip
In CHPT, the structure of any particle is made up by pion loops and
higher resonance contributions encoded in the low--energy constants. As an
example, consider the pion charge form factor. To lowest (tree) order, it is
simply equal to unity as demanded by gauge invariance. At next order in the
chiral expansion, loops and counterterms build up the pion radius, with the
lions share due to one counterterm ($L_9$) which is saturated by vector meson
exchange. At yet higher orders, one consistently sees more of the energy
dependence of the pion form factor. However, in this perturbative approach one
does not get the $\rho$
resonance (or similar effects in other channels). That is the
reason why we argued that the scale of the resonance masses sets a natural cut
off to the range of applicability of CHPT in the meson sector. A more detailed
account of this and the many applications of CHPT can be found in the reviews
[2.2,2.3,2.18,2.22] and the connection of the effective Lagrangian to the QCD
Ward identities is elaborated on in ref.[2.24]. Before considering now the
inclusion of matter fields in CHPT, let us briefly discuss the structure of the
pion from a quark model point of view.
\bigskip
\noindent{\bf II.4. \quad MODELLING THE PION}
\medskip
To investigate the formation of vacuum condensates
and the generation of mass, Nambu
and Jona-Lasinio [2.25] proposed a model with a Heisenberg--type four nucleon
interaction in close analogy to developments in superconductivity. One can
extend this approach to QCD where in the phase of the spontaneously
broken chiral symmetry a scalar quark condensate forms and the quarks acquire a
finite constituent mass of the order of a few hundreds of
MeV.\footnote{$^*$}{These constituent masses should not be confused with the
fundamental mass parameters (current quark masses) entering the QCD Lagrangian.
It should be stressed that the notion of a constituent quark
is model--dependent
but helps to understand qualitatively many features of the hadron properties.}
The pion as the Goldstone boson appears as a collective quark--antiquark mode.
To discuss these features in some detail, we follow closely the work of
ref.[2.26]. Consider the two--flavor NJL Lagrangian for massless quarks
interacting via a contact force,
$${\cal L}_{\rm NJL} = {\bar \psi} i \gamma_\mu \partial^\mu \psi  +
G \, [ (\bar \psi \psi )^2 - (\bar \psi \gamma_5 \svec \tau \psi )^2 ]
\eqno(2.53)$$
Here, $G$ is a positive coupling strength with the dimension of a squared
length. The Lagrangian (2.53) is obviously invariant under chiral $SU(2)_L
\times SU(2)_R$. It can be thought as a minimal effective Lagrangian mimicking
some basic properties of non--perturbative QCD in the long--wavelength limit.
One now solves the Dyson equation for the self--energy $\Sigma$ and identifies
$\Sigma$ with the  mass $M$ which is dynamically generated by the
self--interactions. The resulting self--consistent equation relates $M$ and the
coupling $G$. It has a trivial solution $M = 0$ which corresponds to the
ordinary perturbative result. However, for $G$ above some critical value, it
has also a non--trivial solution which is determined by
 a self--consistency equation of the form
$${4 N_f N_c + 1 \over ( 2 \pi )^4} \, G \, \int d^4 p {1 \over p^2 - M^2 + i
\epsilon} = 1                                                  \eqno(2.54)  $$
with $N_f \,  (N_c)$ the number of flavours (colors). The integral in Eq.(2.54)
diverges quadratically due to the zero range interaction. One therefore has to
regularize the integral. Doing this e.g. by a covariant momentum cut--off
$\theta (\Lambda^2 - p^2)$, one obtains
$${ 4 \pi^2 \over 13 G \Lambda^2} = 1 - {M^2 \over \Lambda^2} \ln \biggl[
{\Lambda^2 \over M^2} +1 \biggr]  \quad .                        \eqno(2.55)$$
Eq.(2.55) clearly exhibits that spontaneous symmetry breaking only occurs for
 values of  $G \ge G_{\rm crit}$. For such values, the mass $M$
starts to deviate from zero and increases with $G$. The scalar quark condensate
acquires a non--vanishing vev which can be interpreted as the probability of
finding $\bar q q$ pairs in the vacuum,
$$<\bar \psi \psi > = -{3 \over 4 \pi^2} M^3 \biggl[{\Lambda^2 \over M^2} -
\ln \biggl({\Lambda^2 \over M^2} + 1 \biggr) \biggr]       \eqno(2.56)$$
which  shows the intimate relation
 between the constituent quark mass $M$ and the
quark condensate in this schematic model of chiral symmetry violation.
\midinsert
\smallskip
\hskip 1in
\epsfysize=1in
\epsfxsize=3in
\epsffile{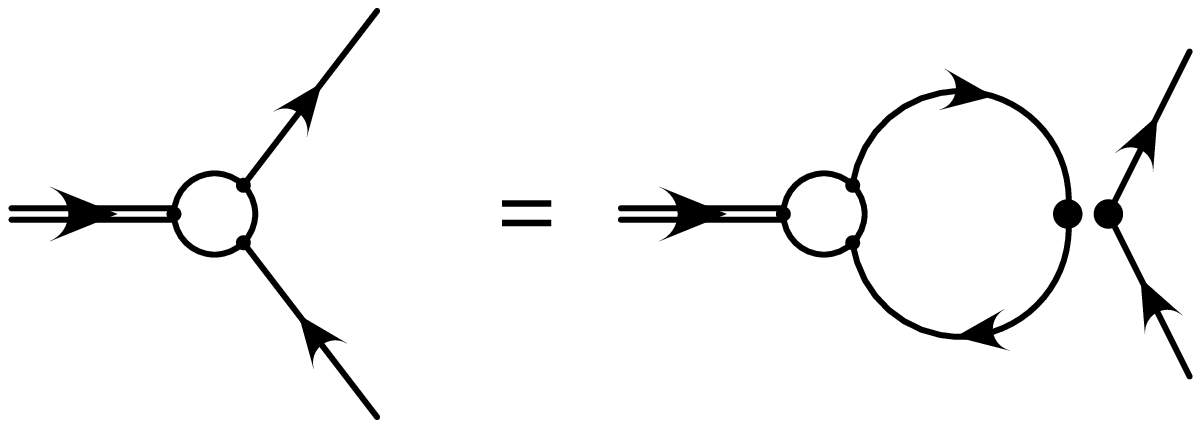}
\smallskip
{\noindent\narrower \it Fig.~2.1: \quad Bethe--Salpeter
equation in the Hartree--Fock approximation.
The double line represents the pion, the solid lines constituent quarks. The
exchange (Fock) diagram is not shown.
\smallskip}
\vskip -0.5truecm
\endinsert
\noindent In studying
the bound--state problem, one finds that the Bethe--Salpeter
equation for the vertex function $\Gamma_5 (p)$ in the pseudoscalar
isovector channel (shown in fig.2.1) is equivalent to the condition (2.54)
when the total four--momentum of the quark--antiquark pair is zero,
$p^2 = 0$. This means
that there exists a massless pseudoscalar isovector particle (the pion) related
to the spontaneous chiral symmetry breaking. This nicely illustrates the
Goldstone theorem in a microscopic picture. As mandated by Goldstone theorem,
the pion has non--vanishing transition matrix--element into the vacuum via the
axial current which defines the pion decay constant. In the NJL model, it is
related to the constituent quark mass via
$$F_\pi = {\sqrt{3} \over 2 \pi} \, M \, \biggl[ \ln(1+u^2) - {u^2 \over
1 + u^2} \biggr] \, , \quad u = {\Lambda \over M}            \eqno(2.57)$$
which shows that $F_\pi$ is linked to the collective nature of the pion. The
model can, of course, also be treated in the case of explicit chiral symmetry
breaking by adding the canonical current quark mass term.

With these basic tools, one can now study very different problems related to
the physics of the Goldstone bosons and other mesons (if one extends the
basic Lagrangian accordingly). Some of these are:
\smallskip
\item{1)}The thermodynamics of the constituent quarks and the pions, i.e.
aspects of finite temperature and density (in the approximation that the
baryon density is given by three times the constituent quark density).
 Such character changes of meson
properties play an important role in the nuclear equation of state and in hot
and dense baryon--rich environments as precursors of the transition to the
much discussed (but not yet observed) quark--gluon plasma.
For an early reference see [2.27]
and the recent reviews [2.28,2.29].
\smallskip
\item{2)}The extension of the model to the three--flavor case and the study
of flavor mixing. This was first addressed in a systematic fashion in
the paper by Bernard et al. [2.30]
where it was shown that the $U(1)_A$ anomaly forces the inclusion of terms
with $2N_f$ fermion fields (within the one-loop approximation to the effective
potential). Certain aspects of the physics of flavor mixing are reviewed in
ref.[2.31].
\smallskip
\item{3)}The relation of NJL--type models to CHPT has been discussed early
[2.32]. It has become clear that a direct comparison  is hampered by the fact
that in the NJL model one does not expand in terms of a small parameter. It
can nevertheless serve as a guideline to understand the physics behind the
low--energy constants (the extended NJL model) [2.33] and to get an estimate
of $p^6$ (and higher order) effects [2.34].
\smallskip
\noindent Let me briefly elaborate on the last point, i.e. the work of
ref.[2.34]. There, a consistent bosonization scheme for the NJL model was
developed and the $p^2$ expansion of certain observables was worked out. In
table 2, we show the results for the pion mass, decay constant and the
constituent mass. One sees that the $p^4$ approximation is within 1$\%$ of the
total (Hartree--Fock) result.
\medskip
$$\hbox{\vbox{\offinterlineskip
\def\strut{\hbox{\vrule height  8pt depth  8pt width 0pt}}
\hrule
\halign{
\strut\vrule# \tabskip 0.1in &
\hfil#\hfil  &
\vrule# &
\hfil#\hfil &
\hfil#\hfil &
\hfil#\hfil &
\hfil#\hfil &
\vrule# \tabskip 0.0in
\cr
\noalign{\hrule}
&  Order
&& ${\cal O} (p^2)$  & ${\cal O} (p^4)$   & ${\cal O} (p^6)$ & Total
& \cr
\noalign{\hrule}
& $F_\pi$   &&  88.6  &  93.8 &  93.0 &  93.1 & \cr
& $M_\pi$   && 141.5  & 138.4 & 139.1 & 139.0 & \cr
& $M$       && 221.2  & 243.9 & 241.4 & 248.1 & \cr
\noalign{\hrule}}}}$$
\smallskip
{\noindent\narrower Table 2:\quad Chiral expansion of the pion decay constant,
the pion mass and the constituent quark mass to order $p^2$, $p^4$, $p^6$ in
comparison to the self--consistent result of the bosonized NJL--model. All
numbers are in MeV. \smallskip}
\goodbreak
\bigskip
Clearly, such results always have to be considered indicative since within the
Hartree--Fock (one--loop effective potential) approximation one does not
include pion loops. What is most important here is to have a microscopic
model for the spontaneous chiral symmetry breakdown and its
associated Goldstone bosons.
\bigskip \bigskip
\noindent{\bf REFERENCES}
\medskip
\item{2.1}J. Gasser and H. Leutwyler, {\it Phys. Reports\/} {\bf C87} (1982)
77.
\smallskip
\item{2.2} H. Leutwyler, in: Proc. XXVI Int. Conf. on High Energy Physics,
Dallas, 1992,

edited by J.R. Sanford, AIP Conf. Proc. No. 272, 1993.
\smallskip
\item{2.3}Ulf-G. Mei{\ss}ner, {\it Rep. Prog. Phys.\/} {\bf 56} (1993) 903.
\smallskip
\item{2.4} J.F. Donoghue, TASI lectures, Boulder, USA, 1993;

H. Leutwyler, Bern University preprint BUTP--94/8, 1994.

\smallskip
\item{2.5}G. 't Hooft,  {\it Nucl. Phys.\/} {\bf B72} (1974) 461.
\smallskip
\item{2.6}R. D. Peccei and J. Sola, {\it Nucl. Phys.\/} {\bf
B281} (1987) 1;

C. A. Dominguez and J. Sola, {\it Z. Phys.\/} {\bf
C40} (1988) 63.
\smallskip
\item{2.7}J. Goldstone, {\it Nuovo Cim.\/} {\bf 19} (1961) 154.
\smallskip
\item{2.8} S. Weinberg, {\it Phys. Rev. Lett.\/} {\bf 17} (1966) 616.
\smallskip
\item{2.9}
J. F. Donoghue, E. Golowich and B. R. Holstein, " Dynamics of the  Standard
Model", Cambridge Univ. Press, Cambridge, 1992.
\smallskip
\item{2.10}J. Gasser and H. Leutwyler, {\it Ann. Phys. (N.Y.)\/}
 {\bf 158} (1984) 142; {\it Nucl. Phys.\/} {\bf B250} (1985) 465,539.
\smallskip
\item{2.11}D. G. Boulware and L. S. Brown,  {\it Ann. Phys.\/} (N.Y.) {\bf
138} (1982) 392.
\smallskip
\item{2.12}M. Gell-Mann, R. J. Oakes and B. Renner,  {\it Phys. Rev.\/} {\bf
175} (1968) 2195.
\smallskip
\item{2.13}M. D. Scadron and H. F. Jones, {\it Phys. Rev.\/} {\bf
D10} (1974) 967;

H. Szadijan and J. Stern, {\it Nucl. Phys.\/} {\bf B94} (1975) 163;

R. J. Crewther, {\it Phys. Lett.\/} {\bf B176} (1986) 172.

N.  H. Fuchs, H. Szadijan and J. Stern, {\it Phys. Lett.\/} {\bf
B238} (1990) 380;

{\it Phys. Lett.\/} {\bf B269} (1991) 183.
\smallskip
\item{2.14}S. Weinberg, {\it Physica} {\bf 96A} (1979) 327.
\smallskip
\item{2.15}A. Manohar and H. Georgi, {\it Nucl. Phys.\/} {\bf B234} (1984) 189.
\smallskip
\item{2.16}J. Wess and B. Zumino, {\it Phys. Lett.\/} {\bf 37B} (1971) 95.
\smallskip
\item{2.17}
E. Witten, {\it Nucl. Phys.\/} {\bf B223} (1983) 422.
\smallskip
\item{2.18}J. Bijnens, {\it Int. J. Mod. Phys.} {\bf 8} (1993) 3045.
\smallskip
\item{2.19}G. Ecker, J. Gasser, A. Pich and E. de Rafael,
{\it Nucl. Phys.\/} {\bf B321}
(1989) 311.
\smallskip
\item{2.20}J. F. Donoghue, C. Ramirez and G. Valencia,
{\it Phys. Rev.\/} {\bf D39}
(1989) 1947.
\smallskip
\item{2.21}J. Bijnens and F. Cornet, {\it Nucl. Phys.\/}
 {\bf B296} (1988) 557.
\smallskip
\item{2.22} J. Bijnens, G. Ecker and J. Gasser, in:
The DAFNE Physics Handbook (vol. 1), eds. L. Maiani, G. Pancheri
and N. Paver, INFN Frascati, 1992
\smallskip
\item{2.23}
J. F. Donoghue, in "Medium energy antiprotons and the quark--gluon structure
of hadrons", eds. R. Landua, J. M. Richard and R. Klapish, Plenum Press, New
York, 1992;

G. Ecker, in: Quantitative Particle Physics, eds. M. Levy et al., Plenum, New
York,

1993;

H. Leutwyler, in ``Recent Aspects of Quantum Fields'', eds. H. Mitter and

M. Gausterer, Springer Verlag, Berlin, 1991.

"Effective field theories of the
standard model", ed. Ulf--G. Mei{\ss}ner, World

Scientific, Singapore, 1992;

G. Ecker, ``Chiral Perturbation Theory'', preprint UWThPh-94-49, 1995,

to app. in Progr. Part. Nucl. Phys., hep-ph/9501357.
\smallskip
\item{2.24}H. Leutwyler,  {\it Ann. Phys. (N.Y.)} {\bf 235} (1994) 165;
see also preprint BUTP-94/13, 1994.
\smallskip
\item{2.25}Y. Nambu and G. Jona--Lasinio, {\it Phys. Rev.\/} {\bf 122}
(1961) 345; {\bf 124} (1961) 246.
\smallskip
\item{2.26}V. Bernard, {\it Phys. Rev.\/} {\bf D34} (1986) 1601.
\smallskip
\item{2.27}V. Bernard, Ulf-G. Mei{\ss}ner and I. Zahed,
{\it Phys. Rev.\/} {\bf D36} (1987) 819.
\smallskip
\item{2.28}S. Klevansky, {\it Rev. Mod. Phys.\/} {\bf 64} (1992) 649.
\smallskip
\item{2.29}T. Hatsuda and T. Kunihiro, {\it Phys. Reports} {\bf 247}
(1994) 221.
\smallskip
\item{2.30}V. Bernard, R.L. Jaffe and Ulf-G. Mei{\ss}ner,
{\it Nucl. Phys.\/} {\bf B308} (1988) 753.
\smallskip
\item{2.31}U. Vogl and W. Weise,
{\it Prog. Part. Nucl. Phys.\/} {\bf 26} (1991) 195.
\smallskip
\item{2.32}
T. H. Hansson, M. Prakash and I. Zahed,
{\it Nucl. Phys.} {\bf B335} (1990) 67;

V. Bernard and Ulf--G. Mei{\ss}ner,
{\it Phys. Lett.} {\bf B266} (1991) 403;

C. Sch\"uren, E. Ruiz Arriola and K. Goeke,
{\it Nucl. Phys.} {\bf A547} (1992) 612;

S. Klevansky and J. Muller, {\it Phys. Rev.} {\bf C} (1994) in print.
\smallskip
\item{2.33}J. Bijnens, C. Bruno and E. de Rafael, {\it Nucl. Phys.} {\bf B390}
(1993) 501. \smallskip
\item{2.34}V. Bernard, A.A. Osipov and Ulf-G. Mei{\ss}ner,
{\it Phys. Lett.\/} {\bf B285} (1992) 119.
\smallskip
\bigskip \bigskip
%
\vfill
\eject
\noindent{\bf III. THE PION--NUCLEON SYSTEM}
\medskip
In this section, we will be concerned with the inclusion of baryons in the
effective field theory. We will specialize to the case of two flavors with
the pions and nucleons as the asymptotically observed fields. The
generalization to the case of three flavors will be taken up later. First,
we discuss the relativistic formulation. In that case, however, the
additional mass scale (the nucleon mass in the chiral limit) destroys the
one--to--one correspondence between the loop and the small momentum expansion.
This can be overcome in the extreme non--relativistic limit in which the
nucleon is essentially considered as a static source. We will then turn to
the systematic renormalization of the effective pion--nucleon Lagrangian to
order $p^3$. Finally, we discuss the appearing low--energy constants and
the role of the $\Delta (1232)$ resonance. As applications, elastic
pion--nucleon scattering and the reaction $\pi N \to \pi \pi N$ at threshold
are considered. Reactions involving electroweak probes
 are relegated to section 4.
\medskip
\noindent{\bf III.1. EFFECTIVE LAGRANGIAN}
\medskip
\goodbreak
In this section, we will be concerned with the inclusion of baryons in
the effective field theory. The relativistic formalism dates back to the
early days, see {\it e.g.} Weinberg [3.1], Callan  et al.  [3.2],
Langacker and Pagels [3.3] and
others (for a review, see Pagels [3.4]).
The connection to QCD Green functions
was performed in a systematic fashion by Gasser, Sainio and
${\rm \check S}$varc [3.5] (from here on referred as GSS)
and Krause [3.6]. As done in the GSS paper, we will
outline the formalism
in the two--flavor case, {\it i.e.} for the pion--nucleon ($\pi N$)
system. The extension to flavor SU(3) is spelled out in
section 6.

Following GSS, we now discuss the modifications of the procedure detailed
in section 2 to include the nucleons. The starting point is the observation
that the time--ordered nucleon matrix elements of the quark currents are
generated by the one--nucleon to one--nucleon transition amplitude
$${\cal F} ({\svec p}',{\svec p};v;a;s;p) = <{\svec p}' \, {\rm out}| {\svec p}
\, {\rm in}>_{v,a,s,p}^{\rm connected} \, , \quad {\svec p}' \ne {\svec p}
\eqno(3.1)$$
determined by the Lagrangian (2.22). Here, $|{\svec p} \, {\rm in}>$ denotes an
incoming one--nucleon state of momentum ${\svec p}$ (and similarly
$|{\svec p}' \, {\rm out}>$). The idea is now to construct in analogy with
(2.25) a pion--nucleon field theory which allows to evaluate the functional
${\cal F}$ at low energies.

First, we consider the general structure of the effective pion-nucleon
Lagrangian ${\cal L}_{\pi N}^{\rm eff}$. It contains the pions collected in the
matrix--valued field $U(x)$ and we combine the
proton $(p)$ and the neutron $(n)$ fields in an isospinor $\Psi$
$$ \Psi = \pmatrix { p \cr n \cr}  \quad . \eqno(3.2)$$
There is a variety of ways to describe the transformation properties
of the spin--1/2 baryons under chiral $SU(2) \times SU(2)$. All
of them lead to the same physics. However, there is one most convenient
choice (this is discussed in detail in Georgi's book [3.7]).
In the previous section, we had already seen that the self--interactions
of the pions are of derivative nature, {\it i.e.} they vanish at
zero momentum. This is a feature we also want to keep for the
pion--baryon interaction. It calls for a non--linear realization of the
chiral symmetry. Following Weinberg [3.1] and CCWZ [3.2], we
introduce a matrix--valued function $K$, and the baryon field transforms as
$$ \Psi \to K(L,R,U) \Psi  \quad .  \eqno(3.3) $$
$K$ not only depends on the group
elements $L,R \in SU(2)_{L,R}$, but also on the pion field (parametrized
in terms of $U(x)$) in a highly non--linear fashion, $K = K(L,R,U)$.
Since $U(x)$ depends on the space--time coordinate $x$, $K$ implicitely
depends on $x$ and therefore the transformations related to $K$
are local. To be more specific, $K$ is defined via
$$ R u = u' K \eqno(3.4) $$
with $u^2 (x) = U(x)$ and $U'(x) = R U(x) L^\dagger =
{u'}^2(x)$.\footnote{*}{We adhere here to the notation of [3.5]. The more
obvious one with interchanging $L$ and $R$ is e.g. used in [3.7].}
The transformation properties of the pion field induce a well--defined
transformation of $u(x)$ under $SU(2) \times SU(2)$. This defines $K$
as a non--linear function of $L$, $R$ and $\pi (x)$. $K$ is a realization
of $SU(2) \times SU(2)$,
$$ K = \sqrt{L U^\dagger R^\dagger} R \sqrt U   \eqno(3.5) $$
The somewhat messy object $K \in $~SU(2) can be
understood most easily in terms of
infinitesimal transformations. For $K = \exp (i \gamma_a \tau_a)$,
$L = \exp (-i \alpha_a \tau_a)
     \exp (i \beta_a \tau_a)$ and
$R = \exp (i \alpha_a \tau_a)
     \exp (i \beta_a \tau_a)$ (with $\gamma_a$, $\alpha_a$, $\beta_a$
real) one finds,
$$ \vec \gamma = \vec \beta - [\vec \alpha \times \vec \pi] / 2 F_\pi
+ {\cal O}({\vec \alpha}^2 ,  {\vec \beta}^2, {\vec \pi}^2)
\eqno(3.6) $$
which means that the nucleon field is multiplied with a function
of the pion field. This gives some credit to the notion that chiral
transformations are related to the absorption or emission of pions.
The covariant derivative of the nucleon field is given by
$$ \eqalign{
D_\mu \Psi&= \partial_\mu  \Psi  + \Gamma_\mu \Psi \cr
\Gamma_\mu&= {1 \over 2} [u^\dagger , \partial_\mu u]
- {i \over 2}  u^\dagger (v_\mu + a_\mu) u
- {i \over 2}  u (v_\mu - a_\mu) u^\dagger \cr}
\eqno(3.7) $$
$D_\mu$ transforms homogeneously under
chiral transformations, $D_\mu' = K D_\mu K^\dagger$.
The object $\Gamma_\mu$ is the so--called chiral connection. It is
a gauge field for the local transformations
$$ \Gamma_\mu' = K \Gamma_\mu K^\dagger + K \partial_\mu K^\dagger
\eqno(3.8) $$
The connection $\Gamma_\mu$ contains one derivative. One can also
form an  object of axial--vector type with one derivative,
$$  u_\mu = i (u^\dagger \nabla_\mu u -
u \nabla_\mu u^\dagger) = i \lbrace u^\dagger , \nabla_\mu
u \rbrace = i u^\dagger \nabla_\mu U u^\dagger
\eqno(3.9) $$
which transforms homogeneously, $u_\mu' = K u_\mu K^\dagger $.
The covariant derivative $D_\mu$ and the axial--vector object $u_\mu$
are the basic building blocks for the lowest order effective theory.
Before writing it down, let us take a look at its most general form.
It can be written as a string of terms with an even number of external
nucleons, $n_{ext} = 0, 2, 4, \ldots$. The term with $n_{ext} = 0$
obviously corresponds to the meson Lagrangian (2.28) so that
$${\cal L}_{eff} [\pi, \Psi, \bar \Psi]
= {\cal L}_{\pi \pi}
 + {\cal L}_{\bar \Psi \Psi}
 + {\cal L}_{\bar \Psi \Psi \bar \Psi \Psi}
 + \ldots
\eqno(3.10) $$
Typical processes related to these terms are pion--pion, pion--nucleon
and nucleon--nucleon scattering, in order. In this section, we will
only be concerned with processes with two external nucleons and no
nucleon loops (in section 5, we will also consider terms with $n=4$),
$$ {\cal L}_{\bar \Psi \Psi} =
          {\cal L}_{\pi N}  =  - \bar \Psi (x) {\cal D}(x) \Psi (x)
\eqno(3.11) $$
The differential operator ${\cal D}(x)$ is  subject to a chiral
expansion as discussed below. We now wish to construct the generating
functional for the vacuum--to--vacuum transitions in the presence of nucleons.
For doing that, we add external Grassmann sources for the
nucleon field to the effective Lagrangian,
$${\cal L}_{eff} [\pi, \Psi, \bar \Psi]
= {\cal L}_{\pi \pi}  + {\cal L}_{\pi N} + \bar{\eta}\Psi + \bar{\Psi}\eta
\eqno(3.12) $$
{}From that, one defines the vacuum--to--vacuum transition amplitude via
$$\eqalign{
\exp \lbrace i {\tilde {\cal Z}}[v,a,s,p;\bar{\eta},\eta]  \rbrace
& = N \int [dU] [d\Psi] [d\bar{\Psi}] \exp i \int d^4 x ( {\cal L}_{\pi \pi} +
{\cal L}_{\pi N} + \bar{\eta} \Psi + \bar{\Psi} \eta ) \cr
& = N' \int [dU] \exp \biggl[i \int d^4 x {\cal L}_{\pi \pi} +
i \int d^4 x d^4 y \, \bar{\eta}(x) S(x,y) \eta (y) \biggr]
({\rm det} {\cal D}) \cr}   \eqno(3.13)$$
where $S$ is the nucleon propagator in the presence of external fields,
$$ DS(x,y;U,v,a,s,p) = \delta^{(4)}(x-y)  \quad .      \eqno(3.14)$$
Evaluating the functional $\tilde{{\cal Z}}$  at
$\bar{\eta} = \eta =0, \, {\rm det} D = 1$ (i.e. no nucleon loops)
 one recovers the functional ${\cal Z}$,
eq.(2.25). Furthermore, the leading order terms in the low--energy expansion of
${\cal F}$ is generated by the tree graphs in $\tilde{{\cal Z}}$. However, the
relation between ${\cal F}$ and $\tilde{{\cal Z}}$ beyond leading
order is  much
more complicated due to the fact that the nucleon mass does not vanish in the
chiral limit as discussed below.

Let us first consider the effective pion--nucleon Lagrangian to lowest order.
 Its explicit form  follows simply by
combining the connection $\Gamma_\mu$ and the axial--vector $u_\mu$
(which are the objects with the least number of derivatives)
with the appropriate baryon bilinears
$$ \eqalign{
{\cal L}_{\pi N}^{(1)} &=  - \bar \Psi {\cal D}^{(1)} \Psi \cr
&= \bar \Psi (i \gamma_\mu D^\mu - \krig m + { \krig g_A  \over 2}
\gamma^\mu \gamma_5 u_\mu) \Psi \cr}
\eqno(3.15) $$
The effective Lagrangian (3.15) contains two new parameters. These are
the baryon mass $\krig m$ and the axial--vector coupling $\krig g_A$
in the chiral limit,
$$\eqalign{
m&= \krig m [1 + {\cal O} (\hat m) ] \cr
g_A&= \krig g_A [1 + {\cal O} (\hat m) ] \cr}
\eqno(3.16) $$
Here, $m = 939$ MeV denotes the physical nucleon mass and $g_A$
the axial--vector strength measured in neutron $\beta$--decay,
$n \to p e^- \bar \nu_e$, $g_A \simeq 1.26$. The fact that $\krig m$
does not vanish in the chiral limit (or is not small on the typical
scale $\Lambda \simeq M_\rho$) will be discussed below. Furthermore,
the actual value of $\krig m$, which has been subject to much recent
debate, will be discussed in the context of pion--nucleon scattering.
The occurence of the constant $\krig g_A$ is all but surprising.
Whereas the vectorial (flavor) $SU(2)$ is protected at zero momentum
transfer, the axial current is, of course, renormalized. Together
with the Lagrangian ${\cal L}_{\pi \pi}^{(2)}$ (2.28), our lowest
order pion--nucleon Lagrangian reads:
$$ {\cal L}_1 = {\cal L}_{\pi N}^{(1)}
              + {\cal L}_{\pi \pi}^{(2)}
\eqno(3.17) $$
To understand the low--energy dimension of ${\cal L}_{\pi N}^{(1)}$,
we have to extend the chiral counting rules of section 2 to the
various operators and bilinears involving the baryon fields. These
are:
$$\eqalign{
& \krig m = {\cal O}(1) \,\, , \, \, \Psi, \, \bar \Psi =
{\cal O}(1) \, \, , \, \,
D_\mu \Psi = {\cal O}(1) \, \, , \, \, \bar \Psi \Psi  = {\cal O}(1) \cr
& \bar \Psi \gamma_\mu  \Psi
= {\cal O}(1) \, , \, \, \,
 \bar \Psi \gamma^\mu \gamma_5 \Psi
= {\cal O}(1) \, , \, \, \,
 \bar \Psi \sigma^{\mu \nu} \Psi = {\cal O}(1) \, , \, \, \,
 \bar \Psi \sigma^{\mu \nu} \gamma_5 \Psi = {\cal O}(1) \cr
& (i \barre D - \krig m) \Psi = {\cal O}(p) \, , \, \, \,
 \bar \Psi \gamma_5 \Psi = {\cal O}(p) \cr}
\eqno(3.18) $$
Here, $p$ denotes a generic nucleon $\underline {{\rm three}}$--momentum.
Since $\krig m$ is of order one, baryon four--momenta can never be
small on the typical chiral scale. Stated differently, any time derivative
$D_0$ acting on the spin--1/2 fields brings down a factor $\krig m$.
However, the operator
$(i \barre D - \krig m) \Psi$ counts as order ${\cal O}(p)$. The proof
of this can be found in ref.[3.6] or in the lectures [3.8].
To lowest order, the Goldberger--Treiman relation  is exact,
$\krig g_A \krig m = \krig g_{\pi N} F$, which allows us to write the
$\pi N$ coupling in the more familiar form $\sim \krig g_{\pi N}
\partial_\mu \pi^a$.

It goes without saying that we have to include pion loops, associated
with ${\cal L}_1$ given in (3.17). Omitting closed fermion loops (det $D$ = 1),
the corresponding generating functional reads [3.5]
$$\eqalign{
\exp \lbrace i {\tilde {\cal Z}} \rbrace & = N' \int [dU] \exp \biggl\lbrace
i \int dx {\cal L}_{\pi \pi}^{(2)} + i \int dx \bar \eta S^{(1)}
\eta \biggr\rbrace  \cr
{\cal D}^{(1)ac} S^{(1)cb} & = \delta^{ab} \delta^{(4)}(x - y)
\cr} \eqno(3.19) $$
with $S^{(1)}$
the inverse nucleon propagator related to ${\cal D}^{(1)}$ (3.15).
$a$, $b$ and $c$ are isospin indices. This generating functional
can now be treated by standard methods. The details are spelled out
by GSS [3.5]. Let us  concentrate on the low--energy structure of
the effective theory which emerges. Pion loops generate divergences,
so one has to add counterterms. This amounts to
$$\eqalign{
{\cal L}_1 & \to {\cal L}_1 + {\cal L}_2 \cr
{\cal L}_2 & =  \Delta {\cal L}_{\pi N}^{(0)}
             +  \Delta {\cal L}_{\pi N}^{(1)}
             +         {\cal L}_{\pi N}^{(2)}
             +         {\cal L}_{\pi N}^{(3)}
             +         {\cal L}_{\pi \pi}^{(4)} \cr} \eqno(3.20) $$
The last three terms on the r.h.s. of
eq.(3.20) are the expected ones. The structure of the $\pi N$
interaction allows for odd powers in $p$, so starting from
${\cal L}_{\pi N}^{(1)}$ to one--loop order one expects couterterms
of dimension  $p^3$. A systematic analysis of all these terms
has been given by Krause [3.6]. The first two terms,
$ \Delta {\cal L}_{\pi N}^{(0)}$
and $ \Delta {\cal L}_{\pi N}^{(1)}$, are due to the fact that the
lowest order coefficients $\krig m$ and $\krig g_A$ are renormalized
(by an infinite amount) when loops are considered.
This indicates that the chiral counting is messed up and is completely
different from the meson sector, the constants $B$ and $F$ are
$\underline {{\rm not}}$ renormalized in the chiral limit. The origin of this
complication lies in the fact that the nucleon mass does not vanish in
the chiral limit. To avoid any shift in the values of $\krig m$
and $\krig g_A$ one thus has to add appropriate counter terms
$$ \Delta {\cal L}_{\pi N}^{(0)} = \Delta \krig m
\biggl( { \krig m \over F} \biggr)^2 \bar \Psi \Psi \, \, \, , \, \, \,
   \Delta {\cal L}_{\pi N}^{(1)} = \Delta \krig g_A
\biggl( { \krig m \over F} \biggr)^2 {1 \over 2}
\bar \Psi  \gamma^\mu \gamma_5
u_\mu \Psi  \eqno(3.21) $$
\midinsert
\hskip 1in
\epsfxsize=2in
\epsfysize=2in
\epsffile{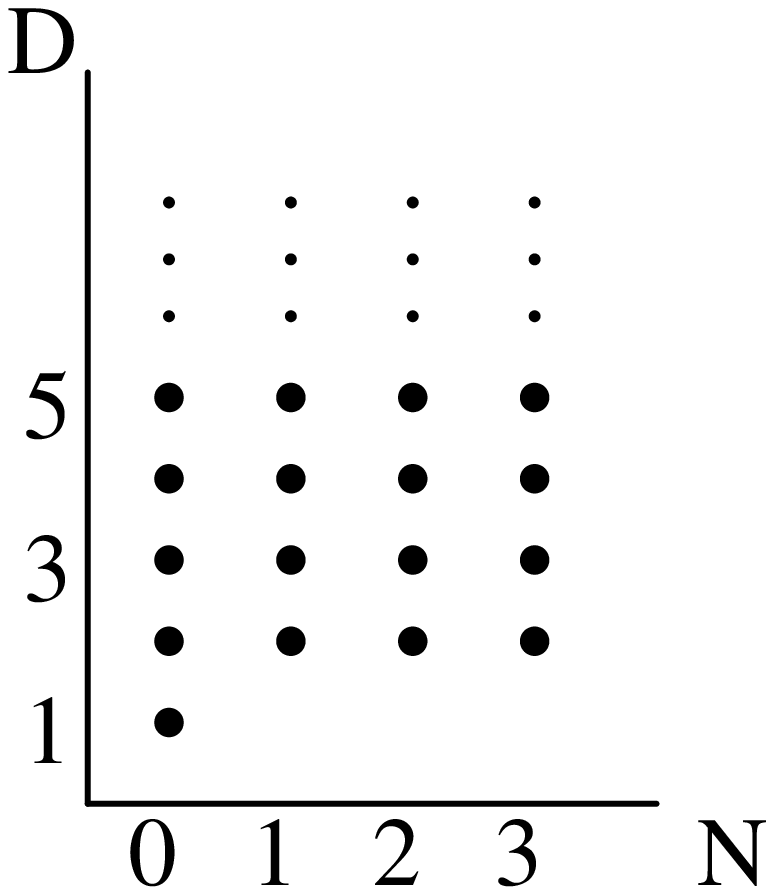}
\smallskip
{\noindent\narrower \it Fig.~3.1:\quad
Chiral expansion for the $\pi N$ scattering amplitude,
$T_{\pi N} \sim p^N = D$. Tree graphs contribute at $N$ = 1, 2, 3 $\ldots$,
$n$--loop graphs at $N$= 2,3, $\ldots$ (after mass and coupling constant
renormalization). The contributions from 2,3,$\ldots$ loops are analytic
in the external momenta at order $p^3$ (here, $p$ is a pion four-,
nucleon three--momentum or the pion mass).
\smallskip}
\endinsert
The first term in (3.21) can be easily worked out when one considers
the nucleon self--energy $\Sigma_{N}(p)$ related to the nucleon
propagator via $S(p) = [\krig m -\barre p - \Sigma_N (p)]^{-1}$
in the one--loop approximation [3.5]. The low--energy structure
of the theory in the presence of baryons is much more complicated
than in the meson sector. This becomes most transparent when one
compares the $\pi \pi$ and $\pi N$ scattering amplitudes, $T_{\pi \pi}$
and $T_{\pi N}$, respectively. While $T_{\pi \pi}^{{\rm tree}}\sim
p^2$ and
$T_{\pi \pi}^{{\rm n-loop}} \sim (p^2)^{n+1}$, the corresponding
behaviour for $T_{\pi N}$ is shown in fig.3.1  [3.5]. Here, $p$ denotes
either a small meson four--momentum or mass or a nucleon three--momentum.
Tree graphs for $T_{\pi N}$ start out at order $p$ followed by a
string of higher order corrections $p^2$, $p^3$, $\ldots$.
One--loop graphs start out at order $p^2$ (after appropriate
mass and coupling constant renormalization) and are non--analytic
in the external momenta at order $p^3$ (in the chiral limit $\hat m =0$).
Higher loops start out at $p^2$ and are analytic to orders
${\cal O}(p^2 \, , \, p^3)$. This again means that the low--energy
constants associated to ${\cal L}_{\pi N}^{(2,3)}$ will
get renormalized.
Evaluation of one--loop graphs associated with ${\cal L}_1$
therefore produces all non--analytic terms in the external momenta
of order $p^3$ like {\it e.g.} leading threshold or branch point
singularities. Let us now consider the case $\hat m \ne 0$.
Obviously, the $\pi N$ amplitude also contains terms which are
non--analytic in the quark masses. A good example is the Adler--Weisberger
relation in its differential form -- it contains a factor $
F_\pi^{-2}$ and therefore a term which goes like
$\hat m \ln \hat m$. Due to the complicated low--energy structure of the
meson--baryon
system, it has never been strictly proven that one--loop graphs
generate all leading infrared singularities, in particular the ones
in the quark masses. However, in all calculations performed so far the
opposite has never been observed. In any case, the exact one--to--one
correspondence between the loop and small momentum expansion is not
valid in the meson--baryon system if one treats the baryons fully
relativistically. This can be overcome, as will be discussed in the
next section, in an extreme non--relativistic limit. Here, however, we
wish to point out that the relativistic formalism has its own advantages.
Two of them are the direct relation to dispersion theory and the
inclusion of the proper relativistic kinematics in certain processes.
These topics will be discussed later on.

The complete list of the  polynomial counter terms
${\cal L}^{(2,3)}_{\pi N}$ can be found in ref.[3.6]. Here, let us just list
the terms which will be used in the calculations of pion photo-- and
electroproduction (see section 4). These are given by
$$
{\cal L}_{\pi N}^{(2)}  = c_1  \bar \Psi \Psi \Tr(\chi_+) +
c_6  \bar \Psi \sigma^{\mu \nu} f_{\mu \nu}^+ \Psi
+ c_7   \bar \Psi \sigma^{\mu \nu} \Tr(f_{\mu \nu}^+) \Psi +\ldots
\eqno(3.22)$$
$$\eqalign{
{\cal L}_{\pi N}^{(3)}  =
&-b_{10}
{1 \over 2 F^2}  \bar \Psi \g5 \gamma_\mu u^\mu \Psi \Tr(\chi_+) +
b_{11}
{\krig{g}_A \krig{m} \over F^2}
\bar \Psi \g5 \chi_- \Psi \cr
&+b_{12}
{1 \over  F^2}  \bar \Psi (i \gamma_\mu D^\mu - \krig{m}) \Psi \Tr(\chi_+)
+d_{1} {F \over 2} \epsilon^{\mu \nu \alpha \beta}
\bar \Psi \gamma_\mu \Psi \Tr(u_\nu f_{\alpha \beta}^{+}) \cr}$$
$$\eqalign{
\qquad \quad & +d_{2}
{F \over 2} \epsilon^{\mu \nu \alpha \beta}
\bar \Psi \gamma_\mu u_\nu \Psi \Tr(f_{\alpha \beta}^{+})
+d_{3}F \bar \Psi \g5 i \sigma^{\mu \nu} \upar{D}^\alpha [u_\alpha\, ,
f_{\mu \nu}^+] \Psi \cr
\qquad \quad &+  d_{4} F
\bar \Psi i \g5  \gamma_\mu [u^\nu \, , f_{\mu \nu}^+] \Psi  \cr}$$
$$\eqalign{
\qquad\quad & + {b_9 \over F^2 } \bar \Psi \gamma^\mu
D^\nu f^+_{\mu\nu} \Psi + {\tilde b_9 \over F^2 } \bar \Psi \gamma^\mu \Psi \,
\Tr( D^\nu f^+_{\mu\nu}) \cr \qquad \quad & + {g_A\over 12} b_{13} \, \bar \Psi
\gamma_5 \gamma^\mu \bigl( [ D^\nu , f^-_{\mu\nu} ] + {i \over 2}
[u^\nu
,f^+_{\mu\nu}]\bigr)  \Psi\,+ \ldots \, . \cr }
\eqno(3.23)$$
where
$$\eqalign{
\chi_{\pm} &= u^\dagger \chi u^\dagger \pm u \chi^\dagger u
\cr
f_{\mu\nu}^\pm &= u^\dagger F_{\mu \nu}^R u \pm u F_{\mu \nu}^L u^\dagger
\cr
F_{\mu \nu}^{L,R} & = \partial_\mu F_\nu^{L,R}- \partial_\nu F_\mu^{L,R}
- i [F_\mu^{L,R} , F_\nu^{L,R}] \cr
F_\mu^R & = v_\mu + a_\mu ,\quad F_\mu^L = v_\mu - a_\mu \cr
\bar{ \Psi} A \upar{D}^\alpha B \Psi &= \bar{ \Psi} A (\partial^\alpha
+ \Gamma^\alpha) B \Psi - \bar{ \Psi} A (\ular{\partial}^\alpha
- \Gamma^\alpha) B \Psi
\cr}
\eqno(3.24) $$
In the case of having only photons as external fields, $f_{\mu \nu}^\pm$
simplifies to
$f_{\mu\nu}^\pm = e (\partial_\mu A_\nu - \partial_\nu A_\mu) (uQu^\dagger \pm
u^\dagger Q u)$
with $A_\mu$ the photon field and $Q = {\rm diag}(1,0)$ the (nucleon) charge
matrix. Let us briefly
discuss the physical significance of the various terms in the pion--nucleon
Lagrangian, eqs(3.22,23).
${\cal L}_{\pi N}^{(2)}$ consists of three terms, the
first ($\sim c_1$) is a mass renormalization counterterm and the
second and third contribute to the anomalous magnetic moments $\kappa_{p,n}$.
In GSS [3.5] it was demonstrated that to a high degree of
accuracy, $c_6 \approx 0$, {\it i.e.} the isovector anomalous
moment of the nucleon is given by the loops (in the one--loop approximation).
The terms of ${\cal L}_{\pi N}^{(3)}$ fall into two types. Let us
first discuss the terms proportional to $b_{10}$, $b_{11}$ and $b_{12}$,
in order. The $b_{10}$--term is needed for the renormalization
of $g_A$. The $b_{11}$--term contributes to the renormalization
of $g_{\pi N}$ and allows to reproduce the empirical deviation
of the Goldberger--Treiman relation from unity. In what follows, we will
always use $g_{\pi N}$ instead of $g_A m / F_\pi$. The term proportional
to $b_{12}$ enters the $Z$--factor which accounts for the
renormalization of the external legs. The  four terms in (3.23)
proportional to $d_i$ ($i=1, \ldots,4$) are finite counterterms which
contribute to pion photo-- and electroproduction.
 The coefficients $d_1, \ldots,d_4$ are not known a priori.
In the $\pi N$ sector, there are three additional terms
contributing to pion electroproduction at order $q^3$, these are the last three
in eq.(3.23).
The two terms in eq.(3.23) proportional to $b_9$ and $\tilde{b}_9$
 are related to the electric mean square charge
radii of the proton and the neutron, see section 4.1.
 The last term in eq.(3.23) is related to the slope of the axial form
factor of the nucleon, $G_A(k^2)$ (see section 4.4). Other terms of order $q^2$
and $q^3$ which enter the calculation of pion--nucleon scattering are discussed
in ref.[3.5].

To end this section, a few remarks concerning the structure
of the nucleons (baryons) at low energies are in order.
Starting from a structureless Dirac field, the nucleon is surrounded
by a cloud of pions which generate {\it e.g.} its anomalous magnetic
moment (notice that the lowest order effective Lagrangian (3.15)
only contains the coupling of the photon to the charge). Besides the
pion loops, there are also counterterms which encode the traces of
meson and baryon excitations contributing to certain properties of the
nucleon. Finally, one point which should be very clear by now: One can
only make a firm statement in any calculation if one takes into account
$\underline {{\rm all}}$ terms at a given order. For a one--loop
calculation in the meson--baryon system, this amounts to the tree terms
of order $p$, the loop contributions of order $p^2$, $p^3$ and the
counterterms of order $p^2$ and $p^3$. This should be kept in mind
in what follows.
\goodbreak \bigskip
\noindent{\bf III.2. EXTREME NON--RELATIVISTIC LIMIT}
\medskip
\goodbreak
As we saw, the fully relativistic treatment of the baryons leads to
severe complications in the low--energy structure of the EFT.
Intuitively,
it is obvious how one can restore the one--to--one correspondence
between the loop and the small momentum expansion. If one considers
the baryons as extremely heavy, only baryon momenta relative to the
rest mass will
count and these can be small. The emerging picture is that of a very
heavy source surrounded by a cloud of light (almost massless) particles.
This is exactly the same idea which is used in the so--called heavy
quark effective field theory methods used in heavy quark physics.
Therefore, it appears natural to apply the insight gained from
heavy quark EFT's to the pion--nucleon sector. Jenkins and Manohar
[3.9,3.10] have given a new formulation of baryon CHPT based on
these ideas. It amounts to taking the extreme non--relativistic
limit of the fully relativistic theory and  expanding in powers of the
inverse baryon mass.
 Notice also that already in the eighties
Gasser [3.11] and Gasser and Leutwyler [3.12] considered a static source
model for the baryons in their determination of quark mass ratios from the
baryon spectrum.

Let us first spell out the underlying ideas before we come back to the
$\pi N$ system. Our starting point is a free Dirac field with mass $m$
$${\cal L} = \bar \Psi (i \barre \partial - m) \Psi    \eqno(3.25)$$
Consider the spin--1/2 particle very heavy. This allows to write its
four--momentum as
$$p_\mu = m v_\mu + l_\mu         \eqno(3.26) $$
with $v_\mu$ the four--velocity satisfying $v^2 = 1$ and $l_\mu$
a small off--shell momentum, $v \cdot l \ll m$.
One can now construct eigenstates of the velocity projection
operator $P_v = ( 1 + \barre v )/2$ via
$$\eqalign{
& \Psi = {\rm e}^{-imv\cdot x} \, (H + h) \cr
& \barre v H = H \, , \, \, \, \ \barre v h = - h
\cr} \eqno(3.27) $$
which in the nucleon rest--frame $v_\mu = (1,0,0,0)$ leads to the
standard non--relativistic reduction of a spinor into upper
and lower components. Substituting (3.27) into (3.25) one finds
$${\cal L} = \bar H(i v \cdot \partial) H
           - \bar h(i v \cdot \partial + 2 m ) h
           + \bar H i \barre \partial^\perp h
           + \bar h i \barre \partial^\perp H
\eqno(3.28) $$
with $\barre \partial^\perp$ the transverse part of the Dirac operator,
$\barre \partial = \barre v (v \cdot \partial) + \barre \partial^\perp$.
{}From eq.(3.28) it follows  that the large component
field $H$ obeys a free Dirac equation (making use of the equation of
motion for $h$)
$$v \cdot \partial H = 0    \eqno(3.29) $$
modulo corrections which are suppressed by powers of  $1/m$. The corresponding
propagator of $H$ reads
$$ S(\omega ) = {i  \over v\cdot k + i \eta} \, , \quad \eta > 0
              \eqno(3.30)$$
with $\omega = v \cdot k$.
The Fourier transform of eq.(3.30) gives the space--time
representation of the heavy baryon propagator. Its explicit form
$\tilde{S} (t, \vec{r} \,)= \Theta(t) \, \delta^{(3)}(\vec r \,)$
illustrates very clearly that the field $H$ represents an (infinitely heavy)
static source.
 The mass--dependence now resides entirely in new
vertices which can be ordered according to their power in $1/m$.
A more elegant path integral formulation is given by Mannel et al. [3.13].
There is one more point worth noticing. In principle, the field $H$
should carry a label '$v$' since it has a definite velocity.
 For the purposes to be discussed we
do not need to worry about this label and will therefore drop it.

Let me now return to the $\pi N$ system. The reasoning is completely
analogous to the one just discussed. We follow here the systematic
analysis of quark currents in flavor $SU(2)$ of Bernard et al. [3.14].
We will derive the effective Lagrangian for heavy nucleons in terms of path
integrals. In this formulation, the $1/m_N$ corrections are easily
constructed.  Consider the generating functional for the chiral
Lagrangian of the  $\pi N$--system
$$ Z[\eta, \bar\eta, v,a,s,p] = \int\! {[d\Psi] [d\bar\Psi] [du] \exp i
\bigl\{S_{\pi N}+S_{\pi\pi} +\int\! d^4x (\bar\eta \psi+\bar\psi \eta) \bigr\}}
\eqno(3.31) $$
where
$$\eqalign{
S_{\pi N} &= \int \Bigl\{  \bar \Psi \bigl( i \barre D - \krig m + {\krig
g_A\over 2} \barre u \gamma_5 \bigr) \Psi +{\cal L}_{\pi N}^{(2)}
+{\cal L}_{\pi N}^{(3)}+ \ldots \Bigr\} \cr
S_{\pi\pi} &= \int d^4x \Bigl\{ {\cal L}_{\pi\pi}^{(2)}+
{\cal L}_{\pi\pi}^{(4)}+ \ldots \Bigr\}\ . \cr}
\eqno(3.32)$$

The aim is to integrate out the heavy degrees of freedom. To this end the
nucleon field $\Psi$ is splitted into upper and lower components with fixed
four velocity $v$
$$\eqalign{
H_v &= e^{i m v\cdot x} {1\over 2} (1+\barre v) \Psi \cr
h_v &= e^{i m v\cdot x} {1\over 2} (1-\barre v) \Psi \ .\cr}
\eqno(3.33)$$
In terms of these fields, the action $S_{\pi N}$ may be rewritten as
$$S_{\pi N} = \int d^4x \Bigl\{\bar H_v {\cal A} H_v+ \bar h_v {\cal B} H_v
+\bar H_v \gamma_0 {\cal B}^\dagger \gamma_0 h_v -\bar h_v {\cal C} h_v
\Bigr\} \ .
\eqno(3.34)$$
The operators ${\cal A}$, ${\cal B}$ and ${\cal C}$ have the low energy
expansions
$${\cal A} = {\cal A}^{(1)}+{\cal A}^{(2)}+ \ldots
\eqno(3.35)$$
where ${\cal A}^{(i)}$ is a quantity of ${\cal O}(q^i)$, $q$ denoting a
low energy momentum. The explicit expressions read
$$\eqalign{
{\cal A}^{(1)} &= i (v\cdot D) + \krig g_A (u\cdot S) \cr
{\cal A}^{(2)} &= {\krig m\over F^2} \Bigl( c_1 \Tr \chi_+  +
c_2(v \cdot u)^2 + c_3 u \cdot u + c_4 \, [S^\mu , S^\nu] u_\mu u_\nu
\cr
& \qquad \quad + c_5 \, (\chi_+ - {1 \over 2}\Tr \chi_+ )
-{i \over 4\krig m}[S^\mu
,S^\nu] ((1+c_6)f_{\mu \nu}^+ + c_7 \, \Tr f_{\mu \nu}^+ ) \Bigr) \cr
{\cal B}^{(1)} &= i \barre D^\bot-{1\over 2} \krig g_A (v\cdot u) \g5 \cr
{\cal C}^{(1)} &= i (v\cdot D)+ 2\krig m +\krig g_A (u\cdot S) \cr
{\cal C}^{(2)} &= -{\cal A}^{(2)} \ . \cr}
\eqno(3.36)$$
$\barre D^\bot= \gamma^\mu(g_{\mu\nu}-v_\mu v_\nu)D^\nu$ is the transverse
part of the covariant derivative which satisfies $\{\barre D^\bot, \barre
v\}=0$.
Here, we have taken advantage of the simplifications for the Dirac
algebra in the heavy mass formulation. It allows to express any Dirac
bilinear $\bar \Psi \Gamma_\mu \Psi$ ($\Gamma_\mu = 1$,
$\gamma_\mu$, $\gamma_5$, $\ldots$)  in terms of the
velocity $v_\mu$ and the spin--operator $2 S_\mu = i \gamma_5
\sigma_{\mu \nu} v^\nu$. The latter obeys the relations (in $d$ space--time
dimensions)
  $$S\cdot v = 0, \, \,  S^2 = {1-d \over 4}, \, \,
                \bigl\{ S_\mu , S_\nu \bigr\}
   = {1\over 2} \bigl( v_\mu v_\nu - g_{\mu \nu}\bigr), \,
  \, \,  [S_\mu , S_\nu]
   = i \epsilon_{\mu \nu \alpha \beta} v^\alpha S^\beta
\eqno(3.37)$$
Using the convention $\epsilon^{0123} = -1$, we can rewrite the
standard Dirac bilinears as:
$$\eqalign{
\quad  &\bar H \gamma_\mu H = v_\mu \bar
  H H, \,\, \bar H \gamma_5 H = 0 ,\,\,
  \bar H \gamma_\mu \gamma_5 H = 2 \bar H
  S_\mu H \cr
\quad & \bar H\sigma^{\mu\nu} H = 2 \epsilon^{\mu\nu\alpha\beta}
 v_\alpha
  \bar H S_\beta H , \, \, \bar H \gamma_5 \sigma^{\mu \nu} H =
  2i(v^\mu \bar H S^\nu H - v^\nu \bar H S^\mu H) \cr}
\eqno(3.38) $$
Therefore, the Dirac algebra is extremely simple in the extreme
non--relativistic limit.

We return to the discussion of the generating functional.
The source term in (3.31) is also rewritten in terms of the fields $H_v$ and
$h_v$
$$
\int\!\! d^4x (\bar\eta \Psi+\bar\Psi \eta)=\int\!\! d^4x(\bar R_v H_v+\bar H_v
R_v+\bar\rho_v h_v+\bar h_v \rho_v)
\eqno(3.39)$$
with
$$\eqalign{
R_v &= {1\over 2} (1+\barre v) e^{i m v\cdot x} \eta \cr
\rho_v &= {1\over 2} (1-\barre v) e^{i m v\cdot x} \eta \ .\cr}
\eqno(3.40)$$
Differentiating with respect to the source $R_v$ yields the Green functions
of the projected fields $H_v$.
The heavy degrees of freedom, $h_v$, may now be integrated out. Shifting
variables $h_v'=h_v-{\cal C}^{-1} ({\cal B} H_v + \rho_v)$
 and completing the square,  the generating functional becomes
$$\eqalign{
 \exp i Z[R_v, \bar R_v, \rho_v, \bar \rho_v ,v,a,s,p]
& = \int\! [dH_v] [d\bar H_v] [du] \Delta_h\exp i
\bigl\{S'_{\pi N}+S_{\pi\pi} + \cr & \qquad \qquad
\int\! d^4x (\bar R_v H_v+\bar H_v R_v)  + \ldots \bigr\} \cr}
\eqno(3.41)$$
where
$$S'_{\pi N} = \int\!d^4x \bar H_v \bigl({\cal A}+(\gamma_0 {\cal B}^\dagger
\gamma_0) {\cal C}^{-1} {\cal B} \bigr) H_v \ ,
\eqno(3.42)$$
and the ellipsis stands for terms with the sources
 $\rho_v$ and $\bar \rho_v$ [3.17].
In (3.42), $\Delta_h$ denotes the determinant coming from the Gaussian
integration over the small component field, i.e.
$$\eqalign{
\Delta_h &= \exp \bigl\{ {1\over 2} \tr \ln {\cal C} \bigr\} \cr
         &= {\cal N} \exp \Bigl\{ {1\over 2} \tr \ln\bigl(1+{\cal C}^{(1)-1}
( i( v \cdot D) + \krig{g}_A (S \cdot u) +
{\cal C}^{(2)}+\ldots )\bigr) \Bigr\} \ . \cr}
\eqno(3.43)$$
As noted in Ref.[3.13], the space time representation of the $h_v$ propagator,
${\cal C}^{(1)-1}$, implies that $\Delta_h$ is just a constant.

The next step consists in expanding the nonlocal functional (3.41) in a series
of operators of increasing dimension. This corresponds to an expansion of the
matrix ${\cal C}^{-1}$ in a power series in $1/\krig m$
$$
{\cal C}^{-1} = {1\over 2 \krig m}-{i(v\cdot D)+\krig g_A (u\cdot S) \over (2
\krig m)^2} + {\cal O}(q^2) \ .
\eqno(3.44)$$
Thus the effective heavy nucleon lagrangian up to ${\cal O}(q^3)$ is given as
$$\eqalign{
S'_{\pi N} &= \int\!d^4x \bar H_v \Bigl\{ {\cal A}^{(1)}+{\cal A}^{(2)}+
{\cal A}^{(3)}+
(\gamma_0 {\cal B}^{(1)\dagger} \gamma_0) {1\over 2 \krig m} {\cal B}^{(1)}\cr
&\qquad\qquad\quad +{(\gamma_0 {\cal B}^{(1)\dagger} \gamma_0) {\cal B}^{(2)}
+(\gamma_0 {\cal B}^{(2)\dagger} \gamma_0) {\cal B}^{(1)} \over 2 \krig m} \cr
&\qquad\qquad\quad
-(\gamma_0 {\cal B}^{(1)\dagger} \gamma_0) {i(v\cdot D)+\krig g_A (u\cdot S)
\over (2 \krig m)^2} {\cal B}^{(1)} \Bigr\} H_v + {\cal O}(q^4) \cr}
\eqno(3.45)$$
Note that the neglected terms of ${\cal O}(q^4)$ may be suppressed by inverse
powers of either $\krig m$ or $\Lambda_\chi=4\pi F_\pi$. These two scales are
treated on the same footing, the only thing which counts is the power of the
low momentum $q$. It is important to note that this expansion of the
non--local action makes the closed fermion loops disappear from the theory
because at any finite order in $1/ \krig m$, $S_{\pi N}'$ is local
(as spelled out in more detail in ref.[3.15]).
To complete the expansion of the generating functional up to order $q^3$, one
has to add the one--loop corrections with vertices from ${\cal A}^{(1)}$
only. Working to order $q^4$ (which still includes only one--loop diagrams),
one also has to include vertices from ${\cal A}^{(2)}$ and from
$(\gamma_0 {\cal B}^{(1)\dagger} \gamma_0) {\cal B}^{(1)} / (2 \krig m)$.

The disappearance of the nucleon mass term to
leading order in $1/ \krig m$ now allows for a consistent chiral power
counting. The nucleon propagator is now of the form (3.30), i.e. has chiral
power $q^{-1}$. Consequently, the dimension $D$ of any Feynman diagram
is given by\footnote{*}{Since this power counting argument is general, we talk
of mesons ($M$) and baryons ($B$) for a while (instead of pions and nucleons).}
$$ D = 4 L - 2I_M -I_B +\sum_d d(N_d^M + N_d^{MB} )  \eqno(3.46)$$
with $L$ the number of loops, $I_M \, (I_B)$ the number of internal meson
(baryon) lines and $N_d^M$ , $N_d^{MB}$ the number of vertices of dimension
$d$ from the meson, meson--baryon Lagrangian, in order. Consider now the
case of a single baryon line running through the diagram [3.15]. In that
case, one has
$$ \sum_d N_d^{MB} = I_B + 1     \qquad .  \eqno(3.47)$$
Together with the general topological relation
$$ L = I_M + I_B - \sum_D (N_d^M + N_d^{MB} ) + 1     \eqno(3.48)$$
we arrive at
$$ D = 2L + 1 + \sum_d (d-2)N_d^M + \sum_d (d-1)N_d^{MB} \quad. \eqno(3.49)$$
Clearly, $D \ge 2L+1$ so that one has  a consistent power counting scheme
in analogy to the one in the meson sector. In particular, the coefficients
appearing in ${\cal L}_{\pi N}^{(1)}$ and ${\cal L}_{\pi N}^{(2)}$ are not
renormalized at any loop order since $D \ge 3$ for $L \ge 1$
(if one uses e.g. dimensional regularization). This is in
marked contrast to the infinite renormalization of ${\krig g}_A$
and $\krig m$ in the relativistic approach, see eq.(3.21). As stated before,
all mass dependence now resides in the vertices of the local pion--nucleon
Lagrangian, i.e.
all vertices now consist of a string of operators with increasing
powers in $1/{\krig m}$. We have for example
$$\eqalign{
&{\rm Photon-nucleon \, \, vertex}:
\quad   i e { 1 + \tau_3 \over 2}
\epsilon \cdot v + {\cal O}(1/\krig m) \cr
&{\rm Pion-nucleon \, \, vertex}:
\quad   (\krig g_A / F) \tau^a S \cdot q
 + {\cal O}(1/\krig m) \cr}
 \eqno(3.50) $$
To summarize, the effective pion--nucleon Lagrangian takes the form
${\cal L}^{\rm eff}_{\pi N} = {\cal L}^{(1)}_{\pi N} + {\cal L}^{(2)}_{\pi N} +
{\cal L}^{(3)}_{\pi N} + {\cal L}^{(4)}_{\pi N} + \ldots$ where the superscript
'(i)' denotes the chiral dimension.
The complete list of terms contributing to
${\cal L}_{\pi N}^{(2)}$ and the corresponding Feynman rules
can be found in appendix A.

Before we turn to the renormalization of the chiral pion--nucleon EFT,
a comment on the heavy fermion formalism is necessary. While it is an
appealing framework, one should not forget that the nucleon (baryon)
mass is not extremely large. Therefore, one expects significant
corrections from $1/m$ suppressed contributions to many observables.
This will become more clear e.g. in the discussion of threshold pion
photo-- and electroproduction. It is conceivable  that going to
one--loop order ${\cal O}(q^3)$ is not sufficient to achieve a very
accurate calculation. Of course, only explicit and complete
calculations can decide upon the quality of the $q^3$ approximation.
 This means that
higher order calculations should be performed to learn about the
convergence of the chiral expansion. For a few selected cases, calculations
including part of or all terms of order $q^4$ have been performed. We will
discuss these in due course. To that accuracy, one has to include the
pertinent contact terms from ${\cal L}_{\pi N}^{(4)}$  and consider one--loop
graphs with exactly one insertion from ${\cal L}_{\pi N}^{(2)}$. Here, let us
note that the calculations which include all terms of order $q^4$ in the chiral
expansion  indeed lead to an improvement for the respective theoretical
predictions. Ultimately, one might want to
include more information in the unperturbed Hamiltonian. At present,
it is not known how to do that but it should be kept in mind.
 \bigskip \goodbreak
\noindent{\bf III.3. RENORMALIZATION}
\medskip
\goodbreak
In this section, we will be concerned with the renormalization of the effective
pion--nucleon Lagrangian to order $q^3$. In the relativistic case, this problem
was addressed for a certain class of divergences in ref.[3.5] and similarly for
the heavy mass formalism in refs.[3.14,3.16]. Ecker [3.17] has recently given
a complete renormalization prescription of the generating functional
at order $q^3$ as discussed below.

Let us first consider the nucleon propagator and mass--shift. The only loop
diagram contributing at order $q^3$ is shown in fig.3.2. and leads to [3.14]
(we have no tadpole contribution since that involves an odd power of the loop
momentum $l$ to be integrated over)
$$\eqalign{
\Sigma_{\rm loop}
(\omega ) & = 3i {{\krig g}_A^2 \over F^2} \int {d^d l \over (2 \pi
)^d} {i \over v \cdot (l-k) + i \eta} {i \over l^2 - M^2_\pi + i \eta}
S \cdot l (-S \cdot l) \cr
& = {3 {\krig g}_A^2 \over 4 F^2} \bigl[ (M^2 - \omega^2 ) J_0(\omega
)  - \omega \Delta_\pi (0) \bigr] \cr}  \eqno(3.51)$$
with $\omega = v \cdot k$ and
the loop functions $J_0 (\omega)$ and $\Delta_\pi (0)$ given in appendix B.
\midinsert
\smallskip
\hskip 1in
\epsfxsize=1.5in
\epsfysize=0.5in
\epsfbox{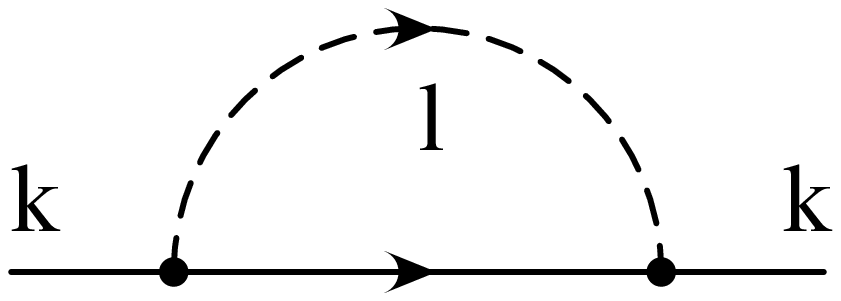}
\smallskip
{\noindent\narrower \it Fig.~3.2:\quad
One--loop contribution to the nucleon self--energy to order $q^3$. The solid
and dashed lines denote nucleons and pions, in order.
\smallskip}
\vskip -0.5truecm
\endinsert
Putting pieces together, we arrive at
$$\eqalign{
\Sigma (\omega ) & = {3 {\krig g}_A^2 \over 4 F^2} \biggl\lbrace
 2 L \omega (2 \omega^2 - 3 M^2) + {\omega \over
8 \pi^2} (2 \omega^2 - 3 M^2) \ln {M \over \lambda}
 + {\omega \over 8 \pi^2} (m^2
- \omega^2) \cr & - {1 \over 4 \pi^2} (M^2 - \omega^2 )^{3/2} \
\arccos {- \omega \over M
} \biggr \rbrace - 4 M^2 \biggl( c_1 + {B_{20} \over 8 \pi^2}
 {\omega \over F^2} \biggr) + {B_{15} \, \omega^3 \over (4 \pi F)^2}
\ldots \cr}  \eqno(3.52)$$
making use of dimensional regularization and separating the infinite from the
finite pieces as in eq.(2.47). The last three terms in eq.(3.52) stem
from  three
contact terms of order $q^2$ and $q^3$, respectively, cf. appendix A
and eqs(3.60),(3.63).
 The coefficient of the
first contact term is obviously finite whereas the other two
low--energy constants
are needed to renormalize the nucleon self--energy. The ellipsis in
(3.52) stands for terms which do not contribute to the mass shift and Z--factor
of the nucleon.  The nucleon propagator now takes the form
$$ S(\omega ) = {i \over p \cdot v - {\krig m} - \Sigma( \omega ) } = { i \over
\omega - \Sigma ( \omega )} \quad .  \eqno(3.53)$$
The propagator develops a pole at $p = m_N v$ with $m_N$ the renormalized
nucleon mass,
$$\eqalign{
m_N & = {\krig m} + \Sigma (0)    \cr
\Sigma(0) & = -4 c_1 M^2 - {3 {\krig g}_A^2   M^3 \over 32 \pi F^2} \quad .
\cr}
\eqno(3.54)$$
As stated in the previous section, the mass shift $\Sigma (0)$ is finite and
vanishes in the chiral limit, quite in contrast to the relativistic approach
(cf. eq.(3.21)). Notice also that the mass--shift contains the non--analytic
piece of order ${\hat m}^{3/2}$ already found in ref.[3]. The nucleon
wave--function renormalization (Z--factor) is determined by the residue of the
propagator at the physical mass pole and given by
$$\eqalign{ S( \omega ) & = {i {\rm Z}_N \over p \cdot v - m_N} \cr
{\rm Z}_N & = 1+ \Sigma ' (0) = 1 - {3 {\krig g}_A^2 M^2 \over 32 \pi^2 F^2}
\biggl[ 3 \ln {M \over \lambda} +1 \biggr] - {M^2 \over 2 \pi^2 F^2}
 B_{20}^{\rm r}
(\lambda ) \quad . \cr} \eqno(3.55)$$
Here, the low energy constant $B_{20}$ has eaten up the infinity in the loop
contribution via the renormalization prescription
$$ B_{20} = B_{20}^{\rm r} (\lambda ) + {\beta_{20} \over 4 \pi^2} \,
L \, ,
 \quad
\beta_{20} = -{9 \over 16} {\krig g}_A^2 \quad  . \eqno(3.56)$$
In a similar fashion, one can renormalize all divergences appearing in the
various Green functions. However, there exists a more systematic method which
we will now turn to.

The starting point for a consistent renormalization scheme is the generating
functional in the presence  of the external sources. In the approximations
described in section 3.2, the fermion determinant is trivial to any finite
order
in $1/m$ and the integration over $H_v$ reduces to completing a square. This
leads to:
$$\eqalign{
\exp \, \lbrace i {\cal Z}[v,a,s,p,\eta, \bar{\eta}] \rbrace & =
N \int [du] \exp i( S_{\pi \pi} + {\cal Z}_{\pi N}[u,v,a,s,p,R_v,\rho_v]  ) \cr
 {\cal Z}_{\pi N}[u,v,a,s,p,R_v,\rho_v] & = \int d^4 x \lbrace
 \bar{R}_v ( {\cal A}
+ \gamma^0 {\cal B}^\dagger \gamma^0 {\cal C}^{-1} {\cal B})^{-1}
R_v + \ldots \rbrace
\cr} \eqno(3.57)$$
where the ellipsis stands for terms linear and quadratic in $\rho_v$
($\bar{\rho}_v$) which we
will not need in what follows, and $U = u^2$.
{}From here on, the standard CHPT procedure as outlined in ref.[3.18] can be
applied. One expands the action in the functional integral around the classical
solution  $U_{\rm cl} [v,a,s,p]
=(u_{\rm cl} [v,a,s,p])^2$ of the lowest oder equations of motion. To calculate
the loop functional to order $p^3$ one has to expand ${\cal L}_{\pi \pi}^{(2)}
+ {\cal L}_{\pi \pi}^{(4)} - {\bar R}_v ({\cal A}^{(1)})^{-1} R_v$ in the
functional integral (3.57) around the classical solution. The divergences are
entirely given by the irreducible diagrams (cf. Fig.1 of ref.[3.17])
corresponding to the generating functional
$$ \eqalign{ {\cal Z}_{\rm irr} [v,a,s,p,R_v] & = \int d^4 x d^4 x' d^4 y d^4
y' \, {\bar R}_v (x) ({\cal A}^{(1)})^{-1}(x,y) [ \Sigma_1 (y,y')
\delta^{(4)}(y-y') \cr & + \Sigma_2 (y,y') ] \, ({\cal A}^{(1)})^{-1}(y',x')
R_v(x') \cr}   \eqno(3.58)$$
with $({\cal A}^{(1)})^{-1}$ the propagator of $H_v$ in the presence of
external
fields. The explicit form of the self--energy functional $\Sigma_{1,2}$ can be
found in ref.[3.17]. Here, it is important to note that these diverge as $y \to
y'$. The divergences can be extracted in a chiral invariant manner by making
use of the heat kernel representation of the propagators in $d-$dimensional
Euclidean space. These divergences will then appear as simple poles in
$\epsilon = (4-d)/2$. After some lengthy algebra as detailed in ref.[3.17] one
arrives at
$$\eqalign{
[ \Sigma_1 (y,y') \delta^{(4)}(y-y') + \Sigma_2 (y,y') ] & =
[ \Sigma_1^{\rm fin} (y,y';\lambda) \delta^{(4)}(y-y') +
\Sigma_2^{\rm fin} (y,y';\lambda) ] \cr & - {2 L \over F^2} \delta^{(4)}(y-y')
[ \hat{\Sigma}_1 (y)  + \hat{\Sigma}_2 (y) ] \cr}    \eqno(3.59)$$
The generating functional ${\cal Z}[v,a,s,p,R_v]$ can now be renormalized by
introducing the local  counterterm Lagrangian
$${\cal L}_{\pi N}^{(3)} (x) = {1 \over (4 \pi F)^2} \sum_i B_i {\bar H}_v (x)
\, O_i (x) \, H_v (x)                          \eqno(3.60)$$
where the coupling constants $B_i$ are dimensionless  and the field monomials
$O_i (x)$ are of order $p^3$.
A minimal set consisting of 22 counterterms has been
given in ref.[3.17].\footnote{*}{For on--shell nucleons, one can further reduce
this number by using the equations of motion for the nucleons.} In complete
analogy to eq.(2.46), one decomposes the low--energy constants $B_i$ as
$$B_i = B_i^r (\lambda ) + (4 \pi )^2 \, \beta_i \, L       \eqno(3.61)$$
The $\beta_i $ depend only on $g_A$ (strictly speaking on ${\krig g}_A$) and
 the corresponding operators $O_i (x)$  are given by
$$\eqalign{
O_1 & = i [u_\mu, v \cdot D u^\mu], \beta_1 = {g_A^4 \over 8} ; \quad
O_2  = i [u_\mu, D^\mu v \cdot u], \beta_2 = -{1+ 5g_A^4 \over 12} ; \cr
O_3 & = i [v \cdot u, v \cdot D v \cdot u], \beta_3 = {4 - g_A^4 \over
8};  \quad
O_4   = S \cdot u \Tr (u \cdot u), \beta_4 = {g_A(4- g_A^4) \over 8} ; \cr
O_5 & = u_\mu \Tr( u^\mu S \cdot u), \beta_5 = {g_A(6 - 6g_A^2+ g_A^4) \over
12}; \quad
O_6   = S \cdot u \Tr (v \cdot u)^2, \beta_6 = -{g_A(8- g_A^4) \over 8} ; \cr
O_7 & = v \cdot u \Tr(S \cdot u v \cdot u), \beta_7 = -{g_A^5 \over 12}; \quad
O_8   = [\chi_- , v \cdot u] , \beta_8 = {1+ 5g_A^2 \over 24} ; \cr
O_9 & = S \cdot u \Tr (\chi_+), \beta_9 = {g_A(4 - g_A^2) \over 8}; \quad
O_{10} =
D^\mu \tilde{f}_{+ \mu \nu} v^\nu, \beta_{10} = -{1+ 5g_A^2 \over 6} ; \cr
O_{11} & = i S^\mu v^\nu [\tilde{f}_{+ \mu \nu}, v \cdot u],
\beta_{11} = g_A ; \quad
O_{12} =  i v_\lambda \epsilon^{\lambda \mu \nu \rho} \Tr (u_\mu u_\nu u_\rho )
, \beta_{12} = -{g_A^3(4 +3 g_A^2) \over 16} \cr}$$
$$\eqalign{
O_{13} & = i v_\lambda \epsilon^{\lambda \mu \nu \rho} S_\rho \Tr [ (v \cdot
D u_\mu) u_\nu ] , \beta_{13} = -{g_A^4 \over 4} ; \quad
O_{14}  = v_\lambda \epsilon^{\lambda \mu \nu \rho}
\Tr (\tilde{f}_{+ \mu \nu} u_\rho )
; \beta_{14}  = -{g_A^3 \over 4} ;  \cr
O_{15} & = i ( v \cdot D)^3 ; \beta_{15} = -3 g_A^2 ; \quad
O_{16}  = v \cdot \larr{D} S \cdot u  v \cdot D ; \beta_{16} =  g_A^3 ; \cr
O_{17} & = \Tr (u \cdot u)i v \cdot D + {\rm h.c.} ; \beta_{17}= -{ 3g_A^2 ( 4
+ 3g_A^2) \over 16} ; \cr
O_{18} & = i \Tr (v \cdot u)^2 v \cdot D + {\rm h.c.} ; \beta_{18}= {  ( 8
+ 9g_A^2) \over 16} ; \cr
O_{19} & = ( v \cdot D S \cdot u ) v \cdot D  + {\rm h.c.} ; \beta_{19} =
{g_A^3 \over 3} ; \quad
O_{20} = \Tr ( \chi_+ ) i v \cdot D + {\rm h.c.} ;
\beta_{20} = -{9 g_A^2 \over 16} ; \cr
O_{21} & =  v_\lambda \epsilon^{\lambda \mu \nu \rho} S_\rho u_\mu u_\nu v
\cdot D + {\rm h.c.} ; \beta_{21} = -{g_A^2 (4 + g_A^2) \over 4} ; \cr
O_{22} & = i v_\lambda \epsilon^{\lambda \mu \nu \rho} S_\rho
\tilde{f}_{+ \mu \nu} v \cdot D + {\rm h.c.} ; \beta_{21} = g_A^2 ;
\, \, \tilde{f}_{+ \mu \nu} = f_{+ \mu \nu} - {1 \over 2} \Tr f_{+ \mu \nu}
  \cr}
\eqno(3.62)$$
The sum of the irreducible one--loop functional (3.58) and the counterterm
functional derived from the Lagrangian (3.60) is finite and scale--independent.
The renormalized low--energy constants $B_i^r (\lambda)$ are
measurable (i.e. they can be determined from a fit to some
observables) and
subject to the followoing renormalization group behaviour under scale changes
$$ B_i^r (\lambda_2) = B_i^r (\lambda_1) - \beta_i \log {\lambda_2 \over
\lambda_1 } \quad.   \eqno(3.63)$$
This completes the  formalism necessary to renormalize the pion--nucleon
(or meson--baryon) Lagrangian to order $q^3$ in heavy fermion CHPT. In what
follows, we will see these renormalization prescriptions being operative for
various physical processes.
 \bigskip \goodbreak \vfill \eject
\noindent{\bf III.4. LOW--ENERGY CONSTANTS AND THE ROLE OF THE $\Delta (1232)$}
\medskip
\goodbreak
As noted in section 2, in the meson sector the low--energy constants $L_i$
could all be fixed from phenomenological constraints (within a certain
accuracy). Furthermore, the actual values of these coefficients could be
understood from a hadronic duality in terms of resonance exchange. We note,
however, that for the non--leptonic weak interactions (which contains 80
new contact terms) this generalized vector meson dominance principle is
not that successful [3.19]. In the nucleon sector, the situation is somewhat
similar to the case of non--leptonic weak interactions of the mesons. At
present, only a subset of the coefficients in ${\cal L}_{\pi N}^{(2)}$ and
${\cal L}_{\pi N}^{(3)}$ (and also in
${\cal L}_{\pi N}^{(4)}$) have been fixed from phenomenology. We will discuss
one example below. In most other cases, one resorts to resonance saturation
which besides meson resonances involves the nucleon excitations, in particular
the $\Delta(1232)$ P--wave resonance. The $\Delta$ plays a particular role for
two reasons. First, its excitation energy is only 300 MeV and second, its
coupling to the $\pi N$ system is very strong, $g_{\Delta N \pi} \simeq 2
g_{\pi N}$. For these reasons and the degeneracy of the $\Delta$ with the
nucleon in the limit of infinite colours, it has been suggested to include the
$\Delta$ from the start in the effective theory [3.20].
We will discuss this below.
Obviously, if one does not want to build in the $\Delta$ in the EFT, it will
feature prominently in the estimation of certain low--energy
constants. We will detail one example which we need for the discussion of
elastic $\pi N$ scattering in the threshold region later on.
\medskip
$$\hbox{\vbox{\offinterlineskip
\def\strut{\hbox{\vrule height  8pt depth  8pt width 0pt}}
\hrule
\halign{
\strut\vrule# \tabskip 0.1in &
\hfil#\hfil  &
\vrule# &
\hfil#\hfil &
\hfil#\hfil &
\hfil#\hfil &
\vrule# \tabskip 0.0in
\cr
\noalign{\hrule}
& &&   occurs in  & determined from & &\cr
\noalign{\hrule}
& $c_1     $            && $m_N$, $\sigma_{\pi N}$& phen. & &\cr
& $c_2, c_3, c_4$       && $\pi N \to \pi N$      & res. exch. + phen. & &\cr
& $c_5     $            && $m_N$, $\sigma_{\pi N}$ $(m_u \ne m_d )$
& phen. & &\cr
& $c_6          $       &&  $\kappa_{p,n}$        & phen. & &\cr
& $c_7          $       &&  $\kappa_{p,n}$        & phen. & &\cr
& $B_1, B_2,B_3, B_8$   && $\pi N \to \pi N$      & res. exch. & &\cr
& $B_{10}       $       &&  $<r^2>_1^V$           & phen. & &\cr
& $B_{20}       $       &&  ${\rm Z}_N$           & unknown & &\cr
\noalign{\hrule}}}}$$
\smallskip
{\noindent\narrower Table 3.1:\quad Occurence of low--energy constants and
their determinations from phenomenological (phen.) constraints or estimation
based on resonance exchange (res. exch.). Note
that $c_5$ is only contributing for $m_u \ne m_d$. For the definition of
the corresponding effective Lagrangian see (3.36) and appendix A1.
\smallskip}
\goodbreak
\bigskip
First, let us tabulate the various low--energy constants from
${\cal L}_{\pi N}^{(2,3)}$ which we will encounter in the following sections
and discuss in which process they can be probed (or determined). As noted
before, while the list for the terms of order $q^2$ is complete, for the
terms of ${\cal O}(q^3)$ we only exhibit the terms which we will use later
on. As becomes clear from table 3.1, certain low--energy constants can only
be probed in the presence of external fields. These are, in turn, the best
determined ones since the nucleon radii and magnetic moments are accurately
known (cf. $c_6, c_7$ or $B_{10}$).
The constants related directly to the $\pi N$
interactions have not yet been determined from a global fit to $\pi N$
scattering data as it was done in the relativistic case. In view of the
present
 discussion about the low--energy $\pi N$ scattering data such a program
has to be performed with adequate care and is not yet available (see section
3.5).
As noted in table 3.1, the low--energy constant $c_i$ can be fixed from
phenomenology. Consider first $c_1$. It is related to the much discussed
pion--nucleon $\sigma$--term, $\sigma_{\pi N}(t) \sim <p'| \hat{m}(\bar{u}u +
\bar{d}d)|p>$ ($t = (p' -p)^2$), via [3.14]
$$c_1 = -{1 \over 4 M_\pi^2} \biggl( \sigma_{\pi N}(0) + {9 g_A^2 M_\pi^3 \over
64 \pi F_\pi^2} \biggr) \quad .  \eqno(3.64)$$
Using the empirical values for $F_\pi$, $M_\pi$ and $g_A$ together with the
recent determination $\sigma_{\pi N}(0) = 45 \pm 8$ MeV [3.21], this amounts to
$$c_1 = -0.87 \pm 0.11 \, \, {\rm GeV}^{-1} \quad . \eqno(3.65)$$
The two constants $c_2$ and $c_3$ are related to the so--called axial
polarizability $\alpha_A$ and the isopin--even $\pi N$ S--wave
scattering length $a^+$ (for the definitions and discussion see section 3.5)
$$\eqalign{
c_3 & = -{F_\pi^2 \over 2} \biggl[ \alpha_A + {g_A^2 M_\pi \over 8 F_\pi^4}
\biggl( {77 \over 48 } + g_A^2 \biggr) \biggr] = -5.25 \pm 0.22 \, {\rm
GeV}^{-1} \cr
c_2 & = { F_\pi^2 \over 2 M_\pi^2 }
\biggl( 4 \pi (1 +{ M_\pi \over m} ) a^+ -
{3 g_A^2 M_\pi^3 \over 64 \pi F_\pi^4} \biggr) + 2c_1 - c_3
+ {g_A^2 \over 8 m} = 3.34 \pm 0.27 \, {\rm GeV}^{-1} \cr}  \eqno(3.66)$$
using the empirical values $\alpha_A = 2.28 \pm 0.10 \, M_\pi^{-3}$ and
$a^+ = -0.83 \pm 0.38 \cdot 10^{-2} \,  M_\pi^{-1}$
(for references, see section
3.5). Note, however, that these observables might not form the best set to
determine the constants $c_{1,2,3}$ since the scattering length $a^+$ is
extremely sensitive to the counter term combination $c_2 + c_3 - 2c_1$ and,
furthermore, there are correlations between the S--wave scattering lengths
and the $\pi N$ $\sigma$--term.
The constants $c_6$, $c_7$ and $B_{10}$ can be determined
from the isovector and isoscalar anomalous magnetic moment of the nucleon and
its isovector charge radius, respectively [3.5,3.14].
The numerical values of the seven low--energy constants in ${\cal
L}_{\pi N}^{(2)}$ are summarized in table 3.2.
The constants $c_2$ and $c_3$ have also been estimated making use of the
resonance saturation hypothesis [3.22]. Consider $c_3$. In that case, the
dominant contribution comes from the $\Delta (1232)$ and there is a small
correction due to the $N^* (1440)$ resonance. In addition, there is a sizeable
contribution due to scalar meson exchange.
The pertinent Lagrangians for the coupling of the mesonic and the
nucleon excitations to the $\pi N$ system read
$$\eqalign{
{\cal L}_{\pi \Delta N}&  = { 3 g_A \over 2 \sqrt{2} } {\bar \Delta}^\mu_a [
g_{\mu \nu} - (Z + {1 \over 2}) \gamma_\mu \gamma_\nu ] u_a^\nu \Psi +
\, {\rm h.c.} \cr
{\cal L}_{\pi N^* N}&  = {1 \over 4} g_A \, R \,{\bar N}^* \barre{u}
\gamma_5 \Psi \,  + {\rm h.c.}  \cr
{\cal L}_{S \pi \pi}&  =  c_m \, \Tr(\chi_+ ) + c_d \, Tr(u \cdot u) \cr
{\cal L}_{S N N}&  =  g_S  \, \bar{\Psi} \Psi
\cr}   \eqno(3.67)$$
where $\Delta_\mu$ denotes the Rarita--Schwinger field and $Z$ parametrizes the
off--shell behaviour of the spin--3/2 field. This parameter is not well known,
the most recent analysis of ref.[3.23] gives $-0.8 \le Z \le 0.3$.
We should stress here that it is mandatory to consider these nucleon
excitations in the relativistic framework. The basic idea is that one starts
from a fully relativistic theory of pions coupled to nucleons and nucleon
resonances chirally coupled. One then integrates out these excitations
from the effective theory which produces a string of pion--nucleon interactions
whose coefficients are given in terms of resonance parameters. Finally, one
defines velocity--dependent nucleon fields eliminating the 'lower component'
 $h(x)$. Using now
the large $N_c$ coupling constant relation $g_{\pi \Delta N} = 3 g_{\pi N} /
\sqrt{2} = 28.42$ (close to the empirical value of 28.37)
and the phenomenological value
$g_{\pi N^* N} = (1/2 \ldots 1/4) g_{\pi N}$ [3.24] (which defines
a parameter $R = 1 \ldots 1/4$), we find
$$\eqalign{
c_3^\Delta & = {g_A^2 \over 8 m_\Delta^2} \biggl( {m_\Delta m - 4 m_\Delta^2 -
m^2 \over m_\Delta - m} + 4 Z [ m_\Delta (2Z+1) + m(Z+1) ] \biggr) \cr
& = -2.54 \ldots -3.18 \, {\rm GeV}^{-1} \cr
c_3^{N^*} & =  {g_A^2 \, R \over 16 ( m - m^* )}
= -0.06 \ldots -0.22 \, {\rm GeV}^{-1} \cr
c_3^S & = 2 c_1 {c_d \over c_m} = -1.33 \, {\rm GeV}^{-1}
  \quad . \cr} \eqno(3.68)$$
using $|c_d| = 32$ MeV and $|c_m| = 42$ MeV [3.31]. In addition, we have
assumed that the value of $c_1$ is saturated by scalar exchange which allows
to eliminate the coupling $g_S$. However, a strongly coupled
scalar--isoscalar with $M_S / \sqrt{g_S} \sim 220$ MeV is needed to saturate
$c_1$ this way.
Altogether, we find that $c_3^{\rm Res} = c_3^\Delta + c_3^{N^*} + c_3^S$
 varies  between -3.6 and -5.0 GeV$^{-1}$,
somewhat smaller than the empirical value discussed above. This demonstrates
that the resonance saturation hypothesis can not yet be considered established
(as it is in the case of the meson sector). However, in the absence of
sufficiently many accurate low--energy data in the meson--baryon sector and
a systematic evaluation of all counterterms up--to--and--including order $q^3$,
it is legitimate to use resonance exchange to
 estimate the low--energy constants
which appear in the processes one considers. The introduction of
this unwanted model--dependence should be considered as a transitional stage
until a complete analysis of the various coupling constants based on fits to
data becomes available.
\medskip
$$\hbox{\vbox{\offinterlineskip
\def\strut{\hbox{\vrule height  8pt depth  8pt width 0pt}}
\hrule
\halign{
\strut\vrule# \tabskip 0.1in &
\hfil#\hfil &
\hfil#\hfil &
\hfil#\hfil &
\hfil#\hfil &
\hfil#\hfil &
\hfil#\hfil &
\hfil#\hfil &
\vrule# \tabskip 0.0in
\cr
\noalign{\hrule}
&  $c_1'$ & $c_2'$  & $c_3'$   & $c_4'$ & $c_5'$  & $c_6'$  & $c_7'$ &
\cr
\noalign{\hrule}
& -1.63   & 6.20  & -9.86 & 7.73 & 0.17 & 11.22 & -2.03 &
\cr
\noalign{\hrule}}}}$$
\smallskip
{\noindent\narrower Table 3.2:\quad Numerical values of the
dimensionless low--energy constants $c_i' = 2 m_N c_i$ $(i=1,\ldots,5$) and
$c_{6,7}' = 2 c_{6,7}$ with $m_N = (m_p + m_n)/2 = 938.92$ MeV the nucleon
mass. $c_4$ is determined from the P--wave $\pi N$ scattering volumes and
$c_5$ follows from the strong $np$ mass difference, $(m_n -m_p)_{\rm str} =
2M_\pi^2 c_5 (m_d -m_u)/ \hat{m}$.
$c_{6,7}$ are determined from the nucleon isovector and isoscalar
anomalous magnetic moments as described in section 4.1.
\smallskip}

\medskip
One particular advantage of the heavy mass formulation is the fact that
it is very easy to include the  baryon decuplet, i.e. the spin--3/2
states. This has been done in full detail by Jenkins and Manohar [3.10,3.20].
The inclusion of the $\Delta(1232)$ is motivated by the arguments given in the
beginning of this section, in particular the fact that
the $N \Delta$ mass--splitting $m_\Delta - m_N$
is only about thrice as much as the pion decay constant,\footnote{*}{Often it
is stated that $m_\Delta - m_N \simeq 2 M_\pi$. While that is numerically true,
the behaviour of these quantities in the chiral limit is very different. While
the former stays constant as $\hat{m} \to 0$, the latter vanishes.}
 so that one expects
significant contributions from this close--by resonance (the same
holds true for the full decuplet in relation to the octet, see section 6). This
expectation is borne out in many phenomenological models and we had
also seen in the discussion of the low--energy constants the prominent role
of the $\Delta$. However,
it should be stressed that if one chooses to include this baryon
resonance (or the full decuplet), one again has to account for all
terms of the given accuracy one aims at, say ${\cal O}(q^3)$ in a
one--loop calculation. This has not been done in the presently
available literature. Furthermore, the mass difference $m_\Delta -
m_N$ does not vanish in the chiral limit thus destroying the consistent
power counting (as it is the case with the baryon mass in the
relativistic formalism discussed in section 3.1). We will come back to this
below. In the extreme non--relativistic limit, the $\Delta$
is described by Rarita--Schwinger spinor $\Delta^\mu_{a}$
with $a \in \lbrace 1$, $2$, $3 \rbrace$. This spinor
contains both spin--1/2 and spin--3/2 components. The spin--1/2 pieces
are projected out by use of the constraint $\gamma_\mu \Delta^\mu_a = 0$.
One then defines a velocity--dependent field via
$$ \Delta^\mu_a = {\rm e}^{-i m v \cdot x} \, (T + t)^\mu_a
\eqno(3.69) $$
In terms of the physical states we have
$$T^1_\mu = {1\over \sqrt 2} \left( \matrix { \Delta^{++} - \Delta^0/\sqrt 3\cr
\Delta^+ / \sqrt 3 - \Delta^- \cr} \right)_\mu , \, \, \,
T^2_\mu = {i\over \sqrt 2} \left( \matrix { \Delta^{++} + \Delta^0/\sqrt 3\cr
\Delta^+ / \sqrt 3 + \Delta^- \cr} \right)_\mu , \, \, \,
T^3_\mu = - \sqrt{{2\over 3}} \left( \matrix { \Delta^{+} \cr \Delta^0 \cr}
\right)_\mu      \eqno(3.70)$$
The effective non--relativistic $\Delta N \pi$ Lagrangian to leading order
reads
$${\cal L}_{\Delta N \pi}^{(1)} = - i \bar T^{\mu a} v \cdot D^{ab} T_\mu^b
+ \Delta \bar T^{\mu a} T_\mu^a + {3 \krig g_A \over 2 \sqrt{2} }
(\bar T^{\mu a} u_\mu^a H + \bar H u_\mu^a T^{\mu a})
\eqno(3.71)$$
with $\Delta = m_\Delta - m_N$ and $u_\mu^a = (i/2) \Tr ( \tau^a u^\dagger
\nabla_\mu U u^\dagger)$. Clearly one is left with some residual
mass dependence. In the language
of ref.[3.20] we have set ${\cal C} = 3 \krig g_A /2 = 1.89$  which is nothing
but the SU(4) coupling constant relation discussed before.
{}From the width of the decay $\Delta \to N \pi$ one has
${\cal C} = 1.8$ [3.10], consistent with the value given before (if one
uses the full decuplet the value of ${\cal C}$ reduces to 1.5). The
propagator of the spin--3/2 fields reads
$$ S_\Delta^{\mu \nu} (\omega) = i { v^\mu v^\nu - g^{\mu \nu} - {4 \over 3}
S^{\mu} S^\nu \over \omega - \Delta}
\eqno(3.72) $$
For all practical purposes, it is most convenient to work in the
rest--frame $v_\mu = (1,0,0,0)$. In that case, one deals with the
well--known non--relativistic isobar model which is discussed in
detail in the monograph by Ericson and Weise [3.24].  Consider now the nucleon
self--energy (i.e. a diagram like in fig.3.2. but with an intermediate $\Delta$
state). Its contribution is non--vanishing in the chiral
limit. Therefore,
a counter term of the following form has to be added [3.25] (like in
the relativistic theory of the nucleon alone)
$$\eqalign{ \delta {\cal L}^{(0)}_{\pi N \Delta} & = - \delta m_0 \Tr (\bar{H}
H ) \cr
\delta m_0 & = {10 \over 3} {C^2 \Delta^3 \over F_\pi^2} \biggl[ L + {1 \over
16 \pi^2} \biggl( \ln \bigl( {2 \Delta \over \lambda } \bigr) - {5 \over 6}
\biggr) \biggr] \quad. \cr} \eqno(3.73)$$
Clearly, such a contribution destroys the consistent power counting. However,
from phenomenological arguments, one might want to consider the quantity
$\Delta$ as a small parameter. While this is not rooted in QCD, it might be
worth
to be explored in a systematic fashion. Such an analysis is, however, not
available at present. Our point of view is that one should not include the
$\Delta$ as a dynamical degree of freedom in the EFT but rather use it to
estimate certain low--energy constants. While this might narrow the range of
applicability of the approach, it at least allows for a  consistent power
counting.
 \bigskip \goodbreak
\noindent{\bf III.5. ASPECTS OF PION--NUCLEON SCATTERING}
\medskip
\goodbreak
Elastic pion--nucleon scattering in the threshold region
can be considered the most basic process to which the CHPT methods can be
applied. This is underlined by the Weinberg's very successfull  current algebra
 prediction [3.26] for the S--wave pion--nucleon scattering
lengths,
$$a_{1/2} = {M_\pi \over 4 \pi F_\pi^2} = -2 a_{3/2} = 0.175 \, M_\pi^{-1}
\eqno(3.74)$$
Tomozawa [3.27] also
derived the sum rule $a_{1/2} - a_{3/2} = 3 M_\pi / 8 \pi F_\pi^2 = 0.263$
$M_\pi^{-1}$. Empirically, the combination $(2 a_{1/2} + a_{3/2}) / 3$ is best
determined from pion--proton scattering.
The Karlsruhe--Helsinki group gives  $0.083 \pm 0.004$ $M_\pi^{-1}$
[3.28] consistent with the pionic atom measurement [3.29] of $0.086 \pm 0.004$
$M_\pi^{-1}$. The value of $a_{1/2} - a_{3/2}$ is more uncertain. The KH
analysis leads to $0.274 \pm 0.005$ $M_\pi^{-1}$ [3.30].
 In what follows, we will use the central values from the work of
Koch [3.28], namely $a_{1/2} = 0.175$ $M_\pi^{-1}$ and $a_{3/2} = -0.100$
$M_\pi^{-1}$. The agreement of the current algebra predictions with these
numbers is rather spectacular.
Therefore, one would like to know what the next--to--leading
order corrections to the original predictions are. This question
was addressed in ref.[3.22]. To be specific,
consider the on--shell $\pi N$ forward scattering amplitude for a nucleon at
rest. Denoting by $b$ and $a$ the
isospin of the outgoing and incoming pion, in order, the scattering amplitude
takes the form
$$T^{ba} = T^+ (\omega) \delta^{ba} + T^- (\omega) i \epsilon^{bac} \tau^c
\eqno(3.75)$$
with $q$ the pion four--momentum and $\omega = v \cdot q $. Under
crossing $(a \leftrightarrow b, \, q \to -q )$ the functions $T^+$ and $T^-$
are even and odd, respectively, $T^\pm (\omega) = \pm T^\pm (-\omega)$. At
threshold one has $\svec q = 0$ and the pertinent scattering lengths are
defined by
$$a^\pm = {1 \over 4 \pi} \bigl( 1 + {M_\pi \over m} \bigr)^{-1}\, T^\pm (
M_\pi ) \eqno(3.76)$$
The S--wave scattering lengths for the total $\pi N$
isospin 1/2 and 3/2 are related to $a^\pm$ via
$a_{1/2} = a^+ + 2 a^- , \quad a_{3/2} = a^+ -  a^- $.
The abovementioned central empirical values translate into  $a^+ = -0.83 \cdot
10^{-2} \, M_\pi^{-1}$ and $a^- = 9.17 \cdot 10^{-2} \, M_\pi^{-1}$. In what
follows, we will not exhibit the canonical units of $10^{-2} \, M_\pi^{-1}$.
The benchmark values are therefore $a^+ = -0.83 \pm 0.38$ and $a^- = 9.17 \pm
0.17$
compared to the current algebra predictions of $a^+ = 0$ and $a^- = 8.76$
(using $M_\pi = 138$ MeV and $F_\pi = 93$ MeV). The empirical values
for the forward amplitudes at threshold follow to be $T^+ (M_\pi) =
-0.17 \pm 0.08$ fm and $T^- (M_\pi) = 1.87 \pm 0.03$ fm.
The four novel counter terms from ${\cal L}_{\pi N}^{(3)}$ which
contribute to $\pi N$ scattering are $O_1, O_2, O_3$ and $O_8$ given
in eq.(3.62).
Due to crossing symmetry, ${\cal L}_{\pi N}^{(2)}$  (these are the terms
proportional to $c_{1,2,3}$, cf eq.(3.36))
contributes only to $T^+(\omega)$ whereas ${\cal L}_{\pi N}^{(3)}$ solely
enters $T^-(\omega)$. For the isospin even/odd threshold amplitude we
derive the following chiral expansion
$$ T^+ (M_\pi) = {2 M_\pi^2 \over F_\pi^2} \, \bigl( c_2 + c_3 - 2c_1
-{g_A^2 \over 8 m} \bigr) + {3 g_A^2 M_\pi^3 \over 64 \pi F_\pi^4} +
{\cal O}(M_\pi^4)  \eqno(3.77)$$

$$ T^- (M_\pi) = { M_\pi \over 2 F_\pi^2} + {M_\pi^3 \over 16 \pi^2
F_\pi^4} \bigl(1 - 2 \ln{M_\pi \over \lambda} \bigr) + {g_{\pi N}^2
M_\pi^3 \over 8 m^4} - 4b^r (\lambda) {M_\pi^3 \over F_\pi^2} +
{\cal O}(M_\pi^4)  \eqno(3.78)$$
with $b^r (\lambda) = -(B_1^r (\lambda ) + B_2^r (\lambda ) +
 B_3^r (\lambda ) +
2 B_8^r (\lambda ) ) / (16 \pi^2 F_\pi^2)$.
$b$ has to be renormalized as follows to render
the isospin--odd scattering amplitude $T^- (\omega)$ finite,
$$b = b^r (\lambda) - { L \over 2 F^2}  \quad , \eqno(3.79)$$
since $\beta_1+\beta_2+\beta_3+2\beta_4 = 1/2$ (cf. eq.(3.62)).
It is remarkable that there are no corrections of order $M_\pi^2$ and
$M_\pi^4$ in $T^-(M_\pi)$. The order $M_\pi^2$ has to be zero since
crossing symmetry forbids any such counter term contribution which
also must be analytic in the quark masses. For the loop contribution
at order $M_\pi^4$ such an argument does not hold (loops can lead to
 non--analyticities), but an explicit calculation of all $q^4$ loop
diagrams shows indeed that they all add up to zero.
The various terms in eq.(3.78) are the
current algebra prediction, the expansion of the nucleon pole term,
the one--loop and the counterterm contribution from ${\cal L}_{\pi
N}^{(3)}$, respectively. $\lambda$ is the scale introduced in
dimensional regularization. In what follows, we
will use $\lambda = m_\Delta = 1.232$ GeV, motivated by the resonance
saturation principle.  Notice that the contact term
contributions are suppressed by a factor $M_\pi^2$ with respect to the leading
current algebra term. Matters are different for the  isospin--even scattering
amplitude $T^+$. It consists of contributions of order $M_\pi^2$ and
$M_\pi^3$.
{}From the form of eq.(3.77) it is
obvious that the contact terms play a more important role in the determination
of $T^+$ than for $T^-$. The most difficult task is to pin down
the various low-energy
constants appearing in eqs.(3.77) and (3.78). In ref. [3.22],
 $c_1$ was fixed as in eq.(3.64). The coefficients $c_2$ and $c_3$ where
estimated from resonance exchange. This induces a dependence on the off--shell
parameter $Z$ as discussed in section 3.4.  From the meson sector, scalar meson
exchange can contribute to $c_1$ and $c_3$,
$$c_1 -{1 \over 2} c_3 \big|_S = c_1 - c_1 {c_d \over c_m} \eqno(3.80)$$
with $2c_d / c_m = L_5 / L_8$. The central values for the parameters $c_d$
and $c_m$ given in ref.[3.31] lead to
$2c_d / c_m = 1.56$. However, within the uncertainty of $L_5$ and $L_8$, this
ratio can vary between 0.75 and 2.25. The $\Delta$ and the $N^* (1440)$
contribute to $c_2 + c_3$ and to $b (\lambda )$
$$\eqalign{
c_2 + c_3 \big|_\Delta & = -{g_A^2 \over 2 m_\Delta^2} ({1\over 2} - Z) \biggl[
2 m_\Delta (1 + Z) + m ( {1 \over 2} - Z) \biggr] \cr
c_2 + c_3 \big|_{N^*} & = -{g_A^2 R \over 16 (m+m^*)} \cr
b^r (\lambda)  \big|_{\lambda = m_\Delta} & = -g_A^2 \biggl[{( Z -
{1 \over 2})^2
\over 8 m^2_\Delta} + {R \over 32 (m + m^* )^2} \biggr] \cr}
\eqno(3.81)$$
Other baryon resonances have been neglected since their couplings are
either very small or poorly (not) known.\footnote{*}{A remark on the
$\rho$--meson is in order. The chiral power counting enforces a $\rho \pi \pi$
vertex of order $q^2$ of the form ${\cal L}^{(2)}_{\rho \pi \pi} =
g_{\rho \pi \pi} \Tr ( \rho_{\mu \nu} [ u^\mu , u^\nu ])$ [3.31]. In forward
direction the contraction of the $\rho$--meson propagator with the
corresponding $\rho \pi \pi$ matrix element vanishes. Therefore, one has no
explicit $\rho$--meson induced contributions to $T^-$ of order $q^2$ and
$q^3$.} Clearly, the contribution of the $N^* (1440)$ is only a small
correction to the $\Delta$--contribution.
The  numerical results are as follows. Consider first the
amplitude $T^-$. Using $M_\pi = 138$ MeV, $F_\pi = 93$ MeV, $m = 938.9$
MeV, $Z = -1/4 $ and $R = 1$, one has
$$T^- (M_\pi) = (1.57+0.24+0.08+0.02) \, \, {\rm fm}
= 1.91 \, \, {\rm fm} \eqno(3.82)$$
where we have explicitely shown the contributions from the current
algebra, the one loop, the nucleon pole and the counter terms.
The total  result is in good agreement with the
empirical value. The largest part of the $M_\pi^3$   term comes from the pion
loop diagrams. We should stress that only this loop contribution can
close the gap between the Weinberg-Tomozawa prediction of 1.57 fm and
the empirical value. As stated before, the uncertainties in $b$ are
completely masked by the small prefactor. If one chooses e.g.
 $\lambda = m$, the loop contribution drops to 0.22 fm.
 The two--loop contribution carries an explicit
factor $M_\pi^5$ and is therefore expected to be much smaller.
 In the case of the isospin--even
scattering amplitude $T^+$, the situation is much less satisfactory.  There are
large cancellations between the loop contribution and the $1/m$ suppressed
kinematical terms of order $M_\pi^2$ and $M_\pi^3$.
 Therefore, the role of the contact terms is even further magnified.
The total result for $T^+$ is very sensitive to some
of the resonance parameters, the empirical value of $T^+$ can, however, be
obtained by reasonable choices of these (cf. figs. 1 and 2 in [3.22]
for the scattering length $a^+$).
A better
understanding of the coefficients of the contact terms appearing at order $q^2$
(and higher) is necessary to further pin down the
 prediction for $T^+ (M_\pi)$.

Another quantity of interest is the so--called nucleon axial polarisability
$\alpha_A$. It is related to the quenching of the axial vector coupling $g_A$
in the nuclear medium as discussed in detail in ref.[3.32].
Consider the standard
helicity non--spin--flip amplitude $C = A + B \nu ( 1 - t /4m)^{-1}$ with
$\nu =( s- u)/4m$ and the  conventional $\pi N$ amplitude is written as $T_{\pi
N} = A + \barre{q} \, B$. Here, $A$ and $B$ are functions of $\nu$ and
the invariant momentum transfer squared $t$. The axial polarisability is then
defined as
$$ \alpha_A = 2 c^+_{01} = 2 {\partial \over \partial t} {\bar A}^+ (m^2 +
M_\pi^2 - t/2, m^2 + M_\pi^2 - t/2 ) \biggl|_{t=0}
\eqno(3.83)$$
where the bar means that the nucleon Born term has been subtracted and we have
also indicated the standard notation which refers to the expansion of $\bar{C}
(\nu, t)$ around $\nu = 0$. Empirically, one has $\alpha_A = 2.28 \pm 0.04 \,
M_\pi^{-3}$ [3.30]. To get $\alpha_A$, we calculate the
on--shell $\pi N$ scattering
amplitude in the cms and subtract the Born term,
$$ \bar{T}^+ ( \omega, {\svec q}' , {\svec q}) = t_0 (\omega ) + {\svec q}'
\cdot {\svec q} \, t_1 (\omega ) + \ldots      \eqno(3.84)$$
with the kinematics $v \cdot q = v \cdot q' = \omega \simeq \nu$
and $t = (q - q')^2
= 2 ( M_\pi^2 - \omega^2 + {\svec q}' \cdot {\svec q} )$ (here, ${\svec q}$ and
${\svec q}'$ are the momenta of the incoming and outgoing pion, respectively).
The axial polarizability is then simply given by:
$$ \alpha_A = t_1 (0) \quad . \eqno(3.85)$$
At order $q^2$, we have the counterterm contribution proportional to $c_3$ and
at order $q^3$, only loops contribute. The possible counterterm of
order $q^3$ proportional to $\omega \vec{q \,}' \cdot \vec{q}$ gives a
vanishing contribution to $\alpha_A$.
The final expression of this calculation
was already given in eq.(3.66). Estimating the value of $c_3$ as discussed
above, we find $\alpha_A = 1.3 \ldots 1.8 \, M_\pi^{-3}$, somewhat below the
empirical value. It is important to stress (see also refs.[3.30,3.32]) that the
$\Delta$ alone is not sufficient to get the empirical value but that one needs
additional scalar exchange (as provided here through the resonance saturation).

For the later discussion in section 6, we will have to
consider the $\pi N$ amplitude for off--shell pions. For doing that, we choose
the pseudoscalar density $P^a = i {\bar q} \gamma_5 \tau^a q$ as the
interpolating pion field. The pion coupling via the pseudoscalar density is
given in terms of $G_\pi$,
$$  < 0 \, |P^b  \, | \pi ^a  > = \delta^{ab} \, G_\pi
\quad , \eqno(3.86)$$
where $G_\pi^2$ is given as the residue of the vacuum correlator
$<0|P^a P^b|0>$ at the pion pole.
The off--shell $\pi N$ amplitude is then defined via
$$ - \int d^4 x {\rm e}^{i q_1 x} <N | T (P^a (x) P^b (0)) | N > =
{G_\pi^2 \, \, i^5
\over (q_1^2 - M_\pi^2 ) (q_2^2 - M_\pi^2 )} \, T^{ab} (q_1 , q_2
) \, . \eqno(3.87)$$
It is now a straightforward exercise to show that the amplitude calculated in
this fashion obeys the Adler conditions,
$$\eqalign{ T^+ (q_1 = q_2 = 0) & = -{\sigma_{\pi N} (0) \over F_\pi^2} \cr
T^+ (q_1^2 = 0,  q_2^2 = M_\pi^2) & = T^+ (q_1^2 = M_\pi^2 , q_2^2 = 0) = 0 \,
.  \cr} \eqno(3.88)$$
\medskip
Finally, we stress that GSS [3.5] have evaluated the full off--shell
pion--nucleon amplitude in the framework of relativistic baryon CHPT and
discussed the so--called remainder of the $\pi N$ $\sigma$--term derived from
it. We will come back to these issues in section 6 because the $\sigma$--term
is intimately related to the strangeness content of the nucleon and the baryon
mass ratios.
 \bigskip \goodbreak
\noindent{\bf III.6. THE REACTION $\pi N \to \pi \pi N$}
\medskip
\goodbreak
Another reaction involving only pions and nucleons is the single pion
production reaction $\pi N \to \pi \pi N$ (for some older references,see
[3.33]).
The interest in this reaction stems mostly from the
fact that it apparently offers a possibility of determining
the low--energy $\pi
\pi$ elastic scattering amplitude whose precise knowledge allows to test our
understanding of the chiral symmetry breaking of QCD. However, at present no
calculation based on chiral perturbation theory  is available which links the
pion production data to the $\pi \pi \to \pi \pi$ amplitude in a {\it
model--independent} fashion. Consequently, all presently available
determinations of the S--wave $\pi \pi$ scattering lengths from the
abovementioned data should be taken {\it cum grano salis}. Over the last years,
new experimental data in the threshold region have become available [3.34-3.38]
which allow for a direct comparison with the CHPT predictions. Beringer [3.39]
has performed a tree calculation in relativistic baryon CHPT. In ref.[3.40]
the chiral expansion of the threshold amplitudes was reanalyzed in terms of the
heavy fermion formalism at next--to--leading order.

To be specific,  consider the process $\pi^a N \to \pi^b \pi^c N$,
with $N$ denoting the nucleon (proton or neutron) and '$a,b,c$' are isospin
indices. At threshold, the transition matrix--element
in the $\pi^a N$ centre--of--mass frame takes the form
$$
T \, = \, i \, {\vec \sigma} \cdot {\vec k} \left[ D_1 (\, \tau^b \delta^{ac} +
\, \tau^c \delta^{ab} ) \, + \, D_2 \, \tau^a \delta^{bc} \right]
\eqno(3.89)
$$
where $\vec k$ denotes the three--momentum of the incoming pion and
the amplitudes $D_1$ and $D_2$
 will be subject to the chiral expansion as discussed below. They are related
to the more commonly used amplitudes ${\cal A}_{2I,I_{\pi \pi}}$, with $I$ the
total isospin of the initial $\pi N$ system
and $I_{\pi \pi}$ the isospin of the two--pion system in the
final state, via
$$
{\cal A}_{32} \, = \, \sqrt{10} \, D_1, \quad {\cal A}_{10} \, = \, -2D_1 \,
- \, 3D_2
\eqno(3.90)$$
Assuming that
the amplitude in the threshold region can be approximated by the exact
threshold amplitude, the total cross section can be written in a compact
form,
$$\eqalign{
\sigma_{\rm tot}(s) & =
 {m^2 \over 2s} \, \sqrt{\lambda(s,m^2,M_\pi^2)} \, \Gamma_3
(s) |\eta_1 D_1 \, + \, \eta_2 D_2 |^2 \, S \cr
\Gamma_3(s) = & {1\over 32 \pi^3 } \int_0^{T_1}dT{\sqrt{T(T+2m)(T_1
- T)(T_2 -  T)} \over T_3 - T}\,,\cr T_1 = & {1\over 2 \sqrt{s}} ( \sqrt{s} -
m -  M_{\pi 1} - M_{\pi2} ) ( \sqrt{s} - m + M_{\pi 1} + M_{\pi 2})\,,\cr T_2 =
& {1\over 2 \sqrt{s}} ( \sqrt{s} -  m -  M_{\pi 1} + M_{\pi2} ) ( \sqrt{s} - m
 + M_{\pi 1} - M_{\pi 2} )\,,\cr T_3 = & {1\over 2 \sqrt{s}} ( \sqrt{s} - m)^2
 \cr
\lambda (x,y,z) & = x^2 + y^2 + z^2 - 2 (xy + xz + yz) \cr}
\eqno(3.91)$$
with $s$ the total
centre--of--mass energy squared. $\Gamma_3 (s)$ denotes the conventional
integrated three--body phase space
where $M_{\pi 1}$ and $M_{\pi 2}$ stand for the masses of the final state pions
and one has the inequality $ 0 \leq T_1 \leq T_{2,3}$.
$\lambda(x,y,z)$ is the K\"all{\'e}n--function. The
$\eta_{1,2}$ are channel-dependent isospin factors and $S$ is a Bose
symmetry factor.
For $\pi^+ p \to \pi^+ \pi^+ n$ and $\pi^- p \to \pi^0 \pi^0 n$ we have
$\eta_1 = 2 \sqrt{2}, \, \eta_2 = 0, \, S = 1/2$ and
$\eta_1 = 0 , \, \eta_2 = \sqrt{2}, \, S = 1/2$, in order. In the threshold
region, one can approximate to  a high degree of accuracy the three--body phase
space and flux factor by analytic expressions as discussed in more detail
in section 4. The chiral expansion of the amplitude functions
$D_1$ and $D_2$ takes the form\footnote{*}{Here, ${\cal D}$ stands as a generic
symbol for $D_{1,2}$.}
$${\cal D} = f_0 \, + \, f_1 \, \mu \, + \ f_2 \, \mu^2 \, + \ldots
\eqno(3.92)$$
 modulo logarithms.
The first two coefficients of this
expansion have been calculated in ref.[3.40]. For that, one needs only
${\cal L}_{\rm eff} = {\cal L}_{\pi N}^{(1)} + {\cal L}_{\pi N}^{(2)}
+ {\cal L}_{\pi \pi}^{(2)}$ and one finds that none of the low--energy
constants $c_i$ will contribute (the sum of the corresponding graphs vanishes).
Notice that the much debated next--to--leading order $\pi \pi$
interaction does not appear at this order in the chiral expansion.
One can therefore write  down low--energy theorems for $D_{1,2}$
which only involve well--known physical (lowest order) parameters,
$$\eqalign{
D_1 & = {g_A \over8 F_\pi^3} \biggl( 1 + {7 M_\pi \over2 m} \biggr)
 + {\cal O}(M_\pi^2) \cr
D_2 &  = -{g_A \over 8 F_\pi^3} \biggl( 3 + {17 M_\pi \over 2 m} \biggr)
 + {\cal O}(M_\pi^2) \cr} \eqno(3.93)  $$
There are  potentially large contributions from diagrams with intermediate
$\Delta(1232)$ states of the type
$M_\pi^2 \, / \, ( \,m_\Delta - m - 2M_\pi \, )$,
which numerically would be of the order $10 \cdot M_\pi$.
As shown in [3.40],  no such terms appear from diagrams involving one or two
intermediate $\Delta$ resonances. Consequently, the chiral expansion is well
behaved but not too rapidly converging. The order $M_\pi$ corrections give
approximatively 50$\%$ of the leading term. However, the
calculations of Beringer [3.39] in relativistic baryon chiral perturbation
theory indicate that further $1/m$ suppressed kinematical corrections are
small. The numerical evaluation of eqs.(3.93) amounts to
$D_1 = 2.4$ fm$^3$ and $D_2 = -6.8$ fm$^3$ or using eq.(3.90)
$${\cal A}_{32} \, = \, 2.7 \, M_\pi^{-3}, \quad {\cal A}_{10} \, = \,
5.5 \, M_\pi^{-3}
\eqno(3.94)$$
which compare fairly with the recent determinations of Burkhardt and Lowe (see
ref.[3.35]), ${\cal A}_{32} \, = \, 2.07 \pm 0.10 \, M_\pi^{-3}$ and
${\cal A}_{10} \, = \, 6.55 \pm 0.16 \, M_\pi^{-3}$. As stressed, however,
in ref.[3.40], one can
confront the LET eqs.(3.93) directly with experimental
data and, furthermore, the global fit to the threshold amplitudes of ref.[3.35]
has to be reexamined critically. The cross sections for
 $\pi^+ p \to \pi^+ \pi^+ n$ and $\pi^- p \to \pi^0 \pi^0 n$
 in comparison to the existing data are shown in fig.3.3. They compare
well to the existing data for the first 30 MeV above threshold.
To get an idea about the higher order corrections, one can calculate
the imaginary parts Im$\, D_{1,2}$. Corrections to Re~$D_{1,2}$ of the
same size
indeed turn to be
 such that they can improve the description of the data since the
first/second reaction allows to test $D_1$/$D_2$, respectively.
\midinsert
\smallskip
\hskip 0.5in
\epsfxsize=3.5in
\epsfysize=5.0in
\epsffile{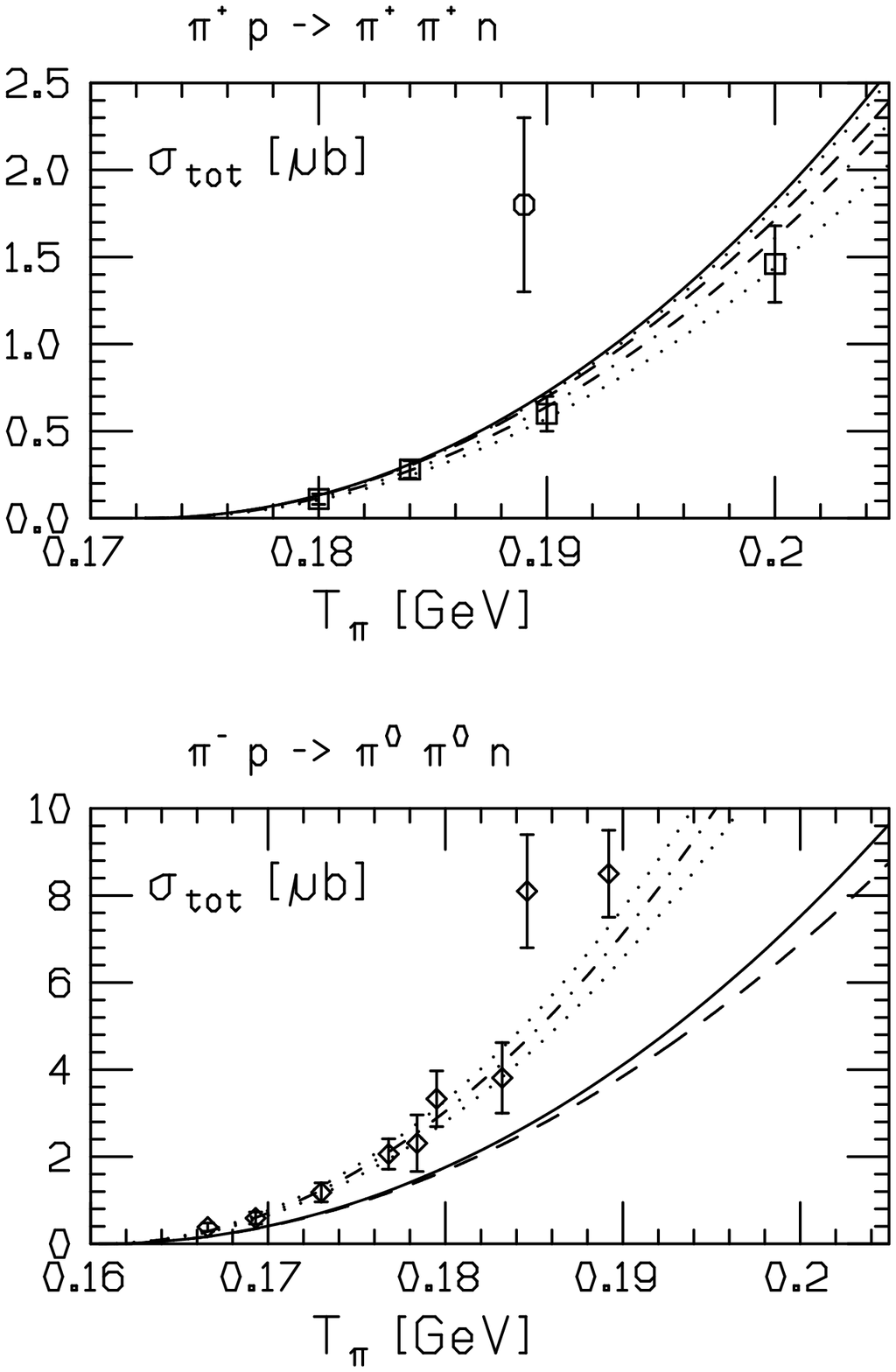}
\smallskip
{\noindent\narrower \it Fig.~3.3:\quad   Total cross sections for
$\pi^+ p \to \pi^+ \pi^+ n$ and $\pi^- p \to \pi^0 \pi^0 n$ in comparison to
the data. Squares: ref.[3.36], octagon: ref.[3.37]
and diamonds: ref.[3.38]. The dashed lines refer to an approximation discussed
in ref.[3.40] and the dash--dotted ones show the best fit to
these data as discussed
in the text (the $1\sigma$--band is indicated by the dotted lines).
\smallskip}
\endinsert
In addition, one finds that a best fit to these data
leads to $D_1 = 2.26$ fm$^3$ and $D_2 = -9.05$ fm$^3$
as indicated by the dotted lines in fig.3.3. Using eq.(3.90), this leads to
${\cal A}_{32} \, = \, 2.5 \, M_\pi^{-3}$ and  ${\cal A}_{10} \,
= \, 8.0 \, M_\pi^{-3}$ somewhat different from the global best fit values
of ref.[3.35]. We believe that the energy range covered by the fit in
ref.[3.35] was too large to reliably extract the threshold amplitudes.
In that fit, the data in the first 30 MeV above
threshold had too little statistical weight.

Another important remark concerns the fashion in which the S--wave
$\pi \pi$ scattering lengths
are in general extracted from the $\pi N \to \pi \pi N$
data [3.34,3.35]. It is based on the old Olsson--Turner model [3.33] which
parametrizes the chiral symmetry breaking in terms of one parameter called
$\xi$. This is, of course, an pre--QCD artefact since we now know that the
breaking via the quark masses is of the $\bar{3} \times 3$ form, i.e. $\xi =
0$. Therefore,
 one can no more accept such a parametrization. The essential
question is now, how can one relate the $\pi \pi$ S-wave scattering
lengths and the
$\pi \pi N$ threshold amplitudes in a model--independent way.
This question is  quite
nontrivial. The Olsson--Turner model with $\xi =0$ only contains the
tree level $\pi \pi$ scattering lengths.
To establish a  firm relationship between the $\pi\pi$ amplitudes and
the $\pi N \to \pi \pi N$ data beyond leading order, one has to perform
a complete one--loop calculation. This has not yet been done. As an
estimate, however, we can combine the  new low energy theorems
 for the $\pi \pi N$
threshold  amplitudes (3.93) with
Weinberg's low energy  theorems for $\pi \pi$ scattering
(i.e. the leading term in the chiral
expansion). This way we derive
$$ {\cal A}_{10} =  4 \pi {g_{\pi N} \over m} \biggl( 1 + {37 M_\pi \over 14 m}
\biggr) \biggl( {a^0_0 \over M_\pi^2 } + {\cal O}(M_\pi^2) \biggr)
\eqno(3.95)$$

$$ {\cal A}_{32} =  -2\sqrt{10} \pi {g_{\pi N} \over m} \biggl( 1 + {7 M_\pi
\over 2 m} \biggr) \biggl( {a^2_0 \over M_\pi^2 } + {\cal O}(M_\pi^2) \biggr)
\eqno(3.96)$$
The corrections of order $M_\pi$ are taken care of by the calculation
to order $p^2$ leading to eqs.(3.93)
and what remains to be done is to systematically work
out the various contributions at $ {\cal O}(M_\pi^2)$. Ignoring these
for the moment and
inserting on the left hand side the present fit value, we extract $a_0^0 = 0.23
\pm 0.02$ and $a^2_0 = -0.042 \pm 0.002 $
 which are quite close to the CHPT prediction
at next-to-leading order. We stress however, the a complete calculation of  the
${\cal O}(M_\pi^2)$ corrections to the above relations is mandatory.
We conclude that the values of the $\pi \pi$ S--wave scattering
lengths can
eventually be infered from the threshold $\pi N \to \pi \pi N$
amplitudes.
The complete
one--loop calculation which would give a sound basis for doing that is,
unfortunately, not yet available. At present, it seems that the most accurate
fashion of determining in particular $a_0^0$ are $K_{\ell 4}$ decays.
 \bigskip \goodbreak
\noindent{\bf III.7. THE PION--NUCLEON VERTEX}
\medskip
\goodbreak
The last topic we want to address in this section is the pion--nucleon vertex,
parametrized in terms of a form factor $G_{\pi N} (t)$. It plays a fundamental
role in many areas of nuclear physics, in particular in the description of the
nucleon--nucleon force via meson--exchange models. Before we discuss the
details, let us stress from the beginning that while the strong pion--nucleon
coupling constant $g_{\pi N} \equiv G_{\pi N} (t = -M_\pi^2 )$
can be unambigously
calculated within CHPT, the $\pi N$ form factor depends on the choice of the
interpolating pion field. Furthermore, if one writes a dispersion relation for
$G_{\pi N} (t)$, one  realizes that the absorptive part starts at $t_0 =
(3 M_\pi)^2$. Therefore, within the context of a one--loop calculation, the
momentum dependence of the form factor will entirely stem from some contact
terms.

After these remarks, consider the Breit frame matrix--element of the
pseudoscalar density between nucleon states\footnote{*}{The Breit frame is most
convenient for the calculation of such matrix--elements since it allows for a
unique translation of Lorentz--covariant matrix--elements into non-relativistic
ones.}
$$<N(p') |i \bar{q} \gamma_5 \tau^a q | N(p)> = 2 i B {\krig g}_A {1 + g(t)
\over M_\pi^2 - t } \, \bar{H} S \cdot (p'-p) \tau^a H  \quad . \eqno(3.97)$$
The form factor $g(t)$ is generated by loop and counterterm contributions. In
fact, the loop contribution is divergent and $t$--independent,
$$g(t) = -{B_{23} \over 8 \pi^2 F^2 } \, t +
{M^2 \over F^2} C(\krig{g}_A^2 , B_{9},
B_{20}, \ldots )     \eqno(3.98)$$
where the constant $C$ sums up all $t$--independent terms. We do not need its
explicit form in what follows. $B_{23}$ is a finite low--energy constant from
${\cal L}_{\pi N}^{(3)}$,
$${\cal L}_{\pi N}^{(3)} = B_{23} \, {\krig{g}_A \over(4 \pi F)^2} \bar{H} i
S \cdot D \chi_- \, H     \quad . \eqno(3.99)$$
The form factor $g(t)$ features in
the so--called Goldberger--Treiman discrepancy (for a review, see [3.41]). To
be specific, let us look at the relation between the divergence of the axial
current and the pseudoscalar density between nucleon states,
$$2 B \krig{m} \krig{g}_A [ 1 + g(0) ] = {m_N g_A \over F_\pi } G_\pi \quad .
\eqno(3.100)$$
On the other hand, the strong pion--nucleon coupling constant
 is defined via the residue of the pole term in (3.97),
$$2 B \krig{m} \krig{g}_A [ 1 + g(M_\pi^2) ] = g_{\pi N} G_\pi \quad ,
\eqno(3.101)$$
which leads to the Goldberger--Treiman discrepancy
$$\Delta_{\pi N} \equiv 1 - {m_N g_A \over F_\pi  g_{\pi N}}
= g(M_\pi^2) - g(0)
= -{M_\pi^2 \over 8 \pi^2 F_\pi^2}\, B_{23} \, \, . \eqno(3.102)$$
Notice that $\Delta_{\pi N}$ is entirely given by the low--energy constant
$b_{11}$. With $m_N = 938.27$ MeV, $F_\pi = 92.5$ MeV, $g_A = 1.257$ and
$g_{\pi N} = 13.3$\footnote{*}{In general, we use the Karlsruhe--Helsinki value
of $g_{\pi N} = 13.4$ [3.30]. In light of the recent discussion about the
actual value of this quantity, we have adopted here the most recent
value proposed by H\"ohler.} we find
$$ \Delta_{\pi N} = 0.04 \, , \quad B_{23} = -1.433 \, \, . \eqno(3.103)$$
If one now describes the whole Goldberger--Treiman discrepancy by a form factor
effect, one identifies the nucleon matrix--element of the pseudoscalar density
with $G_\pi G_{\pi N} (t) / (M_\pi^2 - t )$, so that
$$ G_{\pi N} (t) = g_{\pi N}  [ 1 + g(t) - g(M_\pi^2) ] \quad . \eqno(3.104)$$
Assuming furthermore the standard monopole form, $ G_{\pi N} (t) = ( \Lambda^2
- M_\pi^2 ) / ( \Lambda^2 - t)$, one can calculate the cut--off $\Lambda$,
$$ \Lambda = {4 \pi F_\pi \over \sqrt{ - 2 B_{23} }} = 700 \, {\rm
MeV}
 \, \, , \eqno(3.105)$$
close to the result found by GSS [3.5] in the relativistic calculation.
However, we stress again that this result depends on the choice of the
interpolating field and that it is based on the assumption that the whole
Goldberger--Treiman discrepancy is due to a form factor effect.
\bigskip
\bigskip

\noindent{\bf REFERENCES}
\medskip
\item{3.1}S. Weinberg,
{\it Phys. Rev.\/} {\bf
166} (1968) 1568.
\smallskip
\item{3.2}S. Coleman, J. Wess and B. Zumino,
{\it Phys. Rev.\/} {\bf 177} (1969) 2239;

C. G. Callan, S. Coleman, J. Wess and B. Zumino,
{\it Phys. Rev.\/} {\bf 177} (1969) 2247.
\smallskip
\item{3.3}P. Langacker and H. Pagels,
{\it Phys. Rev.\/} {\bf D8} (1971) 4595.
\smallskip
\item{3.4}H. Pagels, {\it Phys. Rep.\/} {\bf 16} (1975) 219.
\smallskip
\item{3.5}J. Gasser, M.E. Sainio and A. ${\check {\rm S}}$varc,
{\it Nucl. Phys.\/}
 {\bf B 307} (1988) 779.
\smallskip
\item{3.6}A. Krause, {\it Helv. Phys. Acta\/} {\bf
63} (1990) 3.
\smallskip
\item{3.7}H. Georgi,
``Weak Interactions and Modern Particle Physics'',
Benjamin / Cummings, Reading, MA, 1984.
\smallskip
\item{3.8}Ulf-G. Mei{\ss}ner, {\it Int. J. Mod. Phys.}
{\bf E1} (1992) 561.
\smallskip
\item{3.9}E. Jenkins and A.V. Manohar, {\it Phys. Lett.\/} {\bf B255} (1991)
558.
\smallskip
\item{3.10}E. Jenkins and A.V. Manohar, in "Effective field theories of the
standard model", ed. Ulf--G. Mei{\ss}ner, World Scientific, Singapore,
1992.
\smallskip
\item{3.11}J. Gasser, {\it Ann. Phys. (N.Y.)} {\bf 136} (1981) 62.
\smallskip
\item{3.12}J. Gasser and H. Leutwyler,
{\it Phys. Reports\/} {\bf C87} (1982) 77.
\smallskip
\item{3.13}T. Mannel, W. Roberts and Z. Ryzak, {\it Nucl. Phys.\/} {\bf B368}
(1992) 264.
\smallskip
\item{3.14}V. Bernard, N. Kaiser, J. Kambor
and Ulf-G. Mei{\ss}ner, {\it Nucl. Phys.\/} {\bf B388} (1992) 315.
\smallskip
\item{3.15}G. Ecker,  {\it Czech. J. Phys.} {\bf 44} (1994) 405.
\smallskip
\item{3.16}T.--S. Park, D.--P. Min and M. Rho,
{\it Phys. Reports\/} {\bf 233} (1993) 341.
\smallskip
\item{3.17}G. Ecker, {\it Phys. Lett.} {\bf B336} (1994) 508.
  \smallskip
\item{3.18}J. Gasser and H. Leutwyler, {\it Ann. Phys. (N.Y.)\/} {\bf 158}
(1984) 142.
\smallskip
\item{3.19}G. Ecker, J. Kambor and D. Wyler, {\it Nucl. Phys.\/} {\bf
B394} (1993) 101.
\smallskip
\item{3.20}E. Jenkins and A.V. Manohar, {\it Phys. Lett.\/} {\bf B259} (1991)
353.
\smallskip
\item{3.21}J. Gasser, H. Leutwyler and M.E. Sainio, {\it Phys. Lett.\/}
 {\bf 253B} (1991) 252, 260.
\smallskip
\item{3.22}V. Bernard, N. Kaiser and Ulf-G. Mei{\ss}ner, {\it Phys. Lett.\/}
{\bf B309} (1993) 421.
\smallskip
\item{3.23}M. Benmerrouche, R.M. Davidson and N.C. Mukhopadhyay, {\it Phys.
Rev.\/} {\bf C39} (1989) 2339. \smallskip
\item{3.24}T. Ericson and W. Weise,
"Pions and Nuclei", Clarendon Press, Oxford, 1988.
\smallskip
\item{3.25}V. Bernard, N. Kaiser and Ulf-G. Mei{\ss}ner, {\it Z. Phys.\/}
{\bf C60} (1993) 111.
\smallskip
\item{3.26}S. Weinberg, {\it Phys. Rev. Lett.\/} {\bf 17} (1966) 616.
\smallskip
\item{3.27}Y. Tomozawa, {\it Nuovo Cim.\/} {\bf 46A} (1966) 707.
\smallskip
\item{3.28}R. Koch, {\it Nucl. Phys.\/} {\bf A448} (1986) 707.
\smallskip
\item{3.29}W. Beer et al., {\it Phys. Lett.\/} {\bf B261} (1991) 16.
\smallskip
\item{3.30}G. H\"ohler, in Land\"olt--B\"ornstein, vol.9 b2, ed. H. Schopper
(Springer, Berlin, 1983);

M.E. Sainio, private communication.
\smallskip
\item{3.31}G. Ecker, J. Gasser, A. Pich and E. de Rafael,
{\it Nucl. Phys.\/} {\bf B321} (1989) 311.
\smallskip
\item{3.32}M. Ericson and A. Figureau, {\it J. Phys.} {\bf G7} (1981) 1197.
\smallskip
\item{3.33}S. Weinberg, {\it Phys. Rev.} {\bf 166} (1968) 1568;

M.G. Olsson and L. Turner, {\it Phys. Rev. Lett.} {\bf 20} (1968) 1127;
{\it Phys. Rev.} {\bf 181}

(1969) 2142;

L.--N. Chang, {\it Phys. Rev.} {\bf 162} (1967) 1497;

R. Rockmore, {\it Phys. Rev. Lett.} {\bf 35} (1975) 1409;

M.G. Olsson, E.T. Osypowski  and L. Turner,
{\it Phys. Rev. Lett.} {\bf 38} (1977) 296.
\smallskip
\item{3.34} D. Po${\check{\rm c}}$ani{\'c} et al.,
{\it Phys. Rev. Lett.} {\bf 72} (1994) 1156.
\smallskip
\item{3.35} H. Burkhardt and J. Lowe, {\it Phys. Rev. Lett.}
{\bf 67} (1991) 2622.
\smallskip
\item{3.36} M.E. Sevior et al., {\it Phys. Rev. Lett.} {\bf 66} (1991) 2569.
\smallskip
\item{3.37} G. Kernel et al., {\it Z. Phys}. {\bf C48} (1990) 201.
\smallskip
\item{3.38} J. Lowe et al., {\it Phys. Rev.} {\bf C44} (1991) 956.
\smallskip
\item{3.39} J. Beringer, {\it $\pi N$ Newsletter} {\bf 7} (1993) 33.
\smallskip
\item{3.40}V. Bernard, N. Kaiser and Ulf-G. Mei{\ss}ner,
{\it Phys. Lett.\/} {\bf B332} (1994) 415.
\smallskip
\item{3.41}C.A. Dominguez, {\it Riv. Nuovo Cim.} {\bf 8} (1985) N.6.
\smallskip

\vfill \eject

\noindent{\bf IV. NUCLEON STRUCTURE FROM ELECTROWEAK PROBES}
\medskip
In this section, we will mostly be concerned with the nucleon structure
when real or virtual photons are used as probes. This is of particular interest
for the physics program of  the existing CW electron machines and intense light
facilities. Topics included are Compton scattering (spin-averaged and
spin--dependent) and the classical field of single and double pion  production
by real or virtual photons (see e.g. the monograph [4.1]). Another
well--understood probe are the $W$--bosons. Their interactions with the hadrons
lead to the axial form factors and can also be used to produce pions. These
topics will be discussed at the end of this section.
\medskip
\noindent{\bf IV.1. ELECTROMAGNETIC FORM FACTORS OF THE NUCLEON}
\medskip
\goodbreak
The coupling of the photon to the nucleon has an isoscalar and an isovector
component. The chiral expansion of the electric and the magnetic form factors
of the neutron and the proton amounts to a calculation of the corresponding
radii, magnetic moments and so on. Evidently, the further one goes in the
loop expansion,  higher  moments of these form factors are tested. Here, we
will concentrate on the form factors at small momentum transfer. As it was
already mentioned in section 3, the existing precise data on these nucleon
properties are mostly used to fix the values of some $\len$ constants.
However, it is important to understand the interplay of the loop and the
counter term contributions and also to critically examine the absorptive parts
of the isovector form factors.

First, let us consider the
matrix--element of the isovector--vector quark current,
$$<p'|\bar q \gamma_\mu {\tau^a \over 2} q |p> = \bar{u}(p') \biggl[ \gamma_\mu
\, F_1^V (t) + {i \sigma_{\mu \nu} k^\nu \over 2 m} \, F^V_2 (t) \biggr]
{\tau^a \over 2} u(p)          \eqno(4.1)$$
with $k = p' - p$ and $t = k^2$. This defines the so--called Dirac ($F_1^V$)
and
the Pauli ($F_2^V$) form factors.  These are related
to the proton and neutron form factors $F_{1,2}^p$ and $F_{1,2}^n$ via
$$\eqalign{
F_1^V (k^2) & = F_1^p (k^2) - F_1^n (k^2)  \cr
F_2^V (k^2) & = F_2^p (k^2) - F_2^n (k^2)  \, . \cr} \eqno(4.2)$$
At zero momentum transfer, we have $F_1^V (0) = 1$ and
$F_2^V (0) = \kappa_p - \kappa_n = 3.706$.
In relativistic baryon CHPT, these form factors have been discussed by Gasser
et al. [4.2]. Here, we will elaborate on the heavy fermion approach following
ref.[4.3]. For that, one rewrites eq.(4.1) in the Breit frame as
$$\eqalign{
<N(p') |\bar q \gamma_\mu {\tau^a \over 2} q |N(p)> & = \biggl[ F_1^V (t) +
{t \over 4 m^2_N} F_2^V (t) \biggr] \, v_\mu \bar{H} {\tau^a \over 2} H \cr
& + {1 \over m_N} \bigl[ F_1^V (t) + F_2^V (t)\bigr] \bar{H} [S_\mu, S \cdot
(p'-p)] {\tau^a \over 2} H \quad . \cr} \eqno(4.3)$$
This  corresponds to the standard decomposition into the electric and magnetic
form factors $G_E (t) = F_1 (t) + \tau F_2(t)$ and $G_M (t) = F_1 (t) + F_2
(t)$, with $\tau = t / 4 m_N^2$. The Dirac form factor $F_1^V (t)$ is readily
evaluated,
$$F_1^V (t) = 1 + {t \over 6} <r^2>_1^V + {{g}_A^2 -1 \over F_\pi^2} J(t) +
{{g}_A^2 \over F_\pi^2} \biggl[ t \xi (t)  - 2 M_\pi^2 \bar{\xi} (t) \biggr]
\eqno(4.4)$$
with the loop functions $J(t)$ and $\xi (t)$ given in appendix B,
and $\bar{\xi}(t) = \xi (t) - t \xi ' (t)$. It can be
shown analytically that the sum of all loop diagrams does not modify the tree
level result, $F_1^V (0) = 1$. This is, of course, nothing but the charge
non--renormalization by the strong interactions. The isovector charge radius
$$<r^2>_1^V = 6 {d F_1^V (t) \over dt}\biggr|_{t=0}     \eqno(4.5)$$
diverges logarithmically in the chiral limit,
$$<r^2>_1^V = -{5 {g}_A^2 +1 \over 8 \pi^2 F_\pi^2} \ln\bigl({M_\pi \over
\lambda}\bigr) - {7 {g}_A^2 +1 \over 16 \pi^2 F_\pi^2} +
 {3 \over 4 \pi^2 F_\pi^2} B_{10}^r (\lambda )      \eqno(4.6)$$
where the last term stems from a counterterm of order $q^3$ (cf.
eq.(3.62)). It is worth to stress [4.2] that the coefficient of the
logarithm in eq.(4.6) is nine times bigger than in the corresponding expression
for the pion charge radius and therefore this term contributes significantly
even for the physical value of the pion mass. This poses a severe constraint on
any serious attempt of modelling the nucleon (say from a quark model point of
view). To reproduce the empirical value $<r^2>_1^V   =0.578$ fm$^2$, one has to
set $B^r_{10} (\krig{m}) = -0.13$.\footnote{*}{We choose here
$\lambda = \krig{m}$ because of the matching conditions discussed in ref.[4.3].
Naturally, any other choice of $\lambda$ would do as well since physical
observables do not depend on the renormalization scale.} For
$B^r_{10} (\krig{m}) = 0$, one would get $<r^2>_1^V = 0.62$ fm$^2$, 8 $\%$
above the  empirical value.

The Pauli form factor $F_2^V (t)$ takes the very simple form
$$F_2^V (t) = c_6 - {{g}_A^2 m \over 4 \pi F_\pi^2} \int_0^1 dx
\sqrt{ M_\pi^2 + t x (x-1)}   \eqno(4.7)$$
which involves the low--energy constant $c_6$ to be identified with the
isovector anomalous magnetic moment in the chiral limit, $c_6 =
\krig{\kappa}_V$. To order $q^3$, we find for the isovector anomalous magnetic
moment,
$$\kappa_V = c_6 - {{g}_A^2 M_\pi m \over 4 \pi F_\pi^2 }   \eqno(4.8)$$
where the second term is the leading non--analytic piece proportional to
$\sqrt{\hat m}$ first found by Caldi and Pagels [4.5]. Setting $c_6 = 5.62$,
one reproduces the empirical value given after eq.(4.2). The loops generate a
correction of about $-34$ $\%$. The value of $c_6 \simeq 6$ is not quite
surprising if one thinks of generating the corresponding contact term via
$\rho$-meson exchange. The tensor coupling of the $\rho$ to the nucleon is
$\kappa_\rho \simeq 6$. However, we should point out that such an estimate
depends cucially on how one chooses the $\rho NN$ and $\rho \gamma$ couplings.
The isovector magnetic radius,
$$<r^2>_2^V = {6 \over \kappa_V } {d F_2^V (t) \over dt}\biggr|_{t=0}
\eqno(4.9)$$
explodes like $1/M_\pi$ in the chiral limit [4.4] and is not affected by any
counter term contribution to order $q^3$ [4.2],
$$<r^2>_1^V = {g_A^2 m \over 8 \pi F_\pi^2 \kappa_V}
{1 \over M_\pi} = 0.50 \, \, {\rm fm}^2     \eqno(4.10)$$
to be compared with the empirical value of $<r^2>_2^V = 0.77$ fm$^2$. It is
interesting to compare these results to the ones of the relativistic
calculation [4.2]. It becomes obvious that the role of the loop versus the
counter term contributions is rather different. In [4.2] it was shown that one
has to choose $c_6 \simeq 0$ to get the empirical value of $\kappa_V$. This is
due to the additional terms generated by the
relativistic one loop diagrams beyond the
leading non--analytic term in eq.(4.8) at ${\cal O}(q^2)$
 (compare the discussion about the power
counting in section 3.1 and fig.3.1). In the heavy fermion formalism, one loop
diagrams with insertions solely from the lowest order effective Lagrangian only
generate the term $\sim \sqrt{\hat m}$ in the isovector magnetic moment. A
similar statement also holds for the isovector radius, the heavy mass
calculation to order $q^3$ just gives the leading singularity.

\midinsert
\smallskip
\hskip 1in
\epsfxsize=2in
\epsfysize=1in
\epsffile{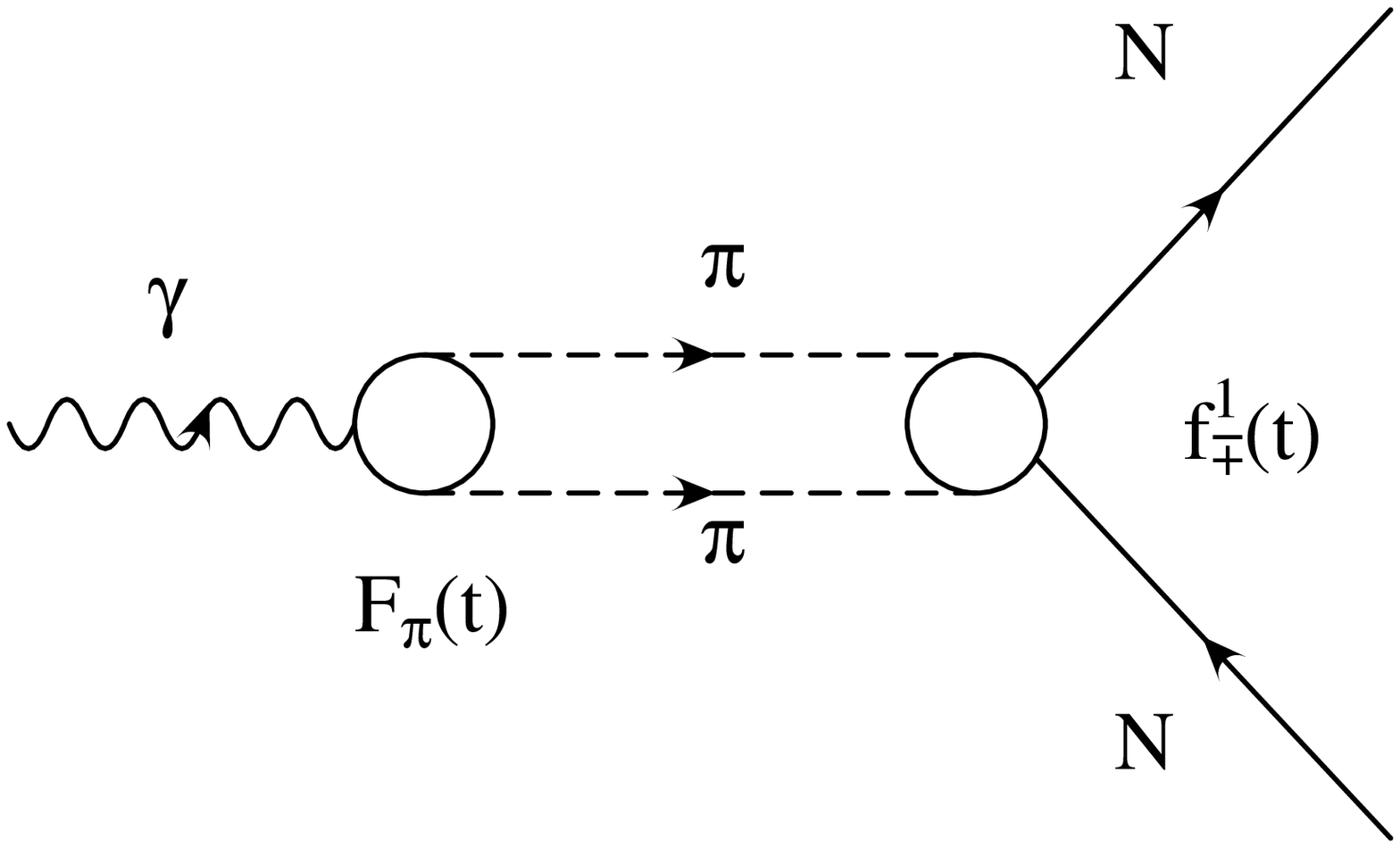}
\smallskip
{\noindent\narrower \it Fig.~4.1:\quad Two--pion cut contribution to
the nucleon isovector electromagnetic form factors.
\smallskip}
\vskip -0.5truecm
\endinsert
Let us now take a closer look at the imaginary parts of the isovector form
factors $F_{1,2}^V (t)$ (for similar discussions, see refs.[4.2,4.6,4.7]). As
first observed by Frazer and Fulco [4.8] and discussed in detail by H\"ohler
and Pietarinen in connection with the nucleon electromagnetic radii [4.9], Im
$F_{1,2}^V (t)$ exhibits a very strong enhancement close to threshold, $t_0 = 4
M_\pi^2$ (the two--pion cut). To be specific, consider $F_2^V (t)$. At low
momentum transfer, its absorptive part is dominated by diagrams like the one
shown in fig.4.1, i.e.
$${\rm Im} \, F_2^V (t) = {2q_t^3 \over (1 - \tau )
\sqrt{t} } F_\pi^* (t) \biggl[
-{1 \over m} f_+^{1} (t) + {1 \over \sqrt{2}}  f_-^{1} (t) \biggr]
\eqno(4.11)$$
with $q_t = \sqrt{t/4-M_\pi^2}$, $F_\pi (t)$ the pion charge form factor and
$f^1_{\pm} (t)$ the P--wave $\pi N$ partial waves in the $t$--channel.
 The
latter are calculated from the standard $\pi N$ amplitudes $A^\pm$ and $B^\pm$
via projection involving ordinary Legendre polynomials [4.10]. In this
procedure, the nucleon pole term of the $\pi N$ amplitudes
proportional to $1/(s-m^2)$ gives rise to Legendre functions
of the second kind, $Q_L
(Z)$, which have logarithmic singularities and
a cut along $-1 < Z < 1$. Consequently, if one continues the
partial waves $f_\pm^1 (t)$ to the second sheet, they have a
logarithmic branch point
at $Z =  \cos \theta_t = m \nu / (p_t q_t) = -1$,
with $p_t = \sqrt{t/4 - m^2}$ and $\nu = (t - 2M_\pi^2)/4 m $.
This translates into
$$t_c = 4M_\pi^2 - M_\pi^4 / m^2 = 3.98 \, M_\pi^2    \eqno(4.12)$$
very close to the physical threshold at $t_0 = 4 M_\pi^2$.  The isovector
form factors $F_{1,2}^V (t)$ inherit this logarithmic singularity
(branch point) on the second Riemann sheet. Actually, the same
phenomenon occurs in the scalar form factor of the
nucleon\footnote{$^*$}{In case of the nucleon scalar form factor, this
 singularity at $t_c$ stems from the partial wave amplitude $f^0_+ (t)$.}
 (for more
details, see ref.[4.10]).
Naturally, one asks the question whether this phenomenon shows up in the chiral
expansion.
Let us first consider  Im~$F_2^V (t)$ in the
relativistic formulation of baryon CHPT. Following Gasser et al. [4.2], one has
$${\rm Im} F_2^V (t) = {8 g_A^2 \over F_\pi^2} m^4 \biggl[ 4 \,
{\rm Im} \, \gamma_4
(t) + {\rm Im} \, \Gamma_4 (t) \biggr]             \eqno(4.13)$$
where the loop functions and their imaginary parts
$\gamma_4$ and $\Gamma_4$ are given in ref.[4.2]. For our purpose, we only need
Im~$\gamma_4 (t)$ since its threshold is the two--pion cut whereas Im~$\Gamma_4
(t)$ only starts to contribute at $t = 4 m^2$. The resulting imaginary part
for Im~$F_2^V (t) / t^2$ is shown in fig.4.2 (solid line). One sees that the
strong increase at threshold is reproduced
since the chiral representation of Im $\gamma_4 (t)$ indeed has the proper
analytical structure, i.e. the singularity on the second sheet at $t_c$ [4.2].
The chiral representation of Im~$F_2^V (t)/ t^2$ does not stay constant on the
left shoulder of the $\rho$--resonance but rather drops.
This is due to the fact that
in the one loop approximation, one is only sensitive to the first  term in
the chiral expansion of the pion charge form factor $F_\pi^V (t)$. Indeed, if
one sets $F_\pi (t) \equiv 1$ in eq.(4.11),
the one loop result reproduces nicely
the empirical curve (as discussed in more detail in ref.[4.7]).
This particular example shows
that in the relativistic version of baryon CHPT the pertinent analytical
structures of current and S--matrix elements  are given correctly.
\midinsert
\smallskip
\hskip 0.5in
\epsfxsize=3.5in
\epsfysize=3in
\epsffile{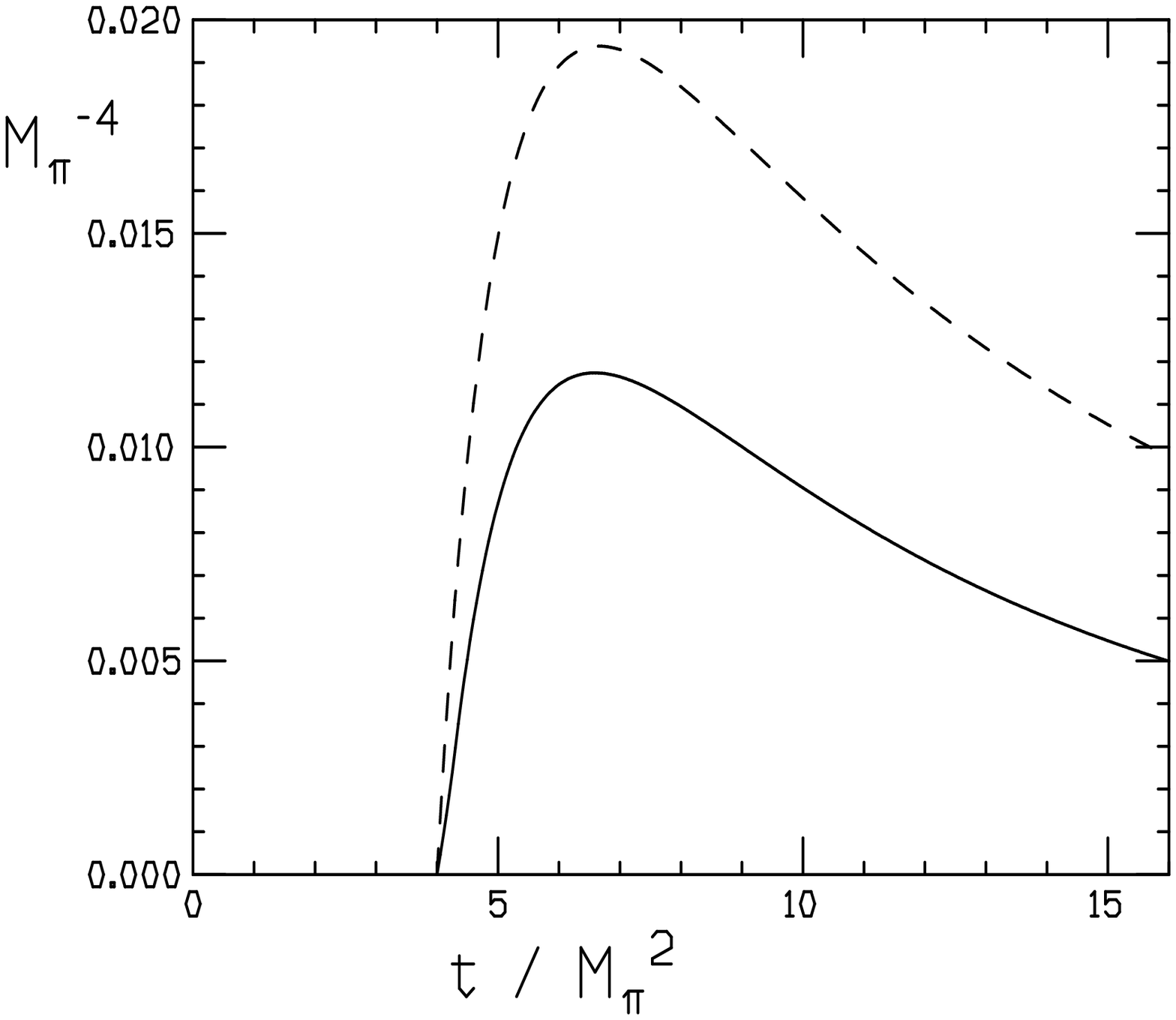}
\smallskip
\vskip -1.5truecm
{\noindent\narrower \it Fig.~4.2:\quad
Chiral representation  of Im~$F_2^V (t)/ t^2$ ($t$ in units of the pion mass)
in the relativistic
formulation of baryon CHPT [4.2] (solid line) and in the heavy mass approach
(dashed line) [4.3,4.6].
\smallskip}
\vskip -0.5truecm
\endinsert
Let us now turn to the heavy mass approach. The corresponding imaginary part
follows from ref.[4.3],
$${\rm Im} F_2^V (t) = {g_A^2 m \over 32 F_\pi^2 \sqrt{t} }
\biggl( t - 4 M_\pi^2 \biggr)  \, \,\Theta (t - 4 M_\pi^2 )   \eqno(4.14)$$
Here, the imaginary part behaves as $q_t^2$ close to threshold and not
as $q_t^3$ as demanded by (4.11). One furthermore finds that $F_2^V(t)$
goes like $\ln (2M_\pi - \sqrt{t})$ and
$t = 4 M_\pi^2$ is a logarithmic branch point in the heavy mass
approach. This incorrect analytic structure is an unavoidable
consequence of the heavy mass limit ($m = \infty$), in which the
square root branch point $t_0$ and the logarithmic branch point $t_c$
(on the second sheet) coincide. Nevertheless
this leads to an enhancement of the
imaginary part of $F_2^V (t) /t^2$ as shown by the dashed line in fig.4.2.
The enhancement is stronger than in the relativistic case and stronger than the
empirical one (for $F_\pi (t) \equiv 1$). In order to get the proper
separation of the singularities $t_0$ and $t_c$ one should therefore
perform an order $q^5$ calculation in
the heavy mass approach.

We also would like to discuss the imaginary part of the isovector Dirac
form factor $F_1^V (t)$. In the heavy mass approach, it reads
$$ {\rm Im} F_1^V(t) = {\sqrt{1-4M_\pi^2/t} \over 96 \pi F_\pi^2}
\biggl[ t - 4
M_\pi^2 + g_A^2 ( 5 t -8 M_\pi^2) \biggr] \eqno(4.15)$$
One can show, that this form is exactly the $m=\infty$
 limit of the corresponding
expression given in [4.2]. The imaginary part
 Im~$ F_1^V(t)$ in the heavy mass limit shows an
abnormal threshold behaviour, close to threshold it grows linear in $q_t$ and
not like $q_t^3$ as demanded by eq.(4.11).  At the moment we are not able to
give a precise explanation for this phenomenon, but certainly it must have to
do with the coalescence of the two singularities $t_0$ and $t_c$ and the
behaviour of the t-channel amplitudes $f^1_{\pm}$ in the heavy mass
 limit.
The main lesson to be learned from this investigation of the imaginary parts is
that the heavy mass formulation also has its own disadvantages. In the infinite
nucleon mass limit the analytical structure (poles and cuts) of certain
amplitudes may be disturbed and this may be a hindrance in order to make
contact to the dispersion theory.

Finally, a few words about the isoscalar form factors are in order. To one loop
accuracy, they are determined mostly by some contact terms. For example, the
isoscalar magnetic moment, $\kappa_S = \kappa_p + \kappa_n$ follows to be
$\krig{\kappa}_S = 2 c_7 +\krig{\kappa}_V $
(as defined in ${\cal L}_{\pi N}^{(2)}$ )
and the isoscalar charge radius $<r^2>_1^S$ is determined by the finite $\len$
constant $b_9 +
2 \tilde{b}_9$ (cf. eq.(3.23)). To access the three pion--cut, at which
the absorptive parts of the isoscalar form factors start, one has to perform a
two loop calculation. Such a two loop calculation
will also answer the yet unresolved question whether or not in the isoscalar
channel there is an enhancement around $t = 9 M_\pi^2$. State of the art
dispersion theoretical analyses of the nucleon form factors assume only a set
of poles in the corresponding spectral distributions [4.10]. Finally, we wish
to stress that in this context
the matching formalism discussed in ref.[4.3] starts to play a role (which
allows to relate matrix--elements in the heavy mass and relativistic
formulation of CHPT) since ultimately one might want to combine the chiral
constraints with dispersion theory.
\goodbreak \bigskip
\noindent{\bf IV.2. NUCLEON COMPTON SCATTERING}
\medskip
\goodbreak
 \bigskip \goodbreak
Low energy Compton scattering of the nucleon is particularly well suited
to investigate the structure of the nucleon since the electromagnetic probe in
the initial and in the final state is well understood. In this section, we will
first discuss the general formalism of Compton scattering and then elaborate on
the nucleon structure as encoded in the so--called electromagnetic (Compton)
polarizabilities ($\alpha, \beta$) and the spin-dependent polarizability
($\gamma$).
We will only consider  Compton scattering which allows for a unique
field--theoretical definition and empirical extraction of these quantities. We
eschew here the commonly used but theoretically uncertain non--relativistic
treatment of these nucleon structure constants.\footnote{*}{We thus drop the
overbar which is frequently used to denote the Compton pola-rizabilities.}

The T-matrix for the process $\gamma (k) + p(p) \to \gamma(k')  + p(p') $
 in the gauge $ \epsilon_0 = 0 =
\epsilon_0' $ for real photons , $ \epsilon\cdot k = 0 = \epsilon'\cdot k' $,
and in the centre--of--mass system $k_0 = k'_0 = \omega$ and
$t = (k - k')^2 = -2 \omega^2 ( 1 - \cos \theta )$ takes the form [4.11]
$$\eqalign{
T  = & e^2  \bigl\{ \svec \epsilon\,' \cdot \svec \epsilon \, A_1 + \svec
\epsilon\,' \cdot \svec k \,\, \svec \epsilon\cdot \svec k\,'  A_2  + i \svec
\sigma \cdot (\svec \epsilon\,' \times  \svec \epsilon)\, A_3  + i \svec
\sigma \cdot (\svec k\,' \times  \svec k)\,\svec \epsilon\,' \cdot\svec
\epsilon \, A_4 \cr &  +i \svec \sigma \cdot \bigl[ (\svec \epsilon\,' \times
\svec k)\,\svec \epsilon \cdot\svec k\,'- (\svec \epsilon \times  \svec
k\,')\,\svec \epsilon\,' \cdot\svec k \bigr]\, A_5   + i \svec \sigma \cdot
\bigl[ (\svec \epsilon\,' \times  \svec k\,')\,\svec \epsilon \cdot\svec k\,'-
(\svec \epsilon \times  \svec k)\,\svec \epsilon\,' \cdot\svec k \bigr] \, A_6
\bigr\} \cr } \eqno(4.16)$$
using the operator basis of ref.[4.12].
The $A_i$ are real below the pion production threshold, $\omega < M_\pi$.
{}From these, one can directly calculate physical obervables like the
unpolarized
differential cross section as well as a set of asymmetries for scattering
polarized photons on polarized protons.
The unpolarized differential cross section in the cm system is
$$\eqalign{  {d \sigma \over d\Omega_{\rm cm}} & = {\alpha^2 m \over m +
2 E_\gamma} \biggl\{ {1\over 2} A_1^2  ( 1 + \cos^2\theta) + {1\over2} A_3^2
(3-\cos^2\theta)  + \omega^2 \sin^2 \theta  \bigl[ 4 A_3 A_6 \cr & +( A_3 A_4 +
2 A_3 A_5 -A_1 A_2)\cos \theta\bigr] + \omega^4 \sin^2 \theta \bigl[{1\over 2}
A_2^2 \sin^2 \theta + {1\over 2}A_4^2(1+\cos^2 \theta)\cr & + A_5^2
(1+2\cos^2 \theta) +3A_6^2  + 2A_6 (A_4 + 3 A_5)\cos \theta + 2 A_4 A_5
\cos^2 \theta\bigr] \biggr\} \cr } \eqno(4.17)$$
 with $\alpha = e^2/4\pi$.
The asymmetry for scattering circular polarized photons on polarized protons
${\cal A}_\parallel$ (i.e. the proton spin parallel or antiparallel
to the photon direction $\svec k$) is given by
$$\eqalign{{\cal A}_\parallel= &  {d \sigma_{\uparrow\uparrow} \over
d\Omega_{\rm cm}} - {d
\sigma_{\uparrow\downarrow}\over d\Omega_{\rm cm}}  = {2\alpha^2 m \over m +
2 E_\gamma} \biggl\{- A_3^2  \sin^2\theta -  A_1 A_3 (1+\cos^2\theta)
 \cr & -\omega^2 \sin^2\theta \bigl[A_6(A_1+3A_3)  +(3 A_3 A_5 -A_1 A_5 + A_3
A_4  - A_2 A_3)\cos \theta\bigr] \cr & - \omega^4  \sin^2 \theta \bigl[A_5( A_2
- A_4)\sin^2 \theta +4 A_5 A_6 \cos\theta + 2 A_6^2 + 2 A_5^2 \cos^2
\theta)\bigr] \biggr\} \cr } \eqno(4.18)$$
Furthermore, we define the perpendicular asymmetry ${\cal A}_\perp$ by
considering right-circularly polarized photons moving in the z-direction
scattering on protons with their spin pointing in positive or
 negative x-direction,
$\svec k\,' = \omega(\sin\theta\cos \phi, \sin\theta\sin\phi, \cos \theta)$,
$$\eqalign{{\cal A}_\perp = &  {d \sigma_{\uparrow\rightarrow} \over
d\Omega_{\rm cm}} -  {d
\sigma_{\uparrow\leftarrow}\over d\Omega_{\rm cm}}  ={2\alpha^2 m \over m + 2
E_\gamma} \biggl\{A_3(A_3-A_1)\cos\theta  + \omega^2
\bigl[(A_1A_5+A_2A_3)\sin^2\theta \cr & +A_3 A_4(1+\cos^2\theta) +A_3
A_5(3\cos^2\theta-1)  + 2A_3 A_6\cos \theta\bigr] \cr & + \omega^4  \sin^2
\theta \bigl[A_6( A_2 + A_4-2A_5) + A_5 (A_2 -A_4 - 2A_5)\cos\theta\bigr]
 \biggr\}  \sin \theta \cos \phi\cr } \eqno(4.19)$$
$\phi$ is (in coordinate-free language) the azimuthal angle
(around the axis defined by the photon momentum) measured with respect to the
plane spanned by the photon momentum and the nucleon spin.
Clearly, if one changes the difference in eqs.(4.18,4.19) to a sum
 one gets in both case just
twice the unpolarized cross section. Letting the nucleon spin point in
y-direction results in $\cos \phi \to \sin \phi$ in (4.19).
If one uses left circular polarized photons instead of right circular polarized
ones, then both asymmetries eqs.(4.18,4.19) change sign.
One can also define a general asymmetry, which means
 right--circularly polarized photons moving in z-direction scatter on
polarized protons and the proton spin lies in the xz-plane inclining an angle
$\delta $ with the z-axis. We consider the difference of cross section for this
configuration and the one with reversed proton spin.  The corresponding
asymmetry reads
$$ {\cal A}(\delta) = \cos\delta \, {\cal A}_\parallel + \sin \delta \,
{\cal A}_\perp  \quad . \eqno(4.20)$$
$ {\cal A}(\delta)$  gives the asymmetry for the most
general spin alignment configuration.

In forward direction, the Compton scattering amplitude takes the form
$$ {1 \over 4 \pi} T (\omega ) = f_1 (\omega^2 ) \, {\svec {\epsilon}'}^* \cdot
 \svec {\epsilon} + i \, \omega \, f_2 (\omega^2 ) \, \svec \sigma \cdot
( {\svec {\epsilon}'}^* \times \svec {\epsilon} )  \eqno(4.21)$$
where the spin--nonflip ($f_1 (\omega )$) and the spin--flip ($f_2 (\omega )$)
amplitudes have the low energy expansions,
$$\eqalign{
f_1 (\omega^2 ) & = f_1 (0) +  (\alpha + \beta ) \omega^2 +
 {\cal O}(\omega^4 ) \cr
f_2 (\omega^2 ) & = f_2 (0) +  \gamma \omega^2 + {\cal O}(\omega^4 ) \cr}
\eqno(4.22)$$
in terms of the electric ($\alpha$), the magnetic ($\beta$)  and the so--called
"spin--dependent" ($\gamma $) polarizabilities. The Taylor coefficient $f_1
(0)$ is given by gauge invariance,
$$ f_1 (0) = - {e^2 Z^2 \over 4 \pi m}               \eqno(4.23)$$
which means that very soft photons only probe global properties like the charge
($Z$) and the mass $m$ of the spin--1/2 particle they scatter off. Eq.(4.23)
constitutes a venerable low--energy theorem (LET).
There exists also a LET for the
Taylor coefficient $f_2 (0)$ due to Low, Gell--Mann and Goldberger
[4.13]. Using gauge invariance, Lorentz invariance and crossing symmetry, they
proved that
$$ f_2 (0) = - {e^2 \kappa^2 \over 8 \pi m^2}               \eqno(4.24)$$
where $\kappa$ denotes the anomalous magnetic moment of the spin--1/2 particle
the photon scatters off. The nucleon structure first shows up in the the
next--to--leading order terms parametrized in terms of the various
polarizabilities. Using the optical theorem, one can derive the following
sum rules
$$\eqalign{
{\rm Re} \, f_1 (\omega ) & = -{e^2 Z^2 \over 4\pi m} -{\omega^2 \over 2 \pi^2}
\, {\cal P} \int_{\omega_0}^\infty d\omega' {\sigma_{\rm tot} (\omega ') \over
{\omega '}^2 - \omega^2} \cr
\alpha + \beta & = {1 \over 4 \pi^2} \int_{\omega_0}^\infty {d\omega \over
\omega^2} [ \sigma_+ (\omega ) + \sigma_- (\omega )] \cr
\gamma & = -{1 \over 4 \pi^2} \int_{\omega_0}^\infty {d\omega \over
\omega^3} [ \sigma_+ (\omega ) - \sigma_- (\omega )] \cr} \eqno(4.25)$$
with $\sigma_{\rm tot} (\omega) = (\sigma_+ (\omega)+\sigma_-(\omega))/2$
 the total photo--nucleon absorption cross
section and $\sigma_\pm (\omega )$ the photoabsorption cross section for
scattering circularly polarized photons on polarized nucleons for total $\gamma
N$ helicity 3/2 and 1/2, respectively. $\omega_0 = M_\pi + M_\pi^2 /2m$ is the
single pion production threshold. Assuming furthermore that the amplitude
$f_2 (\omega )$ satisfies an unsubtracted dispersion relation, Drell, Hearn and
Gerasimov (DHG) derived the sum rule
$$I(0)  =  \int_{\omega_0}^\infty {d\omega \over
\omega} [ \sigma_- (\omega ) - \sigma_+ (\omega )] = -{\pi e^2 \kappa^2 \over 2
m^2} \quad . \eqno(4.26)$$
One can generalize this to the case of virtual photons ($k^2 < 0)$ via
$$I(k^2)  =  \int_{\omega_0}^\infty {d\omega \over
\omega} [ \sigma_- (\omega ,k^2 ) - \sigma_+ (\omega ,k^2) ]
\eqno(4.27)$$
with $\omega_0 = M_\pi + (M_\pi^2 - k^2) /2m$. This extended DHG sum rule will
be discussed later on.
\medskip
We turn now to the calculation of the cms amplitudes in heavy baryon
CHPT (the polarizabilties to this order are discussed for the relativistic
approach in ref.[4.15] and for the heavy mass calculation in [4.3]).
For that, one has to consider nucleon-pole graphs ( expanded up to $1/m^2$),
$\pi^0$  - exchange and  pion loop diagrams. It is important to note that
to this order the only contact terms entering are the ones which generate the
anomalous magnetic moment. The prediction for the various nucleon
polarizabilities will therefore be given entirely in terms of lowest order
parameters. Consequently, one has a particularly sensitive test of the chiral
dynamics in the presence of nucleons. For the invariant functions $A_i $ one
finds (we only give the results for the proton)
$$\eqalign{
A_1  = -{1\over m} + {g_A^2 \over 8 \pi F_\pi^2} \biggl\{ &  M_\pi -
\sqrt{M_\pi^2 - \omega^2} + {2M_\pi^2 - t \over \sqrt{-t}}  \biggl[
{1\over 2} \arctan{\sqrt{-t} \over 2 M_\pi}  \cr & - \int_0^1 dz
\arctan{(1-z)\sqrt{-t}\over 2\sqrt{M_\pi^2 - \omega^2z^2}} \biggr] \biggr\}
\cr}  \eqno(4.28a)$$

$$
A_2  = {1 \over m \omega^2 } +
{g_A^2 \over 8 \pi F_\pi^2} {t- 2M_\pi^2 \over (-t)^{3/2}}\int_0^1 dz
\biggl[\arctan{(1-z)\sqrt{-t}\over 2\sqrt{M_\pi^2 - \omega^2z^2}} -
{2(1-z)\sqrt{t(\omega^2z^2-M_\pi^2)} \over 4M_\pi^2 - 4 \omega^2z^2
- t (1-z)^2} \biggr]\eqno(4.28b)$$

$$\eqalign{ A_3 = & {\omega \over 2m^2}\bigl[1 + 2 \kappa-(1+\kappa)^2 \cos
\theta\bigr] + {g_A t \omega\over 8\pi^2 F_\pi^2(M_\pi^2 -t)} + {g_A^2 \over 8
\pi^2 F_\pi^2} \biggl[ {M_\pi^2 \over \omega} \arcsin^2 {\omega\over M_\pi} -
\omega\biggr] \cr & + {g_A^2 \over 4 \pi^2 F_\pi^2 } \omega^4 \sin^2\theta
\int_0^1 dx \int_0^1 dz {x(1-x)z(1-z)^3\over W^3 } \biggl[ \arcsin
{\omega z\over R} + {\omega z W \over R^2} \biggr] \cr}\eqno(4.28c) $$

$$A_4 = - {(1+ \kappa)^2 \over 2m^2 \omega} + {g_A^2 \over 4 \pi^2 F_\pi^2}
\int_0^1 dx \int_0^1 dz {z(1-z)\over W} \arcsin{\omega z\over R} \eqno(4.28d)$$

$$\eqalign{ A_5 =  {(1+\kappa)^2 \over 2 m^2 \omega} & - {g_A \omega \over 8
\pi^2 F_\pi^2 (M_\pi^2 - t)}  + {g_A^2 \over 8 \pi^2 F_\pi^2} \int_0^1 dx
\int_0^1 dz \biggl[ - {(1-z)^2 \over W} \arcsin{\omega z \over R} \cr & + 2
\omega^2 \cos \theta {x(1-x) z (1-z)^3\over W^3} \biggl( \arcsin{\omega z
\over R} + {\omega z W \over R^2}\biggr) \biggr] \cr } \eqno(4.28e)$$

$$\eqalign{ A_6 =  - {1+\kappa \over 2 m^2 \omega} & + {g_A \omega \over 8
\pi^2 F_\pi^2 (M_\pi^2 - t)}  + {g_A^2 \over 8 \pi^2 F_\pi^2} \int_0^1 dx
\int_0^1 dz \biggl[{(1-z)^2 \over W} \arcsin{\omega z \over R} \cr & -2\omega^2
{x(1-x) z(1-z)^3\over W^3}\biggl( \arcsin{\omega z \over R} + {\omega z
W \over R^2}\biggr)\biggr] \cr } \eqno(4.28f)$$
with
$$ W = \sqrt{M_\pi^2 - \omega^2z^2 + t(1-z)^2 x(x-1)} , \quad R = \sqrt{
M_\pi^2 + t (1-z)^2 x(x-1)} \eqno(4.28g)$$
{}From this, one can read off the polarizabilties as [4.3,4.15]
$$\eqalign{
\alpha_p & = \alpha_n = 10 \beta_p = 10 \beta_n =
{5 e^2 g_A^2 \over 384 \pi^2 F_\pi^2 M_\pi}      \cr
\gamma_p & = \gamma_n = {e^2 g_A^2 \over 96 \pi^3 F_\pi^2 M_\pi^2} \cr}
\eqno(4.29)$$
since the isospin factors in the diagrams contributing to $\alpha$, $\beta$ and
$\gamma$ are the same for the proton and the neutron. Before discussing the
numerical results for the cross sections, asymmetries and polarizabilities, let
us compare to the recent enumeration of third--order spin polarizabilities by
Ragusa [4.16].
Denoting by $\bar A_i$ the Compton amplitudes with the Born terms subtracted,
we can identify
$$a_{1,1} + a_{1,2} + a_3(0) = {e^2\over 8 \pi} {\partial^2\over \partial
\omega^2} \bar A_1(0,0) = \alpha_p + \beta_p = {11e^2 g_A^2\over 768\pi^2
F_\pi^2 M_\pi} \eqno(4.30a)$$
$$a_{1,1} + a_3(0) = {e^2\over 2 \pi} {\partial\over \partial t } \bar A_1(0,0)
= \beta_p = {e^2 g_A^2\over 768\pi^2 F_\pi^2 M_\pi} \eqno(4.30b)$$
$$- a_3(0) = {e^2\over 4 \pi} \bar A_2(0,0) = -\beta_p = -{e^2 g_A^2\over
768\pi^2 F_\pi^2 M_\pi} \eqno(4.30c)$$
 $$a_{2,2} + \gamma_1- \gamma_2-2\gamma_4 = {e^2\over 24 \pi} {\partial^3\over
 \partial \omega^3} \bar A_3(0,0) = \gamma = {e^2 g_A^2\over
 96\pi^3 F_\pi^2 M^2_\pi} \eqno(4.30d)$$
$$a_{2,2} -\gamma_2 - 2 \gamma_4 = {e^2\over 2 \pi} {\partial^2\over \partial
\omega \partial t} \bar A_3(0,0) =  {e^2 g_A\over 16\pi^3 F_\pi^2
M^2_\pi}\eqno(4.30e)$$
$$\gamma_3 = {e^2\over 4\pi} {\partial\over \partial\omega} \bar A_6(0,0)  =
{e^2 g_A(12+g_A)\over 384\pi^3 F_\pi^2 M^2_\pi} \eqno(4.30f)$$
$$\gamma_2 = {e^2\over 4\pi} {\partial\over \partial\omega} \bar A_4(0,0)  =
{e^2 g_A^2\over 192\pi^3 F_\pi^2 M^2_\pi} \eqno(4.30g)$$
$$\gamma_4 = {e^2\over 4\pi} {\partial\over \partial\omega} \bar A_5(0,0)  =
-{e^2 g_A(12+g_A)\over 384\pi^3 F_\pi^2 M^2_\pi} \eqno(4.30h)$$
\medskip
\noindent Of particular interest
is the so-called Compton amplitude $f^{1-}_{EE}$
recently studied in [4.89] (and references therein)  which displays a strong
unitarity cusp. We like to discuss it briefly here. The set of Compton
functions introduced in (4.16) is complete, consequently one can project out
$f^{1-}_{EE}$ via

$$
\eqalign{f^{1-}_{EE}(\omega) & = {e^2 m \over 32 \pi
(\omega+\sqrt{m^2+\omega^2})}
\int_{-1}^{+1} d\xi \cr & \times
\bigl[ (A_1-A_3)(1+\xi^2) +\omega^2(2A_5+A_4 -A_2)
\xi(1-\xi^2) \bigr] \cr}  \eqno(4.31)$$
The soft photon limit in this case is $f^{1-}_{EE}(0) = -e^2/(12 \pi m)$.
Evidently, the one-loop representation has a branch point at $\omega = M_\pi$
and therefore also a cusp. The numerical investigation indeed shows this cusp
but it turns out to be  rather weak at one-loop order ${\cal O}(q^3)$.

We now discuss the numerical results based on the one--loop (order $q^3$)
invariant amplitudes, eqs.(4.28). First, if we set $g_A = 0$, only the nucleon
Born graphs expanded in powers of $1/m$ contribute. For energies below the
pion production threshold, the corresponding differential cross section is
within a few percent of the exact Powell cross section [4.17],
$$\eqalign{
{d \sigma \over d\Omega_{\rm lab}}\biggr|_{\rm Powell} & = {1 \over 2}
\biggl( {\alpha E_\gamma ' \over m E_\gamma} \biggr) \biggl\lbrace {E_\gamma
\over E_\gamma '} + { E_\gamma ' \over E_\gamma} - \sin^2 \theta_{\rm L} +
{2 \kappa E_\gamma E_\gamma ' \over m^2} (1 - \cos \theta_{\rm L} )^2 \cr
& + { \kappa^2 E_\gamma E_\gamma ' \over m^2}
 \bigl[ 4(1 - \cos \theta_{\rm L} )
+{1 \over 2} (1 - \cos \theta_{\rm L} )^2 \bigr] \cr
& +
{ \kappa^3 E_\gamma E_\gamma ' \over m^2} \bigl[ 2(1 - \cos \theta_{\rm L} )
+ \sin^2 \theta_{\rm L}  \bigr] + { \kappa^4 E_\gamma E_\gamma ' \over 2 m^2}
(1 + {1 \over 2} \sin^2 \theta_{\rm L} ) \biggr\rbrace \cr
E_\gamma ' &
= {E_\gamma \over 1 + {E_\gamma \over m} ( 1 - \cos \theta_{\rm L})}
\cr}  \eqno(4.32)$$
where the scattering angle $\theta_{\rm L}$ and the photon energy $E_\gamma$ in
the laboratory  system
 are related to the cms quantities $\theta$ and $\omega$ via
$\cos \theta = ( m + (m+ E_\gamma)(\cos \theta_{\rm L}-1)) / (m + E_\gamma (1 -
\cos \theta_{\rm L} ))$ and $ \omega = E_\gamma / \sqrt{ 1 + 2 E_\gamma / m}$,
respectively.
\midinsert
\smallskip
\hskip 0.5in
\epsfxsize=3in
\epsfysize=4in
\epsffile{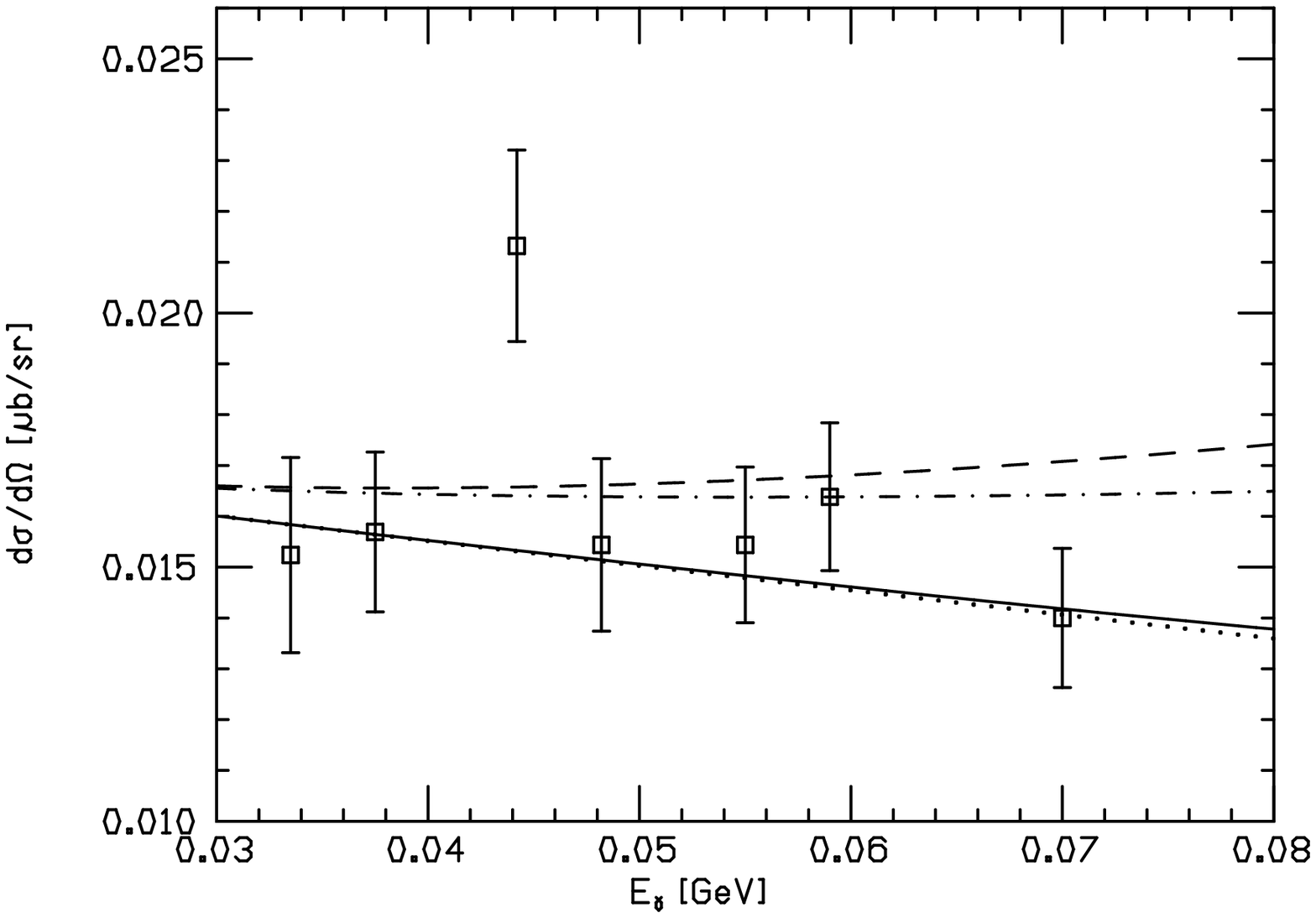}
\smallskip
\vskip -1.5truecm
{\noindent\narrower \it Fig.~4.3:\quad
The unpolarized data from ref.[4.18] in comparison to the chiral expansion of
the Compton amplitude for $\theta_{\rm L} = 135^\circ$.
The dashed, dash--dotted and solid lines
represent the Born, Born + $\pi^0$--exchange and Born + $\pi^0$--exchange +
loop results. If one approximates the loop contribution by the electromagnetic
polarizabilities as described in eq.(4.33), the dotted line results.
\smallskip}
\vskip -0.5truecm
\endinsert
 The chiral expansion to order $q^3$ (solid line)
 reproduces well the differential cross section data of
Federspiel et al.[4.18] as shown in fig.4.3. (the corresponding results for
$\theta_{\rm L} =60^\circ$ look very similar). In this figure it is also shown
that the Born graphs together with the $\pi^0$--exchange contribution are not
sufficient to describe the data.  However, as indicated by the dotted line
in fig.4.3, one is not sensitive
to nucleon structure effects beyond  the electromagnetic polarizabilities
(the $1/M_\pi$-terms). In this latter case, the loop contributions to the
$A_i$ are given by
$$A_1^{\rm loop} = {5  g_A^2 \over 96 \pi F_\pi^2 M_\pi} \omega^2 \bigl( 1
+ {1 \over 10} \cos \theta \bigr) , \, \,
A_2^{\rm loop} = -{  g_A^2 \over 192 \pi F_\pi^2 M_\pi} , \, \,
A_{3,4,5,6}^{\rm loop} = 0 \, \, .  \eqno(4.33)$$
We will discuss these polarizabilities in more detail later on. If one adds the
contribution from static $\Delta$ exchange (which starts at order $q^4$)
the corrections are not dramatic. In any case, for a truely
meaningful comparison one would have to take into account a
 host of other $q^4$  terms.
The parallel asymmetry generically changes sign around 90 degrees
 from negative to
positive values as shown in fig.4.4 for $E_\gamma = 70$ MeV.
This is not at all evident from eq.(4.18)  since there is no overall $\cos
\theta$--factor. For the same photon energy, we also show the perpendicular
asymmetry in fig.4.4. In both cases, the effects from the nucleon structure
encoded in the loop contributions of the $A_i$ are small.
Only for energies $\omega >100$ MeV one is somewhat sensitive to these
structure terms. From an experimental point of view, only an extremely precise
measurement of such asymmetries could shed light on the nucleon structure. An
accuracy as for the unpolarized case [4.17] is certainly insufficient. If one
trusts the $q^3$ approximation up to the pion production threshold, one finds
agreement with the few Saskatoon data [4.18] in this range.
In this  energy range,
loop effects are more significant and could be detected experimentally.
However, the competing contribution from the $\Delta$ (and possible other
$q^4$ effects) becomes appreciable and makes the analysis of such data much
less clear--cut.
\midinsert
\smallskip
\hskip 1in
\epsfxsize=3in
\epsfysize=4.5in
\epsffile{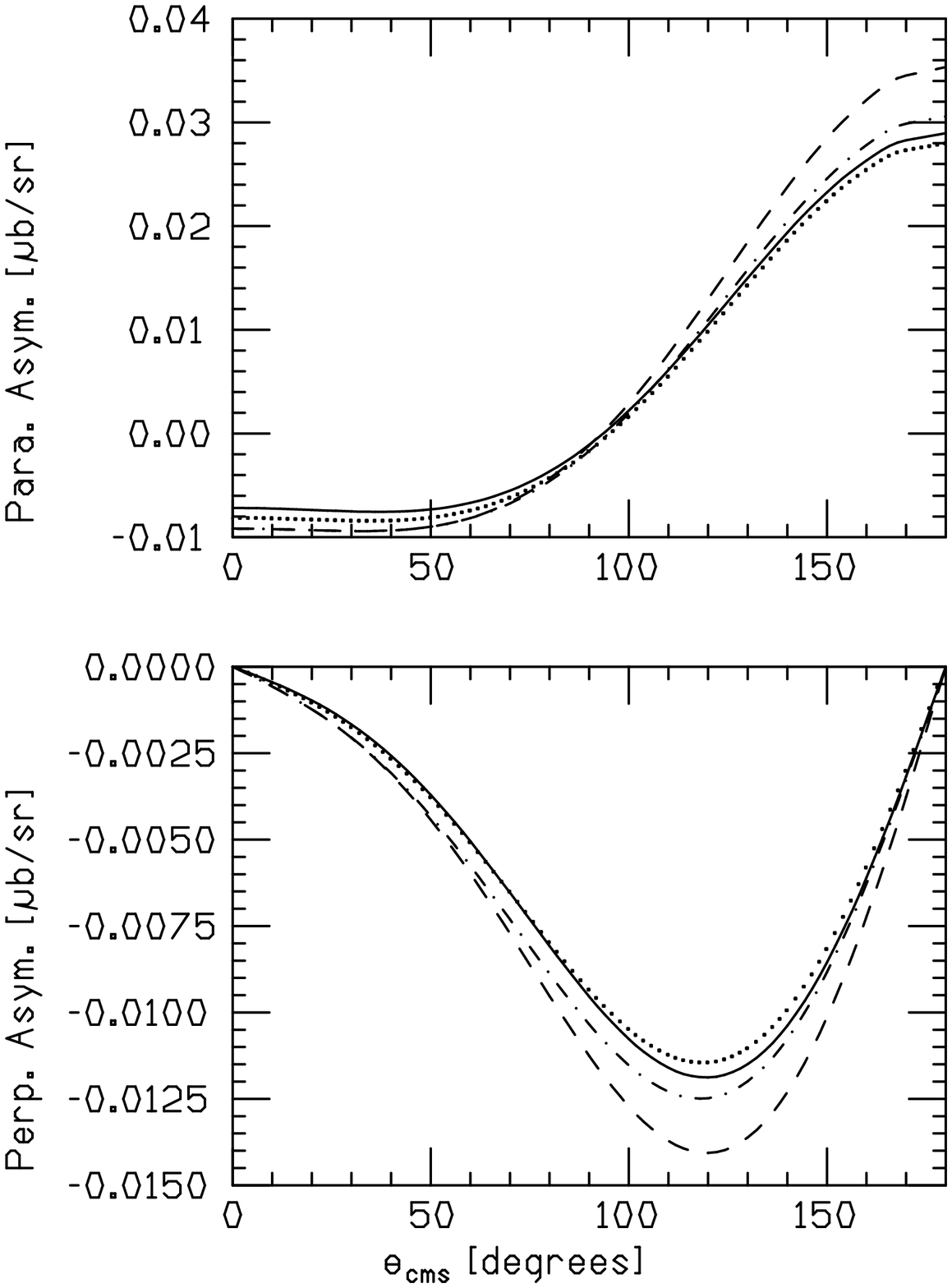}
\smallskip
{\noindent\narrower \it Fig.~4.4:\quad
The parallel and perpendicular asymmetries for $E_\gamma = 70$ MeV. For
notations,see fig.4.3. In case of ${\cal A}_\perp$ we have set $\cos \phi = 1$.
\smallskip}
\vskip -0.5truecm
\endinsert
Let us now take a closer look at the electromagnetic polarizabilities of the
proton and the neutron. These have been rather accurately  determined over
the last years. For the proton, if one combines the Illinois [4.18], Mainz
[4.20] and Saskatoon [4.19] measurements together with the dispersion sum
rule, $( \alpha+  \beta)_p = (14.2 \pm 0.3) \cdot 10^{-4}$ fm$^3$
[4.21], one has
$$ \alpha_p = (10.4 \pm 0.6)
\cdot 10^{-4} {\rm fm}^3 \, , \quad \beta_p   = (3.8 \mp 0.6)
\cdot 10^{-4} {\rm fm}^3  \qquad . \eqno(4.34a)$$
Similarly, the dispersion sum rule
$( \alpha+ \beta)_n = (15.8 \pm 0.5) \cdot 10^{-4}$
fm$^3$\footnote{*}{Notice
that the uncertainty on the sum rule value for the neutron
is presumably underestimated since one has to use
deuteron data to extract the photon-neutron cross section.} [4.22]
together with the recent Oak Ridge [4.23] and Mainz [4.24] measurements leads
to
$$  \alpha_n  = (12.3  \pm 1.3) \cdot 10^{-4} {\rm fm}^3 \, , \quad
\beta_n   = (3.5 \mp 1.3) \cdot 10^{-4} {\rm fm}^3 \qquad .
\eqno(4.34b)$$
Notice that we have added the systematic and statistical errors of
the empirical determinations in quadrature. The salient features of these
experimental results are that both the proton and the neutron behave
essentially as induced electric dipoles and that their respective sums of the
electric and magnetic polarizabilities are almost the same. We should also
point out that due to the strong magnetic M1 $N \Delta$ transition,  one
naively expects a large contribution from the $\Delta$ resonance to the
magnetic polarizabilities. These empirical features are already well
represented by the lowest order ($q^3$) results (4.29), i.e. $\alpha_p =
\alpha_n = 13.6 \cdot 10^{-4} {\rm fm}^3$ and $\beta_p = \beta_n = 1.4
\cdot 10^{-4} {\rm fm}^3$. It is also worth to stress that the
electromagnetic polarizabilities explode as $1/M$ in the chiral limit since
the photon sees the long--ranged pion cloud (compare the chiral limit behaviour
of the isovector formfactors discussed in section 4.1). However, at this order
one has no isospin breaking (if one works in flavor SU(2)) and solely nucleon
intermediate states can contribute in the pertinent graphs. In ref.[4.25],
a systematic analysis of {\it all} ${\cal O}(q^4)$ effects was presented. In
addition to the one--loop diagrams with insertions from ${\cal L}_{\pi
N}^{(1)}$, one also has to include one loop graphs with exactly one insertion
from ${\cal L}_{\pi N}^{(2)}$ and contact terms from ${\cal L}_{\pi N}^{(4)}$,
$$\eqalign{
{\cal L}_{\pi N}^{(4)} & =   {\pi \over 4 e^2}
 (\delta \beta_p - \delta  \beta_n)
\bar H f_{\mu \nu}^+ f^{\mu \nu}_+ H
                        + {\pi \over 4 e^2} \delta  \beta_n
                        \bar H  H \Tr f_{\mu \nu}^+ f^{\mu \nu}_+  \cr
                       &+ {\pi \over 2 e^2} (\delta  \alpha_n + \delta \beta_n
 - \delta  \alpha_p - \delta \beta_p)
    \bar H f_{\mu \nu}^+ f^{\lambda \nu}_+ H v^\mu v_\lambda  \cr
                       &- {\pi \over 2 e^2} (\delta  \alpha_n + \delta \beta_n)
    \bar H H \Tr(f_{\mu \nu}^+ f^{\lambda \nu}_+ ) v^\mu v_\lambda \cr}
 \eqno(4.35)$$
 The unknown coefficients we have to determine
 are the four low-energy constants $\delta
{\alpha}_p$, $\delta {\alpha}_n$, $\delta {\beta}_p$ and $\delta
{\beta}_n$ (see below). The contact terms of order $q^2$ entering ($c_{1,2,3}$)
have already been determined (see section 3).
{}From that, one derives the following formulae for the electric and magnetic
polarizabilities of the proton and the neutron ($i = p,n$)
$$\eqalign{
 {\alpha}_i&= {5 C g_A^2 \over 4 M_\pi} + {C \over \pi} \biggl[ \bigl( {x_i
g_A^2 \over m} - c_2 \bigr) \ln {M_\pi \over \lambda} + {1 \over 4} \bigl(
{y_i g_A^2
\over 2 m} - 6 c_2 + c^+ \bigr) \biggr] +\delta
{\alpha}_i^r(\lambda) \, ,   \cr
  {\beta}_i&= { C g_A^2 \over  8 M_\pi} + {C \over \pi}  \biggl[ \bigl( {3
x_i' g_A^2 \over m} - c_2 \bigr) \ln {M_\pi \over \lambda} + {1 \over 4}
\bigl( {y_i' g_A^2
\over  m} + 2 c_2 - c^+ \bigr) \biggr] + \delta {\beta}_i^r (\lambda)
\, .  \cr} \eqno(4.36)$$
with
$$\eqalign{
 C  &= {e^2   \over 96 \pi^2 F_\pi^2}\, \, = 4.36 \cdot 10^{-4} \, {\rm fm}^2
\, , \cr
  x_p&= 9 \, , \quad x_n = 3 \, , \quad y_p = 71 \, , \quad y_n = 39 \, ,
\cr
  x_p'&= 3 + \kappa_s \, , \quad x_n' = 1 - \kappa_s \, , \quad y_p' = {37
\over
2} + 6 \kappa_s \, , \quad y_n' = {13 \over 2} - 6 \kappa_s \, , \cr
  c^+&= -8 c_1 + 4 c_2 + 4 c_3 - {g_A^2 \over 2 m} \, \, \, . \cr}
\eqno(4.37)$$
At this order, the loop contributions to the polarizabilities contain
divergences which can be absorbed in the $q^4$ counter terms.
The corresponding renormalization prescription reads:
$$\delta  \alpha_i =  {e^2 L \over 6\pi F_\pi^2} \biggl( c_2 -{x_i g_A^2
\over m} \biggr) + \delta  \alpha^r_i(\lambda), \qquad
\delta  \beta_i =  {e^2 L \over 6\pi F_\pi^2} \biggl( c_2 -{3x'_i g_A^2
\over m} \biggr) + \delta  \beta^r_i(\lambda) \eqno(4.38)$$
The results shown in eqs.(4.37) have the following structure.
Besides the leading $1/M_\pi$ term [4.3,4.15],
${\cal O}(q^4)$ contributions from the loops have a $\ln M_\pi$ and a constant
piece $\sim M_\pi^0$. As a check one can recover the coefficient of the
$\ln M_\pi$ term form the relativistic calculation [14.15] if one sets
the new low energy constants $c_i$ and $\kappa_{s,v} = 0$. In that
case only the $1/m$ corrections of the relativistic Dirac formulation
are treated and one necessarily reproduces the corresponding
non--analytic (logarithmic) term of this approach.
The term proportional to $c_2 \, \ln M_\pi$ in eqs.(4.37) represents the
effect
of (pion) loops with intermediate $\Delta (1232)$ states [4.26] consistently
truncated at order $q^4$. We should stress that the decomposition of
the loop and counter term pieces at order $q^4$ has, of course, no deeper
physical meaning but will serve us to separate the uncertainties
stemming from the coefficients accompanying the various contact terms.
Notice that from now on we will omit the superscript '$r$' on
$\delta  \alpha_i^r(\lambda)$ and $\delta  \beta_i^r(\lambda)$
appearing in eqs.(4.36,4.38). The estimation of these low energy constants is
discussed in great detail in ref.[4.27].
The $\Delta(1232)$ enters prominently in the determination of the four
low-energy constants  from ${\cal L}_{\pi N}^{(4)}$.
Therefore, we will determine these coefficients at the scale
$\lambda = m_\Delta$ (this induces some spurious scale--dependence since
we do not treat the remainders as e.g. in eq.(2.52)).
In particular, one gets
a sizeable contribution to the magnetic polarizabilities due to the strong
$N\Delta$ M1 transition.
A crude estimate of this has been given in ref.[4.28]
by integrating the M1 part of the total photoproduction cross section for
single pion photoproduction over the resonance region.
However, this number is afflicted with a large uncertainty.
If one simply uses the  Born
diagrams with an intermediate point--like
$\Delta$, one finds a result which strongly depends
on the strength of the $\gamma N \Delta$ coupling and on the off--shell
parameter $Y$ which is related to
the electromagnetic interaction ${\cal L}^1_{\gamma
N \Delta}$ (we also give ${\cal L}^2_{\gamma N \Delta}$ used below) [4.32]
$$\eqalign{
{\cal L}^1_{\gamma N \Delta} & =
{i e g_1 \over 2 m} \bar{\Delta}^\mu \Theta_{\mu \nu}(Y) \gamma_\nu \gamma_5
T_3 \Psi F^{\nu \lambda} + {\rm h.c.} \cr
{\cal L}^2_{\gamma N \Delta} & =
-{e g_2 \over 4 m^2} \bar{\Delta}^\mu \Theta_{\mu \nu}(X)  \gamma_5
T_3 (\partial_\lambda \Psi) F^{\nu \lambda} + {\rm h.c.} \cr
\Theta_{\mu \nu}(I) & = g_{\mu \nu} + \bigl[ {1 \over 2} (1 + 4 I) A + I \bigr]
\, \, , \quad I = X,Y \cr} \eqno(4.39)$$
where $\Delta_\mu$ denotes the  Rarita--Schwinger (spin--3/2) spinor, the $T's$
are the isospin $1/2 \to 3/2$ transition operators. The parameter $A$ does not
appear in any physical observable and can therefore be chosen to be $A=-1$.
 These parameters ($g_1 , g_2, X, Y$) are not very well determined.
A conservative estimate therefore is
$$\delta {\beta}_p^\Delta (m_\Delta) =
\delta {\beta}_n^\Delta (m_\Delta) \simeq (7.0 \pm 7.0) \cdot 10^{-4}
\, {\rm fm}^3  \eqno(4.40)$$
invoking isospin symmetry. Clearly, the large range in the
value for $\delta {\beta}^\Delta$
is unsatisfactory and induces a major uncertainty in
the determination of the corresponding counterterms.  We choose the
central value of eq.(4.40) as our best determination [4.25,4.27].
In ref.[4.29], the $\Delta(1232)$ was included in the effective field
theory as a dynamical degree of freedom and treated non--relativistically (like
the nucleon). There, it was argued that
the $\Delta$ Born graphs have to be calculated at
the off--shell point $\omega = 0$. This effect can
reduce the large $\delta {\beta}^\Delta$
 by almost an order of magnitude. This is reminiscent of the off--shell
dependence discussed before. Furthermore, as already pointed out in ref.[4.15],
the  relativistic treatment of the $\Delta(1232)$ also induces a finite
electric polarizability at order $q^4$. This contribution depends strongly
on the $\gamma \Delta N$ couplings $g_1$ and $g_2$ as well as the two
off--shell parameters $ X, Y$, cf. eq.(4.39). We thus  assign
an uncertainty of $\pm 2 \cdot 10^{-4}$ fm$^3$  to
the theoretical predictions for the electric polarizabilities.
Another contribution to the coefficients $\delta {\alpha}_i(\lambda)$ and
$\delta {\beta}_i(\lambda)$ comes from loops involving charged kaons [4.30].
Since
we are working in SU(2), the kaons and etas are frozen out and effectively
give some finite contact terms. These have been estimated in refs.[4.25,4.27].
One finds
$\delta {\alpha}_p^K (m_\Delta) = 1.31 \cdot 10^{-4}\, {\rm fm}^3$ and
$\delta {\alpha}_n^K (m_\Delta) = 0.13 \cdot 10^{-4}\, {\rm fm}^3$.
The corresponding numbers for the kaon contributions to the magnetic
polarizabilities are a factor $0.12$ smaller.
These values might, however, considerably overestimate the
kaon loop contribution. Integrating e.g. the data from ref.[4.31] for $\gamma
p \to K \Lambda,K \Sigma^0$, one gets a much smaller contribution since
the typical cross sections are of the order of a few $\mu$barn. This
points towards the importance of a better understanding of SU(3)
breaking effects. At present,  this discrepancy remains to be resolved.
Adding the various theoretical uncertainties in quadrature, one ends up with
$$\eqalign{
{\alpha}_p &= (10.5 \pm 2.0) \cdot 10^{-4} {\rm fm}^3 \, \quad
{\beta}_p = (3.5 \pm 3.6) \cdot 10^{-4} {\rm fm}^3 \, \cr
{\alpha}_n &= (13.4 \pm 1.5) \cdot 10^{-4} {\rm fm}^3 \, \quad
{\beta}_n = (7.8 \pm 3.6) \cdot 10^{-4} {\rm fm}^3 \, \cr}
\eqno(4.41)$$
which with the exception of ${\beta}_n$
agree well with the empirical data (4.33,4.34). The important new effect
is that the loops of order $q^4$  generate a $\ln \,M_\pi$ term with a
large coefficient (for $\beta_p$) which cancels most of the big contribution
from the $\Delta$ encoded in the coefficients of the ${\cal L}_{\pi N}^{(4)}$
contact terms. In case of the neutron, the coefficient in front of the $\ln \,
M_\pi$--term is smaller. This points towards the possible importance of
isospin--breaking in the $p \Delta \gamma$ and $n \Delta \gamma$ couplings or
in the off--shell parameter $Y$
(for which at present we have no empirical indication).
Clearly, an independent determination of the electric and magnetic
nucleon polarizabilities would be needed to further tighten the
empirical bounds on these fundamental quantities. This was also
stressed in ref.[4.29]. It is worth to point out that the uncertainties
given in (4.41) do not include effects of two (and higher) loops which
start out at order $q^5$. We do not expect these to alter the
prediction for the electric polarizabilities significantly [4.25].
Notice also that at present the theoretical
uncertainties are larger than the experimental ones (if one imposes
the sum rules for $( \alpha +  \beta)$).
That there is more spread in
the empirical numbers when the dispersion sum rules are not used
can e.g. be seen in the paper of Federspiel et al. in ref.[4.18]. The role
of dispersion theory and its interplay to the chiral perturbation theory
results is dicussed in refs.[4.33]

In ref.[4.27], the spin--averaged forward Compton amplitude $A_{p,n} (\omega )
= - 4 \pi f_1 (\omega )$ was compared with the available data [4.21,4.34].
To lowest order $q^3$ in the chiral expansion, the expressions for
$A_{p,n}(\omega)$ diverge at $\omega = M_\pi$.
This is an artefact of the heavy mass expansion. The realistic branch point
coincides with the opening of the one--pion channel as given after eq.(4.25).
 To cure this, one introduces the variable
$\zeta = {z \over 1 + M_\pi/2m} = {\omega \over M_\pi ( 1 + M_\pi /2m )}  =
{\omega \over \omega_0}$.
In terms of $\zeta$, the branch point
sits at its proper location and $A_{p,n} (\zeta = 1)$ is finite. We have
$$\eqalign{A_{p,n} (\omega) & =  {e^2 \over 2m} (1 \pm 1)
- 4 \pi ( {\alpha}+ {\beta})_{p,n} \omega^2 \cr
& + {e^2 g_A^2 M_\pi \over 8 \pi F_\pi^2} \biggl\lbrace -{3 \over 2} -{1 \over
 \zeta^2} + \bigl(1+{1 \over \zeta^2}\bigr) \sqrt{1 - \zeta^2} +
{1 \over \zeta} \arcsin\zeta + {11 \over 24} \zeta^2 \biggr\rbrace \cr
 & + {e^2 g_A^2 M_\pi^2 \over 8 \pi^2 m F_\pi^2} \biggl\lbrace -1 +{10\over 3}
\zeta^2 + \biggl[{1 \over \zeta} - 4 \zeta +
{\zeta \over 1 - \zeta^2}\biggr] \sqrt{1 - \zeta^2} \arcsin\zeta \cr
 & + \pi \biggl[ {1 \over \zeta^2}-{1 \over 2 \zeta} \arcsin \zeta
+ {11 \over 24}\zeta^2 - {(1-\zeta^2)^2 + 1 \over 2 \zeta^2 \sqrt{1-\zeta^2}}
\biggr]  \pm \biggl[ -{3 \over 2} + \zeta^2\bigl( {3\over 2} \kappa_s +
{13\over 6} \bigr) \cr & + \bigl({1 \over \zeta} -(2+\kappa_s)
\zeta\bigr) \sqrt{1 -\zeta^2} \arcsin\zeta + {1 \over 2} \bigl({1 \over
\zeta^2}-\kappa_s \bigr) \arcsin^2\zeta \biggr] \biggr\rbrace \cr}
\eqno(4.42)$$
where the '+/-' sign refers to the proton/neutron, respectively. The proper
analytic continuation above the branch point $\zeta = 1$ is obtained through
the substitutions $\sqrt{1-\zeta^2} = -i \sqrt{\zeta^2 -1} $ and $\arcsin \zeta
 = \pi/2 + i \ln(\zeta + \sqrt{\zeta^2-1}) $.
We should stress that in the relativistic formulation of baryon CHPT such
problems concerning the branch point do not occur [4.15].
In the heavy mass
formulation one encounters this problem since the branch point $\omega_0$
itself has an expansion in $1/m$ and is thus different in CHPT at order $q^3$
and $q^4$. As shown in ref.[4.27],
the spin--averaged forward Compton amplitude for the proton is in
agreement with the data up to photon energies $\omega \simeq M_\pi$.
It is dominated by the quadratic contribution, i.e. to order $q^4$
in the chiral expansion the terms of order $\omega^4$ (and higher)
are small. Similar trends are found for the neutron with the exception
of a too strong curvature at the origin.
For the proton, the real and imaginary parts of $A(\omega)$ for $\zeta > 1$
have also been calculated. The imaginary part starts out negative
as it should but becomes positive at $\omega \simeq 180$ MeV. This is
due to the truncation of the chiral expansion and can only be overcome
by a more accurate higher order calculation.
\medskip
The spin--dependent polarizability $\gamma$ has not yet been measured. To
lowest order in the chiral expansion, $\gamma$ explodes like $1/M^2$ in the
chiral limit [4.3],
$$\gamma_p = \gamma_n = {e^2 g_A^2 \over 96 \pi^3 F_\pi^2 M_\pi^2} = 4.4 \cdot
10^{-4} \, {\rm fm}^4 \qquad. \eqno(4.43)$$
The $q^4$ and $q^5$ corrections to this
result have not yet been investigated in a systematic fashion. In ref.[4.3],
the contribution from the $\Delta$ was added (which starts at order $q^5$)
using the off--shell parameters of ref.[4.35] leading to $\gamma^\Delta_{p,n}=
- 3.66 \cdot 10^{-4} \, {\rm fm}^4$, so that
$$\eqalign{\gamma_p & = \gamma_p^{\rm 1-loop} + \gamma_p^\Delta =
-1.50  \cdot 10^{-4} \, {\rm fm}^4 \, , \cr
  \gamma_n & = \gamma_n^{\rm 1-loop} + \gamma_n^\Delta =
-0.46  \cdot 10^{-4} \, {\rm fm}^4 \, , \cr}    \eqno(4.44)$$
which is rather different from the lowest order result, eq.(4.43). One can get
an estimate on the empirical values by use of the dispersion relation (4.25).
Using the latest pion photoproduction multipoles from the SAID data basis, one
arrives at [4.36]
$$\gamma_p = -1.34  \cdot 10^{-4} \, {\rm fm}^4 \, , \quad
  \gamma_n = -0.38  \cdot 10^{-4} \, {\rm fm}^4 \quad . \eqno(4.45)$$
The numbers given in (4.45) differ from the ones in ref.[4.3] because there an
older version of the multipoles from the SAID program was used. The theoretical
estimates (4.44) are amazingly close to the empirical ones, eq.(4.45). In
ref.[4.3], the spin--dependent polarizabilities were also calculated in the
relativistic approach. In that case, even on the level of flavor SU(2), one
finds some isospin--breaking from the one--loop diagrams.

Finally, we turn to a short discussion of the generalized DHG sum rule (4.27).
Models for the photoabsorption cross sections [4.37,4.38,4.39] seem to indicate
the validity of the DHG sum rule (4.26) (on the qualitative level). The direct
experimental test of this sum rule has not yet been performed. Furthermore,
 the recent EMC
measurements in the scaling region $|k^2 | \simeq 10$ GeV$^2$ indicate that the
sum rule behaves a $1 / k^2$ as $|k^2|$ becomes large and that the sign is
opposite to the value at the photon point, $k^2 = 0$. In ref.[4.41], baryon
CHPT was used to investigate the slope of $I(k^2)$ around the origin. It was
found to be negative and of similar size to the recently proposed value by
Soffer and Teryaev [4.42], but opposite to the one of ref.[4.39] (which is due
to the $\Delta$ contribution). At present, experimental data as well as more
detailed theoretical investigation are lacking and we refer the interested
reader to [4.41] for more details.
 \bigskip \goodbreak
\noindent{\bf IV.3. AXIAL PROPERTIES OF THE NUCLEON}
\medskip
\goodbreak
In the previous sections, we were concerned with the coupling of
photons to the nucleon, i.e. pure electromagnetic processes. Within the
framework of the standard model, there also axial currents which can be used as
probes. The structure of the nucleon
 as probed by  charged axial currents is encoded in
two form factors, the axial and the induced pseudoscalar ones. To be specific,
consider the matrix--element of the isovector axial quark
current, $A_\mu^a = \bar{q} \gamma_\mu \gamma_5 (\tau^a / 2) q$,
 between nucleon states
$$<N(p')|\, A_\mu^a \, |N(p)> = \bar{u}(p') \biggl[ \gamma_\mu \, G_A (t) +
{(p' -p)_\mu \over2m} \, G_P(t) \biggr] \gamma_5 {\tau^a \over 2} u(p)
\eqno(4.46)$$
with $t=(p'-p)^2$ the invariant momentum transfer squared.
The form
of eq.(4.46) follows from  Lorentz invariance, isospin conservation and the
discrete symmetries C, P and T.\footnote{*}{We do not consider operators
related to so--called second class currents since these are not observed in
nature.}
$G_A (t)$ is called the nucleon
axial form factor and $G_P (t)$ the induced pseudoscalar form  factor.
We first discuss the axial form factor, which probes the spin--isospin
distribution of the nucleon (since in a non--relativistic language, this is
nothing but the matrix--element of the Gamov--Teller operator $\svec \sigma
\, \ \tau^a$). The small momentum expansion of the axial form factor takes the
form
$$G_A (t) = g_A \, \biggl( 1 + {t \over 6} <r_A^2> + \ldots \biggr)
 \eqno(4.47)$$
with $g_A$ the axial--vector coupling constant, $g_A = 1.2573 \pm 0.0028$
[4.42],
$r_A = <r_A^2>^{1/2}$ the axial root--mean-square (rms) radius and
the ellipsis stands for
terms quadratic (and higher) in $t$. The axial rms radius can be determined
from elastic (anti)--neutrino--proton scattering or from charged pion
electroproduction. While the former method gives $r_A = 0.65 \pm 0.03$ fm
[4.43], the latter leads to somewhat smaller values $r_A = 0.59 \pm 0.05$
fm [4.1,4.44,4.49]. This apparent
 discrepancy will be discussed in the next section.
In any case, we note that the typical size of the nucleon when probed with the
weak charged currents is smaller than the typical electromagnetic size, $r_{\rm
em} =0.8$ fm. There is, of course, no a priori reason why these sizes should
coincide.  This hierachy of nucleon radii finds a natural explanation in the
topological soliton model of the nucleon [4.45] since there the electromagnetic
size is proportional to $r_B^2 + 6/ M_\omega^2 \simeq (0.8)^2$ fm$^2$
 whereas the axial radius is roughly given by
$r_B^2 + 6/ (M_\rho + M_\pi)^2 \simeq (0.7)^2$ fm$^2$, with $r_B \simeq 0.5$ fm
the size related to the distribution of topological charge (baryon number).
 Empirically, the axial
form factor can be rather accurately parametrized by a dipole form
$$G_A (t) =  {g_A \over (1 - t/M_A^2)^2}               \eqno(4.48)$$
and the cut--off mass $M_A$ is thus related to the axial rms radius via
$$ <r_A^2> = {12 \over M^2_A}  \qquad  .                \eqno(4.49)$$

In heavy baryon CHPT and to one--loop accuracy $q^3$, the momentum dependence
of the axial form factor is
essentially given by some contact terms (similar to the isoscalar form factors
discussed in section 4.1) since the absorptive part of $G_A (t)$ starts at the
three--pion cut, $t_0 = 9 M_\pi^2$ (accesible first at two--loop order),
$$\eqalign{
g_A   & = \krig{g}_A \biggl\lbrace Z_N + {M_\pi^2 \over 32 \pi^2 F_\pi^2}
\biggl[ \bigl( \krig{g}_A^2 - 4 \bigr) \ln {M_\pi \over \lambda} + \krig{g}_A^2
\biggr] \biggr\rbrace
+{ M_\pi^2 \over 4 \pi^2 F_\pi^2} \, B^r_{9} (\lambda )  \cr
<r_A^2> & =  {6 \over \krig{g}_A} \, B_{24}          \cr}         \eqno(4.50)$$
with $Z_N$ the nucleon Z--factor (3.55) and the pertinent counter terms
in ${\cal L}_{\pi N}^{(3)}$ are $O_9$ of eq.(3.62) and $B_{24} \bar{H}
S^\mu [D^\nu,f_{\mu \nu}^- ] H$.
 Stated differently, to order $q^4$   in heavy baryon CHPT, $G_A (t)$ is
linear in $t$ since the cut starting at $t_0 = (3 M_\pi)^2$
first shows up  in the chiral expansion at two--loop order ${\cal O}(q^5)$.
Therefore, the contribution to order $q^4$ must be polynomial in $t$.
The  one--loop expression for $G_A (t)$
in the relavistic formulation can be found in ref.[4.46].

Concerning the induced pseudoscalar form factor, it is generally believed to
be understood well in terms of pion pole dominance
as indicated from ordinary muon capture experiments,
$\mu^- + p \rightarrow \nu_\mu + n$ (see e.g. refs.[4.47,4.48,4.49]).
However, it now seems feasible to measure the induced pseudoscalar
coupling constant (the form factor evaluated at $t = -0.88M_\mu^2$) within
a few percent accuracy via new techniques which allow to minimize the
uncertainty in the neutron detection [4.50]. In fact, one is able
to calculate this fundamental quantity within a few
percent accuracy by making use of the chiral Ward identities of QCD.
The pseudoscalar coupling constant as measured in ordinary muon capture is
defined via
$$g_P = {M_\mu \over 2 m} G_P(t = -0.88 M_\mu^2 ) \qquad . \eqno(4.51)$$
The value of $t$ can be understood as follows. If the muon and the proton are
initially at rest, energy and momentum conservation lead to
$$t = - M_\mu^2 \biggl[ 1 - {(M_\mu + m_p)^2 - m_n^2 \over M_\mu (M_\mu + m_p)}
\biggr] = -0.88 \, M_\mu^2 \quad .   \eqno(4.52) $$
To accurately predict $g_P$ in terms of
well--known physical parameters, we exploit the chiral Ward identity of QCD,
$$\partial^\mu [ \bar{q} \gamma_\mu \gamma_5 {\tau^a \over 2} q ] =
\hat{m} \bar{q} i \gamma_5 \tau^a q  \quad . \eqno(4.53)$$
Sandwiching this between nucleon states, one obtains [4.2,4.3]
$$mG_A(t) + {t \over 4 m} G_P(t) = 2 \hat{m} \, B \, \krig{m} \,
\krig{g}_A {1 + g(t) \over M_\pi^2 - t} \eqno(4.54)$$
The pion pole in eq.(4.54) originates
from the direct coupling of the pseudoscalar density  to the pion,
eq.(3.86).
The residue at the pion pole $t = M_\pi^2$ is
$\hat{m} \, G_\pi \, g_{\pi N} = g_{\pi N} \, F_\pi \, M_\pi^2$ as discussed in
section 3.7. Furthermore, to order $q^4$, $G_A (t) $ as well as $g(t)$ are
linear in $t$ as discussed above. Therefore,
$$m \, g_A + m \, g_A {r_A^2 \over 6} t + {t \over4m} G_P (t) =
{g_{\pi N} F_\pi \over M_\pi^2-t} t + g_{\pi N} F_\pi + {2 B_{23} M_\pi^2
g_{\pi N} \over F_\pi}                \eqno(4.55)$$
where we have used $2 \hat{m} B \krig{g}_A \krig{m} = M_\pi^2 ( g_{\pi N}
F_\pi + {\cal O}(M_\pi^2) )$. At $t=0$, eq.(4.55) reduces to the
Goldberger--Treiman discrepancy discussed in section 3.7.
$G_P (t)$ can now be isolated,
$$G_P (t) = {4 m g_{\pi N} F_\pi \over M_\pi^2 - t} \, - {2 \over 3} \, g_A \,
m^2 \, r_A^2 \, + {\cal O}(t, M_\pi^2)          \eqno(4.56)$$
A few remarks are in order. First, notice that only physical and
well--determined parameters enter in eq.(4.56). Second, while the first
term on the right--hand--side of eq.(4.56) is of order $q^{-2}$, the second
one is ${\cal O}(q^0)$ and the corrections not calculated are of order $q^2$.
For $g_P$, this leads to  [4.51]
$$g_P  = {2 M_\mu g_{\pi N} F_\pi \over M_\pi^2 + 0.88M_\mu^2} \,
 - {1 \over 3} \, g_A \, M_\mu \, m \, r_A^2 \quad .
\eqno(4.57)$$
Indeed, this relation has been derived long time ago by
Adler and Dothan [4.52] with the help of PCAC and by
Wolfenstein [4.53]  using a once--subtracted dispersion relation for the
right--hand--side of eq.(4.54) (weak PCAC). It is gratifying that the
result of refs.[4.52,4.53] can be firmly based on
the systematic chiral expansion of
low energy QCD Green functions. In chiral perturbation theory, one could in
principle calculate the corrections to (4.57) by performing a two--loop
calculation while in Adler and Dothan's or
Wolfenstein's method these either depend
(completely) on the PCAC assumption or
could only be estimated.
To stress it again, the main ingredient to arrive at eq.(4.57) is the
linear $t$--dependence of $G_A (t)$ and $g(t)$. Since we are interested here
in a very small momentum transfer, $t = -0.88M_\mu^2 \simeq -0.5 M_\pi^2$,
curvature terms of order $t^2$ have to be negligible. If one uses for example
the dipole parametrization for the axial form factor, eq.(4.48),
the $t^2$--term amounts to a 1.3$\%$ correction to the one
linear in $t$. Consequently, our results can also be used in radiative muon
capture off hydrogen where the four--momentum transfer varies between $-M_\mu^2
\ldots +M_\mu^2$. Using now the PDG values [4.42] for
 $m$, $M_\mu$, $M_\pi = M_{\pi^+}$, $F_\pi$  and $g_A$  together with
 $g_{\pi N} = 13.31 \pm 0.34$\footnote{*}{See the discussion on this in
ref.[4.51]} and $r_A$ from the (anti)neutrino--nucleon scattering experiments,
we arrive at
$$g_P  = (8.89 \pm 0.23)  - (0.45 \pm 0.04) = 8.44 \pm 0.23  \qquad .
\eqno(4.58)$$
The uncertainties in eq.(4.58) stem from the range of $g_{\pi N}$
and from the one for $r_A$ for the first and second term, in order. For the
final result on $g_P$, we have added these uncertainties in quadrature. A
measurement with a 2$\%$ accuracy of $g_P$ could therefore cleanly separate
between the pion pole contribution and the improved CHPT result. This would
mean a significant progress in our understanding of this fundamental
low--energy parameter since the presently available determinations have too
large error bars to disentangle these values (see e.g. refs.[4.47,4.49]).
In fact,
one might turn the argument around and eventually use a precise determination
of $g_P$ to get an additional determination of the strong pion--nucleon
coupling
constant which has been at the center of much controversy over the last years.
The momentum dependence of $G_P (t)$ for $t$ between $-0.07$ and $-0.18$
GeV$^2$ has recently been measured [4.49]. The error bars are, however, too
large to disentangle between the pion pole prediction and the one given in
eq.(4.56). A more accurate determination of the induced pseudoscalar form
factor would therefore help to clarify our understanding of the low--energy
structure of QCD.
\bigskip \goodbreak
\noindent{\bf IV.4. THRESHOLD PION PHOTO-- AND ELECTROPRODUCTION}
\medskip \goodbreak
In this section, we will be concerned with reactions involving photons,
nucleons and pions, i.e. the interplay between vector and axial--vector
currents. This has been a topical field in particle physics in the late sixties
and early seventies before the advent of scaling in deep inelastic scattering,
see e.g. ref.[4.1]. However, over the last few years renewed interest in the
production of pions by real or virtual photons in the threshold region has
emerged. This was first triggered through precise new data on neutral pion
photoproduction [4.54,4.55], which lead to a controversy about their
theoretical interpretations. Furthermore, new precise data on $\pi^0$
electroproduction [4.56] have given further constraints on the understanding
of these fundamental processes in the non--perturbative regime of QCD. In
fact, as we will demonstrate, chiral perturbation theory methods are best
suited to analyze these reactions in the threshold region.

First, we have to supply some basic definitions. For more details, we refer
to refs.[4.1,4.46,4.57,4.58].
Consider the process $\gamma^*(k) + N(p_1) \to \pi^a(q) + N(p_2)$, with $N$
denoting a nucleon (proton or neutron), $\pi^a$ a pion with an
isospin index $a$ and
$\gamma^*$ the virtual photon with $k^2 < 0$. In the case of photoproduction
(real photons), we have $k^2= 0$ and $\epsilon \cdot k = 0$.
 A detailed exposition of the corresponding
kinematics can e.g. be found in ref.[4.59]. The pertinent Mandelstam variables
are $s=(p_1+k)^2, \,\, t=(p_1-p_2)^2$ and $u=(p_1-q)^2$ subject to the
constraint $s+t+u=2m^2+M_\pi^2 +k^2$.
Using Lorentz invariance and the discrete symmetries
P, C and T, the transition current matrix element
can be expressed in terms of six
independent invariant functions, conventionally denoted by
$A_i(s,u),\,\,(i=1,...,6)$, when one makes use of gauge invariance,
$$J^\mu = i \bar u_2 \g5 \sum_{i=1}^6 {\cal M}^\mu_i \, A_i(s,u)\, u_1
\eqno(4.59)$$
with
$$\eqalign{ &
{\cal M}_1^\mu = {1\over 2} (\gamma^\mu \ks - \ks \gamma^\mu), \qquad {\cal
M}_2^\mu = P^\mu ( 2q\cdot k - k^2) - P\cdot k (2 q^\mu - k^\mu),
\cr
& {\cal M}_3^\mu = \gamma^\mu q\cdot k - \ks  q^\mu , \qquad {\cal M}_4^\mu = 2
\gamma^\mu P\cdot k - 2 \ks P^\mu - m \gamma^\mu \ks + m\ks \gamma^\mu, \cr
& {\cal M}_5^\mu =  k^\mu  q \cdot k - q^\mu \, k^2, \qquad
{\cal M}_6^\mu = k^\mu\, \ks - \gamma^\mu \, k^2 \cr } \eqno(4.60)$$
and $P= (p_1+p_2)/2$. The amplitudes $A_i(s,u)$ have the
conventional isospin decomposition (to first order in electromagnetism),
$$A_i(s,u) = A_i^{(+)}(s,u)\,\, \delta_{a3} + A_i^{(-)}(s,u)\,\, {1\over 2}
[\tau_a, \tau_3] + A_i^{(0)}(s,u)\,\, \tau_a \,\,.\eqno(4.61)$$
Under $(s \leftrightarrow u)$ crossing the amplitudes $A^{(+,0)}_{1,2,4},\,
A^{(-)}_{3,5,6}$ are even, while $A^{(+,0)}_{3,5,6},\, A^{(-)}_{1,2,4}$ are
odd. For photoproduction, the number of independent amplitudes is further
reduced to four. In terms of the isospin components, the physical channels
 under consideration are
$$\eqalign{
J_\mu(\gamma^* p \to \pi^0 p ) & = J_\mu^{(0)} + J_\mu^{(+)} \cr
J_\mu(\gamma^* n \to \pi^0 n ) & = J_\mu^{(+)} - J_\mu^{(0)} \cr
J_\mu(\gamma^* p \to \pi^+ n ) & = \sqrt 2 [J_\mu^{(0)} + J_\mu^{(-)}] \cr
J_\mu(\gamma^* n \to \pi^- p ) & = \sqrt 2[ J_\mu^{(0)} - J_\mu^{(-)}]\,.
\cr}\eqno(4.62)$$
Having constructed the current transition matrix element $J_\mu$ it is then
straightforward to calculate observables. The pertinent kinematics and
definitions are outlined in refs.[4.57,4.58].

For the discussion of the low energy theorems , we have to spell
out the corresponding multipole decomposition of the transition current matrix
element at threshold. In the $\gamma^* N$ center of mass system at threshold
$i.e.$ $q_\mu = (M_\pi, 0,0,0)$ one can express the current matrix element in
terms
of the two S--wave multipole amplitudes, called $E_{0+}$ and $L_{0+}$,
$$\vec J = 4\pi i(1+\mu)\, \chi_f^\dagger \bigl\{ E_{0+}(\mu,\nu) \,\vec \sigma
+ \bigl[L_{0+}(\mu,\nu) - E_{0+}(\mu,\nu)\bigr] \, \hat k \, \vec \sigma \cdot
\hat k  \bigr\} \chi_i \eqno(4.63)$$
with $\chi_{i,f}$ two component Pauli-spinors for the nucleon and we
chose the Coulomb gauge $\epsilon_0 = 0$.
For the later
discussion we have introduced the dimensionless quantities
$$\mu = {M_\pi \over m} , \qquad \nu = {k^2 \over m^2}\,. \eqno(4.64)$$
The multipole $E_{0+}$ characterizes the transverse and $L_{0+}$ the
longitudinal coupling of the virtual photon to the nucleon spin.
Alternatively to $L_{0+}$, one also uses the scalar multipole
$S_{0+}$ defined via, $S_{0+} (s, k^2 ) = (|\svec{k}| / k_0) \, L_{0+}
(s, k^2 )$.
At threshold, we can express $E_{0+}$ and $L_{0+}$ through the invariant
amplitudes $A_i(s,u)$ via (suppressing the isospin indices)
$$\eqalign{
E_{0+} & = {m\sqrt{(2+\mu)^2 - \nu} \over 8 \pi (1+\mu)^{3/2} }\biggl\{ \mu A_1
+ \mu m {\mu(2+\mu) + \nu \over 2(1+\mu)} A_3 + m {\mu(\mu^2 - \nu) \over
2(1+\mu)} A_4 - \nu m A_6\biggr\}\,, \cr
L_{0+} & = E_{0+} + {m\sqrt{(2+\mu)^2 - \nu} \over 16 \pi (1+\mu)^{5/2}}
(\mu^2 - \nu)\biggl\{ -A_1 + m^2 {(2-\mu )(\mu (2+\mu ) + \nu ) \over 4 (1+\mu
)} A_2 - \mu m A_4 \cr  & + m^2 \mu {\mu (2+ \mu ) + \nu \over2 (1+\mu )} A_5
 - m(2+\mu) A_6\biggr\} \cr }\eqno(4.65)$$
with the $A_i(s,u)$  evaluated at threshold $s_{th} = m^2(1+\mu)^2$ and
$u_{th} = m^2(1-\mu -\mu^2 + \mu \nu)/(1+\mu)$. In case of photoproduction,
only the electric dipole amplitude $E_{0+}$ survives. Finally, one can
define the S--wave cross section,
$$ a_0 = |E_{0+}|^2 - \epsilon {k^2 \over k_0^2} |L_{0+}|^2
\eqno(4.66)$$
where $\epsilon$ and $k_0 = (s-m^2+k^2)/2 \sqrt{s}$ represent, respectively, a
measure of the transverse linear polarization and the
energy of the virtual photon in the
$\pi N$ rest frame. For $k^2 = 0$, this means in particular that
$(|\svec{k}| / |\svec{q}|) (d \sigma / d\Omega )= (E_{0+})^2$ as $\vec q$
tends to zero. This completes the necessary formalism.
\medskip
We discusss now the electric dipole amplitude $E_{0+}$ as measured in neutral
pion photoproduction, $\gamma + p \to \pi^0 + p$.  This multipole is of
particular interest since in the early seventies a low--energy theorem
(LET)  for
neutral pion production was derived [4.60].\footnote{*}{For the sake of
brevity, we denote $E_{0+}^{\pi^0 p}$ by $E_{0+}$.} The recent measurements
at Saclay and Mainz [4.54,4.55] seemed to indicate a gross violation of
this LET, which predicts $E_{0+} = -2.3 \cdot 10^{-3} / M_{\pi^+}$
at threshold. However, the LET was reconsidered (and rederived)
and the data were reexamined, leading to
$E_{0+} = (-2.0 \pm 0.2) \cdot 10^{-3} / M_{\pi^+}$ in agreement
with the LET prediction. These developments have been subject
of a recent review by Drechsel and Tiator [4.61].
Therefore, we will focus here  on the additional insight gained from CHPT
calculations. In fact, the ``LET'' derived in [4.60] for neutral pion
photoproduction at threshold
is an expansion in powers of $\mu = M_\pi / m \sim 1/7$ and predicts the
coefficients of the first two terms in this series, which are of order $\mu$
and $\mu^2$, respectively, in terms of measurable quantities like the
pion--nucleon coupling constant $g_{\pi N}$, the nucleon mass $m$ and the
anomalous magnetic moment of the proton, $\kappa_p$,
$$E_{0+} (s_{\rm thr})
= -{e g_{\pi N} \over 8 \pi m} \mu \biggl[ 1 - {1 \over 2} (3 +
\kappa_p) \mu + {\cal O}(\mu^2 ) \biggr] \quad . \eqno(4.67)$$
In ref.[4.62] it was, however, shown that a certain class of loop
diagrams modifies the LET at next--to--leading order ${\cal O}(\mu^2 )$.
It is instructive to rederive this result in heavy baryon CHPT [4.3].
Insertions from ${\cal L}_{\pi N}^{(2)}$ and ${\cal L}_{\pi N}^{(3)}$
lead to eq.(4.67) when
the corresponding quantities are given by their chiral limit values. However,
to order $q^3$, one also has to consider one--loop graphs. The standard
derivation of eq.(4.67) is based on the consideration of nucleon pole
graphs (supplemented by form factors). We stress
that such considerations are not
based on a systematic chiral counting.
In the threshold region, only the
so--called triangle and rescattering diagrams are
non-vanishing (compare the detailed discussion of selection rules in ref.[4.3])
leading to
$$\delta E_{0+}^{\rm loop} = {e g_A \over 8 \pi F_\pi^3} \, v \cdot k \,
\biggl[ J_0 (v \cdot k ) +  J_0 (-v \cdot k ) + 2 \gamma_3 (v \cdot
k) + 2 \gamma_3 (-v \cdot k) \biggr]     \eqno(4.68)$$
with the loop functions $J_0$ and $\gamma_3$ given in appendix B. At
threshold, $v \cdot k = M_\pi$, so that $J_0 (M_\pi) +J_0 (-M_\pi)
 = 0$  and $\gamma_3
(M_\pi) + \gamma_3 (-M_\pi) = M_\pi/32$, i.e. only the triangle diagram and its
crossed partner contribute. Therefore, these
particular one loop diagrams contribute at order $\mu^2$,
$$\delta E_{0+}^{\rm loop} = {e g_A M_\pi^2 \over 128 \pi F_\pi^3}
 = {e g_A m^2 \over 128 \pi F_\pi^3} \ \mu^2
\eqno(4.69)$$
Let us look closer at the origin of the finite contribution proportional to
$M_\pi^2$. We follow the argument of ref.[4.62]. In the relativistic
calculation, the loops lead to an expression of the form
$$\delta E_{0+}^{\rm loop}  = {e g_A M_\pi^2 \over (4 \pi
F_\pi )^3 } \bigl[ f(\mu ) - f (-\mu ) \bigr] \eqno(4.70)$$
with
$$f( \mu ) = \int_0^1 dx \int_0^1 dy {2 x y \over \mu^2 + y^2 - 2 \mu x y}
\quad . \eqno(4.70a)$$
Naively, one would argue $\delta E_{0+}^{\rm loop} \sim M_\pi^3$, since the
formula (4.70) is manifestly odd under
 $M_\pi \to - M_\pi$ and this would forbid
an even term proportional to $M_\pi^2$. However, in this argumentation one has
already made an assumption, namely that the function $f(\mu) $ is analytic at
$\mu=0$. The explicit form (4.70a) shows that this is not true, $f(\mu)$ has a
logarithmic singularity at $\mu =0$ and the correct
expansion around this point reads
$$f(\mu ) = -\ln \mu + {3 \pi^2 \over 8} + {1 \over 2} + {\cal O}(\mu), \quad
f(-\mu ) = -\ln \mu - {\pi^2 \over 8} + {1 \over 2} + {\cal O}(\mu) \, .
\eqno(4.70b)$$
Consequently, we find that the odd combination $f(\mu)-f(-\mu)$ has a
nonvanishing limit  $\lim_{\mu \to 0^+} [f(\mu) - f(-\mu)] = \pi^2/2 \ne
0$, which is quite astonishing. Of course, without explicit knowledge of
$f(\mu)$ from the complete loop-calculation in CHPT one would hardly find the
peculiar properties of $f(\mu)$. In the standard derivation (and all
rederivations) of the incomplete LET one tacitely assumes (without better
knowledge) $\lim_{\mu \to 0^+} [ f(\mu) - f(-\mu) ] = 0$.
The
additional term is non-analytic in the quark mass $\hat m$ since $\delta
A_{1,\rm thr}^{\rm loop} \sim M_\pi \sim \sqrt{\hat m}$ [4.62].
Consequently, the correct $\mu$ expansion of $E_{0+}$ in QCD takes the
form\footnote{$^*$}{The meaning of low--energy theorems in the
framework of the Standard Model is discussed in ref.[4.90].}
$$E_{0+} (s_{\rm thr})
= -{e g_{\pi N} \over 8 \pi m} \mu \biggl\lbrace 1 - \biggl[{1 \over 2} (3 +
\kappa_p) + \bigl( {m \over 4 F_\pi} \bigr)^2 \biggr]   \mu +
{\cal O}(\mu^2 ) \biggr\rbrace \quad . \eqno(4.71)$$
One immediately notices that the term of order $\mu^2$ is
even bigger (+4.35) than the leading order one (-3.46)
and of opposite sign. This makes the $\mu$--expansion of $E_{0+}$
truncated at order $\mu^2$ practically useless for a direct comparison
with the data. We also stress that the closeness of the
prediction based on the incomplete expansion (4.67) and the reexamined data
has to be considered accidental and is devoid of any physical significance.
In fact, in ref.[4.59] it was argued that the
$\mu$--expansion of $E_{0+}$ is slowly converging. This has been even further
quantified in a calculation [4.63] in the framework
 of heavy baryon CHPT including
all terms of order $q^4$ and including isospin--breaking by differentiating
between the charged and neutral pion masses as proposed in ref.[4.64].
Furthermore, the theoretical prediction of the
electric dipole amplitude in $\pi^0$ production off protons is afflicted with
some uncertainty related to the $\Delta$ contribution to estimate the appearing
contact terms. At present, it does not seem to serve as a stringent test
of the chiral pion--nucleon dynamics. More accurate data close to threshold are
needed to clarify these questions and also the energy--dependence of $E_{0+}$
from the $\pi^0 p$ threshold ($E_\gamma = 144.68$ MeV) to the $\pi^+ n$
 threshold  at $E_\gamma = 151.44$ MeV. While below this threshold the
multipoles are real, above it they in general become complex.
To one--loop accuracy, one expects a cusp at the $\pi^+ n$ threshold with
$E_{0+} (E_\gamma = 151.44 \, {\rm MeV}) - E_{0+}(E_\gamma = 144.68 \,
{\rm MeV}) \simeq
0.7 \cdot 10^{-3}/M_{\pi^+}$ [4.63]. The various analysis of the Mainz data
[4.55] give very different results, e.g. while Bergstrom [4.65] finds a very
steep energy dependence, the analysis of Bernstein leads to an essentially
flat $E_{0+} (E_\gamma )$ [4.66].
The situation is different for the P--waves. From the
two magnetic multipoles $M_{1+}$ and $M_{1-}$ and the electric $E_{1+}$ one
forms the combinations
$$\eqalign{
P_1 & = 3 E_{1+} + M_{1+} - M_{1-} \cr
P_2 & = 3 E_{1+} - M_{1+} + M_{1-} \cr
P_3 & = 2 M_{1+} + M_{1-} \cr}           \eqno(4.72)$$
To order $q^3$ and including the pion mass difference in the loops,
 one finds [4.63]
$$\eqalign{
P_1 & = \sqrt{\omega^2 - M_{\pi^0}^2}
\biggl\lbrace {e g_{\pi N} \over 8 \pi m^2}
\biggl[1 - {6 \omega \over 5 m} + {M_{\pi^0}^2 \over 5 m \omega} + \kappa_p
\bigl( 1 - {\omega \over  2m} \bigr) \biggr]   \cr
& + {e g_A^3 \over 64 \pi^2 F_\pi^3} \biggl[ {2 \over 3 \omega^2} \bigl(
M_{\pi^+}^3 - ( M_{\pi^+}^2 - \omega^2 )^{3/2} \bigr) + M_{\pi^+} -
 \sqrt{M_{\pi^+}^2 - \omega^2} - {M_{\pi^+}^2 \over \omega} \arcsin {\omega
\over M_{\pi^+} } \biggr] \biggr\rbrace \cr
P_2 & = \sqrt{\omega^2 - M_{\pi^0}^2} \biggl\lbrace
{e g_{\pi N} \over 8 \pi m^2}
\biggl[-1 + {13 \omega \over 10 m} + {M_{\pi^0}^2 \over 5 m \omega} + \kappa_p
\bigl( 1 - {\omega \over  2m} \bigr) \biggr]   \cr
& + {e g_A^3 \over 64 \pi^2 F_\pi^3} \biggl[ {2 \over 3 \omega^2} \bigl(
M_{\pi^+}^3 - ( M_{\pi^+}^2 - \omega^2 )^{3/2} \bigr) - M_{\pi^+}
\biggr] \biggr\rbrace \cr}
\eqno(4.73)$$
with $\omega = (s-m^2+M_{\pi^0}^2)/(2\sqrt{s})$
 the pion energy in the cm system.\footnote{*}{$\omega$ is related
to the frequently used photon energy in the lab system via $\omega = (2 m
E_\gamma + M_{\pi^0}^2) / (2 \sqrt{m^2 + 2 m E_\gamma})$.}
 A closer look at (4.73) reveals
that the terms of order $q^3$ in the threshold region are very small
compared to the leading ${\cal O}(q^2)$ ones, i.e. one can derive the LETs
$$\eqalign{
{1 \over q} P_1\biggl|_{\rm thr} & = {e g_{\pi N} \over 8 \pi m^2}
\biggl\lbrace 1 + \kappa_p + \mu \biggl[ -1 - {\kappa_p \over 2} + {g_{\pi N}^2
(10 - 3\pi ) \over 48 \pi} \biggr] \biggr\rbrace \cr
{1 \over q} P_2\biggl|_{\rm thr} & = {e g_{\pi N} \over 8 \pi m^2}
\biggl\lbrace -1 - \kappa_p + {\mu \over 2} \biggl[ 3 + \kappa_p
 - {g_{\pi N}^2  \over 12 \pi} \biggr] \biggr\rbrace \cr}  \eqno(4.73a)$$
and similarly for the reaction $\gamma n \to \pi^0 n$,
$$\eqalign{
{1 \over q} P_1\biggl|_{\rm thr} & = {e g_{\pi N} \over 8 \pi m^2}
\biggl\lbrace - \kappa_n + {\mu \over 2}
 \biggl[ \kappa_n + {g_{\pi N}^2
(10 - 3\pi ) \over 24 \pi} \biggr] \biggr\rbrace \cr
{1 \over q} P_2\biggl|_{\rm thr} & = {e g_{\pi N} \over 8 \pi m^2}
\biggl\lbrace  \kappa_n + {\mu \over 2} \biggl[ \kappa_n
 + {g_{\pi N}^2  \over 12 \pi} \biggr] \biggr\rbrace \cr}   \eqno(4.73b)$$
with $\kappa_n = -1.913$ the anomalous magnetic moment of the neutron.
 These are examples of  quickly
converging $\mu$ expansions. In fact, the corresponding $P_1$ and $P_2$ of the
relativistic calculation agree quite nicely with the LET (remember that in the
relativistic formulation some higher order terms are included). For example,
at $E_\gamma = 151$ MeV, the LET predicts $P_1 = 2.47$ and $P_2 = -2.48$ while
the P--wave multipoles of ref.[4.59] lead to $P_1 = 2.43$ and $P_2 = -2.60$
(all in units of $10^{-3} / M_{\pi^+}$) (for $\gamma p \to \pi^0 p$). In
contrast, $P_3$ is completely dominated by the $\Delta$ contribution. Its
generic form is $P_3 = q \omega * {\rm const}$, where the constant can either
be fitted from the bell--shaped differential cross sections about the $\pi^+ n$
threshold or by using resonance saturation. Both ways lead to essentially the
same number. Alternatively, one could fit the coefficient in $P_3$ from the
total cross section data if one excludes the very first few MeV above
threshold.  For more details on this, we refer the  reader to ref.[4.63].

The total cross section for $\gamma p \to \pi^0 p$ [4.53,4.54] is only
sensitive to the value of $E_{0+} (s_{\rm thr} )$ very close to threshold
and then dominated by the P-wave combination $(|P_1|^2 +|P_2|^2 + |P_3|^2)/3$.
This is
shown in figure 4.5 for the calculation of refs.[4.59,4.64]. In fact, the
 estimate of the
 low--energy constant $d_4$ in [4.59] should be considered as a
best fit to these total cross section data. The corresponding differential
cross sections calculated in ref.[4.59] do not agree well with the
data since $E_{0+}$ was essentially energy independent $\sim -1.3$,
too large in magnitude to what
is needed to produce the bell--shaped  angular distributions.
 A more detailed account of these topics will be given in ref.[4.63].
Finally, we point out that new data for $\gamma p \to \pi^0 p$ have
been taken at MAMI (Mainz) and SAL (Saskatoon). These are presently in
the process of being analyzed.
\midinsert
\smallskip
\hskip 0.5in
\epsfxsize=3in
\epsfysize=3.5in
\epsffile{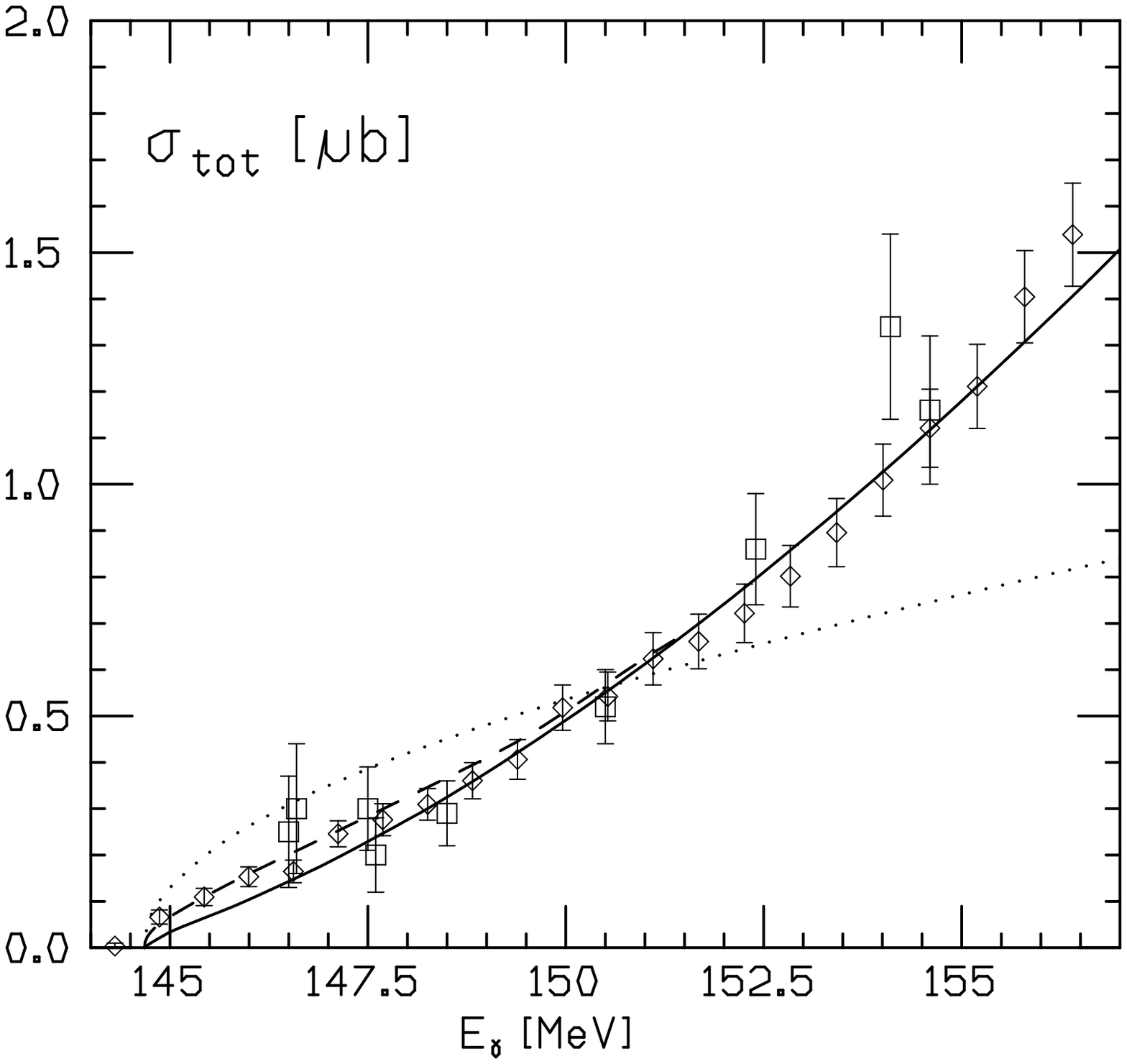}
\smallskip
\vskip -1.5truecm
{\noindent\narrower \it Fig.~4.5:\quad
Total cross section for $\gamma  p \to \pi^0 p$. The solid and dashed curve
represent the one--loop CHPT predictions in the isospin limit and with
isospin--breaking in the class I diagrams, respectively [4.59,4.64]. The dotted
line is the tree level predictions. The Mainz [4.54] and the Saclay [4.53] data
are represented by diamonds and squares, respectively.
\smallskip}
\vskip -0.5truecm
\endinsert
\medskip

The situation is different in the case of charged pion production, $\gamma p
\to \pi^+ n$ and $\gamma n \to \pi^- p$. Here, there exists a
famous LET due to
Kroll and Ruderman [4.67], which states that the corresponding electric dipole
amplitudes do not vanish in the chiral limit and, furthermore, that this
leading
term of order $\mu^0$ is dominant. The chiral corrections do not affect this
result as shown in ref.[4.59]. In fact, the quark mass expansion of $E_{0+}$
for these two channels takes the form (at threshold)
$$\eqalign{
E_{0+}^{\pi^+ n} & = \sqrt{2} {e g_{\pi N} \over 8 \pi m} \biggl[ 1 - {3 \over
2} \mu + {\cal O}(\mu^2 , \mu^2 \ln \mu ) \biggr] = 26.6 \cdot
10^{-3}/M_{\pi^+} \quad , \cr
E_{0+}^{\pi^- p} & = \sqrt{2} {e g_{\pi N} \over 8 \pi m} \biggl[ -1 + {1 \over
2} \mu + {\cal O}(\mu^2 , \mu^2 \ln \mu ) \biggr] = -31.5 \cdot
10^{-3}/M_{\pi^+} \quad . \cr}               \eqno(4.74)$$
The full one--loop corrections (i.e. no expansion in $\mu$)
have been worked out with the result [4.59]
$$E_{0+}^{\pi^+ n}  = 28.4 \cdot 10^{-3}/M_{\pi^+} \quad ,
  E_{0+}^{\pi^- p}  = -31.1 \cdot 10^{-3}/M_{\pi^+} \quad ,  \eqno(4.75)$$
which compare, naturally, well with the empirical data $E_{0+}^{\pi^+ n} = 27.9
\pm 0.5$ [4.68], $E_{0+}^{\pi^+ n} = 28.8 \pm 0.7$ [4.69], $E_{0+}^{\pi^- p} =
-31.4 \pm 1.3$ [4.68] and $E_{0+}^{\pi^- p} = -32.2 \pm 1.2$ [4.70] (all in
canonical units). The numbers in (4.75) should not be considered as
rigorous predictions of CHPT since they depend to some extent on the
assumptions made on the unknown counter terms. One should perform a
similar calculation in HBCHPT to order $q^4$.
 A more accurate determination of these threshold multipoles
would give a further constraint on the pion--nucleon coupling constant via
the Goldberger--Miyazawa--Oehme sum rule [4.10,4.71]
$$ J + {g_{\pi N}^2 \over 2 m^2 - M_\pi^2/2} = {m+M_\pi \over m M_\pi } a^-
\eqno(4.76)$$
with $a^-$ the isospin--odd $\pi N$ scattering length
and $J$ a dispersion integral over the hadronic $\pi^\pm p$ total cross
sections. The integral $J$ can be calculated either from the pertinent
Karlsruhe--Helsinki cross sections or the ones from the SAID data basis.
 One possibility of obtaining the difference $a_{1/2} -a_{3/2}$, which is
most uncertain at present, is via
the Panofsky ratio $P = \sigma (\pi^- p \to n \pi^0 ) / \sigma ( \pi^- p \to n
\gamma )$  [4.10],
$$(a_{1/2} -a_{3/2})^2 = (9k/q) P |E_{0+}^{\pi^- p}|^2 =
 (9k/q) P R |E_{0+}^{\pi^+ n}|^2                  \eqno(4.77)$$
with $R =\sigma (\gamma n  \to p \pi^- ) / \sigma (\gamma p  \to n \pi^+ )$.
To make use of eqs.(4.76,4.77), one needs a very accurate understanding and
determination of the electric dipole amplitude at threshold for charged pion
photoproduction (for further details, see e.g. refs.[4.10,4.72,4.73]). This
concludes our discussion of threshold photopionproduction.
\medskip
We now turn to a short discussion of some topics related to pion
electroproduction. A much more detailed account of these topics can be
found in the recent review [4.46]. There, one can find a thorough discussion of
the pertinent low--energy theorems in the various channels. In particular, it
is stressed (see also ref.[4.74]) that in a systematic chiral expansion one is
only sensitive to the first few moments of the pertinent nucleon form factors,
in contrast to the commonly used practise of supplementing the photon--nucleon
and pion--nucleon vertices with the corresponding  full form factors. Also, in
the loop expansion there is no need for equating the pion charge
form factor and
the isovector nucleon charge form factor to maintain gauge invariance as it is
often done. The chiral expansion keeps gauge invariance at any step of the
calculation and thus allows naturally for $F_\pi (t) \ne F_1^V (t)$. Here, let
us briefly discuss the axial rms radius of the nucleon as measured in
charged pion electroproduction.
 The starting point is the venerable LET due to Nambu, Luri{\'e}
and Shrauner [4.75] for the isospin--odd electric dipole amplitude
$E_{0+}^{(-)}$ in the chiral limit,
$$
E_{0+}^{(-)}(M_\pi=0, k^2) ={e g_A \over 8 \pi
F_\pi} \biggl\lbrace 1 +{k^2 \over 6} r_A^2 + { k^2 \over 4m^2} (\kappa_V
+ {1 \over 2}) + {\cal O} (k^3) \biggr\rbrace
\eqno(4.78)$$
Therefore, measuring the reactions
$\gamma^\star p \to \pi^+ n$ and
$\gamma^\star n \to \pi^- p$ allows to extract
$E_{0+}^{(-)}$ and one can determine the axial radius of the nucleon, $r_A$.
In section 4.3, we had pointed out that the determinations of the axial radius
from electroproduction data and from (anti)neutrino--nucleon scattering show
a small discrepancy. This discrepancy is usually not
taken seriously since the values overlap within the error bars. However, it
was shown in ref.[4.76] that pion loops modify the LET (4.78) at order $k^2$
for finite pion mass. In the heavy mass formalism, the coefficient of the
$k^2$ term reads
$$ {1 \over 6} r_A^2 + {1 \over 4 m^2}(\kappa_V +{1 \over2}) +
{1 \over 128 F_\pi^2} (1 - {12 \over \pi^2})
\eqno(4.79)$$
where the last term in (4.79) is the new one. This means that previously one
had extracted a modified radius, the correction being $3 (1 - 12/\pi^2 ) / 64
F_\pi^2 \simeq -0.046$ fm$^2$. This closes the gap between the values of
$r_A$ extracted from electroproduction and neutrino data.
As detailed in appendix C, the $1 / m$ suppressed terms (i.e. of order
$q^4$) modifying the result (4.79) are small [4.91].

Another interesting quantity is the S--wave cross section defined in eq.(4.66).
The most precise measurement of it for neutral pion production off the proton
close to the photon point was presented in ref.[4.56]. In fig.4.6 we show the
data of ref.[4.56] at $k^2 = -0.042 , \, -0.0501$ and $-0.0995$ GeV$^2$ in
comparison to the one--loop CHPT result and the corresponding tree level
prediction [4.46,4.77]. The most interesting feature of the data is the
flatness of $a_0$ as $|k^2|$ increases.
\midinsert
\smallskip
\hskip 0.75in
\epsfxsize=3.5in
\epsfysize=2.5in
\epsffile{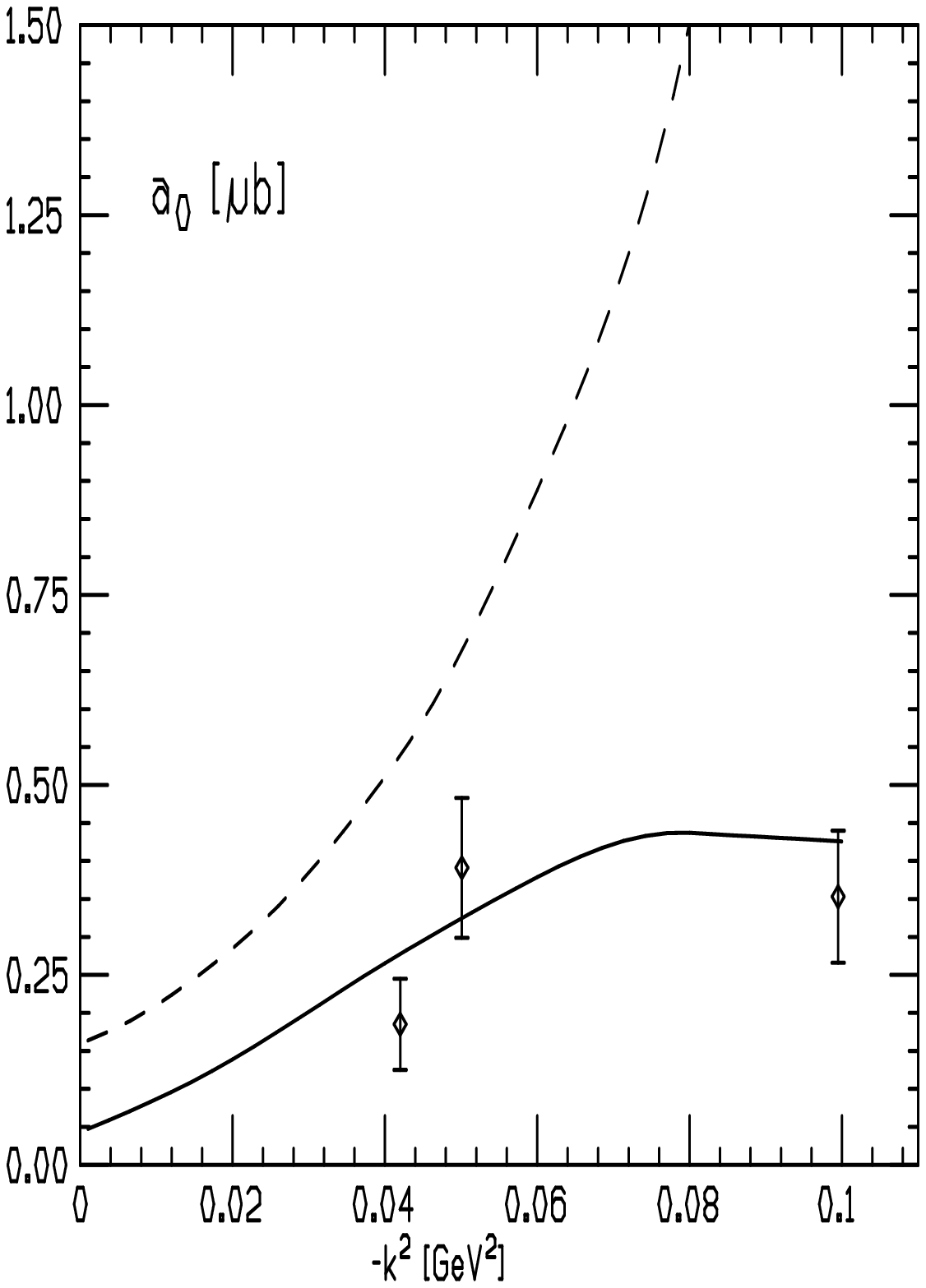}
\smallskip
\vskip -1.5truecm
{\noindent\narrower \it Fig.~4.6:\quad
The $S$--wave component of the neutral pion electroproduction
cross section, calculated from one--loop CHPT
(solid line) and tree graphs (dotted line). The kinematics is $W = 1074$ MeV
and $\epsilon = 0.58$ [4.77].  The data extracted in ref.[4.56] are also shown.
\smallskip}
\vskip -0.5truecm
\endinsert
This trend is also exhibited by the
one--loop CHPT result but not by the tree graphs (or by tree graphs
supplemented with form factors). Chiral loops are required to explain the trend
of the data. We should stress that the calculation of $a_0$ to one loop
accuracy did not involve any new adjustable counter terms (all low--energy
constants were previously fitted in photoproduction [4.59] or via nucleon
radii). Clearly, to have a better test of the chiral dynamics, one should
measure at smaller values of $|k^2|$ since there the loop corrections are
sizeable but not as large as at $k^2 = -0.1$ GeV$^2$. In ref.[4.77], it was
further stressed that to test the chiral predictions, one should investigate in
more detail the angular distributions. The most striking feature is that the
CHPT predictions for the transverse differential cross sections become
 forward peaked as $|k^2|$ becomes larger than $0.04$ GeV$^2$. Indeed, a group
at MAMI (Mainz) has measured this differential cross section at $k^2 = -0.1$
GeV$^2$ and the shape agrees very nicely with the CHPT prediction [4.78].
Clearly, the wide field of single
pion electroproduction in the threshold region
is a good testing ground of the
chiral dynamics and just begins to play again an important role. For a more
detailed account of the existing predictions and limitations within CHPT, we
refer the reader to ref.[4.46].
 \bigskip \goodbreak
\noindent{\bf IV.5. TWO--PION PRODUCTION}
\medskip
In the previous section, we considered single pion photo-- and
electroproduction.  Complementary information
can be gained from the two pion production process $\gamma N \to \pi \pi N$,
with  $\gamma$ a real or virtual photon. The two
pions in the final state can both be charged, both neutral or one charged and
one  neutral.
Here, we will be concerned with the threshold region, i.e. the photon in the
initial state has just enough energy to produce the two pions (and the outgoing
nucleon) at rest. This energy is very close to the first strong resonance
excitation of the nucleon, the $\Delta (1232)$. In fact, presently
available data focus on the resonance region and above. In that case, a
two--step reaction mechanism of the form $\gamma N \to \pi \Delta \to \pi \pi
N$ is appropriate to describe these data as detailed in refs.[4.1,4.79,4.80].
As we will argue, there is however a narrow window above threshold which is
particularly sensitive to chiral loops, i.e. to the strictures of the
spontaneously broken chiral symmetry. First
measurements of
two--pion production at low energies have been performed at MAMI
and we expect that the theoretical predictions discussed below will give
additional motivation to perform yet more detailed measurements of this
particular reaction. The CHPT calculation presented in ref.[4.81]
extends the one of Dahm and Drechsel [4.82] who
discussed certain aspects of two--pion photoproduction in the framework of
Weinberg's chiral pion--nucleon Lagrangian [4.83].

First, we must outline the formalism necessary to treat two-pion photo-
and electroproduction in the threshold region. We will only be concerned with
the kinematics close to or at threshold, the corresponding amplitudes
and the total cross sections. For a more general discussion we refer the reader
to ref.[4.82].  To be specific,
consider the process $\gamma(k) + N(p_1) \to \pi^a(q_1) + \pi^b(q_2) + N(p_2)$.
The corresponding  current transition matrix element is
$$T\cdot \epsilon = \quad <\pi^a(q_1),\pi^b(q_2),N(p_2) \, \, {\rm out}|
 J^{\rm em}_\mu(0)
\epsilon^\mu| N(p_1) \, \, {\rm in}> \eqno(4.80)$$
with $J^{\rm em}_\mu$ the electromagnetic current operator and  $\epsilon_\mu$
the polarization vector of the photon.  From the two initial
states $\gamma p$ and $\gamma n$ we can form in total six final states
$$\eqalign{ \gamma p & \to \pi^+ \pi^- p\,, \quad \gamma p   \to \pi^+ \pi^0 n
\,, \cr \gamma p & \to \pi^0 \pi^0 p\,, \quad \gamma n  \to \pi^+ \pi^- n\,,
\cr  \gamma n & \to \pi^0 \pi^- p\,, \quad \gamma n  \to \pi^0 \pi^0 n\,. \cr}
\eqno(4.81)$$
In what follows we will  concentrate on the
channels with a proton in the initial state.
To first order in the electromagnetic coupling $e$ the threshold
amplitudes for $\gamma p \to \pi^+ \pi^0 n$ and $\gamma n \to \pi^- \pi^0 p$
are equal.
In general, one can form five/six Mandelstam variables for the two-pion
photo/electroproduction process from the independent four-momenta. For our
purpose, it is most convenient to work in the photon-nucleon center-of-mass
frame. At threshold, the real or virtual photon has just enough energy to
produce the two pions  at rest.
 The threshold center-of-mass energy squared is
$s_{\rm thr} = (p_1 + k)^2_{\rm thr} =
(m + 2M_\pi)^2 = m^2 ( 1 + 4 \mu + 4 \mu^2)$.
The photon center-of-mass energy can be expressed in terms of $s$ and the
photon virtuality $k^2$ as $ k_0 = ( s - m^2 + k^2 ) / (2 \sqrt{s})$
with its threshold value
$$k_0^{\rm thr} ={2m\over 1+ 2 \mu}
\biggl[ \mu + \mu^2 + {\nu \over 4} \biggr] \eqno(4.82)$$
in terms of the small parameters $\mu$ and $\nu$. In the lab system,
the threshold value for two pion-photoproduction is given by
$E_\gamma^{\rm thr} = 2 M_\pi( 1 + \mu)$. The kinematics above threshold
is discussed in more detail in ref.[4.81].

At threshold in the center-of-mass frame ($i.e.\,\,\vec q_1 = \vec q_2 = 0$),
the two-pion electroproduction current matrix element can be decomposed into
 amplitudes as follows if we work to first order in the
electromagnetic coupling $e$,
$$\eqalign{ T \cdot \epsilon  = \chi_f^\dagger \bigl\{ & i \vec \sigma\cdot (
\vec \epsilon \times \vec k) \bigl[ M_1 \delta^{ab} + M_2 \delta^{ab} \tau^3 +
M_3 (\delta^{a3} \tau^b + \delta^{b3} \tau^a) \bigr] \cr  & + \vec \epsilon
\cdot \vec k  \bigl[ N_1 \delta^{ab} + N_2 \delta^{ab} \tau^3 + N_3
(\delta^{a3} \tau^b + \delta^{b3} \tau^a) \bigr] \bigr\} \chi_i \cr }
\eqno(4.83)$$
in the gauge $\epsilon_0 = 0$. Clearly, for real photons only the $M_{1,2,3}$
 can contribute. For virtual photons, gauge invariance $T\cdot k = 0$ allows to
reconstruct $T_0$ as $T_0 = \vec T \cdot \vec k /k_0$. The amplitudes
$M_{1,2,3}$ and $N_{1,2,3}$ encode the information about the structure of the
nucleon as probed in threshold two pion photo- and electroproduction. The
physical channels listed in eq.(4.81) give rise to the following linear
combination of $M_{1,2,3}$ (and $N_{1,2,3}$ for $k^2 <0$).
$$\eqalign{ \gamma p & \to \pi^+ \pi^- p: \,\,M_1 + M_2, \cr
\gamma p & \to \pi^+ \pi^0 n: \,\,\sqrt{2} M_3, \cr \gamma p & \to
\pi^0 \pi^0 p:\,\,M_1 + M_2 + 2 M_3, \cr \gamma n & \to \pi^+ \pi^- n:\,\,
M_1 - M_2,\cr  \gamma n & \to \pi^0 \pi^- p:\,\,\sqrt{2} M_3, \cr
\gamma n & \to \pi^0  \pi^0 n:\,\, M_1 - M_2 - 2 M_3 \cr} \eqno(4.84)$$
Close to threshold, the invariant
matrix element squared averaged over nucleon spins and photon polarizations
takes the form
$|{\cal M}_{fi}|^2 = \vec k^2 \,| \eta_1 M_1 + \eta_2 M_2 + \eta_3 M_3|^2$
with the isospin factors $\eta_{1,2,3}$ given in eq.(4.84).
The main dynamical assumption in this relation is that the two-pion
photoproduction amplitude in the threshold region can be approximated by the
amplitude at threshold.
Expressing $\vec
k^2$ in terms of $s$ and supplementing $|{\cal M}_{fi}|^2$ by the photon
flux factor $m^2/ p_1\cdot k= 2m^2 /(s-m^2)$, we find for the unpolarized total
cross section
$$\sigma_{\rm tot}^{\gamma N \to \pi\pi N}(s) = {m^2 \over 2s} (s-m^2)
\Gamma_3(s)\, |\eta_1 M_1 + \eta_2 M_2 + \eta_3 M_3|^2 \, S\,. \eqno(4.85)$$
Here, $\Gamma_3(s) $ is the integrated three-body phase space, eq.(3.91),
 and $S$ a Bose
symmetry factor, $S = 1/2$ for the $\pi^0 \pi^0$ final state and $S = 1$
otherwise.   For equal pion masses an excellent approximation to the
integrated three--body phase space is given by [4.81]
$$\Gamma_3(s) \approx { M_\pi m^{5/2} \over 64 \pi^2 ( m + 2 M_\pi)^{7/2}} \,
[E_\gamma - 2 M_\pi( 1 + \mu) ] ^2\,. \eqno(4.86)$$
Of course, an analogous approximation can be derived for unequal pion masses.
Consequently, the unpolarized total cross section can be approximated within a
few percent by the handy formula
$$\sigma_{\rm tot}^{\gamma N \to \pi \pi N}(E_\gamma)
\approx { M_\pi^2 ( 1 + \mu)
\over 32 \pi^2 ( 1 + 2\mu)^{11/2}} |\eta_1 M_1 + \eta_2 M_2 + \eta_3 M_3 |^2
\,S \, (E_\gamma - E_\gamma^{\rm thr})^2\,. \eqno(4.87)$$
For electroproduction, the prefactor in eq.(4.87) has
to be modified slightly to account for the virtual photon flux normalization
and then it gives the transverse total electroproduction cross section.
In general above threshold the total cross section is given by a
four--dimensional integral over e.g. the two pion  energies and two angle
variables (for details, see ref.[4.81]).
One remark on isospin breaking is in order. To one--loop accuracy ${\cal
O}(q^3)$, it is legitimate to work with one nucleon and one pion mass.
 However, the  pion mass difference
$M_{\pi^\pm} - M_{\pi^0} = 4.6$ MeV in reality leaves a 11.9 MeV gap between
the production threshold of two neutral versus two charged pions. While we are
not in position of performing a calculation including all possible
isospin-breaking effects, a minimal procedure  to account for the mass
difference of the physical particles is to put in these by hand in the
pertinent kinematics, such that the thresholds open indeed at the correct
energy value. To be specific, for the $\pi^+ \pi^- p$ final state the threshold
photon energy is
$E_\gamma^{\rm thr}(\pi^+ \pi^- p) = 320.66$ MeV
whereas for $\pi^0 \pi^0 p$ it is
$E_\gamma^{\rm thr}(\pi^0 \pi^0 p) = 308.77$ MeV.
Therefore, in the pertinent three--body phase space integrals
 we will differentiate
between neutral and charged pion mass when we present results incorporating
the correct opening of the thresholds.

Consider now the chiral expansion of the
threshold amplitudes $M_{1,2,3}$ and $N_{1,2,3}$. In each case we will give two
complete chiral powers, the leading and next-to-leading term. It is worth to
elaborate a bit on the chiral counting here. The S--matrix elements are
calculated up--to--and--including order $q^3$. This means that the threshold
amplitudes are given to order $q$ since two chiral powers are factored out,
$\epsilon \sim {\vec k} \sim q$. Due to the various selection rules which
apply for heavy baryon CHPT and additional ones due to the threshold
kinematics,
only a few diagrams are contributing. These are discussed in detail in [4.81].
Here, we just mention that the leading nonzero contributions
comes from tree graphs with one insertion from ${\cal L}^{(2)}_{\pi
N}$. At next order, one has a plethora of possible contributions.  Four loop
graphs (plus their crossed partners) remain and the only contact
 terms which survive are the ones with
one insertion from ${\cal L}^{(3)}_{\pi N}$.\footnote{*}{Within our
approximation, the contribution from
${\cal L}^{(4)}_{\pi \pi} $ containing the Wess-Zumino term incorporating
the anomalous (natural parity violating) vertex $\gamma \to 3 \pi$ vanishes.}
 The corresponding low--energy
constants are estimated via resonance saturation, i.e. the $\Delta$
contribution.  One expects sizeable effects from the
$\Delta(1232)$ since first it is quite close to threshold and second its
couplings to the $\gamma \pi N$ system are very large (about twice the nucleon
couplings). On first sight the distance of only $14.6 $ MeV of the
$\Delta(1232)$ from threshold seems to give rise to overwhelming contributions
since one naively expects that the very small denominator
$1 / ( m_\Delta^2- s_{\rm thr}) = 1 / ( m_\Delta - m - 2 M_\pi)\cdot 1 / (
m_\Delta + m + 2 M_\pi)$ enters the result. However, as shown in ref.[4.81],
this dangerous denominator always gets cancelled by exactly the same term
in the numerator in the corresponding diagrams. Therefore, the expansion in
$M_\pi$ is not a priori useless. For
the transverse threshold amplitude,  the resulting chiral expansion takes
the form
$$M_1 = {e g_A^2 M_\pi \over 4 m^2 F_\pi^2} + {\cal O}(q^2) \eqno(4.88a)$$
$$\eqalign{
M_2 = & {e \over 4 m F_\pi^2} (2g_A^2 - 1 - \kappa_v) + {e M_\pi \over 4 m^2
F_\pi^2} (g_A^2 - \kappa_v) + {e g_A^2 M_\pi \over 8 m m_\Delta^2 F_\pi^2}
B_\Delta \cr & + {e g_A^2 M_\pi \over 64 \pi F_\pi^4} \biggl\{ {8 + 4r \over
\sqrt{1+r}} \arctan\sqrt{1+r} - {r \over 1 + r} -{ 1+ r+r^2 \over
(1+r)^{3/2}}\biggl[ {\pi \over 2 } + \arctan{r \over \sqrt{1 + r}} \biggl]
\cr & + i \biggl[ {\sqrt{3}(2+r) \over 1 + r} - {1 + r + r^2 \over (1+r)^{3/2}}
\ln { 2 + r + \sqrt{3(1+r)} \over \sqrt{1 + r +r^2}} \biggr] \biggr\} + {\cal
O}(q^2)\,, \cr} \eqno(4.88b)$$
$$\eqalign{
M_3 = & {e \over 8 m F_\pi^2} (1 + \kappa_v- 2 g_A^2) + {e M_\pi \kappa_v\over
8 m^2 F_\pi^2} - {e g_A^2 M_\pi \over 16 m m_\Delta^2 F_\pi^2}
B_\Delta \cr & + {e g_A^2 M_\pi \over 256 \pi F_\pi^4} \biggl\{6- {4 + 2r \over
\sqrt{1+r}} \arctan\sqrt{1+r} - {r \over 1 + r} -{ 1+ r+r^2 \over
(1+r)^{3/2}}\biggl[ {\pi \over 2 } + \arctan{r \over \sqrt{1 + r}} \biggl]
\cr & + i \biggl[ {\sqrt{3}(2+r) \over 1 + r} - {1 + r + r^2 \over (1+r)^{3/2}}
\ln { 2 + r + \sqrt{3(1+r)} \over \sqrt{1 + r +r^2}} \biggr] \biggr\} + {\cal
O}(q^2) \cr } \eqno(4.88c)$$
with the ratio $r = -k^2/4M_\pi^2$
and
$$B_\Delta = {2m_\Delta^2 + m_\Delta m - m^2 \over m_\Delta - m} + 4Z [
m_\Delta(1+2Z) + m(1+Z)] \eqno(4.88d)$$
which involves the off-shell parameter $Z$ of the $\pi N \Delta$ vertex. In
fact, taking the allowed range of $Z$ given in ref.[4.43],
one finds a weak $Z$-dependence, $i.e.$ 9.9 GeV $< B_\Delta < 15.1$ GeV.
Furthermore, from the isospin factors of eq.(4.88) we see that to order $M_\pi$
the $\Delta$ contributions are absent in the $\pi^0 \pi^0$ channels. We also
note that to lowest order, $M_1 =0$ and $M_2 = -2 M_3$ so that the production
of two neutral pions is strictly suppressed.
Another point worth mentioning is that the transverse amplitudes
$M_{2,3}$ are $k^2$-dependent only through their loop contribution. This can
be understood from the fact that the tree graphs have to be polynomial in both
$M_\pi$ and $k^2$ and that a term linear in $k^2$ is already of higher order in
the chiral expansion. It is also not possible to further expand the
$r$-dependent functions since $r = -k^2/4 M_\pi^2$ counts as order one and all
terms have to be kept.   We also notice that
the amplitudes $M_{2,3}$ have a smooth behaviour in the chiral limit.
Finally, we note that
the loop contribution to the transverse amplitudes of two pion
production as given in eq.(4.88) have a nonzero imaginary part even at
threshold. This comes from the rescattering type graphs. Due
to unitarity the pertinent loop functions have a right hand cut starting at $s
= (m+ M_\pi)^2$ (the single pion production threshold) and these functions are
here  to be evaluated at $s = (m+2M_\pi)^2$ (the two-pion production
threshold).
In the electroproduction case, we also have the longitudinal threshold
amplitudes $N_{1,2,3}$. Since we can no more exploit the condition $\epsilon
\cdot k = 0$ the photon coupling to an out-going pion line is non-vanishing and
therefore we obtain a nonzero contribution already at leading order ${\cal
O}(q)$ involving a pion propagator (for charged pions).
 Adding up all terms which arise at order
$q$ and $q^2$ we find the following results
$$\eqalign{N_1 & = {\cal O}(q)\,, \cr N_2 & = {e M_\pi( 1 + \mu)
\over F_\pi^2 ( 4 M_\pi^2 - k^2)} + { e(2g_A^2 - 1) \over 4 m F_\pi^2} + {\cal
O}(q)\,, \cr N_3 & = - {1\over 2} N_2 + {\cal O}(q)\,. \cr} \eqno(4.89)$$
It is interesting to note that none of the low energy constants
$c_1,c_2,c_3,c_4$ and the anomalous magnetic moments which enter ${\cal
L}^{(2)}_{\pi N}$ show up in the final result.
\medskip
We now turn to the numerical results for the threshold
amplitudes and total cross sections. The isospin symmetric case is discussed
in great detail in ref.[4.81].
To connect to the experimental situation,  consider the three--body
phase space with the physical masses for the corresponding pions. This
automatically takes care of the various threshold energies. In the loops
we work, however, with one pion mass. This effect is small as discussed in
ref.[4.81]. In  fig.4.7 we show the calculations with the correct phase--space
 and  using the threshold matrix--elements.
\midinsert
\smallskip
\vskip -0.8truecm
\hskip 0.5in
\epsfysize=3.9in
\epsfxsize=3in
\epsffile{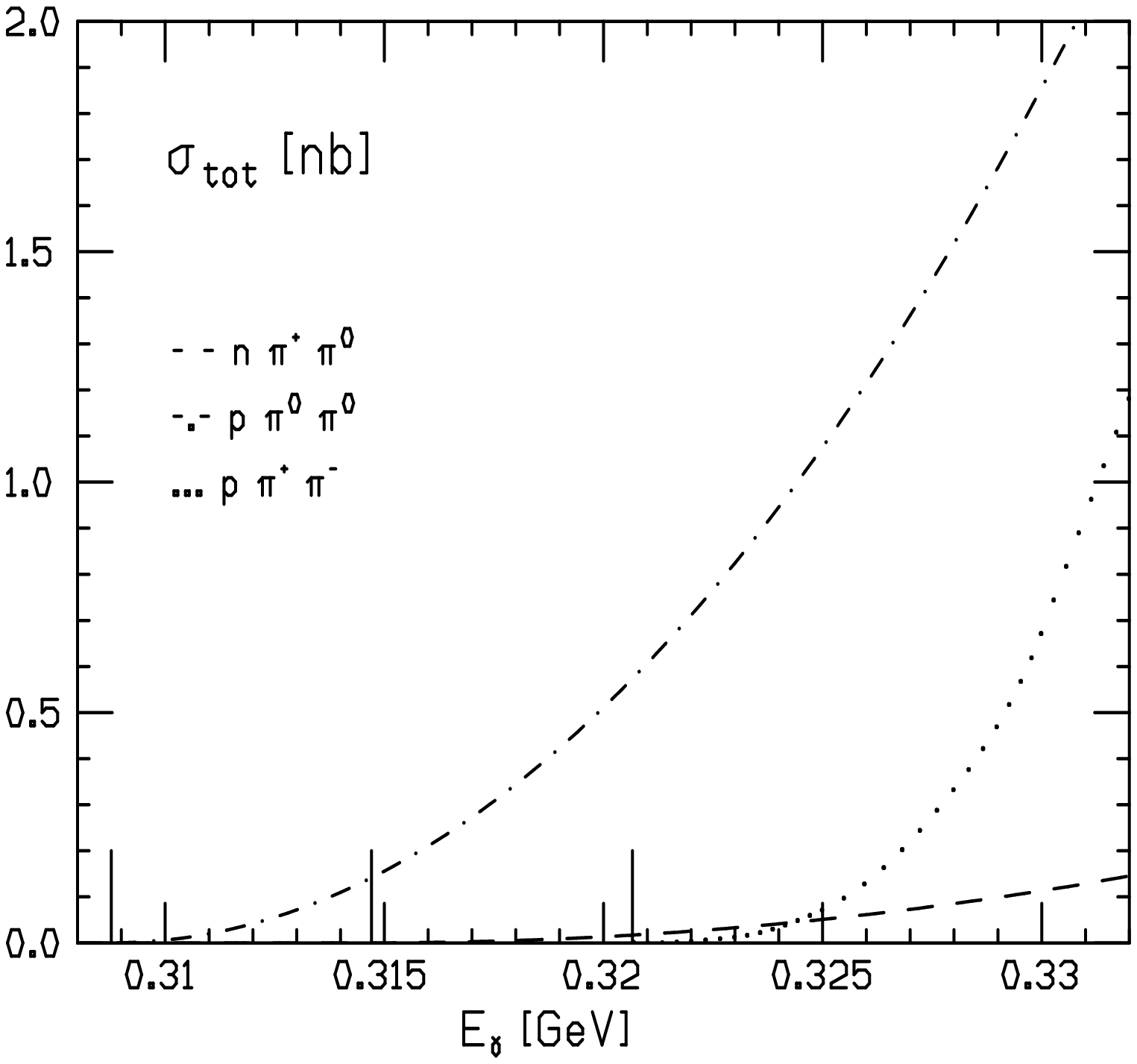}
\smallskip
\vskip -2truecm
{\noindent\narrower \it Fig.~4.7:\quad
Total cross sections (in nb) for the $\gamma p$
initial state ($k^2 =0$) with the correct three--body phase space. The dotted,
dashed and dashed--dotted lines refer to the $\pi^+ \pi^- p$, $\pi^+ \pi^0 n$,
and the $\pi^0 \pi^0 p$ final state, in order.
For $\gamma p \to \pi^+ \pi^- p$ we show the first correction as
discussed in ref.[4.81]. The various thresholds are indicated.
\smallskip}
\vskip -0.3truecm
\endinsert
For $\gamma p \to \pi^+ \pi^- p$, we show the first correction above
threshold from ${\cal L}_{\pi N}^{(1)}$ which is bigger than the
matrix--element calculated with the threshold amplitudes already 3 MeV
above threshold in this particular channel (for details, see [4.81]).
 At $E_\gamma = 320$ MeV, the total
cross section for $\pi^0 \pi^0$ production is 0.5 nb  whereas the competing
$\pi^0 \pi^+ n$ final state has $\sigma_{\rm tot} = 0.013$ nb.
 Double neutral pion
production reaches $\sigma_{\rm tot} = 1.0$  nb at $E_\gamma = 324.3$ MeV
in comparison to $\sigma_{\rm tot} (\gamma p \to \pi^0 \pi^+ n) < 0.05$ nb and
$\sigma_{\rm tot} (\gamma p \to \pi^+ \pi^- p) = 0.4$ nb.
 This means that for the
first 10...12 MeV above $\pi^0 \pi^0$ threshold, one has a fairly clean signal
and much more neutrals than expected.
We stress that the enhancement of the total cross section in the $\pi^0 \pi^0$
channel is a chiral loop effect. Consider the corresponding threshold matrix
element $M_1 + M_2 + 2M_3$. Whereas the Born graphs contribute only 4.9
GeV$^{-3}$, the loop contribution at the same chiral power is much larger (26.6
+ 16.0 i ) GeV$^{-3}$ and enhances the total cross section by a factor 50.
Already the knowlegde of the order of magnitude of the
experimental $\pi^0 \pi^0$
cross sections allows to test the enhancement effect of the chiral
loops.\footnote{$^*$}{We have been informed by Th. Walcher that a
first analysis of double neutral pion production making use of the
TAPS detector seem to indicate an even stronger increase of the $p
\pi^0 \pi^0$ cross section at threshold as indicated by the
calculation of ref.[4.81].}
Remember that to leading order
in the chiral expansion, the production of two neutral pions is
completely suppressed. Of course, the above threshold correction for this
channel, which comes from ${\cal L}_{\pi N}^{(2)}$ (and higher orders)
should be calculated systematically. The first correction, which vanishes
proportional to $|{\svec q}_i|$ ($i=1,2$) at threshold, has been calculated
and found to be very small. The corresponding cross section
at $E_\gamma =320$, $325$ and $330$ MeV is $\sigma_{\rm tot}^{\rm
first
\, corr}
= 0.009, \, 0.026$ and $0.056$ nb, i.e a few percent of the leading order
result.
It is, therefore, conceivable that the qualitative features
described above will not change if even higher order corrections are taken
into account. A more detailed account of these topics can be found in
ref.[4.81].
 \bigskip \goodbreak
\noindent{\bf IV.6. WEAK PION PRODUCTION}
\medskip
As discussed in the preceeding sections,
 single and double pion production off nucleons by real or virtual photons
gives important information about the structure of the nucleon. As stressed
in particular by Adler [4.84], weak pion production involves the isovector
axial
amplitudes and  a unified treatment of pion photo-, electro- and weak
production allows to relate information from electron--nucleon and
neutrino--nucleon scattering experiments. In this spirit, we will consider
here  pion production through the isovector axial current in the
threshold region, extending the classical
work of Adler [4.84], Adler and Dothan [4.52] and
Nambu, Luri{\'e} and Shrauner [4.75],
who have considered soft pion emission induced by weak interactions making
use of PCAC and gauge invariance, relating  certain electroweak
form factors of the nucleon to particular threshold multipole amplitudes.
The corrections beyond this were considered in ref.[4.85], were
novel relations between
various axial threshold multipole amplitudes and physical observables
like electroweak form factors, S--wave pion--nucleon scattering lengths and,
in particular, the nucleon scalar form factor, $\sigma(t) \sim\,<N|\hat m (\bar
u u + \bar d d)|N>$ are given.

We consider processes of the type $\nu(k_1) + N(p_1)
\to l(k_2) + N(p_2) + \pi^a(q)$, which involve the isovector vector
and axial--vector currents. Here, we will focus on the pion production
induced by the axial current, $A_\mu^b = \bar q \gamma_\mu \gamma_5 (
\tau^b / 2) q$ in terms of the $u$ and $d$ quark fields. Denoting by
$k = k_1 - k_2$ the four--momentum of the axial current, the pertinent
Mandelstam variables are $s=(p_1+k)^2$, $t=(q-k)^2$ and $u=(p_1 -q)^2$
subject to the constraint $s+t+u = 2m^2 + M_\pi^2 +k^2$.
The pertinent matrix element decomposes into an isospin--even and
an isospin--odd part (analogous to the $\pi N$ scattering amplitude),
$$< N(p_2), \pi^a(q) \, \, {\rm out} | A^b \cdot \epsilon | N(p_1) \, \,
{\rm in} >
\, = \, i \delta^{ab}\, T^{(+)} \cdot \epsilon \, \,- \, \, \epsilon^{abc} \,
 \tau^c \, T^{(-)} \cdot \epsilon    \eqno(4.90)$$
where $\epsilon_\mu$ is the axial polarization vector, $\epsilon_\mu
\sim \bar{u}_l \gamma_\mu \gamma_5 u_\nu$. Notice that one can use the
Dirac equation to transform terms of the type $\epsilon \cdot k$ into
lepton mass terms via
$\epsilon \cdot k \sim \bar{u}_l \barre{k} \gamma_5 u_\nu = - m_l
\bar{u}_l  \gamma_5 u_\nu$.
This means that in the approximation of zero lepton mass, one has
$\epsilon \cdot k = 0$ and all diagrams where the axial source  couples
directly to a pion line vanish. The general Dirac structure for the transition
current involves the eight operators ${\cal O}_1 = (\barre \epsilon\,
\barre q - \barre q\,\barre \epsilon)/2 , \, {\cal O}_2 = \epsilon \cdot q,\,
{\cal O}_3 = \barre \epsilon,\, {\cal O}_4 = \epsilon \cdot (p_1 + p_2)/2,\,
{\cal O}_5 = \barre k\,\epsilon \cdot q,\, {\cal O}_6 = \barre k\,\epsilon\cdot
(p_1 + p_2) /2 ,\, {\cal O}_7 = \epsilon \cdot k,\, {\cal O}_8 = \barre k\,
\epsilon \cdot k$ which are accompanied by invariant functions denoted
$A_i^{(\pm )} (s,u)$ $(i=1, \ldots ,8)$ [4.84]. At threshold
in the $\pi N$ center of mass frame, one can express the
pertinent matrix element in terms of six S--wave multipoles, called
$L_{0+}^{(\pm )}$, $M_{0+}^{(\pm )}$ and $H_{0+}^{(\pm )}$,
$$T^{(\pm)}\cdot \epsilon = \bar{u}_2 \sum_{i=1}^8 {\cal O}_i A_i^{(\pm)}
(s_{\rm th}, u_{\rm th}) u_1 = 4 \pi ( 1 + \mu ) \chi_2^\dagger \bigl[
\epsilon_0 L_{0+}^{(\pm )} + \epsilon \cdot k H_{0+}^{(\pm )} + i \vec{\sigma}
\cdot (\hat{k} \times \vec{\epsilon} ) M_{0+}^{(\pm )} \bigr] \chi_1 \, \, .
\eqno(4.91)$$
At threshold, one can express
$L_{0+}$, $M_{0+}$ and $H_{0+}$ (suppressing isospin indices) through
the invariant amplitudes $A_i (s,u)$ via
$$\eqalign{
M_{0+} &= {\sqrt{\mu^2 - \nu} \over 8 \pi (1+ \mu )^{3/2}}\bigl\lbrace
m \mu A_1 - A_3 \bigr\rbrace \cr
L_{0+} &= {\sqrt{(2+\mu)^2 - \nu} \over
8 \pi (1+ \mu )^{3/2}} \biggl\lbrace -{\mu ( 2 + \mu ) + \nu \over
(2+\mu)^2 - \nu} \mu m A_1 + m \mu A_2 \cr
&\quad + {2(1+\mu)(2+\mu) \over (2+\mu)^2 - \nu} A_3 + m\bigl(1+ {\mu \over 2}
\bigr)A_4 + \mu^2 m^2 A_5 + m^2 \mu \bigl( 1+ {\mu \over 2} \bigr) A_6
\biggr\rbrace \cr
H_{0+} &= {\sqrt{(2+\mu)^2 - \nu} \over
8 \pi (1+ \mu )^{3/2}} \biggl\lbrace {2 ( 1 + \mu )  \over
(2+\mu)^2 - \nu} \bigl(\mu  A_1 - {1 \over m} A_3 \bigr) - {1 \over 2} A_4
- {m \over 2} \mu A_6 + A_7 + \mu m A_8
\biggr\rbrace
 \cr} \eqno(4.92)$$
where the $A_i (s,u)$ are evaluated at threshold.

 We seek an expansion of these
threshold multipoles in powers of $\mu$ and $\nu$ up to and including
order ${\cal O}(\mu^2 ,\nu )$ (modulo logarithms).
To work out the corrections at order ${\cal O}(q^3)$, it it mandatory to
perform a complete one--loop calculation with insertions from
${\cal L}_{\pi N}^{(1)}$ and the tree diagrams with exactly one insertion
from ${\cal L}_{\pi N}^{(3)}$. One also has to consider tree graphs with
two insertions from ${\cal L}_{\pi N}^{(2)}$ with a nucleon propagator,
which scales as $1/q$, in between.
The resulting low--energy theorems for the various S--wave multipoles are
$$\eqalign{
  M_{0+}^{(+)}
 &= {\sqrt{M_\pi^2 - k^2} \over 16  \pi m F_\pi}
\biggl\lbrace g_A^2 + C_M^{(+)} M_\pi \biggr\rbrace + {\cal O}(q^3 ) \cr
M_{0+}^{(-)}
 &= {\sqrt{M_\pi^2 -k^2} \over 16 \pi m F_\pi} \biggl\lbrace G_M^V(k^2-M_\pi^2)
 - g_A^2 + C_M^{(-)} M_\pi \biggr\rbrace + {\cal O}(q^3) \cr} \eqno(4.93a) $$
$$\eqalign{
L_{0+}^{(+)} &= {1 \over 3 \pi M_\pi F_\pi} \biggl\{ \sigma(k^2 - M_\pi^2) -
{1 \over 4} \sigma (0) \biggr\} - {a^+ F_\pi \over M_\pi} - {g_A^2
M_\pi \over 16 \pi m F_\pi} + C^{(+)}_L M_\pi^2 + {\cal O}(q^3) \cr
L_{0+}^{(-)} &= { 1 \over 8 \pi F_\pi} \biggl\lbrace -G_E^V (k^2 - M_\pi^2)
+{M_\pi\over2m}  (g_A^2 + 1) - {k^2 \over 8m^2} \biggr\rbrace + C_L^{(-)}
M_\pi^2 +  {\cal O}(q^3 )  \cr} \eqno(4.93b)$$
$$\eqalign{
H_{0+}^{(+)} &= {a^+ F_\pi \over k^2 - M_\pi^2} + {\sigma(0) - \sigma(k^2 -
M_\pi^2 ) \over 12 \pi F_\pi (k^2 - M_\pi^2) } + C_H^{(+)}M_\pi +{\cal O}(q^2)
\cr  H_{0+}^{(-)} &= {a^- F_\pi \over k^2 - M_\pi^2} + {M_\pi [G_E^V (k^2 -
M_\pi^2 ) - 1 ] \over 8 \pi F_\pi (k^2 - M_\pi^2) } +{1 \over 16 \pi m F_\pi} +
C_H^{(-)}M_\pi + {\cal O}(q^2) \cr} \eqno(4.93c)$$
with $a^{\pm}$ the isopin--even and odd S--wave $\pi N$ scattering lengths.
The form of the pion pole term in $H_{0+}^{(\pm )}$ can easily be understood
from the fact that as $k \to q$  one picks up as a residue the forward $\pi N$
scattering amplitude which at threshold is expressed in terms of the two
S--wave scattering lengths. The relation between axial pion production and
the $\pi N$ scattering amplitude has also been elucidated by Adler in his
seminal work [4.84].
The constants $C^{(\pm )}_{H,L,M}$ subsume numerous $k^2$--independent
kinematical, loop and counterterm corrections (the latter ones stem mainly
from
${\cal L}_{\pi N}^{(3)}$) which we do not need for the following discussion and
which are difficult to pin down exactly. There is, however, one exception to
this. The chiral Ward identity
$ \partial^\mu A^b_\mu = \hat{m} \, \bar{q} i \tau^b \gamma_5 q \sim M_\pi^2$
demands that $k_0 L_{0+} + k^2 H_{0+} \sim M_\pi^2$ and thus with $k_0 = m(2\mu
+ \mu^2 + \nu)/2(1+\mu)$  we have
$$\eqalign{
C_H^{(+)} &= { a^+ F_\pi \over 2m M_\pi^2} - {\sigma(0) \over 8 \pi
M_\pi^2 m F_\pi} + {g_A^2 \over 32 \pi m^2 F_\pi} = { c_2 + c_3  \over 4 \pi m
F_\pi} \cr C_H^{(-)} &= -{2 g_A^2 + 5 \over 64 \pi  m^2 F_\pi}\,\,. \cr}
\eqno(4.94)$$
The numerical values of the constants are $C_H^{(+)} = -1.0$ GeV$^{-3}$ and
$C_H^{(-)} = 0.5$ GeV$^{-3}$.
The argument of the various nucleon form factors in (4.94) is
 the threshold value of the invariant momentum transfer squared
$ t_{\rm thr} = (q-k)^2_{\rm thr} = (k^2 - M_\pi^2 )/( 1+ \mu) = k^2 -
M_\pi^2 + {\cal O}(q^3)$.
Of particular interest is the low--energy theorem for $L_{0+}^{(+)}$
where one has the following slope at the photon point $k^2 = 0$
$$ {\partial L_{0+}^{(+)} \over \partial k^2}\bigg|_{k^2 = 0} =
{\sigma'(-M_\pi^2) \over 3 \pi M_\pi F_\pi} + {\cal O}(M_\pi) = {g_A^2 \over
128 \pi^2 F_\pi^3 } \biggl( {6\over 5} - \arctan{1\over 2} \biggr) + {\cal
O}(M_\pi)\,.   \eqno(4.95)$$
It is very interesting to note that although $L_{0+}^{(+)}$ vanishes
identically in the chiral limit $M_\pi = 0$ the slope at $k^2 = 0$ stays
finite. The formal reason for this behaviour is the
non--analytic dependence of $L_{0+}^{(+)}$ on $M_\pi$ which does not allow to
interchange the order of taking the derivative with respect to $k^2$ at $k^2 =
0$ and the chiral limit.
Notice also that for $k^2 \simeq 0$ and assuming that $C_L^{(+)}$ of
the order of 1 GeV$^{-3}$, the term proportional to the scalar form factor
$\sigma(-M_\pi^2) - \sigma(0) / 4$
dominates the behaviour of $L_{0+}^{(+)}$ using the numbers from
the recent analysis of Gasser,
Leutwyler and Sainio [4.86].
In principle, an accurate measurement of this
particular multipole in weak pion production allows for a new determination of
the elusive nucleon scalar form factor and the $\pi N$ $\sigma$--term.
This might open the possibility of another determination of this fundamental
quantity. In the standard model, the axial part of the weak neutral
current is the third component of the isovector axial current. To see
this most interesting correction, one should therefore consider neutral
neutrino  reactions like $\nu p \to \nu p \pi^0$ (in that case the
zero lepton mass approximation is justified).
First, however, a complete calculation involving also the
isovector vector current
has to be performed to find out how cleanly one can
separate this multipole in the analysis of neutrino--induced single
pion production. For that, it will be mandatory to include the
$\Delta$ resonance since the presently available  data are concentrated
around this mass region [4.87]. In parity--violating electron scattering, the
interference of this axial current with the vector one is suppressed by the
factor $(1- 4 \sin^2 \theta_{\rm W})$, with $\sin^2 \theta_{\rm W} \simeq 0.23$
the Weinberg (weak mixing) angle.

In contrast to this, the behaviour of $H^{(+)}_{0+}$, which also contains
the scalar form factor, is dominated by the pion
pole term proportional to $a^+$. At $k^2 = 0$, one finds $H^{(+)}_{0+} =
32.8 \, {\rm GeV}^{-2} + ( \sigma(0) - \sigma (-M_\pi^2) ) \cdot 14.1
\, {\rm GeV}^{-3} - 0.14 \, {\rm GeV}^{-2}$. The uncertainty in $a^+$,
$\delta a^+ = \pm 0.38 \cdot 10^{-2}/M_\pi$, gives as large a contribution
as the term proportional to the scalar form factor.

Finally, we point out that Adler's
relation between weak single pion production and the elastic neutrino--nucleon
cross section at low energies [4.88] is also modified by the novel term
proportional to the scalar form factor of the nucleon.
\bigskip
\bigskip
\noindent{\bf REFERENCES}
\medskip
\item{4.1}E. Amaldi, S. Fubini and G. Furlan, {\it Pion Electroproduction},
Springer Verlag, Berlin, 1979. \smallskip
\item{4.2}J. Gasser, M.E. Sainio and A. ${\check {\rm S}}$varc,
{\it Nucl. Phys.\/} {\bf B307} (1988) 779.
\smallskip
\item{4.3}V. Bernard, N. Kaiser, J. Kambor
and Ulf-G. Mei{\ss}ner, {\it Nucl. Phys.\/} {\bf B388} (1992) 315.
\smallskip
\item{4.4}M.A.B. B\'eg and A. Zepeda, {\it Phys. Rev.\/} {\bf D6} (1972) 2912.
\smallskip
\item{4.5}D.G. Caldi and H. Pagels, {\it Phys. Rev.\/} {\bf D10} (1974) 3739.
\smallskip
\item{4.6}Ulf-G. Mei{\ss}ner, {\it Int. J. Mod. Phys.} {\bf E1} (1992) 561.
\smallskip
\item{4.7}Ulf-G. Mei{\ss}ner, Review talk at WHEPP--III, Madras, India, 1994,
preprint CRN--94/04.
\smallskip
\item{4.8}W.R. Frazer and J. Fulco, {\it Phys. Rev.} {\bf 117} (1960)
1603, 1609. \smallskip
\item{4.9}G. H\"ohler and E. Pietarinen, {\it Phys. Lett.} {\bf B53} (1975)
471. \smallskip
\item{4.10}G. H\"ohler, in Land\"olt--B\"ornstein, vol.9 b2, ed. H. Schopper
(Springer, Berlin, 1983).
\smallskip
\item{4.11}R.E. Prange, {\it Phys. Rev.} {\bf 110} (1958) 240. \smallskip
\item{4.12}A.C. Hearn and E. Leader, {\it Phys. Rev.} {\bf 126} (1962) 789.
\smallskip
\item{4.13}F. Low, {\it Phys. Rev.} {\bf 96} (1954) 1428;

M. Gell-Mann and M.L. Goldberger, {\it Phys. Rev.} {\bf 96} (1954) 1433.
\smallskip
\item{4.14}S. Drell and A.C. Hearn, {\it Phys. Rev. Lett.} {\bf 16} (1966) 908;

S.B. Gerasimov, {\it Sov. J. Nucl. Phys.} {\bf 2} (1966) 430.
\smallskip
\item{4.15}V. Bernard, N. Kaiser and Ulf-G. Mei{\ss}ner,
{\it Phys. Rev. Lett.\/} {\bf 67} (1991) 1515;
{\it Nucl. Phys.\/} {\bf B373} (1992) 364.
\smallskip
\item{4.16}S. Ragusa, {\it Phys. Rev.} {\bf D49} (1994) 3157.
\smallskip
\item{4.17}J.L. Powell, {\it Phys. Rev.} {\bf 75} (1949) 32. \smallskip
\item{4.18}F.J. Federspiel et al., {\it Phys. Rev. Lett.\/} {\bf 67}
(1991) 1511.  \smallskip
\item{4.19}E.L. Hallin et al., {\it Phys. Rev.} {\bf C48} (1993) 1497.
\smallskip
\item{4.20}A. Zieger {\it et al.}, {\it Phys. Lett.\/} {\bf B278}
(1992) 34. \smallskip
\item{4.21}M. Damashek and F. Gilman, {\it Phys. Rev.\/} {\bf D1} (1970) 1319;
\smallskip
\item{4.22}V.A. Petrunkin, {\it Sov. J. Nucl. Phys.\/} {\bf 12} (1981) 278.
\smallskip
\item{4.23}J. Schmiedmayer {\it et al.}, {\it Phys. Rev. Lett.\/} {\bf 66}
(1991) 1015.
\smallskip
\item{4.24}K.W. Rose {\it et al.}, {\it Phys. Lett.\/} {\bf B234}
(1990) 460. \smallskip
\item{4.25}V. Bernard, N. Kaiser, A. Schmidt
and Ulf-G. Mei{\ss}ner, {\it Phys. Lett.\/} {\bf B319} (1993) 269.
\smallskip
\item{4.26}M. N. Butler and M. J. Savage, {\it Phys. Lett.} {\bf B294} (1992)
369.
\smallskip
\item{4.27}V. Bernard, N. Kaiser, A. Schmidt
and Ulf-G. Mei{\ss}ner, {\it Z. Phys.\/} {\bf A348} (1994) 317.
\smallskip
\item{4.28}N.C. Mukhopadhyay, A.M. Nathan and L. Zhang,
{\it Phys. Rev.\/} {\bf D47} (1993) R7.
\smallskip
\item{4.29}N.M. Butler, M.J. Savage and R. Springer,
{\it Nucl. Phys.\/} {\bf B399} (1993) 69.
\smallskip
\item{4.30}V. Bernard, N. Kaiser, J. Kambor and Ulf-G. Mei{\ss}ner,
{\it Phys. Rev.\/} {\bf D46} (1992) 2756.
\smallskip
\item{4.31}R. Erbe et al., {\it Phys. Rev.\/} {\bf 188} (1969) 2060.
\smallskip
\item{4.32}M. Benmerrouche, R.M. Davidson and N.C. Mukhopadhyay,
{\it Phys. Rev.\/} {\bf C39} (1989) 2339.
\smallskip
\item{4.33}A. L'vov, {\it Phys. Lett.} {\bf B304} (1993) 29;

B.R. Holstein and A.M. Nathan,
{\it Phys. Rev.\/} {\bf D49} (1994) 6101.
\smallskip
\item{4.34}T.A. Armstrong et al., {\it Phys. Rev.\/} {\bf D5} (1970) 1640;

T.A. Armstrong et al., {\it Nucl. Phys.\/} {\bf B41} (1972) 445.
\smallskip
\item{4.35}R.D. Peccei, {\it Phys. Rev.\/} {\bf 181} (1969) 1902.
\smallskip
\item{4.36}A.M. Sandorfi et al., {\it Phys. Rev.\/} {\bf D50} (1994) R6681.
\smallskip
\item{4.37}I. Karliner, {\it Phys. Rev.\/} {\bf D7} (1973) 2717.
\smallskip
\item{4.38}R.L. Workman and R.A. Arndt, {\it Phys. Rev.\/} {\bf D45} (1992)
1789. \smallskip
\item{4.39}V. Burkert and Z. Li, {\it Phys. Rev.\/} {\bf D47} (1993) 46.
\smallskip
\item{4.40}J. Ashman et al., {\it Nucl. Phys.} {\bf B328} (1989) 1.
\smallskip
\item{4.41}V. Bernard, N. Kaiser and Ulf-G. Mei{\ss}ner,
{\it Phys. Rev.\/} {\bf D48} (1993) 3062.
\smallskip
\item{4.42} Particle Data Group, {\it Phys. Rev.} {\bf D45} (1993) S1.
\smallskip
\item{4.43}T. Kitagaki
{\it et al.}, {\it Phys. Rev.\/} {\bf D28}
(1983) 436;

L.A. Ahrens
{\it et al.}, {\it Phys. Rev.\/} {\bf D35}
(1987) 785;

L.A. Ahrens
{\it et al.}, {\it Phys. Lett.\/} {\bf B202}
(1988) 284.
\smallskip
\item{4.44}A. del Guerra  {\it et al.}, {\it Nucl. Phys.\/}
 {\bf B107} (1976) 65;

M.G. Olsson, E.T. Osypowski and E.H. Monsay, {\it
Phys. Rev.\/} {\bf D17} (1978) 2938.
\smallskip
\item{4.45}Ulf-G. Mei{\ss}ner, {\it Phys. Reports\/} {\bf 161} (1989) 213.
\smallskip
\item{4.46}V. Bernard, N. Kaiser, T.-S. H. Lee and Ulf-G. Mei{\ss}ner,
{\it Phys. Reports\/} {\bf 246}  (1994) 315.
\smallskip
\item{4.47} G. Bardin et al., {\it Phys.Lett.}  {\bf B104} (1981) 320.
\smallskip
\item{4.48} J. Bernabeu, {\it Nucl. Phys.} {\bf A374} (1982) 593c.
\smallskip
\item{4.49}S. Choi et al., {\it Phys. Rev. Lett.} {\bf 71} (1993) 3927.
\smallskip
\item{4.50}
D. Taqqu, contribution presented at the International Workshop on
"Large Experiments
at Low Energy Hadron Machines", PSI, Switzerland, April 1994;
and  private communication.
\smallskip
\item{4.51}V. Bernard, N. Kaiser and Ulf-G. Mei{\ss}ner,
{\it Phys. Rev.\/} {\bf D50} (1994) 6899.
\smallskip
\item{4.52} S.L. Adler and Y. Dothan, {\it Phys. Rev.} {\bf 151} (1966) 1267.
\smallskip
\item{4.53} L. Wolfenstein, in: High-Energy Physics and Nuclear Structure,
ed. S. Devons (Plenum, New York, 1970) p.661.
\smallskip
\item{4.54}E. Mazzucato et al., {\it Phys. Rev. Lett.\/} {\bf 57} (1986) 3144.
\smallskip
\item{4.55}R. Beck et al., {\it Phys. Rev. Lett.\/} {\bf 65} (1990) 1841.
\smallskip
\item{4.56}T. P. Welch et al., {\it Phys. Rev. Lett.\/}
{\bf 69} (1992) 2761.
\smallskip
\item{4.57}S. Nozawa and T.-S. H. Lee, {\it Nucl. Phys.} {\bf A513} (1990) 511,
544. \smallskip
\item{4.58}F.A. Berends, A. Donnachie and D.L. Weaver,
{\it Nucl. Phys.\/} {\bf B4} (1967) 1.
\smallskip
\item{4.59}V. Bernard, N. Kaiser and Ulf-G. Mei{\ss}ner,
{\it Nucl. Phys.\/} {\bf B383} (1992) 442.
\smallskip
\item{4.60}I.A. Vainshtein and V.I. Zakharov,
{\it Sov. J. Nucl. Phys.\/} {\bf 12} (1971) 333;
{\it Nucl. Phys.\/} {\bf B36} (1972) 589;

P. de Baenst, {\it Nucl. Phys.\/} {\bf B24} (1970) 633.
\smallskip
\item{4.61}D. Drechsel and L. Tiator,
{\it J. Phys. G: Nucl. Part. Phys.\/} {\bf 18} (1992) 449.
\smallskip
\item{4.62}V. Bernard, J. Gasser, N. Kaiser and Ulf-G. Mei{\ss}ner,
{\it Phys. Lett.\/} {\bf B268} (1991) 291.
\smallskip
\item{4.63}V. Bernard, N. Kaiser and Ulf-G. Mei{\ss}ner,
``Neutral Pion Photoproduction off Nucleons Revisited'', preprint CRN
94-62 and TK 94 18, 1994, hep-ph/9411287.
\smallskip
\item{4.64}V. Bernard, N. Kaiser and Ulf-G. Mei{\ss}ner,
$\pi N$ {\it Newsletter} {\bf  7} (1992) 62.
\smallskip
\item{4.65}J.C. Bergstrom, {\it Phys. Rev.} {\bf C44} (1991) 1768.
\smallskip
\item{4.66}A.M. Bernstein, private communication.
\smallskip
\item{4.67}N.M. Kroll and M.A. Ruderman, {\it Phys. Rev.\/} {\bf 93} (1954)
233.
\smallskip
\item{4.68}J.P. Burg, {\it Ann. Phys. {\rm (Paris)}} {\bf 10} (1965) 363.
\smallskip
\item{4.69}M.J. Adamovitch et al.,
{\it Sov. J. Nucl. Phys.} {\bf 2} (1966) 95.
\smallskip
\item{4.70}E.L. Goldwasser et al., Proc. XII Int. Conf. on High Energy Physics,
Dubna, 1964, ed. Ya.-A. Smorodinsky (Atomizdat, Moscow, 1966).
\smallskip
\item{4.71}M.L. Goldberger, H. Miyazawa and R. Oehme, {\it Phys. Rev.} {\bf 99}
(1955) 986.
\smallskip
\item{4.72}R.L. Workman, R.A. Arndt and M.M. Pavan, {\it Phys. Rev. Lett.} {\bf
68} (1992) 1653.
\smallskip
\item{4.73}G. H\"ohler, Karlsruhe University preprint TTP 92--21, 1992.
\smallskip
\item{4.74}V. Bernard, N. Kaiser and Ulf-G. Mei{\ss}ner,
{\it Phys. Lett.\/} {\bf B282} (1992) 448.
\smallskip
\item{4.75}Y. Nambu and D. Luri\'e, {\it Phys. Rev.\/} {\bf 125}
(1962) 1429;

Y. Nambu and E. Shrauner, {\it Phys. Rev.\/} {\bf 128}
(1962) 862.
\smallskip
\item{4.76}V. Bernard, N. Kaiser and Ulf-G. Mei{\ss}ner,
{\it Phys. Rev. Lett.\/} {\bf 69} (1992) 1877.
\smallskip
\item{4.77}V. Bernard, N. Kaiser, T.--S. H. Lee and Ulf-G. Mei{\ss}ner,
{\it Phys. Rev. Lett.} {\bf 70} (1993) 387.
\smallskip
\item{4.78}M. Distler and Th. Walcher, private communication.
\smallskip
\item{4.79}J.M. Laget, {\it Phys. Rep.\/} {\bf 69} (1981) 1.
\smallskip
\item{4.80}P.W. Carruthers and H.W. Huang, {\it Phys. Lett.} {\bf B24} (1967)
464;

H.W. Huang, {\it Phys. Rev.} {\bf 174} (1968) 1799;

S.C. Bhargava, {\it Phys. Rev.} {\bf 171} (1968) 969.
\smallskip
\item{4.81}V. Bernard, N. Kaiser, Ulf-G. Mei{\ss}ner and A. Schmidt,
{\it Nucl. Phys.} {\bf A580} (1994) 475
\smallskip
\item{4.82}R. Dahm and D. Drechsel, in Proc.
 Seventh Amsterdam Mini--Conference,
eds. H.P. Blok, J.H. Koch and H. De Vries, Amsterdam, 1991. \smallskip
\item{4.83}S. Weinberg, {\it Phys. Rev.\/} {\bf 166} (1968) 1568.
\smallskip
\item{4.84}S.L. Adler, {\it Ann. Phys. {\rm (N.Y.)}\/} {\bf 50}
(1968) 189. \smallskip
\item{4.85}V. Bernard, N. Kaiser, and Ulf-G. Mei{\ss}ner,
{\it Phys. Lett.\/} {\bf B331} (1994) 137.
\smallskip
\item{4.86}J. Gasser, H. Leutwyler and M.E. Sainio, {\it Phys. Lett.\/}
{\bf B253} (1991) 252,260.
\smallskip
\item{4.87}S.J. Barish et al., {\it Phys. Rev.\/} {\bf D19}
(1979) 2521;

M. Pohl et al., {\it Lett. Nuovo Cimento}  {\bf 24} (1979) 540;

N.J. Baker et al, {\it Phys. Rev.\/} {\bf D23}
(1981) 2495.
\smallskip
\item{4.88}S.L Adler, {\it Phys. Rev. Lett.\/} {\bf 33} (1974) 1511;
{\it Phys. Rev.\/} {\bf D12} (1975) 2644.
\smallskip
\item{4.89}J.C. Bergstrom  and E.L. Hallin, {\it Phys. Rev.} {\bf C48}
(1994)  1508. \smallskip
\item{4.90}G. Ecker and Ulf-G. Mei{\ss}ner, ``What is a Low--Energy
Theorem ?'', {\it Comments. Nucl. Part. Phys.} (1995) in print.
\smallskip
\item{4.91}A. Schmidt, Thesis TU M\"unchen, 1995 (unpublished).
\smallskip
\vfill \eject
\noindent{\bf V. THE NUCLEON--NUCLEON INTERACTION}
\medskip
One of the best studied objects in nuclear physics is the interaction
between two nucleons. It is well--known that to a high degree of
accuracy one can consider nuclei as made of nucleons which behave
non--relativistically and interact pair--wise. Furthermore, three--
and many--body forces are believed to be small. This has lead to the
construction of semi--phenomenological boson exchange
potentials. These
describe accurately deuteron properties and low--energy
nucleon--nucleon phase shifts. The salient feature of these potentials
[5.1--5.9] can be summarized as follows.
At large separation, there is one--pion exchange first introduced by
Yukawa [5.1]. The intermediate--range attraction between two nucleons
can be understood in terms of a {\it fictitious} scalar--isoscalar
$\sigma$--meson with a mass of approximatively 550
MeV. $\omega$--meson exchange gives rise to part of the short--range
repulsion and the $\rho$ features prominently in the isovector--tensor
channel, where it cuts down most of the pion tensor potential.
There are, of course, differences in the various potentials but these
will not be discussed here. As we will show in what follows, the
effective chiral Lagrangian approach of QCD can be used to gain some
insight into the question why these potentials work after all. One can
also extend these considerations to many--nucleon forces as well as
meson--exchange currents. The latter are the cleanest signal of
non--nucleonic degrees of freedom in nuclei, in particular of pions
[5.10,5.11]. First, however, we have to discuss some technical
subtleties related to the appearance of different energy scales in two
(and many) nucleon systems.
\medskip
\noindent{\bf V.1. GENERAL CONSIDERATIONS}
\medskip
\goodbreak
The consequences of the spontaneous chiral symmetry breakdown for the
problem of the forces between nucleons were first discussed by
Weinberg [5.12,5.13]. Since in his papers and the subsequent ones of
the Texas/Seattle group [5.14,5.15,5.16,5.17,5.18] another language than the
previously discussed one is used, we first have to review the
construction of the chiral Lagrangian and the power counting in this
scheme. We will then address the problem of small energy scales (small
energy denominators) related to the nuclear binding. Since
SU(2)$\times$SU(2) is locally isomorphic to SO(4) and SU(2)$\sim$SO(3), one
can use stereographic coordinates to describe the Goldstone bosons
living on the three sphere $S^3 \sim$SO(4)/SO(3). The covariant
derivative of the pions  is
$$ \vec{D}_\mu = {\partial_\mu {\vec \pi} \over 2 D F_\pi} \, \, ,
\quad D = 1 + {{\vec \pi \,}^2 \over 4 F_\pi^2} \quad . \eqno(5.1)$$
Notice that we use $F_\pi = 93$ MeV in contrast to the conventions of
refs.[5.12-5.18] which have $F_\pi = 186$ MeV. Nucleons are decribed
by a Dirac spinor $N$, which is also a Pauli spinor in isospace. The
effective chiral Lagrangian is constructed out of the fields
$\vec{D}_\mu $, $N$ and their covariant derivatives,
$$\eqalign{
{\cal D}_\mu  \vec{D}_\nu & = \partial_\mu \vec{D}_\nu  + i {\vec
E}_\mu \times \vec{D}_\nu \cr
{\cal D}_\mu  N & = ( \partial_\mu   +  {\vec
t} \cdot \vec{E}_\mu ) N \cr}   \eqno(5.2)$$
where ${\vec E}_\mu = i({\vec \pi} \times {\vec D}_\mu)/F_\pi$ and
${\vec t} ={\vec \tau}/2$ the isospin generators in
the ${1 \over 2} $ representation (for more details,
see appendix C. There it is also shown how to include the $\Delta$
resonance in this framework). The most general effective chiral
Lagrangian follows by considering all possible isoscalar terms and
imposing proper Lorentz, parity and time--reversal invariance and
hermiticity. The explicit chiral symmetry breaking is due to the
fourth component of a chiral four vector with coefficient $(m_u +
m_d)/2$. The construction of these terms is also discussed in appendix C.
\medskip
Consider now the S--matrix for the scattering process of N incoming
and N outgoing nucleons, all with momenta smaller than some scale $Q$,
say $Q \ll M_\rho \simeq m$. This means that the nucleons are
non--relativistic and it thus is appropriate to use old--fashioned
time--ordered perturbation theory. In that case, one deals with energy
denominators for the intermediate states instead of the usual particle
propagators. The idea is now to order all contributions in powers of
$Q/m$. This is, however, not straightforward. In fact, as will become
clear later, one is dealing with a three scale problem,
$$m \gg Q \gg {Q^2 \over 2m} \qquad .   \eqno(5.3)$$
The appearance of the nucleon mass $m$ is obvious and the related
scale can be removed by either defining velocity--dependent fields
(cf. section 3.3) or using the equation of motion to eliminate the
large time derivatives, $\partial_0 N \sim m \, N$, as described below [5.13].
The occurence of the third (small) scale $Q^2 / 2m$ is related to the
presence of shallow nuclear bound states. In fact, consider a
time--ordered diagram with only N nucleons in the intermediate state. The
energy denominator associated to such a diagram is of order $Q^2 /
2m$, whereas all other diagrams contain at least one pion in the
intermediate state and have energy denominators of order $Q$. The appearance
of this small scale causes the perturbation theory to diverge and
leads to the formation of nuclear bound states. It is instructive to
understand this in more detail from conventional Feynman diagram
techniques. For that, consider the box graph (called $I$) shown in fig.5.1 for
static nucleons. For a nucleon at rest, the propagator takes the form
$$S_N (q) = -{\Lambda \over q^0 + i \epsilon }   \eqno(5.4)$$
with $\Lambda$ the projection matrix onto positive energy, zero
momentum Dirac wave functions.
\midinsert
\hskip 1.6in
\epsfxsize=1.5in
\epsfysize=1in
\epsffile{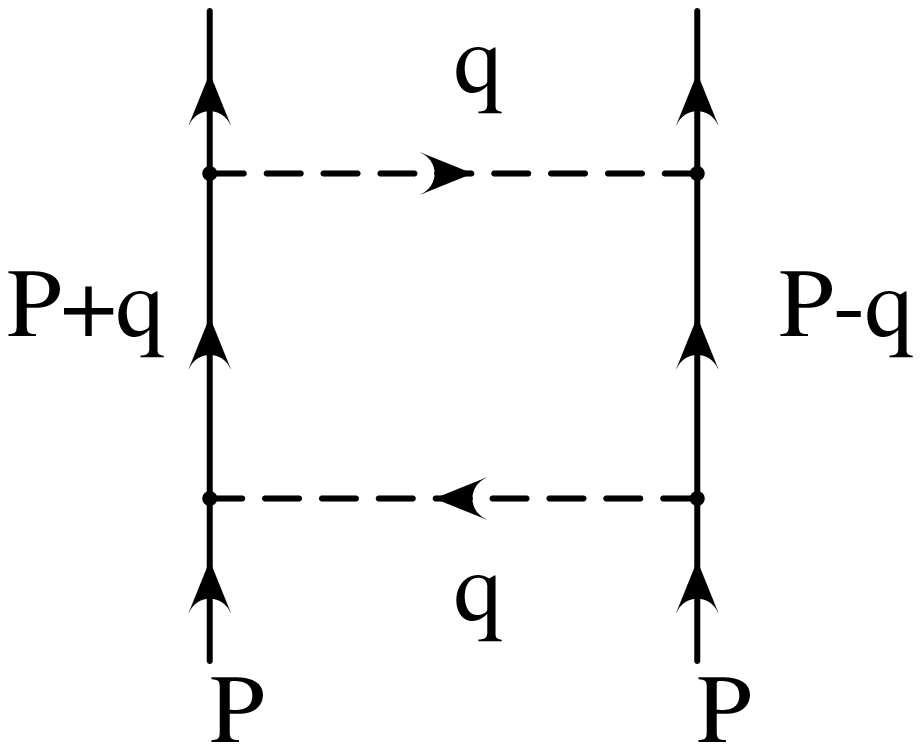}
\smallskip
{\noindent\narrower \it Fig.~5.1:\quad
The box diagram for NN scattering discussed in the text. Solid (dashed) lines
give nucleons (pions) and the pertinent momenta are exhibited.
\smallskip}
\vskip -0.5truecm
\endinsert
\noindent One finds for $I$
$$I \sim \int d^4q {1 \over q^0 + i \epsilon}{1 \over q^0 - i \epsilon}
{P(q) \over (q^2 + M_\pi^2)^2} \sim \int dq^0 {1 \over q^0 + i \epsilon}
{1 \over q^0 - i \epsilon}     \eqno(5.5)$$
where the polynomial $P(q)$ includes terms that are non--vanishing as $q^0$
goes to zero. Consequently, the integral over $q^0$ in $I$ has an infrared (IR)
divergence. The contour of integration is pinched between the two poles at $q^0
= \pm i \epsilon$ and it is therefore impossible to distort it in such a way to
avoid these singularities. This is distinctively different from the
single--nucleon case discussed in the previous sections. Of course, this IR
divergence is an artefact of the approximation (5.4), i.e. treating the
nucleons as static. Indeed, if one includes the nucleon kinetic energy, ${\cal
L}_{\rm kin} = {\bar N} \nabla^2 N / 2m$, the poles are shifted to $q^0 \simeq
\pm[\vec{q \,}^2 /2m - i \epsilon]$ and
the $q^0$ integral has the finite value $2i
\pi m / |{\vec q} \,|^2$. However, from the counting of small momenta one would
expect this integral to scale as $Q^{-1}$ (since each propagator scales as
$Q^{-1}$), i.e. it is enhanced by a large factor $m/Q$. This enhancement is at
the heart of the nuclear binding. Such small scales can only come from
reducible diagrams and to avoid these, one defines an effective potential as
the sum of time--ordered perturbation theory graphs for the T--matrix excluding
those with pure nucleon intermediate states [5.12,5.13]. The full machinery of
expanding in powers of $Q$ is therefore only applied to the reducible diagrams
and the full S--matrix is obtained by solving a Lippmann--Schwinger or
Schr\"odinger equation with the effective potential. This will be discussed in
more detail when we consider the NN--potential. At this point we should stress
that this separation of reducible versus irreducible diagrams is unavoidable
but still poses some concern to the purist since in the process of solving such
bound state problems, one can not completely exclude some large momentum
components.

To remove the scale $m$ from the problem, one can make use of a field
redefinition and replace the time derivative of the nucleon field by the
nucleon field equations [5.13]
$$\bigl[ i \partial_0 - {1 \over 2 D F_\pi^2} \vec{t} \cdot ( \vec{\pi}
\times \partial_0 \vec{\pi} ) \bigr] \, N = \bigl[ m + {g_A \over D
F_\pi} \vec{t}
\times ( \vec{\sigma} \cdot \vec{\nabla} ) \vec{\pi} + \ldots \bigr] \, N
\eqno(5.6)$$
So the chiral invariant time derivative of the nucleon field in the interaction
Lagrangian simply changes the coefficients of other terms  allowed (and
required) by chiral symmetry. Therefore, one can simply adopt a definition of
the fields and the constants in ${\cal L}_{\rm eff}$ such that no time
derivatives appear. Alternatively, one could use the methods described in
section 3.3.

In summary, once the scales $m$ and $Q^2 /2m$ are removed by considering
irreducible diagrams and using appropriate field definitions, one can order
all remaining contributions to the N--nucleon forces in powers of $Q/m \sim
Q/M_\rho$. To do that, we have to extend the power counting scheme discussed in
section 3.2 (since there it was assumed that only one nucleon line runs through
a given diagram). Let us do that in time--ordered perturbation theory.
Derivatives are counting as order $Q$, pion fields as $Q^{-1/2}$ (using the
conventional normalization $\sim 1 / \sqrt{2 \omega}$ for pion fields),
intermediate nucleons or $\Delta$'s\footnote{*}{We include here the $\Delta$
since that has also been done in ref.[5.17] which reported first full scale
numerical results. We remind the reader here of the reservations made in
section 3.4.} as $Q^{-1}$ and loop integrals as $\int d^3k \sim Q^3$. The
chiral dimension of a graph with $E_n$ external nucleon lines, $D$ intermediate
states, $L$ loops, $C$ connected pieces and $V_i$ vertices of type $i$ (with
$d_i$ derivatives or pion masses and $n_i \, (p_i)$ nucleon (pion) fields)
follows to be (we set $C=1$ for the moment) [5.12,5.13,5.14]
$$\nu = 3L - D + \sum_I V_i (d_i - {p_i \over 2} )         \eqno(5.7)$$
and using the topological identities
$$D  = \sum_i V_i - 1 \eqno(5.7a) $$
$$L  = I - \sum_i V_i + 1 \eqno(5.7b)$$
$$2I + E_n  = \sum_i V_i (p_i + n_i)  \eqno(5.7c)$$
with $I$ the total number of internal lines, one arrives at
$$\eqalign{
\nu & = 2L + 2 - {1 \over 2}E_n + \sum_I V_i \Delta_i \cr
\Delta_i & = d_i + {1 \over 2} n_i - 2   \quad . \cr}     \eqno(5.8)$$
In case of $C > 1$, this generalizes to
$$\nu  = 2(L-C) + 4 - {1 \over 2}E_n + \sum_I V_i \Delta_i \quad .\eqno(5.9)$$
It is now important to notice that chiral symmetry demands
$$ \Delta_i \ge 0      \qquad . \eqno(5.10)$$
This can easily be understood. Operators involving pions only have at least two
derivatives or two powers of $M_\pi$ and nucleon bilinears have at least one
derivative. As before, to lowest order one calculates {\it tree} diagrams with
$\Delta_i = 0$. Loop diagrams are suppressed by powers of $Q^2$. We have now
assembled all tools to take a closer look at the nucleon--nucleon potential and
the problem of many--nucleon forces.
\bigskip \goodbreak
\noindent{\bf V.2. THE NUCLEON--NUCLEON POTENTIAL}
\medskip
\goodbreak
{}From eq.(5.8) it follows that for $E_n =4$ ($C=1$), the minimum value of
$\nu$
is given by $L=0$ (tree diagrams) with $\Delta_i = 0$. The latter condition can
either be fulfilled having diagrams with $d_i =1$ and $n_i =2$ (one--pion
exchange) or having $d_i =0$ and $n_i =4$ (four--nucleon contact terms) (see
fig.5.2).
\midinsert
\hskip 1in
\epsfxsize=2.5in
\epsfysize=1.2in
\epsffile{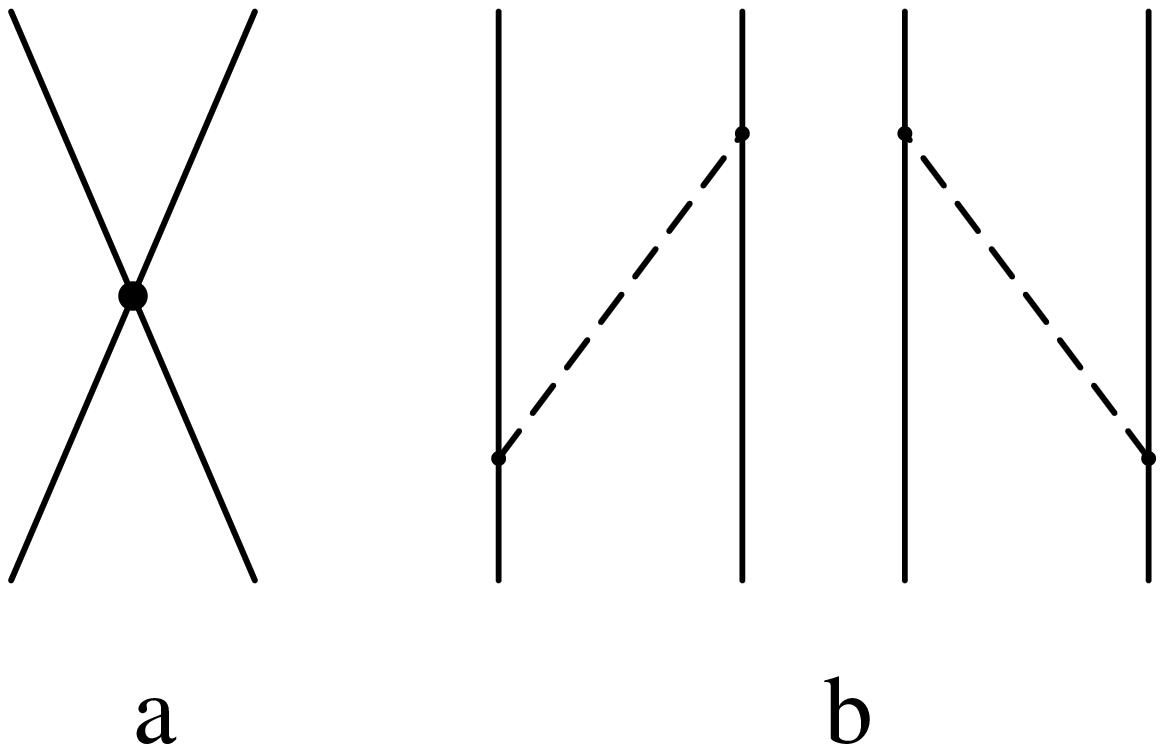}
\smallskip
{\noindent\narrower \it Fig.~5.2:\quad
Lowest order diagrams contributing to the NN interaction.
 (a) is a set of four--nucleon contact terms and (a) is the one--pion
exchange.
\smallskip}
\vskip -0.5truecm
\endinsert
\noindent The corresponding Lagrangian reads
$$\eqalign{ \quad
{\cal L}^{(0)}  =&{\cal L}_{\pi \pi} +
{\cal L}_{\pi N} + {\cal L}_{\bar N N} = -{1 \over 2 D^2}
(\partial_\mu \vec \pi)^2 - {M_\pi^2 \over 2D} \vec \pi ^2 \cr
&
- \bar N \biggl[ i \partial_0 - m
-{ g_A \over 2 D F_\pi} \vec t \cdot (\vec \sigma \cdot \vec \nabla )
\vec \pi
- {1 \over 2 D F_\pi^2} \vec t \cdot (\vec \pi \times \partial_0
\vec \pi) \biggr] N \cr
              & - {1 \over 2} C_S (\bar N N) (\bar N N)
    +{1 \over 2}  C_T (\bar N \vec \sigma N) \cdot (\bar N \vec \sigma N) \cr}
\eqno(5.11)$$
where $C_S$ and $C_T$ are new low--energy  constants related to ${\cal L}_{\bar
\Psi \Psi \bar \Psi \Psi}$ of eq.(3.10).
Notice that because of Fermi statistics (Fierz rearrangement)
one can rewrite a non--derivative four--nucleon contact term involving $\vec t$
as a combination of the last two terms in (5.11). It is straightforward to
construct the interaction Hamiltonian related to ${\cal L}^{(0)} $ as detailed
in refs.[5.12,5.13]. For the two--nucleon case, the effective potential derived
then is simply the sum of one--pion exchange and a contact interaction arising
from the two last terms in eq.(5.11). One finds in coordinate space
$$\eqalign{
V_{12} (\vec{r}_1 - \vec{r}_2 ) & = [ C_S + C_T \vec{\sigma}_1 \cdot
\vec{\sigma}_2 ] \delta^{(3)} (\vec{r}_1 - \vec{r}_2 )
\cr & - \bigl( {g_A \over F_\pi}
\bigr)^2 (\vec{t}_1 \cdot \vec{t}_2 ) (\vec{\sigma}_1 \cdot \vec{\nabla}_1 )
(\vec{\sigma}_2 \cdot \vec{\nabla}_2 ) \, Y ( | \vec{r}_1 - \vec{r}_2 | )
- (1' \leftrightarrow 2' ) \cr} \eqno(5.12)$$
with $Y(r) = \exp (-M_\pi r ) / 4 \pi r$ the standard Yukawa function. Clearly,
the potential (5.12) is only a crude approximation to the NN forces. In
particular, the correlated $J=T=0$ pion pair exchange that is believed to
furnish the intermediate range attraction is hidden in the constant $C_S$. As
stressed by Weinberg [5.12,5.13], the constanst $C_S$ has to be "unnaturally"
large to lead to shallow nuclear bound states. If one considers e.g. the $L=0$
spin singlet state and approximates the potential by $C \delta^{(3)}
(\vec{r}_1 - \vec{r}_2 )$, with $C = C_S - 3C_T + g_A^2 / 4 F_\pi^2$, one can
solve the Lippmann--Schwinger equation in momentum space and finds after
renormalization ($C \to C_R$) a bound state with binding energy
$$B = {16 \pi^2 \over m^3 C_R^2}   \qquad . \eqno(5.13)$$
According to naive dimensional analysis [5.19] one expects $C_R / 2 \pi^2 \sim
1 / \Lambda_\chi^2 \sim 1$ GeV$^{-2}$. However, to get
the deuteron binding energy
of $B =2.22$ MeV, $C_R / 2 \pi^2$ must have the large value of (260
MeV)$^{-2}$. This then suggests that $Q / m \sim (Q / \Lambda_\chi)^2$ and one
adopts the rule that a pure nucleon intermediate state counts as if it
contributes {\it two} more powers of $1/Q$ than any other intermediate state.
Notice also that the lowest order NN potential leaves no room for the
short-range repulsion or the spin--orbit forces and alike. On the other hand,
one--pion exchange is known to describe well the higher partial waves in NN
scattering at low energies.

In the work of refs.[5.16-5.18], the $\Delta$ was also put in the
effective theory based on the closeness of this resonance to the
nucleon ground state [5.20], i.e. $m_\Delta - m \ll M_\rho$. In that
case, one has additional lowest order terms, collected in
${\cal L}^{(0)}_\Delta$
$$\eqalign{
{\cal L}^{(0)}_\Delta & = \bar{\Delta} [ i \partial_0 - {1 \over 2F_\pi^2
D} {\vec t \, \,}^{(3/2)} \cdot (\vec \pi \times \partial_0 \vec \pi ) -
m_\Delta ] \Delta \cr
& - {h_A \over 2F_\pi D} [\bar N \vec T \cdot ( \vec S
\cdot \nabla ) \vec \pi \Delta + {\rm h.c.} ] - D_T \bar N \vec \sigma
\vec t N \cdot \cdot [\bar N \vec S \vec T \Delta +{\rm h.c.} ] +
\ldots \cr} \eqno(5.14)$$
where the ellipsis stands for terms involving more $\Delta$'s which
are irrelevant for the NN potential. The constant $h_A$ can be
calculated e.g. from the decay width $\Gamma (\Delta \to N \pi )$,
$h_A \simeq 2.7$. $D_T$ is a new low--energy constant and only enters
the calculation of 3N (or more) forces.

To calculate corrections, one also has to consider the terms with
$\Delta_i=1$ and $\Delta_i=2$. These are discussed in detail in
refs.[5.14,5.16,5.17]. We only give a short outline of the pertinent
effective Lagrangians here. For ${\cal L}^{(1)}$, one finds
$$\eqalign{
{\cal L}^{(1)} & = -{B_1 \over 4F_\pi^2 D^2} \bar N N [ (\nabla \vec
\pi )^2 - (\partial_0 \vec \pi )^2]  - {B_2 \over 4F_\pi^2 D^2} \bar N
\vec t \vec \sigma N \cdot ( \nabla \vec \pi \times \nabla \vec \pi )
- {B_3 M_\pi^2 \over 4 F_\pi^2 D} \bar N N {\vec \pi \,}^2 \cr
& -{D_1 \over 2 F_\pi D} \bar N N \bar N (\vec t \cdot \vec \sigma
\cdot \nabla \vec \pi \, )N - {D_2 \over 2 F_\pi D} ( \bar N \vec t \,
\vec \sigma \, N \times \bar N \vec t \, \vec \sigma \, N) \cdot
\nabla \vec \pi \cr
& - {1 \over 2} E_1 \bar N N \bar N \vec t \, N \cdot \bar N \vec t \, N
- {1 \over 2} E_2 \bar N N \bar N \vec t \, \vec \sigma \, N \cdot
\bar N \vec t \, \vec \sigma \, N \cr &- {1 \over 2} E_3 ( \bar N \vec t \,
\vec \sigma \, N \times \bar N \vec t \, \vec \sigma \, N) \cdot \bar
N \vec t \, \vec \sigma \, N  + \ldots  \cr}
\eqno(5.15)$$
where the $B_i$, $D_i$ and $E_i$ are new parameters. $B_3$ is
obviously related to the $\pi N$ $\sigma$--term and $B_{1,2}$ could be
determined from $\pi N$ scattering (cf. section 4.3). At present, this
has not been done but all $B_i$, $D_i$ and $E_i$ are left free. The
six--fermion terms proportional to $E_{1,2,3}$ do not enter the NN
potential. Also, terms with explicit $\Delta$'s are not shown. The
terms with $\Delta_i =2$ take the form
$$\eqalign{ {\cal L}^{(2)}  &= {1 \over 2m} \bar N \nabla^2 N
-{A_1' \over 2 F_\pi} [ \bar N ( \vec t \cdot \vec \sigma \cdot \nabla
\vec \pi \, ) \nabla^2 N + \overline{\nabla^2 N} (\vec t \cdot \vec
\sigma \nabla \vec \pi \, )] N \cr & - {A_2' \over 2 F_\pi} \overline{\nabla
N} ( \vec t \cdot \vec \sigma \cdot \nabla \vec \pi \, ) \cdot \nabla
N  - C_1'[(\bar N \nabla N)^2 + (\overline{\nabla N} N)^2] -
C_2' (\bar N \nabla N) \cdot (\overline{\nabla N} N)
 + \ldots \cr}
\eqno(5.16)$$
where the $A_i'$ and $C_i'$ are undetermined coefficients and the
ellipsis denotes other terms with two derivatives or more pion
fields. Because only the term proportional to $B_1$ contains a time
derivative, the corresponding interaction Hamiltonian can be taken as
$ -{\cal L}^{(1)}-{\cal L}^{(2)}$ up to terms with more pion fields.

We are now in the position to systematically discuss the corrections
to the lowest order potential $V_{12}^{(0)}$, eq.(5.12). As already
noted in [5.13], the first corrections arise from the same graphs as
in fig.5.2 with exactly one insertion from ${\cal L}^{(1)}$. However,
since all time derivatives have been eliminated, one would have to
construct a vertex with an odd number of three--momenta. This clashes
with parity and one therefore concludes that
$$V_{12}^{(1)} = 0 \quad.    \eqno(5.17)$$
The second corrections fall into two classes. The first one are tree
graphs with exactly one vertex or kinetic energy insertion from
${\cal L}^{(2)}$, leading to
$$\eqalign{
 & V_{12}^{(2)}  (\vec q , \vec k \,)  = -{2 g_A \over F_\pi} \vec t_1 \cdot
\vec t_2  {\vec \sigma_1 \cdot \vec q \, \vec \sigma_2 \cdot \vec q \over
  {\vec q}^2 + M_\pi^2} \biggl[ A_1 \vec{q \,}^2 + A_2 \vec{k}^2 +
{2 g_A \over
 ({\vec q}^2 + M_\pi^2)^{1/2} } \bigl(E - {\vec{q \,}^2 \over 4 m} -
{\vec{k}^2 \over  m} \bigr) \biggr] + C_1 \vec{q \,}^2  \cr & +
C_2 \vec{k}^2 +(C_3 \vec{q \,}^2 + C_4 \vec{k}^2 ) \vec \sigma_1
\cdot \vec \sigma_2 + i{C_5  \over 2} (\vec \sigma_1 + \vec \sigma_2)
\cdot (\vec{q} \times \vec k )
+ C_6 \vec \sigma_1 \cdot \vec{q} \, \vec \sigma_2 \cdot \vec{q}
+ C_7 \vec \sigma_1 \cdot \vec{k} \, \vec \sigma_2 \cdot \vec{k}
\cr}  \eqno(5.19)$$
with $\vec{q} = \vec{p} - \vec{p} \,'$ the transferred momentum,
$\vec{k} = (\vec{p} + \vec{p}\, ')/2$, $2m+E$ the energy in the center
of mass and $\vec p \,(')$ is the initial (final) cm momentum.
The $A_i, C_i$ are
combinations of the $A_i', C_i'$ in (5.16). Second, there are the one--loop
contributions, i.e. the two--pion exchange, of the form
$$  V_{12, {\rm loop}}^{(2)} = V_{12, {\rm no} \Delta}^{(2)} +
 V_{12, {\rm one} \Delta}^{(2)} +  V_{12, {\rm two} \Delta{\rm 's}}^{(2)}\, \,
\eqno(5.20)$$
corresponding to no, one and two isobars in the intermediate states.
The first term on the r.h.s. of (5.20) reads [5.14]
$$\eqalign{
 & V_{12, {\rm no} \Delta}^{(2)} = -{1 \over 32F_\pi^4} \vec t_1 \cdot
 \vec t_2 \int {d^3 l \over (2\pi)^3}{1 \over \omega_+ \omega_-}
{(\omega_+ - \omega_-)^2 \over \omega_+  + \omega_-} -
\bigl({g_A \over 2F_\pi^2}\bigr)^2 \vec t_1 \cdot
 \vec t_2 \int {d^3 l \over (2\pi)^3}{1 \over \omega_+ \omega_-}
{\vec{q \,}^2 - \vec{l \,}^2 \over \omega_+  + \omega_-} \cr &
\qquad \qquad \qquad
 -  {1 \over 4} \bigl({g_A \over 2F_\pi^2}\bigr)^4
\int {d^3 l \over (2\pi)^3}{1 \over \omega_+^3 \omega_-} \biggl\lbrace
\biggl( {3 \over \omega_-} + { 8 \vec t_1 \cdot \vec t_2
\over \omega_+ + \omega_-} \biggr) (\vec{q \,}^2 - \vec{l \,}^2 )^2 \cr &
\qquad \qquad \qquad + 4
\biggl( {3 \over \omega_+ +\omega_-} + { 8 \vec t_1 \cdot \vec t_2
\over \omega_-} \biggr) \vec \sigma_1 \cdot (\vec{q \,} \times
\vec{l}) \, \vec \sigma_2 \cdot (\vec{q \,} \times \vec{l} \,) \biggr\rbrace
\cr} \eqno(5.20a)$$
with $\omega_{\pm} = \sqrt{ (\vec{q \,} \pm \vec{l}\,)^2 + 4 M_\pi^2}$ and
the explicit expressions for the contributions with one or two
intermediate isobars can be found in ref.[5.16]. Finally, the third
 corrections have also been evaluated. For the same reason as
 discussed above, one finds
$$V_{12,{\rm tree}}^{(3)} = 0 \quad ,    \eqno(5.21)$$
and in the one--loop graphs one has  exactly one insertion from
eq.(5.15) leading to
$$  V_{12, {\rm loop}}^{(3)} = V_{12, {\rm no} \Delta}^{(3)} +
 V_{12, {\rm one} \Delta}^{(3)} \, \,
\eqno(5.22a)$$
with
$$\eqalign{
V_{12, {\rm no} \Delta}^{(3)}  =
 -  {1 \over 4} \bigl({g_A \over 2F_\pi^2}\bigr)^2
\int {d^3 l \over (2\pi)^3}{1 \over \omega_+^2 \omega_-^2}& \biggl\lbrace
3 (\vec{q \,}^2 - \vec{l \,}^2 )^2 \bigl[4M_\pi^2B_3 - B_1
(\vec{q \,}^2 - \vec{l \,}^2 ) \bigr] \cr
& + 16 B_2 \vec \sigma_1 \cdot (\vec{q \,} \times
\vec{l}) \, \vec \sigma_2 \cdot (\vec{q \,} \times \vec{l} \,)
\vec t_1 \cdot \vec t_2 \biggr\rbrace \cr}
\eqno(5.22b)$$
and the corresponding expression with one intermediate $\Delta$ is
given in ref.[5.16]. We are now at the point to discuss the structure
of the momentum space potentials. The term proportional to $A_1$ in
(5.19) can be considered as coming from the expansion of the
pion--nucleon form factor in powers of momenta over the
cut--off. Indeed,
a typical monopole form factor
 $F_M(\vec{q \,}^2 ) = \Lambda^2 /(\Lambda^2 +\vec{q \,}^2)$
would amount to $A_1 = 2 /\Lambda^2 = 2$ GeV$^{-2}$ for $\Lambda=1$
GeV. The $A_2$--term is a so--called non-adiabatic correction and the last
term in the square brackets in (5.19) is the energy--dependent recoil
correction. This modified one--pion exchange gives the long--range
part of the potential. At intermediate distances, the two--pion
exchange generated from the one--loop diagrams comes in. Many of the
box and crossed box diagrams are well-known from the work of Br\"uckner
and Watson [5.21], Sugawara and von Hippel [5.22] and Sugawara and
Okubo [5.23]. However, the diagrams with one $NN\pi\pi$--vertex have
coefficients which are either fixed by chiral symmetry (like
e.g. $g_A^4$ or $(g_A h_A)^2$), or are in principle determined from
$\pi N$ scattering (the $B_{1,2,3}$). We come back to these later on.
Furthermore, at this order there is no correlated two--pion exchange,
it only shows up at order $(Q/M_\rho)^4$ and higher. This is consistent
with the analysis of the intermediate--range attraction made in
ref.[5.24] based on the spectral analysis of the scalar pion form factor.
All physics of shorter ranges is buried in the various contact terms,
i.e. the coefficients $C_i$. The various loop integrals like (5.20a) or
(5.22b) are all divergent. At present, this is treated by a momentum
space cut--off. The form of the cut--off function is chosen to be
gaussian as in the Nijmegen approach [5.9]. Specifically, all loop
momenta $l$ are cut off by $\exp( - \vec l \,^2 /
\Lambda^2)$. Furthermore, since as argued before all momenta should be
smaller than some scale $\Lambda$, the transferred momentum $\vec q$
is also damped with the same type of cut--off, $\exp( - \vec q \,^2 /
\Lambda^2)$. In practise, $\Lambda = M_\rho$ is chosen. Of course, one
would like to see a more elegant regularization employed such as
dimensional regularization. The potential is then transformed into
coordinate space, where it is energy-dependent and takes the form
$$ V = \sum_{p=1}^{20} V_p (r, {\partial \over \partial r},
{\partial^2 \over \partial r^2}; E) \, {\cal O}^p  \, \, \, ,
\eqno(5.23)$$
with
$$V_p (r, {\partial \over \partial r}, {\partial^2 \over \partial r^2}; E)
= V_p^0(r;E) + V_p^1(r;E){\partial \over \partial r} + V_p^2(r;E)
{\partial^2 \over \partial r^2} \, \, , \eqno(5.23a)$$
and the ${\cal O}^{p=1,\ldots,20}$ are a complete basis of operators
made of the $\vec \sigma_i$, the $\vec \tau_i$ $(i=1,2)$, the tensor operator
$S_{12}$, the total spin operator $\vec S = (\vec \sigma_1 + \vec
\sigma_2)/2$ as well as the angular momentum operator $\vec L = -i
\vec r \times \vec \nabla$. Alltogether, the potential contains 26
parameters, but it should be stressed that some of these are indeed
not free but given by constraints from $\pi N$ scattering.
Let us briefly dicuss the connection between the $B_i$ $(i=1,2,3)$ and
the various $c_i$ discussed in section 3.4. One finds [5.25]
$$B_1 = 4 c_3 = 13.6 \, {\rm GeV}^{-1} , \, \,
B_2 = 8 c_1 = -7.0 \, {\rm GeV}^{-1} , \, \,
B_3 = -4 c_4 - {1 \over m} = -17.5 \, {\rm GeV}^{-1} , \eqno(5.24)$$
for the central values of the $c_i$ from section 3.4. These
constraints have not yet been implemented in the numerical
calculations. Furthermore, in the fitting procedure of ref.[5.17],
even the fundamental parameters $F_\pi$, $g_A$ and $h_A$ were left free.

The parameters are fixed from a best fit to deuteron properties
(binding energy, magnetic moment and electric quadrupole moment)  and
the $np$ and $pp$ phase shifts with $J \le 2$ and $T_{\rm lab} \le
100$ MeV. The higher partial waves are supposedly dominated by one
pion exchange and were therefore not used to constrain the fit [5.17].
The results for the deuteron properties are summarized in table 5.1
and some typical phases are shown in fig.5.3 (more of these can be
found in ref.[5.17]). The resulting values for $F_\pi$, $g_A$ and
$h_A$ are 86 MeV, 1.33 and 2.03, respectively, not too far from their
empirical values. Using the Goldberger--Treiman relation, this
corresponds to a pion--nucleon coupling constant of 14.5.
\bigskip
$$\hbox{\vbox{\offinterlineskip
\def\strut{\hbox{\vrule height  8pt depth  8pt width 0pt}}
\hrule
\halign{
\strut\vrule# \tabskip 0.1in &
\hfil#\hfil  &
\vrule# &
\hfil#\hfil &
\hfil#\hfil &
\vrule# \tabskip 0.0in
\cr
\noalign{\hrule}
&  Observable
&& Fit       & Exp.   & \cr
\noalign{\hrule}
& B [MeV]        && 2.18    & 2.224579(9) & \cr
& $\mu_d$ [n.m.]   && 0.851   & 0.857406(1) & \cr
& Q [fm$^2$]     && 0.231   & 0.2859(3)   & \cr
& $\eta$         && 0.0239  & 0.0271(4) & \cr
\noalign{\hrule}}}}$$
\smallskip
{\noindent\narrower Table 1:\quad Deuteron properties: binding energy
B, magnetic moment $\mu_d$, quadrupole moment Q and the asymptotic D/S
ratio $\eta$ [5.17]. The data are from ref.[5.26].
\smallskip}
\goodbreak
\medskip
\midinsert
\hskip 1in
\epsfxsize=3in
\epsfysize=4in
\epsffile{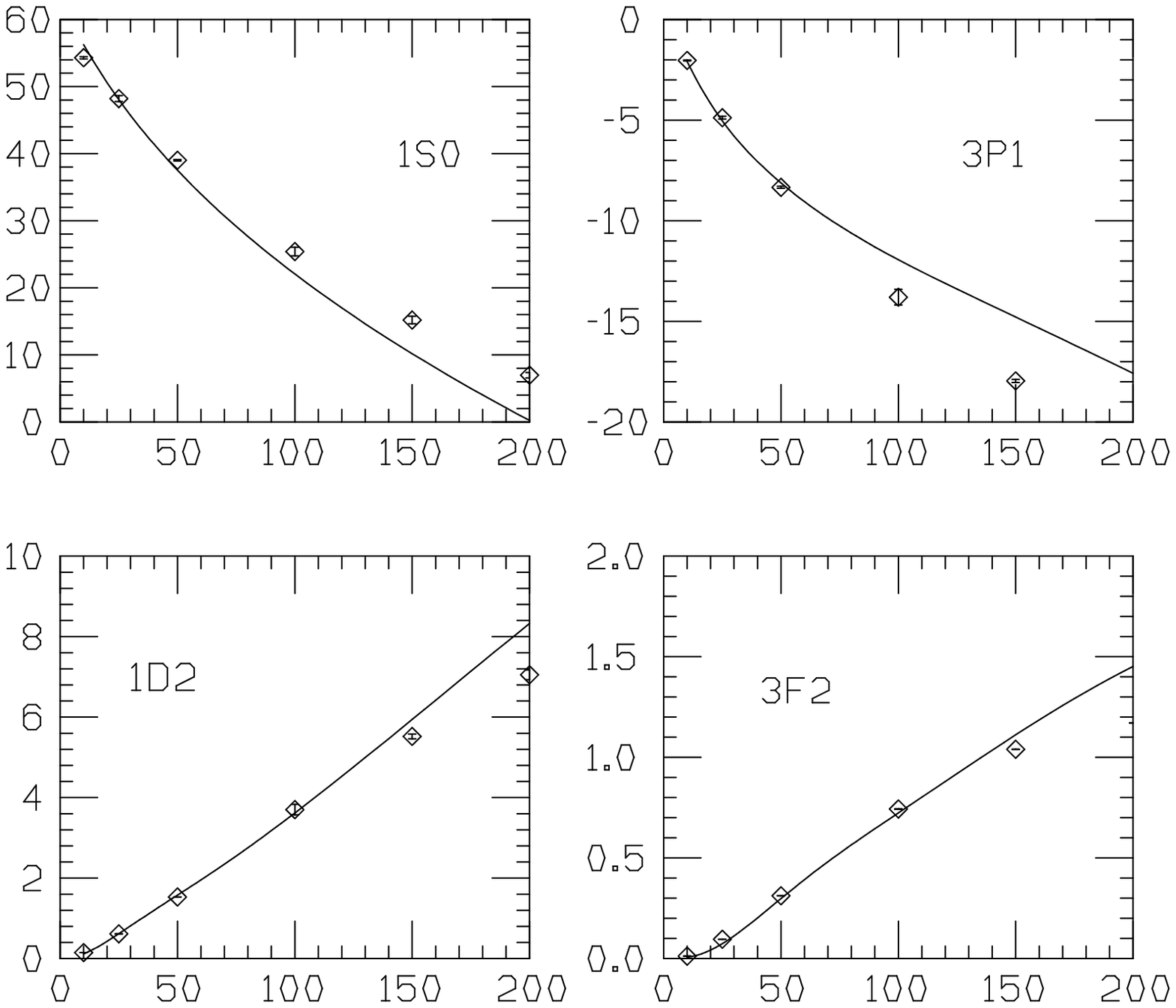}
\vskip -0.5truecm
{\noindent\narrower \it Fig.~5.3:\quad
Best fit to some partial waves. Phase shifts in degrees versus $T_{\rm
lab}$ in MeV.  We thank L. Ray and U. van Kolck for
supplying us with the pertinent numbers.
\smallskip}
\vskip -0.5truecm
\endinsert
\noindent The calculated D--state probability is 5$\%$, i.e. of
comparable size to what one gets from the Bonn or Paris potential. The
$L=0$ singlet and triplet scattering lengths are predicted to be -15.0
fm and 5.46 fm,  in fair agreement with the empirical
values of -16.4(1.9) fm and 5.396(11) fm [5.27], respectively. A close
look at table 1 and the phase shifts reveals that the fit is not
too satisfactory, in particular the deviation in the deuteron
quadrupole moment as well as in certain P--waves are quite substantial
for the accuracy one is used from the semi-phenomenological
potentials. Holinde has argued [5.28]  that some of these discrepancies
reside in the asymmetric treatment of the pions and rho mesons. In
particular, the fine cancellations between the tensor forces from the
the $\pi$ and the $\rho$ are unbalanced here. This in turn leads to an
overall unsatisfactory tensor force which mostly shows up in the
before mentioned observables. At present, it is not clear how one has
to go about these problems. Clearly, more detailed fits allowing also
for variations in the cut-off function and its associated cut off are
called for and as already stressed a few times, the strictures from
the single nucleon sector on some of the parameters should be
enforced. On the positive side, it is worth pointing out that such a
straightforward potential based solely on chiral symmetry constraints
can describe the low--energy NN phases and deuteron properties within
some accuracy.
\bigskip \goodbreak
\noindent{\bf V.3. MORE THAN TWO NUCLEONS}
\medskip
\goodbreak
Nucleons interact mainly via two--body forces. However, there is some
indication of small three--nucleon forces (for a recent review see
ref.[5.29]). Standard two--nucleon potentials when employed to $^3$H
and $^4$He tend to lead to an  underbinding of typically 0.5 to 5 MeV.
This in turn means that if this discrepancy is due to a three--nucleon
force, it has to be small, typically a few percent of the $2N$ force.
However, recent calculations of the Bochum group [5.30] have indicated
that a fine--tuning of the NN--potential can lead to a satisfactory
description of the three--nucleon system. This, however, involves
rather large charge--symmetry breaking effects. The chiral Lagrangian
analysis can shed some light about the size of three (and many)
nucleons forces to be expected as will be discussed in this section.
The role of chiral symmetry, i.e. the use of the pseudovector $\pi N$
coupling leading to small $3N$ forces has long been conjectured
[5.31,5.32]. As will be shown, this can now be put on firmer grounds.

To leading order, the potential between $A$ nucleons is simply given
by a pair--wise sum of the lowest order two--nucleon potential (5.12)
since decreasing the number of connected pieces $C$ costs powers of
$Q$, cf. eq.(5.9). Therefore, $\nu_{\rm min} = 6 - 3A$ leading to
$$V^{(0)}(\vec{r}_1 , \ldots ,\vec{r}_A \,) = \sum_{(ij)} \, V^{(2)}
(\vec{r}_i - \vec{r}_j \,) \quad , \eqno(5.25)$$
where the sum runs over all nucleon pairs.
This is a rather crude approximation. Therefore, one has to consider
corrections [5.13,5.18]. We follow here the recent analysis by van
Kolck [5.18]. To order $\nu = \nu_{\rm min} + 2$, one has the
following form for the  potential between $A$ nucleons:
$$\eqalign{
& \sum_{n=0}^3 V^{(n)}(\vec{r}_1 , \ldots ,\vec{r}_A \,)
 = \cr & \sum_{(ij)} \sum_{n=0}^3 V_2^{(n)} (\vec{r}_i ,\vec{r}_j  \, )
+ \sum_{(ijk)} \sum_{n=2}^3 V_3^{(n)} (\vec{r}_i ,\vec{r}_j ,\vec{r}_k
\, )
+ \sum_{(ij;kl)} \sum_{n=2}^3 V_{2,2}^{(n)} (\vec{r}_i -\vec{r}_j ;
\vec{r}_k - \vec{r}_l \, ) \, , \cr} \eqno(5.26)$$
At this order, the $2N$ potential contains the one--pion exchange
recoil (5.19) among other terms. The $3N$ potential consists of the
three types of terms shown in fig.5.4. and the double pair potential
$V_{2,2}$ is made of two sets of diagrams, the first being two
disconnected OPE graphs and the second one one OPE graph separated from a
lowest order two--nucleon contact term. It was first shown by Weinberg
[5.12] that all diagrams containing the non--linear $\pi \pi N$ vertex
add up to zero to lowest order. Furthermore, as detailed in [5.18],
the remaining three--body forces and double---pair forces are canceled
by the energy-dependence of the two--body potential when the latter is
iterated in the Lippmann-Schwinger equation. Such kind of cancellation
had been noticed before [5.33] and it means that if one chooses to work
with an energy--dependent $NN$--potential, one has to include at the
same time $3N$ and double--$2N$ forces calculated consistently in the
same framework.
\midinsert
\hskip 1in
\epsfxsize=3in
\epsfysize=1in
\epsffile{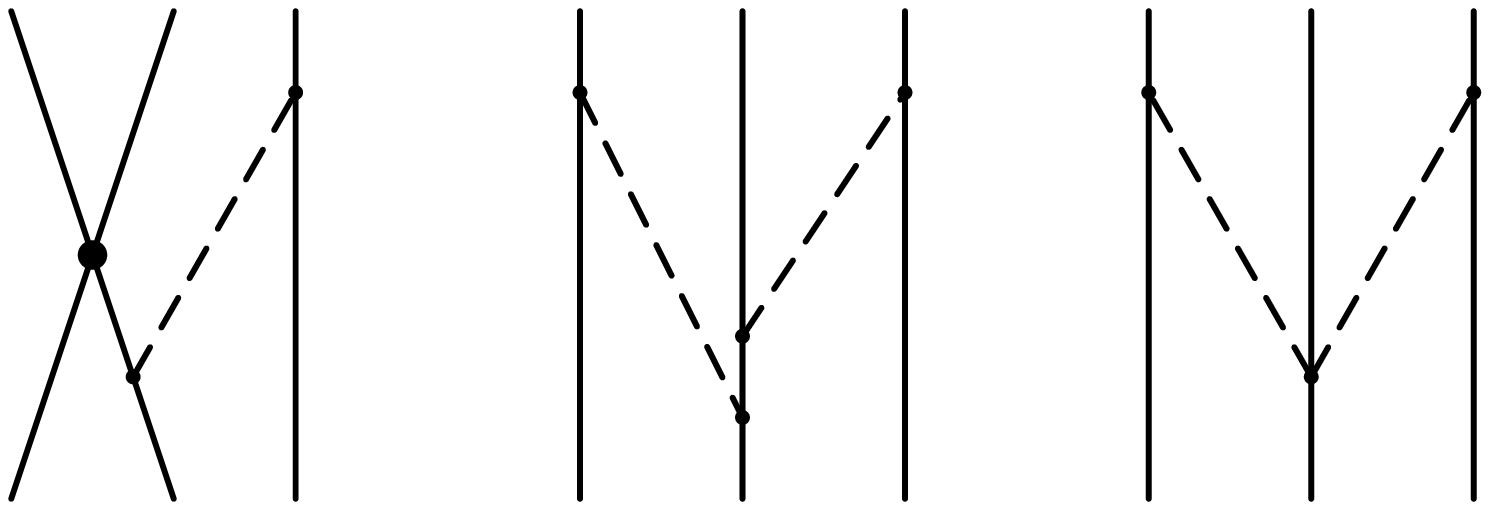}
\smallskip
{\noindent\narrower \it Fig.~5.4:\quad
Tree graphs contributing to the three--nucleon potential. For each
class of contributions, one typical diagram is shown. All other
irreducible time orderings have to be considered.
\smallskip}
\vskip -0.5truecm
\endinsert

The corrections at next order, $\nu = \nu_{\rm min}+3$ are discussed
in [5.18]. The correction to the double--pair potential vanishes for
the same reasons discussed before (5.17) and the remaining $3N$
potential takes the form
$$\eqalign{
V^{(3)}_3 (\vec{q}_{ij} ; \vec{q}_{jk})  & = E_1 \vec{t}_i \cdot \vec{t}_k
+ E_2  \vec{\sigma}_i \cdot \vec{\sigma}_k \, \vec{t}_i \cdot \vec{t}_k
+ E_3 \vec{\sigma}_j \cdot ( \vec{\sigma}_i \times \vec{\sigma}_k ) \,
\vec{t}_j \cdot (\vec{t}_i \times \vec{t}_k \,) \cr - &
{g_A \over 2 F_\pi^2} {1 \over \omega_{ij}} \vec{\sigma}_k \cdot
\vec{q}_{jk} \, \bigl[ D_1 \, (\vec{t}_i \cdot \vec{t}_k \,
\vec{\sigma}_i + \vec{t}_j \cdot \vec{t}_k \, \vec{\sigma}_j \,) -
2D_2 \, \vec{t}_j \cdot (\vec{t}_i \times \vec{t}_k \,) \,
\vec{\sigma}_i \times \vec{\sigma}_j \, \bigr] \cdot \vec{q}_{jk} \cr
+ & 2 \biggl( {g_A \over 2 F_\pi^2} \biggr)^2 {1 \over \omega_{ij}^2
\omega_{jk}^2} \vec{\sigma}_i \cdot \vec{q}_{ij} \,
\vec{\sigma}_k \cdot \vec{q}_{jk} \, \bigl[ \vec{t}_i \cdot \vec{t}_k
\, (B_1 \, \vec{q}_{ij} \cdot \vec{q}_{jk} + B_3 \, M_\pi^2 ) \cr - & B_2 \,
\vec{t}_j \cdot (\vec{t}_i \times \vec{t}_k \,) \, \vec{\sigma}_j
\cdot ( \vec{q}_{ij} \times \vec{q}_{jk} \, ) \bigr]
+  {\rm two} \, \, {\rm cyclic} \, \, {\rm permutations} \, \, {\rm
of} \, \, (ijk) \, \, , \cr}
\eqno(5.27)$$
which contains 8 parameters, three of which are in principle fixed by
$\pi N$ scattering, cf. (5.24), and the $D_i$ could  be
determined form $\pi$--deuteron scattering or pion production on
two--nucleon systems. The three $E_i$ can only be fixed from data on
$3N$ systems. Such an analysis is not yet available. A simplification
arises if one includes the $\Delta (1232)$ in the effective
Lagrangian. In that case, one has an additional $3N$ force of order
$\nu = \nu_{\rm min}+2$ which has the form (5.27) and the
corresponding low--energy constants can be expressed in terms of
$\Delta$ properties,
$$\eqalign{
E_1 & \to 0 , \quad E_2 \to {1 \over 9}{D_T^2 \over m_\Delta - m_N} ,
\quad E_3 \to -{1 \over 18}{D_T^2 \over m_\Delta - m_N} , \cr
D_1 & \to -{4 \over 9}{D_T h_A \over m_\Delta - m_N} , \quad
D_2 \to {2 \over 9}{D_T h_A \over m_\Delta - m_N} , \cr
B_1 & \to -{4 \over 9}{ h_A^2 \over m_\Delta - m_N} , \quad
B_2 \to -{2 \over 9}{ h_A^2 \over m_\Delta - m_N} ,
\quad B_3 \to 0 \cr} \eqno(5.28)$$
with $D_T$ a new low--energy constant. The terms proportional to
$h_A^2$, i.e. the two--pion exchange pieces,
 are nothing but the old Fujita--Miyazawa force [5.34] which is
accompanied here by a shorter--range contribution proportional to the
parameter $D_T$. The relation  of these results to existing
three--nucleon force models is discussed in [5.18]. No explicit $3N$
calculation has yet been performed in the framework outlined here.

To summarize, the chiral Lagrangian approach implies that few--nucleon
forces are generically smaller than the dominant two--nucleon forces.
There are strong cancellations between the leading (static) $3N$ force,
the double--pair forces and the iterated leading energy--dependence of
the two--nucleon force. The remaining $3N$ force is  expected to be
dominated by the Fujita--Miyazawa force plus a shorter--range term
involving one new parameter, $D_T$. These $3N$ forces are expected to
be of order ${\cal O}(M_\pi^2 / M_\rho^2)$, i.e. some 5$\%$ of the
$NN$ contribution. By a similar argument, one expects even smaller
$4N$ forces of order ${\cal O}(M_\pi^4 / M_\rho^4)$ (less than 1$\%$
compared to the $NN$ contribution). Consequently, this analysis leads
one to expect that four--nucleon systems are underbound by roughly
four times the triton underbinding when pure $NN$ forces are
used. These dimensional arguments have yet to be substantiated by a
quantitative calculation in the framework outlined here.
\bigskip \goodbreak
\noindent{\bf V.4. THREE--BODY INTERACTIONS BETWEEN NUCLEONS, PIONS
AND PHOTONS}
\medskip
\goodbreak
In the previous sections we saw that the calculation of the two--body
interactions between nucleons involves a large number of free
parameters and that the resulting potential does not yet have the
accuracy of the standard semi--phenomenological ones. It was therefore
proposed by Weinberg [5.15] to use the empirical knowledge about the
two--body interactions between nucleons as well as pions and nucleons and
combine these with the remaining contributions from the potential of
the same power in small momenta, which are graphs with three particles
(or two pairs of particles) interacting. Stated differently, if one
looks at any nuclear process like elastic pion scattering, pion
photoproduction and so on, the calculation of the S--matrix element
$<\Psi_A | {\cal I} | \Psi_A>$ is split into two parts. On one hand,
chiral perturbation theory is used to calculate the irreducible
kernel ${\cal I}$ to a certain power in $Q/M_\rho$ and on the other
hand, one uses phenomenological input to construct the nuclear wave--function
$\Psi_A$. The virtue of this method lies in the fact that it orders the
relevant contributions to  ${\cal I}$ in a systematic fashion and
thus can explain the dominance of certain digrams contributing to a
certain process (which is often already known from
models but also often not fully understood). One thus encompasses many
of the problems which arise in the CHPT calculation of the NN interaction.
However, one also looses a certain degree of consistency since one does
not calculate nuclear wave--functions and operators in the same framework.
This has to be kept in mind in what follows.
 How much this could be improved by systematically generating also
the nuclear wave--functions from a potential solely derived from
chiral symmetry as described in section 5.2 is not yet clear.
\medskip
In ref.[5.15], this method is applied to pion scattering on complex
nuclei. To be more specific, the calculation is simplified by
considering the corresponding scattering length, i.e. the reaction with the
in-coming and out-going pion having vanishing three--momentum.
In this process, we have $N_n =A$ external nucleons and $N_\pi
=1$ external pions. The leading irreducible graphs are those in which
the pion scatters off a single nucleon, evaluated using the lowest
order vertices with $\Delta_i =0$ in the tree approximation (this is
what is called the impulse approximation). To second order in small
momenta, the two--body interactions involving loop graphs and tree
diagrams with $\Delta_i =1,2$ are taken from phenomenological models of
$\pi N$ scattering and the remaining three--body  interactions between
two nucleons and the pion (calculated from tree diagrams with
$\Delta_i =0$ vertices) are  shown in fig.5.5 (these are the ones
that contribute to the pion--nucleus scattering length).
\midinsert
\hskip 1in
\epsfxsize=3in
\epsfysize=2.4in
\epsffile{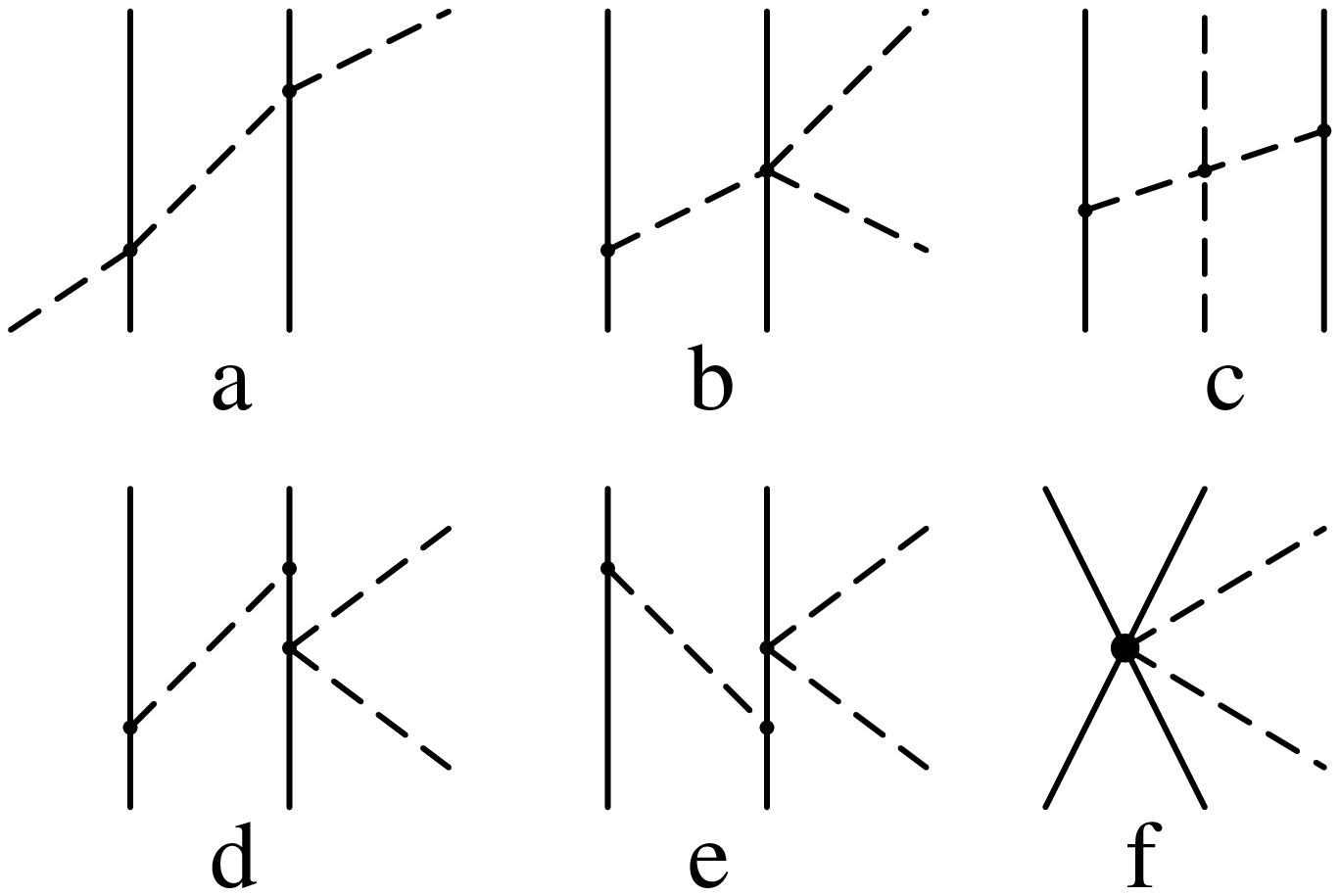}
\smallskip
{\noindent\narrower \it Fig.~5.5:\quad
Irreducible connected graphs for the interaction of a pion with a pair
of nucleons that contribute to the pion--nucleus scattering length.
Only one time--ordered diagram per class is shown.
\smallskip}
\vskip -0.5truecm
\endinsert
The details of the calculation are found in ref.[5.15]. The
 pion--nucleus scattering length takes the form
$$a_{ab} = {1 + M_\pi / m_N \over 1 + M_\pi / A m_N} \, \sum_r \,
a_{ab}^{(r)} \, \, + \, \, a_{ab}^{\rm three-body} \quad ,
\eqno(5.29)$$
where $a$, $b$ are the pion isovector indices, $a_{ab}^{(r)}$ is
the pion scattering length of the $r$th nucleon and $m_N$ is the
nucleon mass. The three--body
contribution stemming from diagrams 5.5a--f takes the form
$$
a_{ab}^{\rm three-body}  = {M_\pi^2 \over 32 \pi^4 F_\pi^4 (1+M_\pi/m_d)}
\sum_{r < s} \biggl< {1 \over \svec{q}_{rs}^2} \bigl( 2 \svec{t}^{(r)} \cdot
\svec{t}^{(s)} \, \delta_{ab} - t_a^{(r)} t_b^{(s)} -  t_a^{(s)} t_b^{(r)}
\bigr) \biggr> $$
$$ - {g_A^2 \delta_{ab} \over 32 \pi^4 F_\pi^4 (1+M_\pi/m_d)}
\sum_{r < s} \biggl< \svec{t}^{(r)} \cdot \svec{t}^{(s)}
\,{\svec{q}_{rs} \cdot \svec{\sigma}^{(r)} \, \svec{q}_{rs}
\cdot \svec{\sigma}^{(s)} \over \svec{q}_{rs}^2 + M_\pi^2} \biggr> $$
$$ + {g_A^2  \over 32 \pi^4 F_\pi^4 (1+M_\pi/m_d)} \sum_{r < s} \biggl<
{[\svec{q}_{rs}^2 \, \svec{t}^{(r)} \cdot \svec{t}^{(s)}
\delta_{ab} + M_\pi^2\,(t_a^{(r)} t_b^{(s)} + t_a^{(s)} t_b^{(r)}]
\svec{q}_{rs} \cdot \svec{\sigma}^{(r)} \, \svec{q}_{rs}
\cdot \svec{\sigma}^{(s)} \over (\svec{q}_{rs}^2 + M_\pi^2)^2} \biggr> $$
$$ +  {g_A^2 M_\pi \over 132 \pi^4 F_\pi^4 (1+M_\pi/m_d)}
\sum_{r < s} \biggl< (\svec{t}^{(r)} + \svec{t}^{(s)}) \cdot
 (\svec{t}^{(\pi)})_{ab}
\,{\svec{q}_{rs} \cdot \svec{\sigma}^{(r)} \, \svec{q}_{rs}
\cdot \svec{\sigma}^{(s)} \over (\svec{q}_{rs}^2 + M_\pi^2)^{3/2}} \biggr>
\eqno(5.30)$$
where $r,s$ label individual nucleons and $(t_c^{(\pi)})_{ab}
=- i \epsilon_{abc}$ is the pion isospin vector. Notice that there is
some cancellation between the second and third term in eq.(5.30) as
${\bf q}_{rs} \to \infty$ so that the result is less sensitive to the
nuclear wave function at small separation (see also ref.[5.35]).
For an isoscalar nucleus like the deuteron, the expressions in (5.30)
simplify considerably  since the last term vanishes and
$t_a^{(r)} t_b^{(s)} + t_a^{(s)} t_b^{(r)}$ can be replaced by
$(2/3)\delta_{ab} \, \svec{t}^{(r)} \cdot \svec{t}^{(s)}$. Even more
important, for an isoscalar nucleus the nominally leading term in
eq.(5.29) are vanishing since they involve an expectation value of
$\sum_r \svec{t}^{(r)} \cdot \svec{t}^{(\pi)}$. So one is left with small
${\cal O}(M_\pi^2)$ contributions of the $\sigma$--term type to the
impulse approximation. This is the reason why it makes sense to
compare the corrections calculated in CHPT directly with the empirical
values of the $\pi d$ scattering length. Using isospin symmetry, one
can now calculate the two--body contribution to the $\pi d$ scattering
length and finds [5.36,5.37]
$${1+M_\pi / m_N \over 1+M_\pi / m_d} \, (a_{\pi p} +a_{\pi n}) =
-(0.021 \pm 0.006) \, M_\pi^{-1} \quad , \eqno(5.31)$$
with $m_d$ the deuteron mass.
The first term in (5.30) gives the well--known and large rescattering
contribution. It is much bigger than the remaining three--body terms
due to the anomalously large radius of the deuteron. Using
empirical information on $\pi N$ scattering to calculate the
rescattering contribution (for details, see e.g. [5.37]),
and  the Bonn potential to produce the
deuteron wave function for the calculation of the remaining
three--body
contributions [5.15],  one finds
$$a^{\rm three-body} = -(0.026 \pm 0.001) \, M_\pi^{-1}
 -  0.0005 \, M_\pi^{-1} \quad , \eqno(5.32)$$
where the first number refers to the first term in (5.30) and the second
one to the remaining three--body contributions. The latter ones are
very small, well within the uncertainties of the other dominant terms.
 This justifies the
final theoretical result of $ -(0.047 \pm 0.006) \, M_\pi^{-1}$ in
good agreement with the empiral value of $ -(0.056 \pm 0.009) \,
M_\pi^{-1}$ [5.37].
This is a good example how the chiral Lagrangian machinery can be used
to explain why one is allowed to take only  certain graphs like the
rescattering contribution but neglect the others which are of the same
order in the expansion in small momenta.
\medskip
A similar calculation has been been performed by Beane et al. [5.38]
for pion photoproduction on nuclei. They have considered all
corrections which are suppressed by two powers in small momenta as
compared to the lowest order impulse approximation using $\Delta_i =0$
vertices. Again, there is a single scattering contribution taken from
phenomenology plus some three--body interactions (and disconnected
graphs involving pairs of nucleons for $A>2$). In the case of the
deuteron and considering neutral pion photoproduction, the calculation
simplifies enormeously.  At threshold, the invariant matrix--element
takes the form
$${\cal M} = 2 i \,  \svec{J} \cdot \svec{\epsilon} \, E_d \quad ,
\eqno(5.33)$$
with $\svec{J}$ the spin of the deuteron. The single scattering
contribution can be compactly written as
$$E^{ss} =  {1+M_\pi / m_N \over 1+M_\pi / m_d} \, \bigl(
E_{0+}^{\pi^0 p} + E_{0+}^{\pi^0 n} \bigr) \, S(k/2) = -1.33 \cdot
10^{-3} \, M_\pi^{-1}
\eqno(5.34)$$
with $S(k/2)$ the deuteron form factor evaluated for the threshold
kinematics ($k = 0.685$ fm$^{-1}$) and using the currently accepted
value of $E_{0+}^{\pi^0 p} = -2.0 \cdot 10^{-3} \, M_\pi^{-1}$ and
taking $E_{0+}^{\pi^0 n} =0.5 \cdot 10^{-3} \, M_\pi^{-1}$ from the incomplete
``LET'' as discussed in section 4.4. Clearly, this prediction hinges on
these particular values for the elementary pion photoproduction
electric dipole amplitudes. Also, the one for the $\gamma n \to \pi^0
n$ is not taken from experiment. The three--body
graphs which contribute are of the exchange current type (see also the
next section). In  class (a), the photon couples to the pion which is
interchanged between the two nucleons and the second class (b) involves the
diagrams with  exactly one $NN\pi\pi$ and one $NN \pi \gamma$ lowest
order vertex. Their contributions to $E_d$ take the form
$$\eqalign{
 E^{(a)} & = -{e g_A M_\pi m_N \over 8 \pi^2 (M_\pi + m_d) F_\pi^3 k}
\, \int_0^\infty \, F(kr) \, dr \cr
 E^{(b)} & = -{e g_A M_\pi m_N \over 16 \pi^2 (M_\pi + m_d) F_\pi^3 }
\,\int_0^1 \, dz \, \int_0^\infty \, G(zkr) \, dr \cr} \eqno(5.35)$$
 with the integrands $F(kr)$ and $G(zkr)$ given in [5.38]. Using again
the Bonn potential to generate the deuteron wave function, one finds
$$E^{(a)} = -2.24 \cdot 10^{-3} \, M_\pi^{-1} \, , \quad
  E^{(b)} = -0.42 \cdot 10^{-3} \, M_\pi^{-1} \, . \eqno(5.36)$$
Summing up (5.34) and (5.36), one finds $E_d = -3.99
\cdot 10^{-3} \, M_\pi^{-1}$ in good agreement with the empirical
value of $(-3.74 \pm 0.25) \cdot 10^{-3} \, M_\pi^{-1}$
[5.39]. However, as already stressed, the single scattering contribution
(5.34) is afflicted by large uncertainties and it remains to be seen
which value for $E_{0+}^{\pi^0 p}$ the new data from Mainz and
Saskatoon  will favor and how accurate the guess for the
electric dipole amplitude $E_{0+}^{\pi^0 n}$ will turn out to be
(once measured).
\bigskip \goodbreak
\noindent{\bf V.5. EXCHANGE CURRENTS}
\medskip
\goodbreak
Meson exchange currents arise naturally in the meson--exchange picture of the
nuclear forces. An external electromagnetic or axial probe does not only
couple to the nucleons (impulse approximation, one--body operators)
 but also to the mesons in
flight or leads to nonlinear seagull-type vertices (these are typical
two--body operators), cf. fig.5.6.
\midinsert
\hskip 0.9in
\epsfxsize=3in
\epsfysize=1.2in
\epsffile{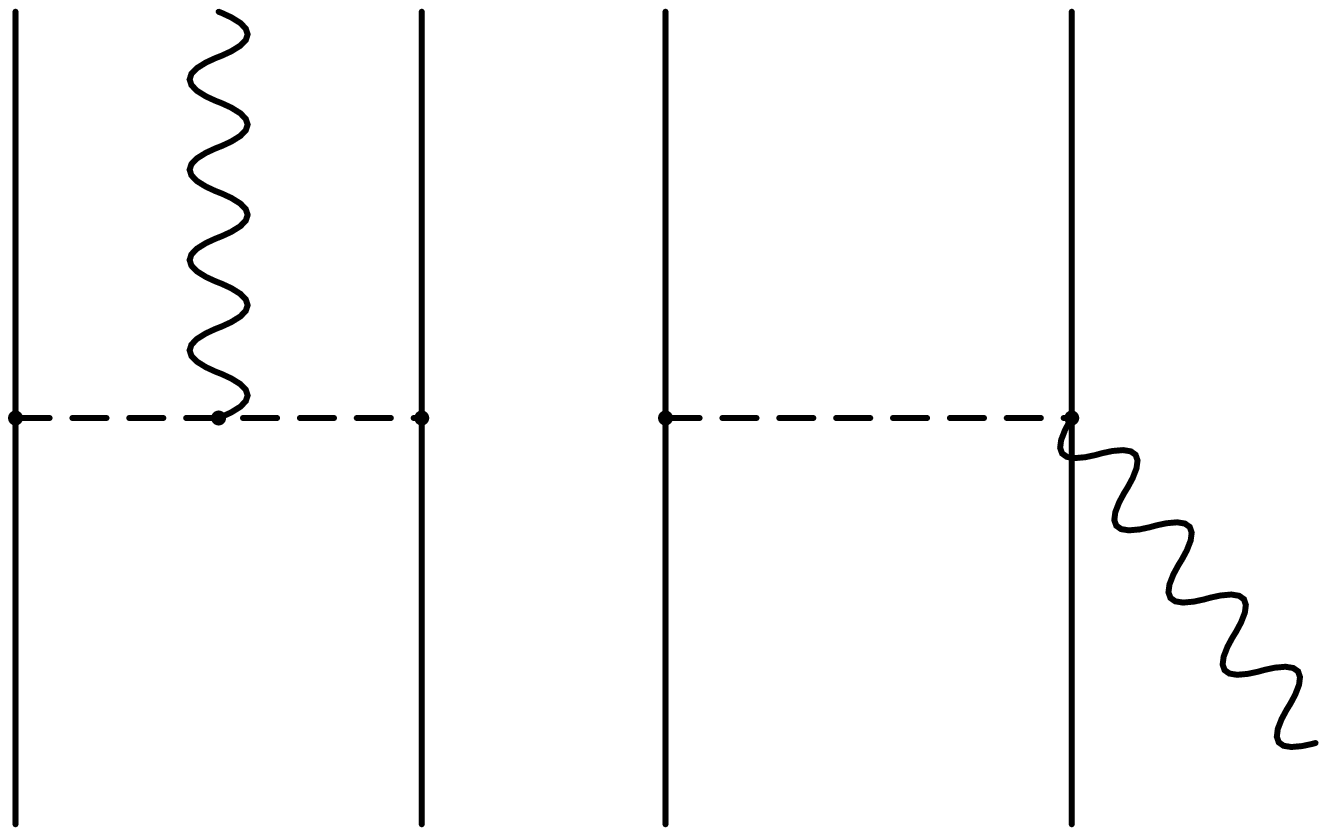}
\smallskip
{\noindent\narrower \it Fig.~5.6:\quad
The lowest order and dominant pion exchange current diagrams. The
wiggly line denotes an electroweak probe.
\smallskip}
\vskip -0.5truecm
\endinsert
\noindent  Also known since many decades [5.40], the first
compelling evidence for meson exchange currents came from the
calculation of neutron radiative capture at threshold and deuteron
photodisintegration [5.10,5.11] (for reviews, see [5.19,5.20]). By now,
 the existence of these two--body operators can be considered verified
experimentally [5.41]. More than 10 years ago,
the so--called "chiral filter hypothesis" was introduced [5.42]. It states
that the response of a nucleus to a long--wavelength electroweak probe is
given solely by the soft--pion exchange terms dictated by chiral
symmetry. Consider any exchange current contribution ${\cal X}$, this
means
$$ {\cal X} = {\cal X}_\pi + {\cal X}_{2 \pi} + {\cal X}_R + {\cal X}_{N^*}
= {\cal X}_{\rm soft-pion} ( 1 + C ) \eqno(5.37)$$
where $C$ is a generally small correction to the leading one
($C \ll 1$), $R$
denotes the effects of heavier meson exchange and $N^*$ the excitation
of nucleon resonances. Stated differently,
 all the heavier mesons and nucleon excitations, multi--pion exchanges
and form factor effects are not
seen, even up to energies of the order of 1 GeV (although individual
contributions can be large). Why this holds true at such
energies has not yet been explained.
\medskip
 Rho [5.43] has given a simple argument
how the "chiral filter" can occur in nuclei for small and moderate
momentum transfer.  His lowest order analysis
follows closely the one of Weinberg [5.12]. Any matrix--element $ME$ of the
effective potential $V$ or of a current $J_\mu$ has the form $ME \sim Q^\nu
\, F(Q/m)$, as discussed before. In the presence of a slowly varying external
electromagnetic field $A_\mu$ (or a weak one), the Hamiltonian takes the form
$$\eqalign{
{\rm H_{\rm eff}} &= {\rm H}_{\pi \pi} + {\rm H}_{\pi N} + {\rm H}_{N N} +
{\rm H_{\rm ext}}    \cr
{\rm H_{\rm ext}} &= {e \over D^2} \bigl[ (\vec{\pi} \times
\partial_\mu \vec{\pi} )_3
+ {i g_A \over 2 F_\pi} \bar{\Psi}_N \gamma_5 \gamma_\mu ( \vec{\tau} \times
\vec{\pi} )_3 \Psi_N \bigr] A_\mu + \ldots \cr} \eqno(5.38)$$
and this additional term H$_{\rm ext}$ modifies the power counting.  Since
one derivative is replaced by the external current, the tree graphs ($L=0$)
with the lowest power $\nu$ must fulfill
$$ d_i + {1 \over 2} n_i - 2 = -1 \eqno(5.39)$$
which leads to $d_i = 0$ and $n_i = 2$ or
$d_i = 1$ and $n_i = 0$. In contrast to the case of the nuclear forces, to
leading order {\it no} four--nucleon contact terms contribute. This means that
there is no short--ranged two--body current (to lowest order), the exchange
current is entirely given in terms of the soft--pion component derived from
(5.38). This justifies the chiral filter hypothesis at tree level.
\medskip Park et
al.[5.44] have also investigated one--loop corrections to the axial--charge
operator, the first correction to the pertinent soft--pion matrix--element is
suppressed by $(Q/m)^2$ . The authors of ref.[5.44] use the heavy mass
formalism which simplifies the calculation considerably. They argue
that $\delta$--function type contact terms are suppressed by the
short--range nuclear correlations. Stated differently, since the
chiral counting is only meaningful as long as $Q^2 /m^2 \ll 1$, one
can not describe processes that involve energy or momentum scales
exceeding this criterion. Short--distance interactions are therefore
not accessible by chiral perturbation theory. This is different in
philosophy from the calculation of the nuclear forces by van Kolck et
al. [5.14,5.17] where it is argued that the four--nucleon contact
terms are smeared out over a distance $\sim 1 / M_\rho$.
In ref.[5.44], it is shown that the
loop corrections are small for distances $r \ge 0.6 $ fm, which means that
the lowest order argument of Rho [5.43] is robust to one--loop order.
To be more precise, the results of ref.[5.44,5.45] can be summarized
as follows. One writes the nuclear matrix element of the axial--charge
operator  as
$$ M^{\rm axial} = M_1 + M_2 \, , \quad M_2 = M_2^{\rm tree} \, (1 +
\delta \,) \eqno(5.40)$$
where the subscript $n=1,2$ refers to one-- and two--body operators and
'tree' correponds to the diagrams shown in fig.5.6 (with renormalized
couplings). One finds almost independently of the mass number that
$\delta < 0.1$, i.e. the tree contribution dominates. In this
particular case, the two--body operator is of comparable size to the
one--body operator, $M_2^{\rm tree} / M_1 \sim 0.7$ [5.45] which is
sufficient to explain the empirical value of $M^{\rm exp} / M_1 \sim
1.6 \ldots 2.0$ [5.46].
\medskip
The thermal $np$ capture has also been considered by Park et
al. [5.47] including terms to order $(Q/m)^3$. Apart from the dominant
one pion exchange diagrams (fig.5.6), there are  additional
graphs corresponding to two--pion exchange as well as counterterm
contributions saturated by $\Delta$ and
$\omega$ meson exchange. While in impulse approximation one finds
$\sigma (np \to d \gamma)=305.6$~mb, the exchange currents calculated
in CHPT together with a short--range correlation cut--off $0 < r_c <
0.7$~fm lead to $\sigma (np \to d \gamma)=334 \pm 3$~mb, in nice
agreement with the experimental value, $\sigma_{\rm exp} (np \to d
\gamma)= 334.2 \pm 0.5$~mb [5.48]. Again, the soft pion contribution
gives the dominant part of the two--body enhancement.
\medskip
The calculations presented here seem to lend credit to the chiral
filter hypothesis and demonstrate once more the importance of chiral
symmetry in nuclear phenomena. Furthermore, in nuclei it appears
natural to make use of the short--range correlations to
suppress operators of the contact term type and alike.
For more details on the calculation of exchange currents from chiral
Lagrangians we refer the reader to refs.[5.45,5.47,5.49].
 What remains mysterious is why the chiral filter
hypothesis works up to so high energies
-- the answer to this lies certainly outside the realm of baryon CHPT.
\bigskip \bigskip
\noindent{\bf REFERENCES}
\medskip
\item{5.1}H. Yukawa, {\it Proc. Phys.-Math. Soc. Jpn.} {\bf 17} (1935)
48. \smallskip
\item{5.2}R. Machleidt, K. Holinde and Ch. Elster, {\it Phys. Rep.}
  {\bf 149} (1987) 1.  \smallskip
\item{5.3}W.N. Cottingham et al., {\it Phys. Rev.\/} {\bf D8} (1973)
  800:

M. Lacombe et al., {\it Phys. Rev.\/} {\bf C21} (1980) 861.
 \smallskip
\item{5.4}M.M. Nagels, T.A. Rijken and J.J. de Swart, {\it
    Phys. Rev.\/} {\bf D17} (1978) 768. \smallskip
\item{5.5}S. Deister et al., {\it Few--Body Systems} {\bf 10} (1991) 1.
\smallskip
\item{5.6}I.E. Lagaris and V.R. Phandharipande, {\it Nucl. Phys.} {\bf
A359} (1981) 331:

R.B. Wiringa, V.G.J. Stoks and R. Schiavilla, ``An accurate
Nucleon--Nucleon

Potential'', ANL preprint 1994, nucl-th/9408016.
 \smallskip
\item{5.7}S.O. B\"ackman, G.E. Brown and J.A. Niskanen, {\it
Phys. Reports} {\bf 124} (1984) 1.
\smallskip
\item{5.8}G.E. Brown and A.D. Jackson, ``The Nucleon--Nucleon
Interaction'', North--Holland, Amsterdam, 1976. \smallskip
\item{5.9}T.A. Rijken, {\it Ann. Phys.} {\bf 164} (1985) 1, 23.
\smallskip
\item{5.10}D.O. Riska and G.E. Brown, {\it Phys. Lett.} {\bf B38}
(1972) 193.  \smallskip
\item{5.11}
J. Hockert, D.O. Riska, M.Gari and A. Huffmann, {\it Nucl. Phys.} {\bf
A217} (1973) 19. \smallskip
\item{5.12}S. Weinberg, {\it Phys. Lett.\/} {\bf B251} (1990) 288.
\smallskip
\item{5.13}S. Weinberg, {\it Nucl. Phys.\/} {\bf B363} (1991) 3.
\smallskip
\item{5.14}C. Ordonez and U. van Kolck, {\it Phys. Lett.\/} {\bf
B291} (1992) 459.
\smallskip
\item{5.15}S. Weinberg, {\it Phys. Lett.\/} {\bf B295} (1992) 114.
\smallskip
\item{5.16}U. van Kolck, Thesis, University of Texas at Austin, 1992
\smallskip
\item{5.17}C. Ordonez, L. Ray and U. van Kolck, {\it Phys. Rev. Lett.\/} {\bf
72} (1994) 1982.
\smallskip
\item{5.18}U. van Kolck, {\it Phys. Rev.\/} {\bf
C49} (1994) 2932. \smallskip
\item{5.19}J.F. Mathiot, {\it Phys. Reports} {\bf 173} (1989) 63.
\smallskip
\item{5.20}D.O. Riska, {\it Phys. Reports} {\bf 181} (1989) 207.
\smallskip
\item{5.21}K.A. Brueckner and K.M. Watson, {\it Phys. Rev.\/} {\bf 92}
(1953) 1023. \smallskip
\item{5.22}H. Sugawara and F. von Hippel, {\it Phys. Rev.\/} {\bf 172}
(1968) 1764. \smallskip
\item{5.23}H. Sugawara and S. Okubo, {\it Phys. Rev.\/} {\bf 117}
(1960) 605, 611. \smallskip
\item{5.24}Ulf-G. Mei{\ss}ner,
{\it Comments Nucl. Part. Phys.\/} {\bf 20} (1991) 119.
\smallskip
\item{5.25}Ulf-G. Mei{\ss}ner, ``Baryon Chiral Perturbation Theory
A.D. 1994'', Bonn preprint TK 94 17, 1994, {\it Czech. J. Phys.}, in print.
\smallskip
\item{5.26}T.E.O. Ericson, {\it Nucl. Phys.\/} {\bf A416} (1984) 281c.
\smallskip
\item{5.27}R.P. Haddock et al., {\it Phys. Rev. Lett.} {\bf 14} (1985)
318;

H.P. Noyes, {\it Phys. Rev.} {\bf 130} (1963) 2025.
\smallskip
\item{5.28}K. Holinde, ``Hadron--Hadron Interactions'', J\"ulich
preprint, 1994.
\smallskip
\item{5.29}J.L. Friar, {\it Czech. J. Phys.} {\bf 43} (1993) 259.
\smallskip
\item{5.30}H. Witala, W. Gl\"ockle and Th. Cornelius,
{\it Few Body Systems\/}
{\bf 6} (1989) 79; {\it Nucl. Phys.} {\bf A496} (1989) 446;
{\it Phys. Rev.} {\bf C39} (1989) 384;

I. Slaus, R. Machleidt, W. Tornow, W. Gl\"ockle and H. Witala,

{\it Comments Nucl. Part. Phys.} {\bf 20} (1991) 85;

W. Gl\"ockle, private communication.
\smallskip
\item{5.31}K.A. Brueckner, C.A. Levinson and H.M. Mahmoud, {\it Phys. Rev.}
{\bf 95} (1954) 217. \smallskip
\item{5.32}G.E. Brown, A.M. Green and W.J. Gerace, {\it Nucl. Phys.}
{\bf A115} (1968) 435;

G.E. Brown, {\it Comments Nucl. Part. Phys.\/} {\bf 5} (1972) 6.
\smallskip
\item{5.33}S.N. Yang and W. Gl{\"o}ckle,
{\it Phys. Rev.} {\bf C33} (1986) 1774;

S.A. Coon and J.L. Friar, {\it Phys. Rev.} {\bf C34} (1986) 1060.
\smallskip
\item{5.34}J. Fujita and H. Miyazawa, {\it Prog. Theor. Phys.}
{\bf 17} (1957) 360. \smallskip
\item{5.35}M.R. Robilotta and C. Wilkin, {\it J. Phys.}
{\bf G4} (1978) L115. \smallskip
\item{5.36}J.M. Eisenberg and D.S. Koltun, ``Theory of Meson Interactions
with Nuclei'', Wiley--Interscience, New York, 1980. \smallskip
\item{5.37}T. Ericson and W.Weise , ``Pions and Nuclei'', Oxford
University Press, Oxford, 1988.  \smallskip
\item{5.38}S.R.Beane, C.Y. Lee and U. van Kolck, Duke preprint, 1995.
\smallskip
\item{5.39}P. Argan et al., {\it Phys. Rev. Lett.}
{\bf 41} (1978) 629; {\it Phys. Rev.} {\bf C21} (1980) 1416.
 \smallskip
\item{5.40}F. Villars, {\it Helv. Phys.  Acta} {\bf 20} (1947) 476;

H. Miyazawa, {\it Prog. Theor. Phys.} {\bf 6} (1951) 801. \smallskip
\item{5.41}
B. Frois and J.--F. Mathiot,
{\it Comments Part. Nucl. Phys.\/} {\bf 18} (1989) 291.
\smallskip
\item{5.42}K. Kubodera, J. Delorme and M. Rho, {\it Phys. Rev. Lett.\/} {\bf
40} (1978) 755

M. Rho and G. E. Brown, {\it Comments Part. Nucl. Phys.\/} {\bf 10} (1981)
201.
\smallskip
\item{5.43}M. Rho, {\it Phys. Rev. Lett.\/} {\bf 66} (1991) 1275.
\smallskip
\item{5.44}T.-S. Park, D.-P. Min and M. Rho, {\it Phys. Reports\/}
{\bf 233} (1993) 341. \smallskip
\item{5.45}T.-S. Park, I.S. Towner
and K. Kubodera, {\it Nucl. Phys.} {\bf A579} (1994) 381.  \smallskip
\item{5.46}E.K. Warburton and I.S. Towner, {\it Phys. Reports\/}
{\bf 242} (1994) 103. \smallskip
\item{5.47}T.-S. Park, D.-P. Min and M. Rho, preprint SNUTP 94--124, 1992,
nucl-th/9412025. \smallskip
\item{5.48}A.E. Cox, S. Wynchank and C.H. Collie, {\it Nucl. Phys.}
{\bf 74} (1965) 497. \smallskip
\item{5.49}M. Rho, {\it Phys. Reports\/} {\bf 240} (1994) 1.
\smallskip
\vfill \eject

\noindent{\bf VI. THREE FLAVORS, DENSE MATTER AND ALL THAT}
\bigskip
In this section, we will first be concerned with the extension to the case of
three flavors and discuss the calculation of baryon masses and $\sigma$--terms.
Then, we will turn to the topic of kaon--nucleon scattering and the behaviour
of pions in dense matter. This latter topic (also extended to kaons)
is a rather new and
rapidly developing field, so we can only provide a state of the art summary.
Finally, various omissions are briefly touched upon.
\bigskip
\noindent{\bf VI.1. FLAVOR $SU(3)$, BARYON MASSES AND $\sigma$--TERMS}
\medskip
\goodbreak
Let us first  provide the necessary definitions for the three flavor
meson--baryon system. It is most convenient to write the eight meson and
baryon fields in terms of $SU(3)$ matrices $\Phi$ and $B$, respectively,
$$ \Phi =  \sqrt{2}  \left(
\matrix { {1\over \sqrt 2} \pi^0 + {1 \over \sqrt 6} \eta
&\pi^+ &K^+ \cr
\pi^-
        & -{1\over \sqrt 2} \pi^0 + {1 \over \sqrt 6} \eta & K^0 \cr
K^-
        &  \bar{K^0}&- {2 \over \sqrt 6} \eta  \cr} \right)
\eqno(6.1a) $$
$$ B =  \left(
\matrix  { {1\over \sqrt 2} \Sigma^0 + {1 \over \sqrt 6} \Lambda
&\Sigma^+ &  p \cr
\Sigma^-
    & -{1\over \sqrt 2} \Sigma^0 + {1 \over \sqrt 6} \Lambda & n \cr
\Xi^-
        &       \Xi^0 &- {2 \over \sqrt 6} \Lambda \cr} \right)
\eqno(6.1b) $$
with
$$ U(\Phi) = u^2 (\Phi) = \exp \lbrace i \Phi / F_p \rbrace
\eqno(6.2) $$
with $F_p$ the pseudoscalar decay constant in the chiral limit.
Of course, beyond leading order, one has to account for the fact that $F_\pi
\not= F_K \not= F_\eta$ [6.1].  The covariant derivative acting on $B$ reads
$$\eqalign{
D_\mu B &= \partial_\mu B + [ \Gamma_\mu , B ] \cr
\Gamma_\mu &= {1 \over 2} \bigl\lbrace u^\dagger [ \partial_\mu - i ( v_\mu +
a_\mu )]u + u [ \partial_\mu - i ( v_\mu - a_\mu )]u^\dagger \bigr\rbrace
\cr}
\eqno(6.3) $$
with $v_\mu$ and $a_\mu$ external vector and axial--vector sources. Under
$SU(3)_L \times SU(3)_R$, $B$ and $D_\mu B$ transform as
$$ B' = K B K^\dagger \, \, \, , \, \,\, (D_\mu B)' = K (D_\mu B) K^\dagger
\eqno(6.4)  $$
It is now straightforward to construct the lowest--order ${\cal O}(p)$
meson--baryon Lagrangian,
$${\cal L}_{\rm MB}^{(1)} = \Tr \bigl\lbrace i \bar{B} \gamma^\mu D_\mu B
- m_0 \bar{B} B + {1 \over 2} D \bar{B} \gamma^\mu \gamma_5 \lbrace
u_\mu , B \rbrace
+ {1 \over 2} F \bar{B} \gamma^\mu \gamma_5 [ u_\mu , B ] \bigr\rbrace
\eqno(6.5)  $$
where $m_0$ stands for the (average) octet mass in the chiral limit. The
trace in (6.5) runs over the flavor indices. Notice that in contrast to the
$SU(2)$ case, one has two possibilities of coupling the axial $u_\mu$ to
baryon bilinears. These are the conventional $F$ and $D$ couplings subject to
the constraint $F + D = g_A = 1.26$. At order ${\cal O}(p^2)$ the baryon mass
degeneracy is lifted by the terms discussed below. However, there are
many other terms at this order. If one works in the one--loop approximation,
one also needs the terms of order ${\cal O}(p^3)$ (or eventually from
${\cal O}(p^4)$).
The complete local
effective Lagrangians ${\cal L}_{\rm MB}^{(2)}$ and ${\cal L}_{\rm MB}^{(3)}$
are given by Krause [6.2]. The extension of this to the heavy mass formalism
is straightforward, as spelled out in detail in the review article by Jenkins
 and Manohar [6.3]. For our purpose, we only give the lowest order Lagrangian
and the three terms of order $q^2$ which account for quark mass insertions
(in the isospin limit $m_u = m_d$),
$${\cal L}^{(1)}_{\rm MB}= \Tr(\bar B \,i v\cdot{\cal D} \,B) + D\, \Tr (\bar B
S^\mu \{ u_\mu , B\} ) + F \, \Tr ( \bar B S^\mu [ u_\mu, B] ) \eqno(6.6)$$
with the baryons considered as static sources and
equivalently their momenta decompose as
$p_\mu = m_0 \, v_\mu + l_\mu$, $v \cdot l \ll m_0$.
Beyond leading order and in the
present context we consider only counter terms of chiral power $q^2$ which
account for quark mass insertions,
$${\cal L}_{\rm MB}^{(2)} = b_D\, \Tr (\bar B \{ \chi_+ , B \} ) + b_F \, \Tr
(\bar B [ \chi_+,B]) + b_0 \, \Tr (\bar BB ) \Tr (\chi_+) \eqno(6.7)$$
with $\chi_+ = u^\dagger \chi u^\dagger + u \chi^\dagger u$ and $\chi = 2 B_0 (
{\cal M} + {\cal S})$ where $\cal S$ denotes the nonet of external scalar
sources. As we will see later on, the constants $b_D$, $b_F$ and $b_0$ can be
fixed from the knowledge of the baryon masses and the $\pi N$ $\sigma$-term (or
one of the $KN$ $\sigma$-terms). The constant $b_0$ can not be determined from
the baryon mass spectrum alone since it contributes to all octet members in the
same way.

We now consider the calculation of baryon masses in CHPT.
Gasser [6.4] and Gasser and Leutwyler [6.5] were the first to systematically
investigate the baryon masses at next--to--leading order. The quark mass
expansion of the baryon masses takes the form
$$m_B = m_0 + \alpha {\cal M} + \beta {\cal M}^{3/2} + \gamma {\cal M}^2 +
\ldots \quad .
\eqno(6.8)$$
The non--analytic piece proportional to ${\cal M}^{3/2}$ was
first observed by Langacker and Pagels [6.6]. If one retains only the terms
linear in the quark masses, one obtains the Gell-Mann--Okubo relation
$m_\Sigma + 3 m_\Lambda = 2 (m_N + m_\Xi )$ (which is fulfilled within 0.6$\%$
 in nature) for the octet and the equal spacing rule for the decuplet,
$m_\Omega- m_{\Xi^*} = m_{\Xi^*} - m_{\Sigma^*}
= m_{\Sigma^*} - m_\Delta$
(experimentally, one has 142:145:153 MeV). However, to extract quark mass
ratios
>from the expansion (6.8), one has to work harder. This was done in
refs.[6.4,6.5]. The non--analytic
terms were modelled by considering the baryons
as static sources surrounded by a cloud of mesons and photons -- truely the
first calculation in the spirit of the heavy mass formalism.  The most
important result of this analysis was the fact
that the ratio $R = (\hat m - m_s ) / (m_u - m_d )$ comes out
consistent with the value obtained from the meson spectrum.  Jenkins [6.7]
has recently repeated this calculation using the heavy fermion EFT of
refs.[6.3,6.8], including also the spin--3/2 decuplet fields in the EFT.
She concludes that the success of the octet and decuplet mass relations is
consistent with baryon CHPT as long as one includes the decuplet. Its
contributions tend to cancel the large corrections from the kaon loops like
$m_s M_K^2 \ln M_K^2$.
The calculation was done in the isospin limit $m_u = m_d =
0$ so that nothing could be said about the quark mass ratio $R$. This latter
question was recently addressed by
Lebed and Luty [6.9] who arrive at a negative
conclusion concerning the possibility of extracting current quark mass ratios
>from the baryon spectrum. We follow here ref.[6.10] in which the whole scalar
sector, i.e the baryon masses and $\sigma$--terms, are considered and which
sheds some doubt on the results obtained so far when the decuplet is included
in the EFT. Following [6.10],  a complete calculation up to order $q^3$
 involves only intermediate octet states.
At this order (one-loop approximation)
one has three counterterms with a priori unknown but finite coefficients. These
can be fixed from the octet masses $(m_N, m_\Lambda, m_\Sigma, m_\Xi)$ and the
value $\sigma_{\pi N}(0) $ since one of the counter terms appears in the baryon
mass formulae in such a way that it always can be lumped together with the
average octet mass in the chiral limit. This allows  to predict
the two $KN$ $\sigma$-terms, $\sigma_{KN}^{(1)}(0)$ and $\sigma^{(2)}_{KN}(0)$
as well as
the $\sigma$-term shifts to the respective Cheng-Dashen points and the
matrix element $m_s <p|\bar ss |p>$.
To this order in the chiral expansion, any baryon mass takes the form
$$m_B = m_0 - {1\over 24 \pi F_p^2}\bigl[ \alpha^\pi_B M_\pi^3 + \alpha^K_B
M_K^3 + \alpha^\eta_B M_\eta^3 \bigr] + \gamma^D_B b_D + \gamma^F_B b_F
-2b_0(M_\pi^2 + 2M_K^2) \eqno(6.9)$$
The second  term on the right hand side
comprises the Goldstone boson loop contributions and the third
term stems from the counter terms eq.(6.7). Notice that the loop
contribution is
ultraviolet finite and non-analytic in the quark masses since
$M_{\pi, K ,\eta}^3 \sim {\cal M}^{3/2}$.
The constants $b_D$, $b_F$ and $b_0$ are therefore finite. The
numerical factors read
$$\eqalign{
\alpha_N^\pi & = {9\over 4}(D+F)^2, \quad
\alpha_N^K = {1\over 2}(5D^2 - 6DF +9F^2), \quad
\alpha_N^\eta = {1\over 4}(D-3F)^2; \cr
\alpha_\Sigma^\pi &  = D^2+6F^2, \quad
\alpha_\Sigma^K = 3(D^2+F^2), \quad
\alpha_\Sigma^\eta = D^2; \cr
\alpha_\Lambda^\pi &  = 3D^2, \quad
\alpha_\Lambda^K  = D^2  +9F^2, \quad
\alpha_\Lambda^\eta   = D^2; \cr
\alpha_\Xi^\pi & = {9\over 4}(D-F)^2, \quad
\alpha_\Xi^K   = {1\over 2}(5D^2 + 6DF +9F^2), \quad
\alpha_\Xi^\eta = {1\over 4}(D+3F)^2; \cr
\gamma_N^D & = -4M_K^2,\quad
\gamma_N^F    =  4M_K^2-4M^2_\pi; \quad \quad
\gamma_\Sigma^D   =  -4M_\pi^2, \quad
\gamma_\Sigma^F  =  0; \cr
\gamma_\Lambda^D & =  -{16\over 3}M_K^2 + {4\over3}M_\pi^2, \quad
\gamma_\Lambda^F = 0; \quad \quad
\gamma_\Xi^D =  -4M_K^2, \quad \gamma_\Xi^F =  -4M_K^2+4M^2_\pi. \cr}
\eqno(6.9a)$$
At this order, the deviation from the Gell-Mann-Okubo formula is given by
$$\eqalign{{1\over4} \bigl[ 3m_\Lambda + m_\Sigma - 2 m_N - 2 m_\Xi\bigr]  & =
{3F^2 - D^2 \over 96 \pi F_p^2}\bigl[ M_\pi^3 - 4 M_K^3 + 3 M_\eta^3 \bigr] \cr
& = {3F^2 - D^2\over 96\pi F_p^2} \bigl[ M_\pi^3 - 4 M_K^3 + {1\over \sqrt 3}
(4 M_K^2 - M_\pi^2)^{3/2}\bigr] \cr} \eqno(6.10)$$
where in the second line we have used the GMO relation for the $\eta$-meson
mass, which is legitimate if one works at next-to-leading order.

Further information on the scalar sector is given by the scalar form factors or
$\sigma$-terms which measure the strength of the various matrix-elements $m_q\,
\bar q q$ in the proton. One defines:\footnote{*}{These quantities are
renormalization-group invariant in a mass-independent subtraction scheme, which
is what one usually employs.}
$$\eqalign{
\sigma_{\pi N}(t) &= \hat m <p'|\bar u u + \bar d d| p>\cr
\sigma_{KN}^{(1)}(t)&={1\over 2}(\hat m + m_s) <p'|\bar u u + \bar s s | p> \cr
\sigma^{(2)}_{KN}(t) &=  {1\over 2}(\hat m + m_s) <p'|-\bar u u + 2 \bar d d +
\bar s s| p> \cr } \eqno(6.11)$$
with $t = (p'-p)^2$ the invariant momentum transfer squared
and $\hat m = (m_u +
m_d)/2$ the average light quark mass. At zero momentum transfer, the strange
quark contribution to the nucleon mass is given by
$$m_s <p|\bar s s |p> = \biggl({1\over 2} - {M_\pi^2 \over 4M_K^2} \biggr)
\biggl[3\sigma^{(1)}_{KN}(0)+ \sigma_{KN}^{(2)}(0) \biggr]+ \biggl({1\over 2} -
{M_K^2 \over M_\pi^2 } \biggr) \sigma_{\pi N}(0) \eqno(6.12)$$
making use of the leading order meson mass formulae $M_\pi^2 = 2 \hat m B_0 $
and $M_K^2 = (\hat m + m_s) B_0$ which are sufficiently accurate to the order
we are working. The chiral expansion at next-to-leading order for the
$\sigma$-terms reads
$$ \sigma_{\pi N}(0) = {M_\pi^2 \over 64 \pi F_p^2}
\biggl[
- 4 \alpha_N^\pi M_\pi
- 2 \alpha_N^K M_K - {4\over3} \alpha_N^\eta M_\eta \biggr]
- 2M_\pi^2 (b_D +
  b_F +   2 b_0)  \eqno(6.13a)$$
$$\eqalign{ \sigma^{(j)}_{K N}(0)= & {M_K^2 \over 64 \pi F_p^2}\biggl[- 2
\alpha_N^\pi
M_\pi -3\xi^{(j)}_K M_K - {10\over3} \alpha_N^\eta M_\eta \cr &\quad - 2
\xi^{(j)}_{\pi \eta}   \alpha_N^{\pi \eta}   {M_\pi^2 + M_\pi M_\eta +
M_\eta^2 \over M_\pi + M_\eta}
\biggr]
+ 4M_K^2 (\xi^{(j)}_D\, b_D + \xi^{(j)}_F\,
b_F - b_0) \cr} \eqno(6.13b)$$
 for $j = 1,2$ with the coefficients
$$\eqalign{\xi^{(1)}_K & = {7\over 3} D^2 - 2DF + 5F^2, \quad \xi^{(2)}_K =
3(D-F)^2, \quad \xi^{(1)}_{\pi \eta} = 1, \quad \xi^{(2)}_{\pi\eta} = -3, \cr
\xi^{(1)}_D  & = -1, \quad \xi^{(2)}_D = 0, \quad \xi^{(1)}_F = 0, \quad
\xi^{(2)}_F =  1; \quad \alpha_N^{\pi \eta} = {1 \over 3} (D+F)(3F-D).
\cr }\eqno(6.13c)$$
This completely determines the scalar sector at next-to-leading order. Note
that the $\pi N$ $\sigma$-term is given as $\sigma_{\pi N}(0) = \hat m\,
(\partial m_N/ \partial \hat m)$ according to the Feynman-Hellman theorem. The
shifts of the $\sigma$-terms from $t=0$ to the respective
Cheng-Dashen points  do  not involve any contact terms,
$$\eqalign{ \sigma_{\pi N}(2M_\pi^2) - \sigma_{\pi N}(0) = & { M_\pi^2 \over 64
\pi F_p^2}\biggl\{ {4 \over 3} \alpha_N^\pi
\, M_\pi \cr  & +  {2\over 3} \alpha_N^K
\biggl[ {M_\pi^2 - M_K^2 \over \sqrt 2 M_\pi } \ln {\sqrt 2 M_K + M_\pi \over
\sqrt 2 M_K - M_\pi} + M_K \biggr] \cr & + {4\over 9} \alpha_N^\eta
\biggl[ {M_\pi^2
- M_\eta^2 \over \sqrt 2 M_\pi} \ln {\sqrt 2 M_\eta - M_\pi \over \sqrt 2
  M_\eta - M_\pi } + M_\eta \biggr] \biggr\} \cr }
 \eqno(6.14a)$$
$$\eqalign{  \sigma&_{KN}^{(j)}(2M_K^2)  - \sigma_{KN}^{(j)}(0) = {M_K^2 \over
128 \pi F_p^2}  \biggl\{ {4 \over 3} \alpha_N^\pi
\biggl[
{M_K^2 - M_\pi^2 \over \sqrt 2 M_K} \biggl( \ln  {M_K +\sqrt 2M_\pi \over
M_K - \sqrt 2 M_\pi} + i \pi \biggr) + M_\pi \biggr] \cr
&+ {20\over 9} \alpha_N^\eta
         \biggl[ {M_K^2 - M_\eta^2 \over \sqrt 2 M_K}\ln
{\sqrt 2 M_\eta +  M_K \over \sqrt 2 M_\eta - M_K}   + M_\eta \biggr]  +
2\xi^{(j)}_K M_K\cr &+ \xi^{(j)}_{\pi \eta}
 \alpha_N^{\pi \eta}   \biggl[{2M_K^2 - M_\pi^2 - M_\eta^2 \over \sqrt 2 M_K}
 \ln  {\sqrt 2M_K +  M_\pi + M_\eta \over
M_\pi +M_\eta - \sqrt 2 M_K}
+ 2{M_\pi^2 + M_\eta^2
\over M_\pi + M_\eta}  \biggr] \biggr\} \cr} \eqno(6.14b)$$
Notice that the shifts of the two $KN$ $\sigma$-terms acquire an imaginary part
since the pion loop has a branch cut starting at $t
= 4M_\pi^2 $ which is below the kaon Cheng--Dashen
point $t = 2M_K^2$\footnote{*}{Since we choose the GMO value for the
$\eta$ mass, $M_\pi + M_\eta > {\sqrt 2} M_K$, the $\pi \eta$ loop does
not contribute to the imaginary part in eq.(6.14b). For the physical
value of the $\eta$ mass this contribution is tiny compared to the pion
loop.}.
In the limit of large kaon and eta mass the result
eq.(6.14a) agrees, evidently, with the ancient
calculation of Pagels and Pardee [6.11]
once one accounts for the numerical error of a
factor 2 in that paper. Clearly, the $\sigma$-term shifts are non-analytic in
the quark masses since they scale with the third power of the pseudoscalar
meson masses. Our strategy will be the following: We use the empirically known
baryon masses  and the recently determined value of $\sigma_{\pi N}(0)$ [6.12]
to fix the unknown parameters $m_0, b_D,b_F$ and $b_0$. This allows us to
predict the two $KN$ $\sigma$-terms $\sigma_{KN}^{(j)}(0)$. The shifts of the
$\sigma$-terms are independent of this fit.

Since we use $\sigma_{\pi N}(0)$ as input in what follows, let us briefly
review the status of this much debated quantity.
The quantity $\sigma_{\pi N} (0)$ can be calculated from the baryon spectrum.
To leading order in the quark masses, one finds
$$ \eqalign{
\sigma_{\pi N}(0) &= { \hat m \over m_s - \hat m} {m_\Xi + m_\Sigma - 2 m_N
\over 1 - y} + {\cal O}({\cal M}^{3/2}) \cr
y&= { 2<p| \bar s s | p> \over <p| \bar u u + \bar d d| p> }
\cr}
\eqno(6.15)$$
where $y$ is a measure of the strange quark content of the proton. Setting
$y=0$ as suggested by the OZI rule, one finds $\sigma_{\pi N} (0) = 26$ MeV.
However,
from the baryon mass analysis it is obvious that one has to include the
${\cal O}({\cal M}^{3/2})$ contributions and estimate the
${\cal O}({\cal M}^2 )$ ones. This was done by Gasser [6.4] leading to
$$\sigma_{\pi N}(0) = {35 \pm 5 \, {\rm MeV} \over 1 - y }
= {\sigma_0 \over 1 - y }
\eqno(6.16)$$
However, in $\pi N$ scattering one does not measure $\sigma_{\pi N}(0)$, but a
quantity called $\Sigma_{\pi N}$ defined via
$$\Sigma_{\pi N} = F_\pi^2 \bar D^+ (\nu = 0, t = 2 M_\pi^2)
\eqno(6.17)$$
with  the bar on $D$ denoting that the pseudovector Born terms have been
subtracted, $\bar D = D - D_{\rm pv}$. The amplitudes $D^\pm$ are related to
the more conventional $\pi N$ scattering amplitudes $A^\pm$ and $B^\pm$ via
$D^\pm = A^\pm + \nu B^\pm$, with $\nu = (s-u) / 4m$. The superscript '$\pm$'
denotes the isospin  (even or odd). $D$ is useful since it is related to the
total cross section via the optical theorem. The kinematical choice $\nu = 0
, t = 2 M_\pi^2$ (which lies in the unphysical region) is called the
Cheng--Dashen point [6.13]. The relation between $\Sigma_{\pi N}$
and the $\pi N$
scattering data at low energies is rather complex, see e.g. H{\"o}hler [6.14]
for a discussion or Gasser [6.15] for an instructive pictorial (given also in
ref.[6.16]). Based on dispersion theory, Koch [6.17] found $\Sigma_{\pi N}
 = 64 \pm 8$
MeV (notice that the error only reflects the uncertainty of the method, not
the one of the underlying data). Gasser et al. [6.12] have recently repeated
this analysis and found $\Sigma_{\pi N} = 60$ MeV
(for a discussion of the errors,
see that paper). There is still some debate about this value, but in what
follows we will use the central result of ref.[6.12].  Finally, we have to
relate $\sigma_{\pi N}(0)$ and $\Sigma_{\pi N}$. The relation of these two
 quantities is based
on the LET of Brown et al. [6.18] and takes the following form at the
Cheng--Dashen point:
$$\eqalign{
\Sigma_{\pi N}&= \sigma_{\pi N} (0) + \Delta \sigma_{\pi N} + \Delta R \cr
\Delta \sigma_{\pi N}&= \sigma_{\pi N}(2 M_\pi^2) - \sigma_{\pi N}(0) \cr}
\eqno(6.18)$$
$\Delta \sigma_{\pi N}$ is the shift due to the
scalar form factor of the nucleon,
and $\Delta R$ is related to a remainder not fixed by
chiral symmetry. The latter was found to be very small by GSS [6.19], $\Delta
R = 0.4$ MeV.  In this case, one is dealing with
strong $S$--wave $\pi \pi$ and $\pi N$ interactions.
Therefore,  the suspicion arises
that the one--loop approximation is not sufficient to give an accurate
description of the scalar form factor (compare e.g. ref.[6.20]).
 Therefore, Gasser et al. [6.12] have performed a
dispersion--theoretical analysis tied together with CHPT constraints for the
scalar form factor $\sigma_{\pi N} (t)$. The resulting numerical value is
$$\Delta \sigma_{\pi N} = ( 15 \pm 0.5 ) \, {\rm MeV}        \eqno(6.19)$$
which is a stunningly large correction to the one--loop result (see below).
If one parametrizes the scalar form factor as
$\sigma_{\pi N} (t) = 1 + <r_S^2> \, t + {\cal
O} (t^2)$, this leads to $<r_S^2> = 1.6$ fm$^2$, substantially larger than the
typical electromagnetic size. This means that the scalar operator
$\hat m (\bar u u + \bar d d )$ sees a
more extended pion cloud. Notice that for the pion, a similar enhancement of
the scalar radius was already observed, $(r_S^2 / r^2_{\rm em})_\pi \simeq
1.4$ [6.20]. Putting pieces together, one ends up with $\sigma_{\pi N}(0)
 = 45 \pm 8$ MeV
[6.12] to be compared with $\sigma_0 / (1-y) = ( 35 \pm 5 ) \, {\rm MeV} /
(1-y)$. This means that the strange quarks contribute approximately 10 MeV to
the $\pi N$ $\sigma$--term and thus the mass shift induced by the
strange quarks is
$m_s <p| \bar s s |p> \simeq (m_s / 2 \hat m ) \cdot 10 \, {\rm MeV} \simeq
130$ MeV. This is consistent  with the estimate made in ref.[6.21] based on
the heavy mass formalism including the decuplet fields. The effect of the
strange quarks is not dramatic. All speculations starting from the first
order formula (6.15) should thus be laid at rest. The lesson to be learned
here is that many small effects can add up constructively to explain a
seemingly large discrepancy like $\Sigma_{\pi N} - \sigma_0 \approx \sigma_0$.
What is clearly needed are more accurate and reliable low--energy pion--nucleon
scattering data to further pin down the uncertainties.

We now return to the order $q^3$ calculation in CHPT.
We use an average value $F_p = (F_\pi + F_K)/2\simeq 100$
MeV, together with $F=0.5$ and $D=0.75$, which leads to $g_A = 1.25$. The
uncertainties in these parameters and how they affect the results are discussed
in [6.10].
The four unknowns, which are the three low-energy constants $b_D,b_F$ and $b_0$
and the average octet mass (in the chiral limit) $m_0$ are obtained from a
least square fit to the physical baryon masses ($N,\Sigma,\Lambda,\Xi$) and the
value of $\sigma_{\pi N}(0) \simeq 45$ MeV. This allows to predict
$\sigma^{(1)}_{KN}(0)$ and $\sigma^{(2)}_{KN}(0)$ and the much discussed matrix
element $m_s<p|\bar ss |p>$, $i.e.$ the contribution of the strange quarks to
the nucleon mass. Typical results are [6.10]: (a)
The strangeness matrix element
in most cases is negative and of reasonable magnitude of about 200 MeV,
(b) within the accuracy of the calculation, the $KN$ $\sigma$-terms
turn out to be
$$\sigma^{(1)}_{KN}(0) \simeq 200 \pm 50 \,{\rm MeV} \, , \quad
\sigma^{(2)}_{KN}(0) \simeq 140 \pm 40 \,{\rm MeV} \eqno(6.20)$$
which is comparable to the first order perturbation theory analysis having no
strange quarks, $\sigma^{(1)}_{KN}(0) = 205$ MeV and $\sigma^{(2)}_{KN}(0) =
63$ MeV [6.22]. These results are indeed most sensitive to
the value of $\sigma_{\pi N}(0)$.
The $\sigma$-term shifts are given by [6.11,6.23]
 $$\sigma_{\pi N}(2M_\pi^2 ) -\sigma_{\pi N}(0) = 7.4\,{\rm MeV}
\eqno(6.21)$$
which is half of the empirical value discussed above, eq.(6.19).
 Furthermore, one finds
$$\eqalign{ \sigma^{(1)}_{KN}(2M_K^2) - \sigma^{(1)}_{KN}(0) & = (271 + i \,
303)\,{\rm MeV}\cr  \sigma^{(2)}_{KN}(2M_K^2) - \sigma^{(2)}_{KN}(0) & = ( 21 +
i \,   303)\,{\rm MeV}\cr } \eqno(6.22)$$
The real part of the first $\sigma$--term
 can be estimated simply via Re$(\sigma_{KN}^{(1)}(2M_K^2) -
\sigma_{KN}^{(1)}(0))
\simeq [\sigma_{\pi N}(2M_\pi^2) - \sigma_{\pi
N}(0)](M_K/M_\pi)^3 = 7.4 \cdot 46.2 $ MeV = 340 MeV. The rather small
real part in $\Delta \sigma_{KN}^{(2)}$ stems from the large negative
contribution of the $\pi \eta$--loop which leads to strong cancellations.
Notice the large
imaginary parts in $\sigma^{(j)}_{KN}(2M_K^2)-\sigma^{(j)}_{KN}(0) $ due to the
two--pion cut. The situation concerning the empirical determinations of the
kaon--nucleon $\sigma$--terms is rather uncertain, see e.g. refs.[6.24,6.25].
Since most of the phase shift data stem from kaon--nucleus scattering, it is
advantegeous to define the $KN$ $\sigma$--terms by means of the nuclear
isospin,
$$\sigma_{KN}' = {1 \over 4} \bigl( 3 \sigma_{KN}^{(2)} + \sigma_{KN}^{(1)}
\bigr) \, , \quad
\sigma_{KN}'' = {1 \over 2} \bigl( \sigma_{KN}^{(2)} - \sigma_{KN}^{(1)}
\bigr) \, . \eqno(6.23)$$
The best available determinations give $\sigma_{KN}'(0) = 599 \pm 377$ MeV and
$\sigma_{KN}''(0) = 87 \pm 66$ MeV which translates into
$\sigma_{KN}^{(1)}(0) = 469 \pm 390$  MeV and $\sigma_{KN}^{(2)}(0) = 643 \pm
 378 $ MeV.
\medskip
As has been argued e.g. in refs.[6.3,6.7,6.8], one should account for the
spin-3/2 decuplet in the EFT since it leads to a natural
 cancellation of the large
kaon cloud contributions of the type $m_s M_K^2 \ln M_K^2$.
However, as shown in
section 3.4, the inclusion of these fields spoils the power counting much like
the nucleon mass term in the relativistic formulation of baryon CHPT. For the
case at hand, one also has an infinite renormalization of the three constants
$b_D$, $b_F$ and $b_0$ [6.10],
$$b_D = b_D^r (\lambda) - {\Delta {\cal C}^2 \over 2 F_p^2} \, L \, , \quad
b_F = b_F^r (\lambda) + {5 \Delta {\cal C}^2 \over 12 F_p^2} \, L \, , \quad
b_0 = b_0^r (\lambda) + {7 \Delta {\cal C}^2 \over 6 F_p^2} \, L \, .
\eqno(6.24)$$
with $\lambda$ the scale of renormalization, $\Delta = 231 \, {\rm MeV}$
 the average octet--decuplet mass splitting,
${\cal C}$ the Goldstone--octet--decuplet coupling as discussed after (3.71)
and $L$ is defined in (2.47). The
graphs with intermediate decuplet--states start to contribute at order $q^4$
(explicit formulae can e.g. be found in ref.[6.10]). If one now assumes that
these contributions dominate at this order, one does not find a consistent
description of the scalar sector, as long as one keeps $F$ and $D$ close to
their physical values. In that case, the $KN$ $\sigma$--terms turn out to be
very small and the strange matrix--element $m_s <p|\bar s s|p>$ very large.
In ref.[6.7], some additional tadpole diagrams with one insertion from ${\cal
L}^{(2)}_{\rm MB}$ were considered, but that does not alter these conclusions.
As stressed before, a complete and systematic $q^4$ calculation has to be
performed before one can draw a definite conclusion on the role of the
intermediate decuplet states. For another critical discussion, see e.g.
ref.[6.26]. However, there is one curious result in the two--flavor sector
we would like to mention [6.10,6.27]. If one considers the shift of the
$\pi N$ $\sigma$--term calculated with intermediate nucleon and $\Delta (1232)$
states and uses the large $N_c$ coupling constant relation, one finds
$$\eqalign{\sigma_{\pi N} (2M_\pi^2) - \sigma_{\pi N} (0) & = {3 g_A^2 M_\pi^2
\over 64 \pi^2 F_\pi^2} \biggl\lbrace \pi M_\pi + (\pi -4 ) \Delta - 4
\sqrt{\Delta^2 - M_\pi^2} \ln \bigl( {\Delta \over M_\pi} +
 \sqrt{{\Delta \over M_\pi} -1} \bigr) \cr
& + {8 \Delta^2 \over M_\pi} \int_\Delta^\infty {dy \over \sqrt{2y^2 + M_\pi^2
- 2\Delta^2}} \arctan {M_\pi \over \sqrt{2y^2 + M_\pi^2 - 2\Delta^2}}
\biggr\rbrace \cr & = 14 \, \, {\rm MeV} \cr}   \eqno(6.25)$$
very close to the empirical value (6.19) (with $\Delta = m_\Delta -m =
293$ MeV).  This means that the graph with the
intermediate $\Delta (1232)$ contributes almost
as much as the lowest order diagram
with a nucleon as intermediate state. However, the spectral distribution ${\rm
Im} \, \sigma_{\pi N} (t) /t^2$ is much less pronounced around $t = 4 M_\pi^2$
as in the analysis of ref.[6.12] but has a longer tail leading to the same
result for the integral. The $\Delta$--contribution mocks up the higher loop
corrections of the dispersive analysis [6.12].
This is similar to the discussion
of the spectral function related to the intermediate range attraction in the
nucleon--nucleon interaction (cf. section 5.2). It remains to be seen how the
result (6.25) will be affected when all $q^4$ (and possibly higher order) terms
will be accounted for. The essential difference to the baryon mass calculation
and the shifts of the $KN$ $\sigma$--terms is that the kaon and eta
contributions to (6.25) are essentially negligible
(they contribute approximately one extra MeV to (6.25)),
i.e. we are dealing with an SU(2) statement so that one does
not expect higher loop diagrams to contribute significantly.
\bigskip
\noindent{\bf VI.2. KAON--NUCLEON SCATTERING}
\bigskip
A topic of current interest is the dynamics of the kaon--nucleon system based
on SU(3) extensions of chiral effective Lagrangians. Such investigation were in
particular triggered by Kaplan and Nelson [6.28]
who proposed a mechanism for kaon
condensation in dense nuclear matter using a (however incomplete) chiral
Lagrangian. Besides this, kaon--nucleon scattering at
low energies is of its own interest as a testing ground for three-flavor
 chiral dynamics. In comparison to the SU(2) sector of pion--nucleon
interaction (discussed to some extend in sect.III.5) the kaon--nucleon dynamics
involves several  complications. First, the pertinent expansion
parameter for the chiral expansion is much larger, namely
$${M_K \over m} \simeq 0.53 \eqno(6.26)$$
in comparison to $\mu = M_\pi/m \simeq 0.14$ for the $\pi N$ system. Therefore
one expects the next-to-leading order corrections to be numerically less
suppressed in  comparison to the leading terms. The $KN$ system with
strangeness $S= +1$ is physically still quite simple at low energies since it
is a purely elastic scattering process with a dominant S--wave contribution.
The analogous $\bar K N$ system with strangeness  $S=-1$  however greatly
differs, mainly because there are a number of baryons and
baryon resonances with $S=-1$ but none with $S=+1$. For the $K^- p $ reaction
there exist inelastic channels down to threshold involving a pion and a
hyperon, $K^- p \to \pi^0 \Lambda, \, \pi^\pm \Sigma^\mp, \pi^0 \Sigma^0$.
Furthermore, there is a subthreshold resonance in the $K^- p$ system, the
$\Lambda(1405)$ of isospin zero. It may be considered as a kind of a
$K^- p$  bound state which can decay into the kinematically
open $\pi \Sigma$ channel and
thus receives its physical width. The dynamical differences in $K N$ versus
$\bar K N$ naturally show up in the values of the corresponding $S$-wave
scattering lengths [6.29]. The
experimental numbers stem from data on kaon--nucleon scattering and
$K^-$-atomic level shifts and we give them here without error bars,
$$\eqalign{ a^{K^+ p} &  = - 0.31 \, {\rm fm} , \qquad \qquad \qquad
\, \, a^{K^+ n}  = - 0.20 \,
{\rm fm},  \cr  a^{K^- p} & = (- 0.67+ i\, 0.63) \, {\rm fm},
 \qquad a^{K^- n}  =
(+ 0.37+ i \, 0.57) \, {\rm fm} \, .\cr}   \eqno(6.27)$$
The experimental values for $K^+N$ are comparable in magnitude to $\pi N$
scattering lengths, $a^{\pi^+ p} = a_{3/2} = a^+ - a^- \simeq -0.14 \, $fm.
The negative signs of the $a^{K^+ N} $ signal that
the $K^+$-nucleon interaction is
repulsive. Characteristic for the $K^- N$ scattering lengths is their very
large imaginary part, which originates from the open inelastic $\pi \Sigma, \pi
\Lambda$ channels  and the subthreshold $\Lambda(1405)$-resonance.

Such inelastic channels are not a problem for CHPT of kaon--nucleon
interaction, since they reflect themselves simply as unitarity cuts in the
amplitudes which extend below the physical threshold. They are mainly of
kinematical origin. On the other hand, the existence of a strong subthreshold
resonance (like $\Lambda(1405)$ in $K^- p$) poses a problem to the
expansion scheme of chiral perturbation theory, since bound states and
(subthreshold) resonances can not be obtained at any finite order of
perturbation theory. They require infinitely many orders and are out of the
scope of pertubation theory. Consequently, the $\Lambda(1405)$ will have to be
added by hand (compare the discussion concerning the decuplet states
in the EFT in section 3.4). In ref.[6.30] a
model has been proposed to generate the $\Lambda(1405)$ dynamically. For that
one solves a Schr\"odinger equation for the coupled $\bar K N, \, \pi \Sigma$
and $\pi \Lambda$ channels with local or separable meson-baryon potentials
linked to an $SU(3)$ chiral Lagrangian. The latter means that the relative
strengths of the transition potentials are fixed by the
Clebsch--Gordan coefficients of a chiral
meson-baryon vertex.  A few finite range parameters for the potentials and a
coupling stength are then adjusted to the mass and width of the $\Lambda(1405)$
and several measured branching ratios.  It turns out that such a few parameter
fit is quite sucessful in describing the low--energy $K^- N$ data. We will
not pursue this approach further here, but outline some aspects of $K N$ and
$\bar K N$ scattering in CHPT.

The leading order Lagrangian is a straightforward generalization of the chiral
$\pi N$-Lagrangian of sect.III to SU(3) as described in section 6.1.
To lowest order, all octet baryon masses equal.
The baryon mass splittings are due to higher
orders in the quark mass expansion. The corresponding
${\cal L}^{(1)}_{M B}$ is given in (6.6) for the heavy fermion
approach. From that, one finds for the  S--wave scattering lengths:
$$ a^{K^+ p} = - \bigl( 1 + {M_K\over m} \bigr)^{-1} {M_K \over 4 \pi F_p^2} =
2 a^{K^+ n} \simeq - 0.6 \, {\rm fm} \eqno(6.28)$$
This current algebra result has the correct negative sign and the order of
magnitude is reasonable. It is about a factor 2, respectively 1.5, too large
for $K^+ p$ and $K^+ n$ if we use $F_p = F_\pi = 93$ MeV. However, there is
quite some theoretical uncertainty in this leading order result. According
to the chiral power counting the prefactor $1/(1+M_K/m)$ could be neglected and
$F_p$ could be taken to be $F_K = 1.22 \, F_\pi$. Such ambiguities point
towards the importance of higher order calculations, at least up to
 order $q^3$.
The corresponding $K^- N$ scattering lengths have the opposite sign of the $K^+
N$  ones to leading order, i.e. the chiral $K^- N$ interaction is attractive.
This feature was considered as quite important for understanding the dynamics
of the $\Lambda(1405)$-resonance formation in ref.[6.30].

At next-to-leading order, ${\cal O}(q^2$), the SU(3)
chiral Lagrangian contains a host
of new terms. We display here only those which contribute to the
S--waves [6.10,6.31]
$$\eqalign{{\cal L}^{(2)}_{M B} = & b_d \Tr(\bar B \{ \chi_+,B\})+b_f
\Tr(\bar B [\chi_+,B])+b_0 \Tr(\bar B B) \Tr (\chi_+) \cr & + d_1 \Tr(\bar B
u_\mu u^\mu B) + d_2 \Tr(\bar B v\cdot u v \cdot u B) + d_3 \Tr(\bar B B u_\mu
u^\mu) + d_4 \Tr(\bar B B v \cdot u v\cdot u)\cr & +d_5 \Tr(\bar BB) \Tr( u_\mu
u^\mu) + d_6 \Tr(\bar BB) \Tr( v \cdot u v \cdot u) + d_7 \Tr(\bar B
u_\mu) \cdot \Tr( u_\mu B)\cr &+ d_8 \Tr(\bar B v \cdot u) \Tr(v \cdot u B) +
d_9 \Tr(\bar B u_\mu B u^\mu ) + d_{10} \Tr(\bar B v \cdot u B v\cdot u) +
\dots \cr } \eqno(6.29)$$
where the first three terms obviously coincide with the ones given in eq.(6.7).
In ref.[6.32] the last two terms have been forgotten.
The complete list of terms at order
$q^2$ and $q^3$  for flavor-SU(3) can be found in ref.[6.2] (for the
relativistic approach).
There are 13 new parameters for chiral SU(3) in comparison
to 3 ($c_1, c_2, c_3$) in flavor SU(2) and some of the $d_i$ contain $1/m$
corrections from the  expansion  of the relativistic leading order Lagrangian
formulated in terms of Dirac-fields. The coefficients of the first three terms
in eq.(6.29) (often named "sigma--terms") can be fixed at this order from the
mass splittings in the baryon octet and the empirical value of the $\pi N$
$\sigma$-term,
$$\eqalign{ & m_\Sigma - m_\Lambda = {16 \over 3 } b_d (M_K^2 - M_\pi^2) =
79 \, {\rm MeV} , \qquad b_d = 0.066 \, {\rm GeV}^{-1} \cr & m_\Xi - m_N = 8
b_f (M_K^2 - M_\pi^2) = 383 \, {\rm MeV} , \qquad b_f = -0.213 \, {\rm
GeV}^{-1} \cr & \sigma_{\pi N}(0) = - 2 M_\pi^2 ( b_d + b_f + 2b_0) = 45 \,
{\rm MeV} , \qquad  b_0 = - 0.517 {\rm GeV}^{-1} \cr } \eqno(6.30)$$
Of course such a fit is somewhat problematic, since one neglects all
higher order in the quark masses, compare the discussion after eq.(6.15).
Consequently, the strangeness content of the proton
$$y  = {2(b_0 +b_d-b_f)
\over 2 b_0 +b_d +b_f} \simeq  0.4 \eqno(6.31)$$
is about twice the value obtained by Gasser, Leutwyler and Sainio
[6.12].  If one
enforces, however, $y \simeq 0$, which is possible
due to the uncertainties going into
the theoretical analysis of $y$, one finds $b_0 = -0.279$ GeV$^{-1}$. This
corresponds to $\sigma_{\pi N}(0) = 26 $ MeV, the usual estimate at linear
order in the quark masses.  The $K^+ N$ and $K^- N$ scattering lengths are now
given as [6.31]:
$$\eqalign{a^{K^\pm p} &  = \bigl( 1 + {M_K\over m} \bigr)^{-1} {M_K \over 4
\pi F_p^2} \biggl[ \mp 1 + M_K(D_s + D_v)\biggr] \cr
a^{K^\pm n} &  = \bigl( 1 +
{M_K\over m} \bigr)^{-1} {M_K \over 4 \pi F_p^2} \biggl[\mp {1\over 2}  +
M_K(D_s - D_v)\biggr] \cr} \eqno(6.32)$$
with $D_s $ and $D_v$ some linear combination of the $b_{d,f,0}$ and
$d_{1,\ldots,10}$. Of course, as long as one has not found a reliable way to
estimate all new coefficients, the expressions given above have not much
predictive  power. A similar situation occurs for the isospin even $\pi N$
scattering length $a^+$. At this order, the $K^- N$ scattering lengths are
still real, since the scattering into the inelastic channels is a loop effect
which first shows up at order $q^3$.

In ref.[6.32] an order $q^3$ calculation for the $K^\pm N$ scattering
lengths has been presented. In this work, however, the loop contribution has
not been separated  cleanly from the counterterms at the same order. The
knowledge of the magnitude of the full loop
 correction (say at a scale $\lambda \simeq$ 1 GeV)
would however be very important in order to get a feeling for the genuine size
of the (non-analytic) corrections in SU(3).
We remind here that for the isospin odd
$\pi N$ scattering length $a^-$ the loop correction at order $M_\pi^3$ just
has the right sign and magnitude to fill the gap between the Weinberg-Tomozawa
prediction and the empirical value. The $K^- N$ scattering lengths given in
ref.[6.32] are still real at order $q^3$. However, at this order the
rescattering processes $K^- N \to \pi \Sigma , \pi \Lambda \to K^- N$ into the
inelastic channels  are possible and they manifest themselves as complex valued
scattering lengths, cf. fig.6.1.
\midinsert
\smallskip
\hskip 1in
\epsfxsize=2in
\epsfysize=1in
\epsfbox{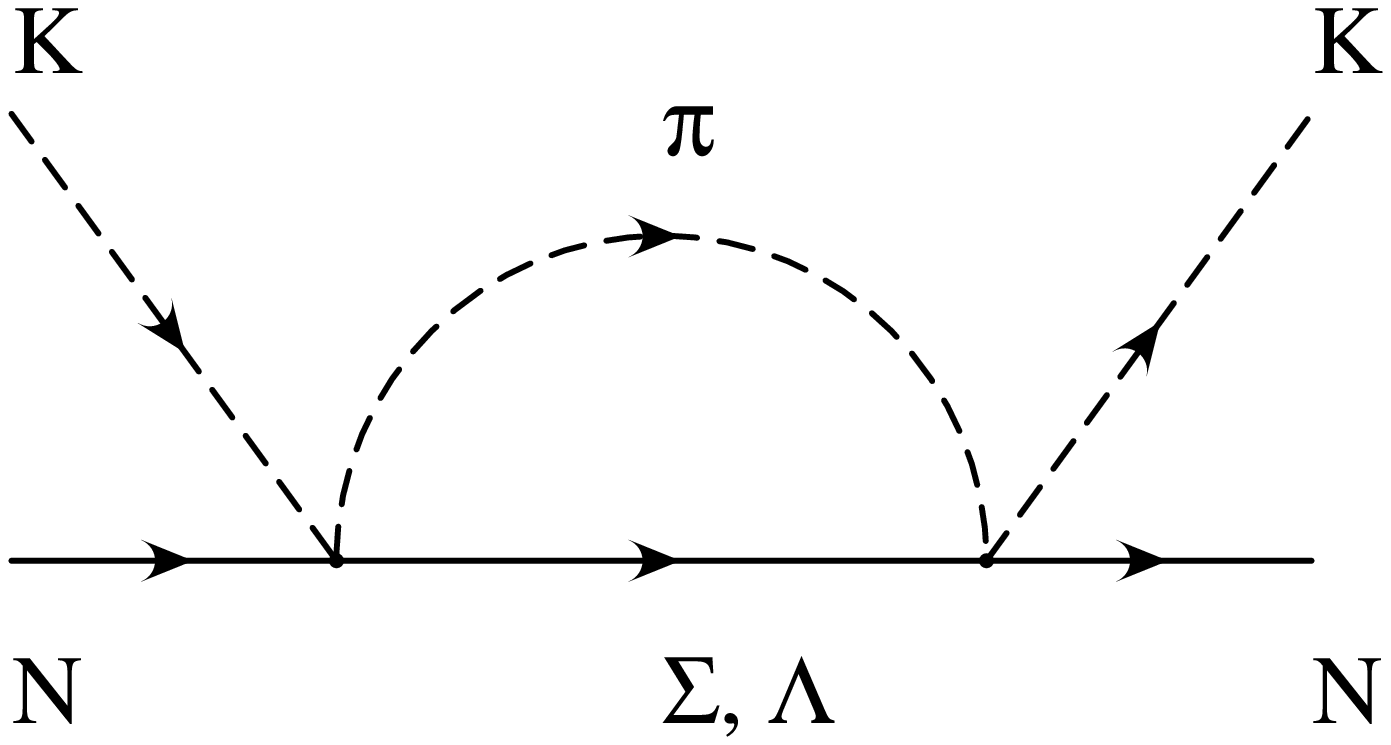}
\smallskip
{\noindent\narrower \it Fig.~6.1:\quad
Rescattering diagram which leads to the imaginary part of the
scattering lengths at order $q^3$ for $K^- N \to K^- N$.
\smallskip}
\vskip -0.5truecm
\endinsert

In the heavy mass limit one calculates the
following nonzero imaginary parts from the rescattering diagrams
$$ {\rm Im} \, a^{K^- p} = \bigl( 1 + {M_K \over m} \bigr)^{-1} { M_K^2 \sqrt
{M_K^2 - M_\pi^2} \over 32 \pi^2 F_p^4} = {4 \over 5} \, {\rm Im } \, a^{K^- n}
\simeq 0.63 \, {\rm fm} \eqno(6.33)$$
which are surprisingly close to the empirical values. However, one should not
put to much emphasis on these numbers since all mass splittings in the baryon
octet have been neglected and this affects the available phase
space. It is interesting to observe that
such a simple rescattering calculation tends to explain the near
equality of the imaginary parts for proton and neutron $K^-$-scattering and at
least gives the correct order of magnitude without an explicit $\Lambda(1405)$.

Clearly, all what has been discussed here points towards the importance of
more systematic
calculations using the complete chiral Lagrangian at a given order. Many of the
present controversies in the literature (in particular concering the in-medium
behavior of pions and kaons) stem from the use of incomplete Lagrangians. It is
also clear that baryon CHPT for flavor SU(3) is just at its beginning and a lot
more work (complete higher order calculations) is necessary in order to judge
the quality of such an approach.
\bigskip
\noindent {\bf VI.3. THE PION IN MATTER}
\medskip
In this chapter we will describe how effective chiral Lagrangians can be used
to get information on the modification of pionic properties in nuclear
matter (so-called medium modifications). The medium modifications of hadron
properties are relevant for a broad class of problems in nuclear physics. Among
these are pion and kaon condensation (in neutron stars), chiral symmetry
restoration in relativistic heavy ion collisions, the "dropping" of hadron
masses in medium, to name a few.
 In the following we will only touch upon part of these
many issues, namely the density dependence of the quark condensate $<\bar u
u >$ and the density dependence of the pion decay constant $F_\pi$ and of the
pion mass $M_\pi$. We will make use of chiral effective Lagrangian techniques
and show that such a method indeed leads to the correct linear terms in
density. The latter are often called "low-density theorems" and can be derived
from a multiple scattering expansion. We follow here closely the ideas
spelled out in ref.[6.33].

If one remembers the additional complications one encounters in each step in
extending chiral perturbation theory from the pure meson sector
(section 2) to
single baryon processes (sections 3 and 4) and further to the $B=2$
(here, $B$ denotes the baryon number)
sector of nucleon-nucleon interaction and exchange currents (section 5)
it is not
surprising that a rigorous formulation of a systematic chiral expansion in
nuclear matter ($i.e.$ at finite baryon density) has not yet been found. Finite
baryon density introduces a new scale parameter, the Fermi momentum of the
nucleons $k_F \sim \rho^{1/3}$ and it is not clear how to deal with it in the
chiral power counting scheme. Furthermore, Lorentz invariance is
broken at finite
density and the effects of nuclear correlations have to be considered. A
rigorous (and still predictive) expansion scheme which can account for all
of these  many-body complexities as well as the chiral structure of QCD has
not yet been formulated and may even be too demanding. A recent
approach due to Shankar [6.34] appears promising but needs further
detailed study. For a general approach to non--relativistic effective
theories, see Leutwyler [6.35].

In a first step, following ref.[6.33] one can simply use the free
space chiral
Lagrangian for the $B=0$ and $B=1$ sectors developped so far and evaluate the
pertinent nucleon operators at the mean field level. Consequently, one works to
linear order in the nuclear matter density,
$$\rho \simeq \bar H H \simeq H^\dagger H \, \, . \eqno(6.34)$$
Formally, such a mean field approximation means that any local
term in the effective $\pi N$ Lagrangian of the  form $\bar H(x) O(x)  H(x)$ is
replaced by ${1\over 2} \Tr[O(x)] \cdot \rho$. The averaging trace here
goes over both spin and isospin coordinates since we consider only a
homogeneous, isospin symmetric and (spin-)unpolarized nuclear matter
distribution of density $\rho$. Of course, such a mean field approximation may
have no rigorous foundation, but intuitively it should be reasonable at least
for low densities. Furthermore, since its starting point is the most general
effective chiral Lagrangian to a given order the information gained this way is
 more general than model calculations of the in--medium properties.

Let us now apply the simple mean field approximation to the effective chiral
Lagrangian ${\cal L}^{(2)}_{\pi\pi} + {\cal L}^{(1)}_{\pi N} + {\cal
L}^{(2)}_{\pi N} $. The averaging procedure going along with the mean field
approximation used to decribe  spin and isospin symmetric nuclear matter
makes the piece ${\cal L}^{(1)}_{\pi N}$ vanishing identically since $i v\cdot
\partial = \partial_0$ gives zero and the free term $- m \, \rho$ is absent
in the heavy mass formulation. The other terms in ${\cal L}^{(1)}_{\pi N}$
 vanish since the isospin
traces $\Tr \Gamma_\mu, \, \Tr u_\mu$ are zero by construction. So we have to
consider only ${\cal L}^{(2)}_{\pi N}$ and here not all terms vanish in the
mean field approximation. These are the ones proportional to $c_1, c_2 -
g_A^2/8m, c_3$ which are of scalar--isoscalar nature. One obtains the following
"finite density chiral Lagrangian" (in the isospin limit $m_u =m_d$)
$${\cal L}(\rho) = \biggl({F_\pi^2 \over 4} + {c_3\over 2} \rho \biggl)\,
\Tr(u_\mu u^\mu) + \biggl({c_2 \over 2} - {g_A^2\over 16m} \biggl) \rho \,
\Tr(v\cdot u v\cdot u) + \biggl({F_\pi^2 \over 4} + c_1 \,\rho \biggl)
\Tr(\chi_+) \eqno(6.35)$$
The terms coming from $[(v\cdot D)^2 - D\cdot D]/2m$ have been neglected.
They either represent nucleon kinetic energies or as the terms $\Tr(\Gamma_\mu
\Gamma^\mu) , \Tr(v\cdot\Gamma \,v\cdot \Gamma)$ start out at order $\pi^4$ in
the expansion in powers of the pion mass, which are not of interest here.
The form of eq.(6.35) is very illustrative  when compared with the free space
Lagrangian ${\cal L}^{(2)}_{\pi \pi}$. The two parameters $F \simeq F_\pi$ and
$B_0 = -<\bar u u >/F_\pi^2$ have become density dependent through the mean
field approximation of the nucleons. One can immediately read off the
corresponding medium modifications. From the last term
in eq.(6.35) we get the density dependence of quark condensate (in the absence
of pions $\chi_+ $ is proportional to quark mass times quark condensate),
$${<\bar u u>(\rho) \over <\bar u u>(0) } = 1 + {4 c_1 \over F_\pi^2} \, \rho =
1 - {\sigma_{\pi N}(0) \over F_\pi^2 M_\pi^2} \, \rho  \eqno(6.36)$$
where we used the leading order relation to the $\pi N$ sigma term $
\sigma_{\pi N}(0) = - 4 c_1 M_\pi^2 $ (see eq.(3.54)). The result eq.(6.36) for
the linear term of the density
 dependence of the quark condensate has been derived
by several authors using
quite different methods [6.36,6.37] and is often called
``low -density theorem''. It
was also  found in calculations using the Nambu-Jona-Lasinio model,
see e.g. ref.[6.38]. It it quite
interesting that the simple mean field approximation to the effective chiral
$\pi N$ Lagrangian very naturally leads to this general result. This gives some
confidence in the approximations one is using. Putting in the empirical value
of $\sigma_{\pi
N}(0) = 45 $ MeV one finds a 30 $\%$ reduction of the quark condensate at
normal nuclear matter density, giving strong hints that the chiral restoration
will take place at about several times nuclear matter density. This is an
important issue for relativistic heavy ion collisions, where one hopes
to reach
such high densities. Of course, in order to make a more quantitative statement
about the actual chiral restoration density one has to know corrections
to eq.(6.36) at higher order in the density $\rho$. There is a certain
 similarity to
the calculation of the temperature dependence of the quark condensate
in CHPT, it allows for an accurate calculations at low $T$ but can not
be used to calculate the critical temperature since then the higher
order corrections have to completely cancel the leading term, i.e. one
has no longer a controlled expansion [6.39].
\medskip
The first and second term in eq.(6.35) take the form of a
 density dependent pion
kinetic term. As finite density breaks Lorentz invariance the time and spatial
components of the pionic gradients are treated now differently. One sees that
at finite density the pion decay constant splits up into a "time component"
$F_t(\rho)$ and a "spatial component" $F_s(\rho)$ which behave differently.
They are given as
$$\eqalign{
F^2_t(\rho) = &  F_\pi^2 + \bigl( 2 c_2 + 2 c_3 - {g_A^2 \over 4m}) \, \rho \cr
F^2_s(\rho) = &  F_\pi^2 +  2 c_3  \, \rho \cr} \eqno(6.37)$$
The phenomenon that the breakdown of Lorentz invariance leads to
different  time and space components
of the pion decay constant has also been discussed
in some model calculations of pion properties [6.40].
 Using the most general effective chiral Lagrangian at
order $q^2$ and the simple mean field approximation to describe
density such a behaviour, i.e. $F_t \ne F_s$,
follows very naturally from the underlying chiral structure.
Therefore one should take care of this possibility in model calculations of the
in-medium effects of the pion.
\medskip
Of particular interest is the density dependence of the pion mass because of
the pions Goldstone boson nature. The inverse
pion propagator is
 $$D^{-1}(\omega, \vec q; \rho) = \omega^2 - {\vec q \,}^2 -
\Pi(\omega, \vec q; \rho) \, \, ,   \eqno(6.38)$$
with $\Pi$ the self--energy correction due to the
interaction with the medium. Performing an expansion of eq.(6.35) to quadratic
order in the pion field one finds
$$D^{-1}(\omega, \vec q; \rho) = \bigl(1+{2c_3\over F_\pi^2}\rho\bigr)
\,(\omega^2 - {\vec q \,}^2) +{\rho\over F_\pi^2}\bigl( 2c_2- {g_A^2\over 4m}
\bigr)\,\omega^2 - \bigl(1+{4c_1\over F_\pi^2}\rho\bigr) M_\pi^2 \, \, .
\eqno(6.39)$$
Evaluating the poles of this propagator one finds for the effective pion mass
$M^{*2}_\pi(\rho) = \omega^2(\vec q =0; \rho) $
$$\eqalign{
M_\pi^{*2}(\rho) & = M_\pi^2 \bigl(1+{4c_1\over F_\pi^2} \rho
\bigr)\bigl[1+{\rho\over F_\pi^2}\bigl( 2c_2 +2c_3- {g_A^2\over 4m}
\bigr)\bigr]^{-1}   \cr
& = M_\pi^2 \bigl[1-{\rho\over F_\pi^2}\bigl( 2c_2 +2c_3-4c_1-
{g_A^2\over 4m} \bigr)\bigr] \cr & = M_\pi^2 - T^+(M_\pi) \cdot \rho = M_\pi^2
- 4 \pi \bigl( 1+ {M_\pi \over m}\bigr) \, a^+ \cdot \rho \cr} \eqno(6.40)$$
with $a^+$ the isospin even $\pi N$ scattering length calculated to lowest
order (cf. eq.(3.66) without the loop contribution $\sim M_\pi^3$).
In ref.[6.41] it was emphasized that the linear term in the density
dependence of the pion mass is
proportional to the isoscalar $\pi N$ scattering length $a^+$.
This fact is a rigorous result from the leading order of a multiple
scattering expansion. The correct coefficient proportional to $a^+$ is indeed
reproduced here using the complete chiral $\pi N$ Lagrangian at order $q^2$ and
the mean field approximation. The statement in ref.[6.41] that the chiral
Lagrangian techniques
can not give this result is therefore wrong. The argumentation of
ref.[6.41]
was based on an incomplete chiral Lagrangian which consists only of the term
proportional to $c_1$ (the "sigma-term"). We would like to stress here again
that the {\it complete} chiral Lagrangian
up to a given order has to be used to automatically
produce the correct results modulo corrections of higher order.
Since the value of $a^+ \simeq -0.01 \, M_\pi^{-1}$ is negative one finds that
the pion mass slightly increases with density. Of course, this statement is
based exclusively on the knowledge of the very small linear term in density and
could be modified by higher orders, ${\cal O}(\rho^2)$ and so on. In
the absence of calculations including higher
orders in the density one even has no control on the range of validity of the
linear density approximation, $i.e.$ to what fraction of nuclear matter density
it is reliable.
\medskip
As a final issue let us address the validity of
the Gell-Mann-Oakes-Renner (GMOR)
relation at finite density. It is often assumed or found to hold in model
calculations, see e.g. ref.[6.42]. At
 zero density, the GMOR relation is well founded in chiral
perturbation theory and takes the form, $F_\pi^2 M_\pi^2 = - \hat m \, <\bar u
 u + \bar d d> + {\cal O}(\hat m^2)$. An extension to finite densities is not
immediately obvious since there is not just a single density dependent pion
decay constant and, also, other quantities are density--dependent.
 The combination of eqs.(6.36,6.37,6.40) yields an
in-medium  version of
the GMOR relation to linear order in the density $\rho$, but only if
one replaces
the free space pion decay constant $F_\pi$ by its density dependent time
component $F_t(\rho)$,
$$F_t^2(\rho) \, M_\pi^{*2}(\rho) = - \hat m <\bar u u + \bar d d> (\rho)
\eqno(6.41)$$
This relation only  holds  modulo corrections which are of higher
 order in the light quark masses and the
density. It is important to note that for the spatial
component $F_s(\rho) $
the in-medium GMOR relation does not hold. In ref.[6.33] functional
methods have been used to show that the in-medium properties of the pion  up to
linear order in the density do not depend on the actual choice of
the interpolating
pion field. This feature is of course quite important in order to make the
concept of density dependent mass, decay constant and so on meaningful
at all.
The independence from the interpolating field becomes
 also clear if one goes back to
eq.(6.35), the "density dependent chiral Lagrangian". Any
parametrization of
the chiral matrix $U(x)$ in terms of some pion field
(exponential, $\sigma$-model
gauge, stereographic coordinates, $\ldots$) gives the same result for
the expansion
truncated at the quadratic order and this is all one needs to read off the
density dependent pion mass and decay constant.
\medskip
The salient features from this study of the pion in nuclear matter can
be summarized as follows. Using the
effective chiral Lagrangian up to order $q^2$ and a simple mean field
approximation to describe the nuclear density, one can easily
reproduce the so-called
``low- density theorems'' for the density variation of the quark condensate and
the pion mass which follow from a multiple scattering expansion.
The pion decay constant $F_\pi$ splits up into a time component $F_t(\rho)$ and
a spatial component $F_s(\rho)$, which do not have the same density dependence,
nevertheless both actually decrease with density. The corresponding linear
coefficients $ 2
c_2 + 2 c_3 - g_A^2/4m$ and $2c_3$ are negative. The results presented here
could be obtained rather easily from lowest order tree level chiral Lagrangian
${\cal L}^{(1,2)}_{\pi N}$, but is seems rather nontrivial to go to higher
orders. For example at order $q^3$ loop diagrams give rise to
non-local  terms of
the form $\int d^4y \bar H(x) O(x,y) H(y)$ and it is not clear how to handle
them in mean field approximation. Furthermore, all four-nucleon terms showing
up in the $B=2$ sector should be considered, since they will give information
on $\rho^2$ correction and eventually nuclear correlations. Presently a
systematic scheme to account for all these complications is unknown and may
only be feasible if one supplies phenomenological information as
discussed in section 5.
\bigskip
\noindent {\bf VI.4. MISCELLANEOUS OMISSIONS}
\medskip
In this section, we want to give a list of topics not covered in
detail. This list is neither meant to be complete or does the order
imply any relevance. The references should allow the interested reader
to further study these topics.
\medskip
\item{$\bullet$}{\it Isospin} {\it violation:} Although the down quark is
almost twice as heavy as the up quark, the corresponding isospin
violations are perfectly masked in almost all observables since $(m_d
- m_u) / \Lambda_\chi \ll 1$. All purely pionic low--energy processes
automatically conserve isospin to order $m_d - m_u$
besides from true electromagnetic
effects. The reason is that G--parity forbids a term of the type $\bar
u u - \bar d d $ using only pion fields with no
derivatives. Therefore, Weinberg [6.43] considered isospin violating
effects in the scattering lengths of neutral pions off nucleons. As
pointed out by Weinberg [6.43] and later quantified by Bernard et
al.[6.44], the absolute values of $a(\pi^0 n \to \pi^0 n)$ and of
$a(\pi^0 p \to \pi^0 p)$ are hard to pin down accurately. However, in the
difference many of the uncertainties cancel and one
expects a sizeable isospin violating effect of the order of 30$\%$
[6.43]. In view of this, Bernstein [6.45] has proposed a
second--generation experiment to accurately measure the phase of the
reaction $\gamma p \to \pi^0 p$ below $\pi^+ n$ threshold and use the
three--channel unitarity to deduce the tiny $\pi^0 p$
phase. More recent discussions of these topics are due to van Kolck
[6.46] and Weinberg [6.47].
\medskip
\item{$\bullet$}{\it Baryon} {\it octet} {\it and} {\it decuplet}
{\it properties:} The high energy hyperon beams at CERN and Fermilab
allow to study aspects of the electromagnetic structure of these
particles. For example, one can make use of the Primakoff effect to
measure the $\Sigma^{\pm}$ polarizabilities. In the quark model, one
expects $\alpha_{\Sigma^+} \gg \alpha_{\Sigma^-}$ since in a system
with like--sign charges like the $\Sigma^- \,(dds)$ dipole excitations are
strongly suppressed [6.48]. This was quantified in a CHPT calculation
to order $q^3$ in ref.[6.49]. To that accuracy, one expects
$\alpha_{\Sigma^+} \simeq 1.5\, \alpha_{\Sigma^-}$. Further studies of
hyperon radiative decays and an analysis of the octet magnetic moments
can be found in refs.[6.50,6.51,6.52]. In the EFT with the spin--3/2
decuplet as active degrees of freedom , one can also address the
properties of the decuplet states. Topics considered include the
$E2/M1$ mixing ratio in the decay $\Delta \to N \gamma$ [6.53], the
decay $\Delta \to N \ell^+ \ell^-$ (where $\ell$ denotes a lepton)
[6.54] or the strong and electromagnetic decays of the decuplet states [6.55].
Furthermore, in the large $N_c$ limit, one needs the spin--3/2 states
to restore unitarity in $\pi N$ scattering (see [6.56] and references therein).
This is often used as a strong support for the inclusion of the
decuplet states in the EFT. However, we would like to stress that since
the chiral and the large $N_c$ limites do not commute, considerable
care has to be taken when such arguments are employed, see e.g.
refs.[6.57,6.58,6.59]).
\medskip
\item{$\bullet$}{\it Kaon}  {\it and} {\it pion} {\it condensation:}
The  work of Kaplan and Nelson [6.28] triggered a flurry of papers
addressing the question whether Bose condensates of charged mesons
may be found in dense nuclear matter formed e.g. in the cores of
neutron stars, the collapse of stars or in the collision of heavy
ions. The physical picture
behind this is the attractive S--wave kaon--nucleon interaction which
could lower the effective mass of kaons to the extent that the kaons
condense at a few times the nuclear matter density. This question, its
consequences for the nuclear equation of state, neutron stars and the
related question of S--wave pion condensation are addressed e.g. in
refs.[6.31,6.32,6.33,6.41,6.60,6.61,6.62,6.63,6.64] (and references given
therein).
\medskip
Finally, let us mention that a state of the art update can be found in
the workshop proceedings [6.65] from which many more references can be
traced back.
\bigskip
\bigskip \noindent
{\bf REFERENCES}
\bigskip
\item{6.1}J. Gasser and H. Leutwyler, {\it Nucl. Phys.\/}
 {\bf B250} (1985) 465.
\smallskip
\item{6.2}A. Krause, {\it Helv. Phys. Acta\/} {\bf 63} (1990) 3.
\smallskip
\item{6.3}E. Jenkins and A.V. Manohar, in "Effective field theories of the
standard model", ed. Ulf--G. Mei{\ss}ner, World Scientific, Singapore,
1992. \smallskip
\item{6.4}J. Gasser, {\it Ann. Phys.\/} (N.Y.) {\bf 136} (1981) 62.
\smallskip
\item{6.5}J. Gasser and H. Leutwyler, {\it Phys. Reports\/} {\bf C87} (1982)
77. \smallskip
\item{6.6}P. Langacker and H. Pagels,
{\it Phys. Rev.\/} {\bf D8} (1971) 4595.
\smallskip
\item{6.7}E. Jenkins, {\it Nucl. Phys.\/} {\bf B368} (1992) 190.
\smallskip
\item{6.8}E. Jenkins and A.V. Manohar, {\it Phys. Lett.\/} {\bf B259} (1991)
353. \smallskip
\item{6.9}R.F. Lebed and M.A. Luty, {\it Phys. Lett.\/} {\bf B329} (1994)
479. \smallskip
\item{6.10}V. Bernard, N. Kaiser
and Ulf-G. Mei{\ss}ner, {\it Z. Phys.\/} {\bf C60} (1993) 111.
\smallskip
\item{6.11}H. Pagels and W. Pardee, {\it Phys. Rev.\/} {\bf D4} (1971) 3225.
\smallskip
\item{6.12}J. Gasser, H. Leutwyler and M.E. Sainio, {\it Phys. Lett.\/}
 {\bf 253B} (1991) 252, 260.
\smallskip
\item{6.13}T.P. Cheng and R. Dashen, {\it Phys. Rev. Lett.\/}
{\bf 26} (1971) 594.
\smallskip
\item{6.14}G. H\"ohler, in Landolt--B\"ornstein, vol.9 b2, ed. H. Schopper
(Springer, Berlin, 1983).
\smallskip
\item{6.15}J. Gasser, in "Hadrons and Hadronic Matter", eds. D. Vautherin et
al., Plenum Press, New York, 1990.
\smallskip
\item{6.16} Ulf-G. Mei{\ss}ner, {\it Int. J. Mod. Phys.} {\bf E1} (1992) 561.
\smallskip
\item{6.17}R. Koch, {\it Z. Phys.\/} {\bf C15} (1982) 161.
\smallskip
\item{6.18}L. S. Brown, W. J. Pardee and R. D. Peccei, {\it Phys. Rev.\/}
{\bf D4} (1971) 2801. \smallskip
\item{6.19}J. Gasser, M.E. Sainio and A. ${\check {\rm S}}$varc,
{\it Nucl. Phys.\/} {\bf B307} (1988) 779.
\smallskip
\item{6.20}J. Gasser and Ulf-G. Mei{\ss}ner, {\it Nucl. Phys.\/}
 {\bf B357} (1991) 90. \smallskip
\item{6.21}E. Jenkins and A.V. Manohar, {\it Phys. Lett.\/} {\bf B281} (1992)
336. \smallskip
\item{6.22}R.L. Jaffe and C. Korpa,
{\it Comm. Nucl. Part. Phys.\/} {\bf 17} (1987) 163.
\smallskip
\item{6.23}V. Bernard, N. Kaiser, J. Kambor
and Ulf-G. Mei{\ss}ner, {\it Nucl. Phys.\/} {\bf B388} (1992) 315.
\smallskip
\item{6.24}P.M. Gensini,
{\it J. Phys. G: Nucl. Part. Phys.\/} {\bf 7} (1981) 1177.
\smallskip
\item{6.25}P.M. Gensini, in $\pi N$ Newsletter no. 6, eds. R.E. Cutkowsky,
G. H\"ohler, W. Kluge and B.M.K. Nefkens, April 1992.
\smallskip
\item{6.26}M.A. Luty and A. White, {\it Phys. Lett.\/} {\bf B319} (1993)
261. \smallskip
\item{6.27}I. Jameson, A.W. Thomas and G. Chanfray,
{\it J. Phys. G: Nucl. Part.
Phys.\/} {\bf 18} (1992) L159.
\smallskip
\item{6.28}D.B. Kaplan and A.E. Nelson, {\it Phys. Lett.} {\bf  B175}
(1986) 57; {\bf B192} (1987) 193. \smallskip
\item{6.29}O. Dumbrais et al., {\it Nucl. Phys.} {\bf  B216} (1982) 277.
\smallskip
\item{6.30}P.B. Siegel and W. Weise, {\it Phys. Rev.} {\bf C38} (1988) 2221.
\smallskip
\item{6.31}G.E. Brown, C.-H. Lee, M. Rho and V. Thorsson,
{\it Nucl. Phys.} {\bf A567} (1993) 937.
\smallskip
\item{6.32}C.-H. Lee, H. Jong, D.-P. Min and M. Rho,
{\it Phys. Lett.} {\bf B326} (1994) 14.
\smallskip
\item{6.33}V. Thorsson and A. Wirzba, ``S--wave Meson--Nucleon
  Interactions and the

Meson Mass in Nuclear Matter from Effective Chiral Lagrangians'',

NORDITA preprint 1995, in preparation..
\smallskip
\item{6.34}R. Shankar, {\it Rev. Mod. Phys.} {\bf 66} (1994) 124.
\smallskip
\item{6.35}H. Leutwyler, {\it Phys. Rev.} {\bf D49} (1994) 3033.
\smallskip
\item{6.36}
E.G. Drukarev and E.M. Levin, {\it Nucl. Phys.} {\bf A511} (1988) 697.
\smallskip
\item{6.37}T.D. Cohen, P.J. Furnstuhl and D.K. Griegel, {\it
    Phys. Rev.} {\bf  C45} (1992) 1881. \smallskip
\item{6.38}
M. Lutz, S. Klimt and W. Weise, {\it Nucl. Phys.} {\bf  A542} (1992) 521.
\smallskip
\item{6.39}P. Gerber and H. Leutwyler, {\it Nucl. Phys.}
{\bf B321} (1989) 387.   \smallskip
\item{6.40}A. Le Yaouanc, L. Oliver, S. Ono, O. P\`ene and
J.-C. Raynal,  {\it Phys. Rev.} {\bf D31} (1985) 137;

V. Bernard, {\it Phys. Rev.} {\bf D34} (1986) 1601.   \smallskip
\item{6.41}J. Delorme, M. Ericson and T.E.O Ericson, {\it Phys. Lett.}
{\bf B291} (1992) 379.   \smallskip
\item{6.42}V. Bernard and Ulf--G. Mei{\ss}ner, {\it Nucl. Phys.}
{\bf A489} (1988) 647.   \smallskip
\item{6.43}S. Weinberg, {\it Trans. N.Y. Acad. Sci.} {\bf 38} (1977)
185. \smallskip
\item{6.44}V. Bernard, N. Kaiser and Ulf--G. Mei{\ss}ner, {\it Phys. Lett.}
{\bf B309} (1993) 421.   \smallskip
\item{6.45}A.M. Bernstein, $\pi N$ {\it  Newsletter} {\bf 9} (1993)
55. \smallskip
\item{6.46}U. van Kolck, Thesis, University of Texas at Austin, 1992.
\smallskip
\item{6.47}S. Weinberg, ``Strong Interactions at Low Energies'',
preprint UTTG-16-94, 1994. \smallskip
\item{6.48}H.J. Lipkin and M.A. Moinester, {\it Phys. Lett.} {\bf B287} (1992)
179. \smallskip
\item{6.49}V. Bernard, N. Kaiser, J. Kambor
and Ulf-G. Mei{\ss}ner, {\it Phys. Rev.\/} {\bf D46} (1992) 2756.
\smallskip
\item{6.50}E. Jenkins et al., {\it Nucl. Phys.} {\bf B397} (1993) 84.
\smallskip
\item{6.51}H. Neufeld, {\it Nucl. Phys.} {\bf B402} (1993) 166.
\smallskip
\item{6.52}E. Jenkins et al., {\it Phys. Lett.} {\bf B302} (1993) 482.
\smallskip
\item{6.53}M.N. Butler, M.J. Savage and R.P. Springer,
{\it Phys. Lett.} {\bf B304} (1993) 353. \smallskip
\item{6.54}M.N. Butler, M.J. Savage and R.P. Springer,
{\it Phys. Lett.} {\bf B312} (1993) 486. \smallskip
\item{6.55}M.N. Butler, M.J. Savage and R.P. Springer,
{\it Nucl. Phys.} {\bf B399} (1993) 69; {\it Phys. Rev.} {\bf D49}
(1994) 3459. \smallskip
\item{6.56}R. Dashen, E. Jenkins and A.V. Manohar, {\it Phys. Rev.}
{\bf D49} (1994) 4713. \smallskip
\item{6.57}J. Gasser and A. Zepeda, {\it Nucl. Phys.} {\bf B174} (1980) 45.
\smallskip
\item{6.58}E. Jenkins and A.V. Manohar, in ``Effective Field Theories
of the Standard Model'', ed. Ulf--G. Mei{\ss}ner (World Scientific,
Singapore, 1992). \smallskip
\item{6.59}N. Kaiser, in ``Baryons as Skyrme Solitons'',
ed. G. Holzwarth (World Scientific, Singapore, 1993). \smallskip
\item{6.60}H.D. Politzer and M.B. Wise,
{\it Phys. Lett.} {\bf B273} (1991) 156. \smallskip
\item{6.61}G.E. Brown, K. Kubodera, M. Rho and V. Thorsson, {\it
Phys. Lett.} {\bf B291} (1992) 355. \smallskip
\item{6.62}V. Thorsson, M. Prakash and J.M. Lattimer,
{\it Nucl. Phys.} {\bf A572} (1994) 693;
{\it Nucl. Phys.} {\bf A574} (1994) 851, \smallskip
\item{6.63}C.-H. Lee, G.E. Brown and M. Rho, ``Kaon condesation in
nuclear star matter'',  Seoul University preprint SNUTP-94-28, hep-ph/9403339,
1994. \smallskip
\item{6.64}H. Yabu, F. Myhrer and K. Kubodera, {\it Phys. Rev.} {\bf
D50} (1994) 3549 (and references therein).
\item{6.65}Proceedings of the workshop on ``Chiral Dynamics: Theory
and Experiment'', B.R. Holstein and A.M. Bernstein (eds), Springer
Lecture Notes in Physics, 1995, in print.
\vfill \eject
\noindent{\bf APPENDIX A: FEYNMAN-RULES}
\medskip
Here, we wish to collect the pertinent Feynman rules which are needed to
calculate tree and loop diagrams. In order to prarametrize  the SU(2) matrix
$U$ in terms of pion fields we use the so-called sigma gauge $U = \sqrt{1-\vec
\pi^2 /F^2 } + i \vec \tau \cdot \vec \pi / F$ which is more convenient than
the exponential parametrization $U = \exp[i \vec \tau \cdot \vec \pi/F]$. For
effective vertices involving 3 and more pions, the Feynman rules differ in the
two parametrizations. Of course the complete S-matrix for a process with a
certain number of on-shell external pions  is the same in either
parametrization. The parametrization dependence of matrix elements for
off-shell pions signals that these are indeed non-unique in CHPT. Physically
this is clear, since in order to calculate $e.g.$ off-shell pion amplitudes,
one has to know the exact pion field of QCD, not just some interpolating field.
\medskip
\noindent We use the following notation:
\smallskip \noindent
\item{$l$}Momentum of a pion or nucleon propagator.
\smallskip \noindent
\item{$k$}Momentum of an external vector or axial source.
\smallskip \noindent
\item{$q$}Momentum of an external pion.
\smallskip \noindent
\item{$\epsilon$}Photon polarization vector.
\smallskip \noindent
\item{$\epsilon_A$}Polarization vector of an axial source.
\smallskip \noindent
\item{$p$}Momentum of a nucleon in heavy mass formulation.
\smallskip \noindent
and pion isopin indices are $a,b,c,d,3$.Furthermore, $v_\mu$ is the nucleon
four-velocity and $S_\mu$ its covariant spin--vector. All parameters
like $Q = F, g_A, m, \ldots$ are meant to be taken in the chiral limit. We
also give the orientation of momenta at the vertices, i.e.
 which are "in"-going or "out"-going.
\medskip
\noindent{\bf Vertices from ${\cal L}^{(2)}_{\pi\pi}$}
\medskip
\noindent
pion propagator: $${ i \delta^{ab} \over l^2 - M_\pi^2 + i0} \eqno(A.1)$$
1 pion, pseudoscalar source:   $$ 2 i B F \delta^{ab} \eqno(A.2)$$
3 pions, pseudoscalar source:  $$0 \eqno(A.3)$$
1 pion, axial source ($k$ in): $$F \delta^{ab} \epsilon_A\cdot k \eqno(A.4)$$
3 pions, axial source (all $q$'s out):
$${1\over F}\epsilon_A\cdot[\delta^{bc}\delta^{de}(q_2+q_3-q_1) +\delta^{bd}
\delta^{ce}(q_1+q_3-q_2) +\delta^{be}\delta^{cd}(q_1+q_2-q_3)] \eqno(A.5)$$
2 pions, photon ($q_1$ in, $q_2$ out): $$e \epsilon^{a3b} \epsilon\cdot
(q_1+q_2) \eqno(A.6)$$
4 pions, photon: $$0 \eqno(A.7)$$
2 pions, 2 photons: $$2ie^2 \epsilon'\cdot \epsilon
(\delta^{ab}-\delta^{a3}\delta^{b3}) \eqno(A.8)$$
2 pions, scalar source: $$i\delta^{ab} M_\pi^2 \eqno(A.9)$$
4 pion vertex (all $q$'s in): $${i\over F^2} \{ \delta^{ab} \delta^{cd}
[(q_1+q_2)^2-M_\pi^2] +\delta^{ac} \delta^{bd} [(q_1+q_3)^2-M_\pi^2]+
\delta^{ad} \delta^{bc} [(q_1+q_4)^2-M_\pi^2] \} \eqno(A.10)$$
\medskip
\noindent{\bf Vertices from ${\cal L}^{(1)}_{\pi N}$}
\medskip
\noindent
nucleon propagator: $$ {i \over v\cdot l + i0} \eqno(A.11)$$
1 pion ($q$ out): $$ {g_A\over F } S\cdot q \tau^a \eqno(A.12)$$
photon: $${ie\over 2}(1+ \tau^3) \epsilon\cdot v \eqno(A.13)$$
2 pions ($q_1$ in,$q_2$ out): $${1\over 4 F^2 } v\cdot (q_1+q_2)
\epsilon^{abc}\tau^c \eqno(A.14)$$
1 pion, 1 photon: $$ {ieg_A\over F} \epsilon\cdot S \epsilon^{a3b}
\tau^b \eqno(A.15)$$
3 pions (all $q$'s out): $${g_A \over 2 F^3} [ \tau^a \delta^{bc}
S\cdot(q_2+q_3)+\tau^b \delta^{ac} S\cdot(q_1+q_3)+\tau^c
\delta^{ab} S\cdot(q_1+q_2)] \eqno(A.16)$$
2 pions, photon: $${ie \over 4F^2} (\tau^a \delta^{b3} + \tau^b \delta^{a3} - 2
\tau^3 \delta^{ab}) \epsilon\cdot v \eqno(A.17)$$
3 pions, photon: $$0 \eqno(A.18)$$
axial source: $$ig_A S\cdot \epsilon_A \tau^b \eqno(A.19)$$
1 pion, axial source: $${i\over 2F} \epsilon_A \cdot v \epsilon^{abc}
\tau^c \eqno(A.20)$$
2 pions, axial source: $${i g_A \over 2 F^2} S\cdot \epsilon_A
(\delta^{ab}\tau^c + \delta^{bc}\tau^a - 2 \delta^{ac} \tau^b) \eqno(A.21)$$
\medskip
\noindent{\bf Vertices from ${\cal L}^{(2)}_{\pi N}$}
$$\eqalign{ {\cal L}_{\pi N}^{(2)} & = \bar H \biggl\{ {1\over 2m} (v\cdot D)^2
- {1 \over 2m} D\cdot D -{ig_A\over 2m} \{S\cdot D ,v\cdot u\} + c_1 {\rm
  Tr}\chi_+ \cr & +(c_2 - {g_A^2 \over 8m})( v\cdot u)^2  + c_3 u\cdot u +(c_4
 +{1\over 4m} ) [S^\mu, S^\nu] u_\mu u_\nu  \cr & + c_5 \Tr( \tilde{\chi}_+ )
 -{i \over 4m} [S^\mu , S^\nu] \bigl(
  (1+\krig \kappa_v) f^+_{\mu\nu} +{1\over2}(\krig \kappa_s- \krig \kappa_v)
\Tr(f_{\mu\nu}^+) \bigr) \biggr\}H \cr} \eqno(A.22)$$
with $\tilde{\chi}_+ = \chi_+ - (1/2)\Tr{\chi}_+$ (this term is only
non--vanishing for $m_u \ne m_d$).
All parameters, $g_A, m, c_i, \krig\kappa_{s,v}$ are understood as the ones
in the chiral limit. $f^+_{\mu\nu} = u^\dagger
F^R_{\mu\nu} u + u F^L_{\mu\nu} u^\dagger$ with $F^{L,R}_{\mu\nu}$ the field
strength tensor corresponding
to external (isovector) left/right vector sources
(isovector photon, W and Z boson). The external vector source in $f_{\mu\nu}^+$
is understood to have also an isoscalar component (isoscalar photon).
Here, we will use the (Coulomb) gauge $\epsilon\cdot v = 0$ for the photon.
The first three terms in ${\cal L}^{(2)}_{\pi N}$ stem from the $1/m$ expansion
of the chiral nucleon Dirac lagrangian. They have no counter part in the
relativistic theory, $i.e.$ no bilinears form involving $\gamma$--matrices.
Their coefficients are fixed in terms of the lowest order parameters and
nucleon mass $m$. The other terms involving the new low--energy constants
come from the most general relativistic lagrangian at order $q^2$ (see
ref.[3.6]) after translation into the heavy mass formalism. One observes, that
there is some overlap between the two types of terms at order $q^2$, namely in
the $(v \cdot u)^2 $ and $[S^\mu, S^\nu] u_\mu u_\nu$  terms and the magnetic
moment couplings. The low--energy constants $c_6$ and $c_7$ which are
discussed in sections three and four are related to the anomalous magnetic
moments of the nucleon (in the chiral limit) via
$$ c_6 = \krig{\kappa}_v \, , \quad c_7 = {1 \over 2} (\krig{\kappa}_s
- \krig{\kappa}_v ) \quad .\eqno(A.23) $$
The pertinent Feynman insertions read ($p_1$ is always ingoing and $p_2$
outgoing):
\medskip

\noindent
nucleon propagator: $${i\over 2m} \biggl[ 1 - {l^2 \over (v\cdot l+i0)^2}
\biggr] \eqno(A.24)$$
2 photons: $${ie^2 \over 2m} (1+\tau^3) \epsilon'\cdot
\epsilon \eqno(A.25)$$
1 pion ($q$ out): $$-{g_A\over 2 m F} S\cdot(p_1+p_2) v\cdot q \tau^a
 \eqno(A.26)$$
1 photon ($k$ in): $${ie\over 4m}(1+ \tau^3) \epsilon\cdot(p_1+p_2) + {ie \over
2m} [S\cdot \epsilon, S\cdot k] ( 1 + \krig\kappa_s + (1+\krig\kappa_v)
\tau^3) \eqno(A.27)$$
1 pion, 1 photon ($q$ out): $$-{eg_A\over 2m F} S\cdot \epsilon v\cdot q
(\tau^a + \delta^{a3}) \eqno(A.28)$$
2 pions ($q_1$ in, $q_2$ out): $$\eqalign{&{i\delta^{ab}\over
F^2}\bigl[-4c_11M_\pi^2 +(2c_2-{g_A^2\over4m})v \cdot q_1 v\cdot q_2 + 2 c_3
q_1\cdot q_2\bigr] \cr & + {1\over 8mF^2}\epsilon^{abc}\tau^c
\bigl[(p_1+p_2)\cdot(q_1+q_2) -v\cdot (p_1+p_2) v \cdot (q_1+q_2) \bigr] \cr
&-{1\over F^2} \biggl(2c_4+{1\over 2m} \biggr) \epsilon^{abc}\tau^c [S\cdot q_1
, S\cdot q_2]  \cr } \eqno(A.29)$$
3 pions (all $q$'s out): $$\eqalign{&-{i g_A\over4 m F^3} \epsilon^{abc}[v\cdot
q_1 S\cdot (q_2-q_3)+v\cdot q_2 S\cdot (q_3-q_1)+v\cdot q_3 S\cdot(q_1-q_2)]\cr
& -{g_A\over 4 mF^3} S\cdot (p_1+p_2) [ \tau^a \delta^{bc} v\cdot (q_2+q_3)+
\tau^b \delta^{ac} v\cdot (q_1+q_3)+ \tau^c \delta^{ab} v\cdot
(q_1+q_2)]\cr} \eqno(A.30)$$
2 pions, 1 photon ($q_1$ in, $q_2$ out): $$\eqalign{ &{ie\over 2
mF^2}\bigl\{2(1+\krig\kappa_v)[S
\cdot\epsilon, S\cdot k]+ \epsilon\cdot (p_1+p_2)\bigr\} (\tau^a \delta^{b3}
+\tau^b \delta^{a3} -2\tau^3 \delta^{ab}) \cr &+{e\over 8 mF^2}
\epsilon\cdot(q_1+q_2) \epsilon^{abc} ( \tau^c + \delta^{3c}) + 2c_3 {e\over
F^2} \epsilon^{a3b}\epsilon\cdot(q_1+q_2) \cr} \eqno(A.31)$$
3 pions, 1 photon (all $q$'s out): $$\eqalign{-{eg_A\over 4 m F^3} S\cdot
\epsilon &\bigl[ 2 \delta^{a3} \delta^{bc} v\cdot(q_2+q_3-q_1) +2\delta^{b3}
\delta^{ac} v\cdot(q_1+q_3-q_2)+2\delta^{c3} \delta^{ab} v\cdot(q_1+q_2-q_3)
\cr & + \tau^a \delta^{bc} v\cdot(q_2+q_3)+\tau^b \delta^{ac}
v\cdot(q_1+q_3)+\tau^c \delta^{ab} v\cdot(q_1+q_2)  \bigr] \cr}\eqno(A.32) $$
2 pions, 2 photons: $${ie^2\over 4 mF^2} \epsilon'\cdot \epsilon [ \tau^a
\delta^{b3}  +\tau^b \delta^{a3} +2\delta^{a3}\delta^{b3}
-2\delta^{ab}(1+\tau^3)] + 4ic_3 {e^2\over F^2} \epsilon'\cdot
\epsilon(\delta^{ab} -\delta^{a3} \delta^{b3}) \eqno(A.33)$$
axial source: $$-i{g_A \over 2mF} S\cdot(p_1+p_2) \epsilon_A\cdot v
\tau^b  \eqno(A.34)$$
1 pion, axial source ($q$ out): $$\eqalign{ &{i\over 4mF} \epsilon^{abc} \tau^c
[ \epsilon_A\cdot (p_1+p_2) - \epsilon_A\cdot v v \cdot (p_1+p_2) + 2 {c_3\over
F}  \delta^{ab} \epsilon_A\cdot q \cr & +{1\over F}(2c_2 - {g_A^2\over 4m} )
\delta^{ab} \epsilon_A\cdot v v \cdot q + {i\over 2mF}(1+\krig\kappa_v)[S\cdot
\epsilon_A , S\cdot k] \epsilon^{abc} \tau^c \cr & +i\biggl( 2c_4 +{1\over 2m}
\biggr)[ S\cdot q , S\cdot \epsilon_A] \epsilon^{abc} \tau^c\cr }\eqno(A.35)$$

\bigskip
\noindent{\bf APPENDIX B: LOOP FUNCTIONS}
\medskip
Here, we will define many of the loop functions which frequently occur in our
calculations and we will give these functions in closed analytical form as far
as possible. Divergent loop functions are regularized via dimensional
regularization and expanded around $d=4 $ space-time dimensions. In the
following all propagators are understood to have an infinitesimal negative
imaginary part.

$$ {1\over i} \int {d^d l \over (2\pi)^d} {1 \over M_\pi^2 - l^2 } = \Delta_\pi
= M_\pi^{d-2} (4\pi)^{-d/2} \Gamma(1-{d\over2}) \eqno(B.1)$$
$$\Delta_\pi = 2 M_\pi^2 \biggl( L + {1\over 16 \pi^2} \ln {M_\pi\over \lambda}
\biggr)  + {\cal O}(d-4) \eqno(B.2)$$
with
$$L = {\lambda^{d -4} \over 16 \pi^2 } \biggl[ {1\over d-4} + {1\over 2} (
\gamma_E -1 - \ln 4\pi) \biggr] \eqno(B.3)$$
containing a pole in $d=4 $ and $\gamma_E = 0.557215...$. The scale $\lambda $
is introduced in dimensional regularization.
$$ {1\over i} \int {d^d l \over (2\pi)^d} {\{1,l_\mu , l_\mu l_\nu\}  \over
(v\cdot l-\omega)  (M_\pi^2 - l^2) } = \{J_0(\omega), v_\mu J_1(\omega),
g_{\mu\nu} J_2(\omega) + v_\mu v_\nu J_3(\omega) \} \eqno(B.4)$$
$$J_0(\omega ) = -4L\omega + {\omega\over 8 \pi^2}\biggl(1-2\ln {M_\pi\over
\lambda} \biggr)- {1\over 4 \pi^2} \sqrt{M_\pi^2 - \omega^2} \arccos{-\omega
\over M_\pi} +{\cal O}(d-4) \eqno(B.5)$$

$$J_1(\omega) = \omega J_0(\omega) + \Delta_\pi, \quad J_2(\omega) ={1\over
d-1}\biggl[ (M_\pi^2-\omega^2) J_0(\omega) - \omega \Delta_\pi\biggr]
\eqno(B.6)$$
$$J_3(\omega) = \omega J_1(\omega) - J_2(\omega) \eqno(B.7)$$

$$ {1\over i} \int {d^d l \over (2\pi)^d }{\{1,l_\mu , l_\mu l_\nu\}  \over
v\cdot l(v\cdot l-\omega)  (M_\pi^2 - l^2) } = \{\tilde\Gamma_0(\omega), v_\mu
\tilde \Gamma_1(\omega),  g_{\mu\nu} \tilde \Gamma_2(\omega) + v_\mu v_\nu
\tilde \Gamma_3(\omega) \} \eqno(B.8)$$
Using the identity
$${1\over v\cdot l ( v\cdot l - \omega)} = {1\over \omega} \biggl( {1\over
v\cdot l - \omega} - {1 \over v\cdot l} \biggr) \eqno(B.9)$$
$$\tilde\Gamma_i(\omega ) = {1\over \omega}\bigl[ J_i(\omega) - J_i(0)\bigr],
\quad(i = 0,1,2,3) \eqno(B.10)$$

$$ {1\over i} \int {d^d l \over (2\pi)^d} {\{1,l_\mu , l_\mu l_\nu\}  \over
(v\cdot l-\omega)^2  (M_\pi^2 - l^2) } = \{G_0(\omega), v_\mu G_1(\omega),
g_{\mu\nu} G_2(\omega) + v_\mu v_\nu G_3(\omega) \} \eqno(B.11)$$
Using the identity
$${1\over ( v\cdot l - \omega)^2} = {\partial\over \partial\omega} \biggl( {1
\over v\cdot l- \omega} \biggr)  \eqno(B.12)$$
$$G_i(\omega ) = {\partial\over \partial\omega}J_i(\omega) , \quad(i =
0,1,2,3) \eqno(B.13)$$

$$ \eqalign{ {1\over i} \int {d^d l \over (2\pi)^d}& {\{1,l_\mu , l_\mu l_\nu\}
\over v\cdot l  (M_\pi^2 - l^2)(M_\pi^2 - (l+k)^2) } = \{\gamma_0(\omega),
k_\mu \gamma_1(\omega) + v_\mu \gamma_2(\omega),\cr
& g_{\mu\nu} \gamma_3(\omega)
+k_\mu k_\nu \gamma_4(\omega) +(k_\mu v_\nu + k_\nu v_\mu) \gamma_5(\omega) +
v_\mu v_\nu \gamma_6(\omega) \} \cr } \eqno(B.14)$$
where $\omega = v \cdot k,\,\, k^2 = 0$ since we consider only real photons.
$$\gamma_0(\omega) = {1\over 16 \pi^2 \omega}\biggl( \pi + \arcsin{\omega\over
M_\pi}\biggr) \arcsin{\omega\over M_\pi} \eqno(B.15)$$
The vector and tensor functions $\gamma_{j}(\omega), \,j = 1,2,3,4,5,6$ can be
obtained by the following procedure. One multiplies the defining equation with
$v $ thereby canceling a factor $v \cdot l$ or one multiplies by $2k$ and uses
the identity
$ 2 k\cdot l = (M_\pi^2 - l^2) -(M_\pi^2 - (l+k)^2))$ for $k^2=0$. This leads
to linear relation among the $\gamma_j(\omega), \, j\ne 0$ where the right hand
sides are loop functions with fewer propagators.  For illustration of the
method, we give  the explicit solution for the functions
$\gamma_j(\omega)$:
$$\gamma_1(\omega) = {1\over 8 \pi^2 \omega^2} \biggl[ \sqrt{M_\pi^2 -
\omega^2} \arccos{-\omega\over M_\pi} - {\pi \over 2} M_\pi -
\omega\biggr]
\eqno(B.16)$$
$$\gamma_2(\omega) = - 2 L + {1 \over 16 \pi^2 } \biggl( 1 - 2 \ln {M_\pi \over
\lambda} \biggl) + {1 \over 8 \pi^2 \omega} \biggl[ {\pi \over 2} M_\pi -
\sqrt{M_\pi^2 - \omega^2} \arccos{-\omega \over M_\pi} \biggr] \eqno(B.17)$$
$$\eqalign{ \gamma_3(\omega) = & L \omega + {\omega \over 16 \pi^2} \biggl( \ln
{M_\pi \over \lambda} -1 \biggr) + {1\over 16 \pi^2 } \sqrt{M_\pi^2 - \omega^2}
\arccos{-\omega \over M_\pi} \cr & + {M_\pi^2 \over 32 \pi^2 \omega }
\biggl( \pi + \arcsin{\omega\over M_\pi}\biggr) \arcsin{\omega\over
M_\pi} \cr} \eqno(B.18)$$
$$\gamma_4(\omega) =  {1 \over 32 \pi^2\omega^3} \biggl[M_\pi^2\biggl( \pi +
\arcsin{\omega\over M_\pi}\biggr) \arcsin{\omega\over M_\pi}-2 \omega
\sqrt{M_\pi^2 - \omega^2} \arccos{-\omega \over M_\pi} +
\omega^2 \biggr] \eqno(B.19)$$
$$\gamma_5(\omega) =  L + {1\over 16 \pi^2}\ln{M_\pi \over \lambda}  + {
\sqrt{M_\pi^2 - \omega^2}\over 16 \pi^2 \omega}  \arccos{-\omega \over M_\pi}-
{M_\pi^2 \over 32 \pi^2 \omega^2 }   \biggl( \pi + \arcsin{\omega\over
M_\pi}\biggr) \arcsin{\omega\over M_\pi} \eqno(B.20)$$
$$\gamma_6(\omega) = -2L\omega +
{\omega\over 16 \pi^2}\biggl(1-2\ln {M_\pi\over
\lambda} \biggr)- {1\over 8 \pi^2} \sqrt{M_\pi^2 - \omega^2} \arccos{-\omega
\over M_\pi} \eqno(B.21)$$

\noindent From now on we give only the scalar loop functions for $d=4$.
$${1\over i} \int {d^4 l \over (2\pi)^4} {1\over v\cdot l (v\cdot l - \omega)
(M_\pi^2 - l^2)(M_\pi^2 - (l-k)^2) } = \Omega_0(\omega) \eqno(B.22)$$
Using the same identity as for $\tilde \Gamma_i(\omega)$
$$\Omega_0(\omega) = {1\over \omega}\bigl[ \gamma_0(\omega) - \gamma_0(-\omega)
\bigr] = {1\over 8 \pi^2 \omega^2} \arcsin^2{\omega\over M_\pi} \eqno(B.23)$$
$${1\over i} \int {d^4 l \over (2\pi)^4} {1\over v\cdot l (M_\pi^2 - l^2)^2
(M_\pi^2 - (l+k)^2) } = \Lambda_0(\omega) \eqno(B.24)$$
$$\Lambda_0(\omega) ={1\over 32 \pi M_\pi^2 \omega^2} \biggl[ M_\pi -
\sqrt{M_\pi^2 - \omega^2} \biggr] +{1\over 16 \pi^2 M_\pi^2 \omega^2} \biggl[
\omega -\sqrt{M_\pi^2 - \omega^2}\arcsin{\omega\over M_\pi} \biggr]
\eqno(B.25) $$
$${1\over i} \int {d^4 l \over (2\pi)^4} {1\over (v\cdot l)^2 (M_\pi^2 - l^2)
(M_\pi^2 - (l+k)^2) } = \psi_0(\omega) \eqno(B.26)$$
$$\psi_0(\omega) = {1\over 8 \pi^2 \omega} \biggl[ (M_\pi^2 -
\omega^2)^{-1/2}\arccos{-\omega \over M_\pi}- {\pi \over 2 M_\pi}
\biggr]  \eqno(B.27)$$

$${1\over i} \int {d^4 l \over (2\pi)^4} {1\over (v\cdot l)^2 (M_\pi^2 - l^2)^2
(M_\pi^2 - (l+k)^2) } = \chi_0(\omega) \eqno(B.28)$$
$$\chi_0(\omega) = {1\over 16 \pi^2 M_\pi^2 \omega}\biggl[ (M_\pi^2 -
\omega^2)^{-1/2}\arccos{-\omega \over M_\pi}- {\pi \over 2 M_\pi}
\biggr] \eqno(B.29)$$
For the calculation of form factor we need $\gamma_j$ with $v\cdot k = 0$ but
$k^2 =t \ne 0$.
$$ {1\over i} \int {d^d l \over (2\pi)^d} {1,l_\mu l_\nu\over v\cdot l
(M_\pi^2 - l^2)(M_\pi^2 - (l+k)^2) } = \gamma_0(t), g_{\mu\nu} \gamma_3(t) +
\dots \eqno(B.30)$$
$$\gamma_0(t) = {1\over 8 \pi \sqrt{-t}} \arctan{\sqrt{-t}\over 2
M_\pi} \eqno(B.31)$$
$$\gamma_3(t) = {1\over 32\pi} \biggl[ M_\pi + \biggl( {1\over 2} - {2M_\pi^2
\over t} \biggr) \sqrt{-t} \arctan{\sqrt{-t}\over 2 M_\pi}
\biggr] \eqno(B.32)$$

The two loop function which enter the nucleon isovector Dirac from factor are:
$$J(t) = -{1\over 16 \pi^2} \biggr\{ {t\over 6} + \int_0^1 dx \bigl[ M_\pi^2 +
t x (x-1)\bigr] \ln\bigl[ 1 + {t\over M_\pi^2} x(x-1) \bigr]
\biggr\} \eqno(B.33)$$
$$\xi(t) = -{1\over 16 \pi^2} \int_0^1 dx \ln\bigl[ 1 + {t\over M_\pi^2} x(x-1)
\bigr] \eqno(B.34)$$
\bigskip
\noindent {\bf APPENDIX C: THE "AXIAL RADIUS DISCREPANCY"}
\medskip
In the end of sect.IV.4 we discussed pion electroproduction at threshold and
found that chiral loops modify the LET of Nambu, Luri\'e and Shrauner. The
conclusion was that an analysis of threshold charged pion electroproduction
data in terms of soft pion theory (in order to link the measured cross sections
to nucleon electromagnetic and axial form factors) does not lead to the nucleon
mean square axial radius $r^2_A$ itself but to a modified quantity $\tilde
r^2_A$ which subsumes the chiral loop corrections.  For this "discrepancy",
$\tilde r^2_A - r^2_A$, the leading term which survives in the chiral limit is
given in eq.(4.79).  Numerically it is a --10$\%$ effect and allows to
understand the systematic discrepancies between present (anti)neutrino
experiments (which measure the true nucleon axial radius) and charged pion
electroproduction experiments [4.76]. (Note that we are considering
here exclusively small values of the momentum transfer $k^2$).

Since the lowest order result of the discrepancy
$\tilde r^2_A - r^2_A$ is quite small,
one should investigate higher order corrections (in $\mu = M_\pi/m$) in order
to see whether the numerical value of the leading order prediction is actually
reliable. Such a complete next-to-leading order calculation has been done in
ref.[4.91].
For that one has to go to order $q^4$ in the chiral expansion, which amounts to
an evaluation of all one loop graphs with a single insertion from ${\cal
L}^{(2)}_{\pi N} $ and possible counter terms. The latter were estimated from
resonance exchange contributions, in the present case from $\rho(770)$ and
$\Delta(1232)$ exchange.

The relevant observable to be studied is the transition matrix element for
$\gamma^* p \to \pi^+ n$ in the center of mass frame at threshold. Only the
transverse part is of interest and it takes the form
$$T\cdot \epsilon = {ie g_A \over \sqrt{2} F_\pi } \vec \sigma \cdot \vec
\epsilon \, E(k^2) = 4 \pi\,i\, ( 1+ \mu)\, \vec \sigma \cdot \vec \epsilon\,
E_{0+,\rm thr}^{\pi^+ n} \eqno(C.1)$$
The auxiliary quantity $E(k^2)$ introduced here is proportional to the
transverse threshold S--wave multipole for $\pi^+$-electroproduction. In the
chiral limit the corresponding current algebra result becomes exact and gives,
when expanded in $k^2$,
$$E(k^2) = 1 + {k^2 \over 6} \krig r^2_A - {k^2 \over 2 m^2} \biggl(\krig
\kappa_n +{1\over 4 }\biggr) +{\cal O}(k^3)\eqno(C.2)$$
with
$$\krig r^2_A = r^2_A +{\cal O}(\mu^2), \qquad \quad \krig \kappa_n = \kappa_n
-{g_A^2 m M_\pi \over 8 \pi F_\pi^2} +{\cal O}(\mu^2)\eqno(C.3)$$
the nucleon mean square axial radius and neutron magnetic moment in the chiral
limit. Note that the axial mean square radius has no non-analytic piece $\sim
\sqrt{\hat m}$ in its quark mass expansion [4.91] (in contrast to the
isovector magnetic moment). The aim is to work out all tree and loop graphs up
to order $q^4$ which contribute to the slope terms
$E'(0) = \partial E(0)/\partial
k^2$  proportional to $M_\pi^0$ and/or $M_\pi^1$. The quantity $6 E'(0)$
is the sum of the mean square axial radius $r^2_A$ and a host of other terms.
Among these other terms the contributions from the relativistic Born graphs
including electromagnetic form factors will not be counted for the axial radius
discrepancy $\tilde r^2_A- r^2_A$, since such effects are taken
 into account in the
standard analysis of the pion electroproduction data.  The discrepancy
therefore subsumes (per definition) all additional loop (and counter term)
effects which go beyond the form factors. Stated differently, the dicrepancy
represents all those $k^2$-pieces which are missing in a tree calculation (with
form factors) of $E(k^2)$. After some lengthy calculation [4.91] one
arrives at
$$\eqalign{ \tilde r^2_A - r^2_A =  & {3\over 64 F_\pi^2}\biggl( 1 - {12 \over
\pi^2} \biggr) + {3 M_\pi \over 64 m F_\pi^2} + {3c^+(\pi - 4) M_\pi \over 32
\pi F_\pi^2} \cr & + {3 g_A^2 \over 8 \pi^2 m F_\pi^2 } \biggl( \ln{M_\pi \over
\lambda} -{\pi^2 \over 16} +{7\pi\over 12} -{1\over 4} \biggr) + 6
E'_\rho(0) + 6 E'_\Delta(0) \cr} \eqno(C.4)$$
The first term in (C.4) is the leading order result given  already in
eq.(4.79). The combination of low energy constants $c^+ = -8c_1+4c_2
+4c_3-g_A^2/2m$ can be related to the isospin even $\pi N$ scattering length
$a^+$ via $c^+ = 8 \pi F_\pi^2 a^+/M_\pi^2$.  The last two terms in (C.3)
represent the counter term contributions at order $q^4$ which have been
identified with $\rho(770)$ and $\Delta(1232)$ exchange contributions,
$$\eqalign{
6E'_\rho(0) & = -{3 (1+\kappa_\rho) M_\pi \over 16 \pi^2 g_A m F_\pi^2}
\cr 6 E'_\Delta(0) & = {\kappa^* M_\pi \over \sqrt{2} m^2 m_\Delta^2} \biggl[
{m_\Delta^2 - m_\Delta m +m^2 \over m_\Delta - m} - 2m (Y+Z+2YZ)
-2m_\Delta(Y+Z+4YZ) \biggr] \cr} \eqno(C.5)$$
Here, $\kappa_\rho \simeq 6$ stands for the tensor-to-vector ratio of the $\rho
NN$ couplings and $\kappa^* = g_1 \simeq 5$ is the $\gamma N \Delta$ coupling
constant. The second $\gamma N \Delta$ coupling $g_2$ of eq.(4.39)
does not contribute at order $q^4$. The off-shell parameters $Y,Z$ have been
estimated roughly in ref.[4.32] as $-0.75 \leq Y \leq 1.67$ and $-0.8
\leq Z \leq 0.3$. For a numerical evaluation of (C.5)  these ranges are much
too large and they should be further constrained. The strategy of
ref.[4.91] was to link them to some nucleon structure parameters. The
off-shell parameter $Y$ enters the $\Delta(1232)$ contribution to the proton
magnetic polarizability
$$\delta \beta^{(\Delta)}_p = {e^2 \kappa^{*2} \over 18 \pi m^2 m_\Delta^2}
\biggl[ {m_\Delta^2 - m_\Delta m +m^2 \over m_\Delta - m} - 4 Y\bigl( m (Y+1)
 +m_\Delta(2Y+1) \bigr)\biggr]  \eqno(C.6)$$
Experimental determinations of this quantity in ref.[4.28] give a
value of $\delta \beta ^{(\Delta)}_p \simeq  7\cdot 10^{-4}$ fm$^3$
corresponding to $Y \simeq 0.12$. It is clear that the wide range of $Y$
mentioned above is inacceptably large, since it also leads to (absurd) negative
magnetic polarizabilities $\delta \beta^{(\Delta)}_p$. Furthermore the
off-shell parameter $Z$ of the $\pi N \Delta$ -vertex has been
constrained. The $\Delta(1232)$ gives a large contribution to the P-wave $\pi
N$ scattering volume $a_{33}$. In the isobar model (an approximation without
the off-shell parameter $Z$) the experimental value $a_{33} = 0.214 /M_\pi^3$
is understood to come in equal shares from nucleon pole graphs and  from
$\Delta(1232)$ excitation. Using a fully relativistic treatment of the
$\Delta$ (Rarita-Schwinger spinors) the maximal value of $a_{33}^{\rm max} =
0.185/M_\pi^3$ is obtained with $ Z\simeq  -0.3$. These values of $Y=0.12$ and
$Z=-0.3$ will now be used to evaluate (C.5). Putting all pieces together, one
finds for the axial radius discrepancy
$$\eqalign{
\tilde r^2_A - r^2_A & = ( -4.6 + ( 3.1 + 1.1 -4.5) + (-7.2 +8.0)) \cdot
10^{-2} \,{\rm fm}^2 \cr & = ( -4.6 +0.5) \cdot 10^{-2}\, {\rm fm}^2 \cr}
\eqno(C.7)$$
The first term (-4.6) gives the leading order contribution and the others are
the order $\mu $ corrections. In (C.6) $a^+ = -0.83 \cdot 10^{-2}/ M_\pi$ and
$\lambda = 1 $GeV has been used. Although there is some numerical uncertainty
in the order $\mu$ correction, one obverves that the individual loop and
counter term contributions cancel each other to a large extent if one makes
reasonable assumptions on the parameters involved. We stress that the
individual terms at order $\mu$ in (C.7) coming from certain classes of loop
diagrams have no physical meaning, only the total sum counts. The latter tends
to be quite small, similar to the sum of $\rho$ and $\Delta$ contributions.
In essence one concludes from this complete order $q^4$ calculation that no
dramatic corrections to the leading order prediction of the discrepancy $\tilde
r^2_A - r^2_A$ are to be expected. It is now the task of future precision
experiments (like $\pi^+$-electroproduction at low $k^2$ close to threshold and
inverse $\beta$-decay) to test the prediction of $\tilde r^2_A-r^2_A$
presented here.
\bigskip
\goodbreak
\noindent{\bf APPENDIX D: STEREOGRAPHIC COORDINATES}
\medskip
In this appendix, we briefly summarize how one can
use stereographic coordinates
to parametrize the non--linearly realized pion and matter fields. This
formalism is used extensively in section 5. The pions inhabit the
three--sphere of radius $F_\pi$,
$$ S^3 \sim {{\rm SO(4)}  \over {\rm SO(3)}} \eqno(D.1)$$
since SU(2)$_L \times$~SU(2)$_R \sim$ SO(4) and SU(2)$_V \sim$~SO(3)
(local isomorphisms).
Embedding the sphere in euclidean space $E^4$ of cartesian coordinates
$\phi = \{ \phi_\alpha \} = \{ \svec{\phi} , \phi_4 = \sigma \}$, the sphere
is defined via
$$ \sum_{\alpha=1}^4 \phi_\alpha^2 = F_\pi^2 \quad . \eqno(D.2)$$
Three pion fields $\svec{\pi}$ can be obtained by applying e.g.
a four--rotation $R (\svec{\pi})$ ($RR^T =1$) to the north pole,
$$ \phi_\alpha (\svec{\pi} ) = R_{\alpha 4} (\svec{\pi}) \, F_\pi
\quad .
 \eqno(D.3)$$
In stereographic coordinates, one has
$$ R_{\alpha \beta}[\svec{\pi}] = \left(
\matrix { \delta_{ij} - {1 \over D} {\pi_i \pi_j \over 2F_\pi^2} &
          {1 \over D} {\pi_i \over F_\pi} \cr
          {1 \over D} {\pi_j \over F_\pi} &
          {1 \over D} \bigl( 1 - {\svec{\pi}^2 \over 4F_\pi^2}
          \bigr) \cr} \right) \quad , \eqno(D.4)$$
with
$$ D \equiv 1 + {\svec{\pi}^2 \over 4F_\pi^2} \quad . \eqno(D.5)$$
The corresponding covariant derivative follows to be as given in
eq.(5.1). It transforms linearly under the unbroken subgroup SU(2)$_V$
but highly non--linear under SU(2)$_A$. Furthermore, it expresses the
Goldstone boson character of the pions, i.e. their interactions vanish
as the momentum transfer goes to zero.
\medskip
Explicit symmetry breaking can be included in the following
way. Rewrite the mass term in two--flavor QCD as
$${\cal L}_{\rm mass} = -{1 \over 2} (m_u + m_d) (\bar q q)
 - {1 \over 2} (m_u - m_d) (2\bar q \, t_3 \, q) \, \quad
q = \left( \matrix{ u  \cr d} \right) \, \, .  \eqno(D.6)$$
The first term is the fourth component of the four--vector
 $S = (2\bar q i \gamma_5 \svec{t} q , \bar q q)$ and the second the third
component of the SO(4) vector
 $P = (-2\bar q  \svec{t} q , \bar q i \gamma_5 q)$ with opposite
transformation properties under parity and time reversal. Both terms
break chiral symmetry, but only the second one breaks isospin (the
invariant SO(3) subgroup does not affect the fourth component). In
terms of the pion fields, one constructs
$$S[\svec{\pi}] = \biggl( {\svec{\pi} \over D F_\pi} , 1-
{\svec{\pi}^2 \over 2D F_\pi^2} \biggr) \, \, , \eqno(D.7)$$
which leads to the canonical pion mass term
$${\cal L}_{\pi , {\rm mass}} =  -{1 \over 2D} M_\pi^2 \svec{\pi}^2 +
{\rm constant} = -{1 \over 2} M_\pi^2  \svec{\pi}^2 + \ldots \, , \eqno(D.8)$$
and the pion mass squared is proportional to $(m_u + m_d)$.
\medskip
To include the matter fields $\Psi = \Psi_N, \Psi_\Delta^\mu , \ldots$,
 one has to furnish a
representation $\svec{t}^{(\Psi)}$ of SO(3). For the nucleon,
 the corresponding $2 \times 2$ matrices
are $\svec{t}^{(N)} \equiv \svec{t} = {1 \over 2} \tau$, with $\tau$
the Pauli matrices. It is then most convenient to use a non--linear
realization
$$ N = \left( \matrix { p \cr n } \right) =
 {1 \over \sqrt{D}} \biggl( 1 + i \gamma_5 {\svec{t} \cdot
\svec{\pi} \over F_\pi} \biggr) \Psi_N \, \, \, , \eqno(D.9)$$
where $\Psi_N$ transforms linearly under the chiral group. The
pertinent covariant derivative is given in eq.(5.2). For the isobar,
$$ \Delta_\mu = \left( \matrix { \Delta_\mu^{++} \cr \Delta_\mu^{+}
\cr \Delta_\mu^{0} \cr \Delta_\mu^{-} } \right)   \eqno(D.10)$$
the construction is the same, only that the
$\svec{t}^{(\Delta)}= \svec{t}^{(3/2)} $ are
hermitean $4 \times 4$ matrices. The covariant derivative reads
$$ {\cal D}_\mu \Delta = (\partial_\mu + \svec{t}^{(3/2)} \cdot
\svec{E}_\mu \, ) \Delta \quad, \eqno(D.11) $$
with $\svec{E}_\mu$ defined after eq.(5.2).
Finally, one also needs the $2 \times 4$ spin ($S_i$) and isospin ($T_a$)
${1 \over 2} \to {3 \over 2}$ transition  matrices satisfying
$$\eqalign{
S_i S^+_j = & {1 \over 3} (2 \delta_{ij} - i \epsilon_{ijk} \sigma_k )
\, ,\cr
T_a T^+_b = & {1 \over 6} ( \delta_{ab} - i \epsilon_{abc} t_c ) \,
. \cr}  \eqno(D.12)$$
\bigskip
\goodbreak
\vfill \eject
\end